\documentclass[review]{elsarticle}

 \journal{TexExchange}
 
    \journal{Journal of \LaTeX\ Templates}
    
    \usepackage{tgtermes}

    \usepackage{hhline}  % http://ctan.org/pkg/hhline
    \usepackage{float}
    \usepackage{subfig}
    \usepackage{multirow}
    \usepackage{graphicx,booktabs}
    \usepackage{xfrac}
    \usepackage{titlesec}
    \usepackage{wrapfig,booktabs}
    \usepackage{mwe}
    \usepackage{adjustbox}
    \usepackage{caption}
    \usepackage{makecell}
    
    \usepackage{amsmath,array}
    \usepackage[usenames, dvipsnames]{color}
    \usepackage[table]{xcolor}
    \usepackage{array}
    \usepackage{longtable}
    \usepackage{ragged2e}
    \usepackage{mathtools}

    \usepackage{MnSymbol}
    \usepackage{xcolor}
    \usepackage{fullpage}

    \usepackage{chngcntr}
    \usepackage{longfbox}
    \usepackage{setspace}

    \usepackage{tikz}
    \usetikzlibrary{plotmarks}
    
    \newcommand{\etal}{\textit{et al}.}

    \arrayrulecolor{red}
    \fboxsep=\tabcolsep\relax
    \fboxrule=\arrayrulewidth\relax

    \newcolumntype{A}[2]{%
    >{\minipage{\dimexpr#1\linewidth-2\tabcolsep-#2\arrayrulewidth\relax}\vspace\tabcolsep}%
    c<{\vspace\tabcolsep\endminipage}}
    
    \usepackage{tikz}
    \usepackage{siunitx}
    \definecolor{scalebgcolor}{rgb}{0.08,0.52,0.80}     %% To change the backgroundcolor, adjust the scalebgcolor variable to use the desired color. Add the following code to your preamble:

    \newcommand{\scalebarbackground}[5][white]{
        \begin{tikzpicture}
            \draw (0,0) node[anchor=south east,inner sep=0] (image) { #2 };
                \begin{scope}[x={(image.south west)},y={(image.north west)}]
                \fill [fill=scalebgcolor, fill opacity=0.75] (0.02,1.3em) rectangle (#5*#4/#3+0.04,0.1em);
                \draw [#1, line width=0.2em] (0.02,0.4em) -- node[above,inner sep=0.1em, font=\footnotesize] {\SI{#5}{\nano \meter}} (#5*#4/#3+0.04,0.4em);
            \end{scope}
        \end{tikzpicture}
    }
    
    \newcommand{\scalebarbackgroundsabc}[5][white]{
        \begin{tikzpicture}
            \draw (0,0) node[anchor=south east,inner sep=0] (image) { #2 };
                \begin{scope}[x={(image.south west)},y={(image.north west)}]
                \fill [fill=scalebgcolor, fill opacity=0.75] (0.04,1.3em) rectangle (#5*#4/#3+0.04,0.1em);
                \draw [#1, line width=0.15em] (0.02,0.4em) -- node[above,inner sep=0.1em, font=\footnotesize] {\SI{#5}{\nano \meter}} (#5*#4/#3+0.02,0.4em);
            \end{scope}
        \end{tikzpicture}
    }

    \graphicspath{{images/}}
     
    \arrayrulecolor{black} % <---

    \usepackage{hyperref}
    
    \usepackage[letterpaper, portrait, margin=0.65in]{geometry}
    \biboptions{numbers,sort&compress}

\begin{document}

    %\twocolumn[{
    
    \begin{frontmatter}

    \title{Exploration of the Microstructure Space in TiAlZrN Ultra-Hard Nanostructured Coatings}
    %\title{Influence of Al on the microstructure, and phase transition of low Zr (c)-Ti$_{1-x-y}$Al$_{x}$Zr$_{y}$N Systems }
    %\title{On the Spinodal Decomposition of Pseudoternary (c)-Ti$_{1-x-y}$Al$_{x}$Zr$_{y}$N Systems }
    \author{Vahid Attari$^{a,}$\fnref{myfootnote1}}
    \author{Aitor Cruzado$^{b,c}$}
    \author{Raymundo Arroyave$^{a,c,d}$}
    \address{$^a$Department of Materials Science and Engineering Department, Texas A\&M University, College Station, TX 77843}
    \address{$^b$Department of Aerospace Engineering, Texas A\&M University, College Station, TX 77843}
    \address{$^c$Center for intelligent Multifunctional Materials and Structures, TEES,  \\  Texas A$\&$M University, College Station, TX 77843}
    \address{$^d$Department of Mechanical Engineering, Texas A\&M University, College Station, TX 77843}

    \cortext[mycorrespondingauthor]{Corresponding author email:attari.v@tamu.edu} 
    %\small
    \begin{abstract} 
       Ti$_{1-x-y}$Al$_{x}$Zr$_{y}$N cubic alloys within the 25-70\% Al composition range have high age-hardening capabilities due to metastable phase transition pathways at high temperatures. They are thus ideal candidates for ultra-hard nano-coating materials. There is growing evidence that this effect is associated with the elasto-chemical field-induced phase separation into compositionally-segregated nanocrystaline nitride phases. Here, we studied the microstructural evolution in this pseudo-ternary system within spinodal regions at 1200 $^\circ$C by using an elasto-chemical phase field model. Our simulations indicate that elastic interactions between nitride nano-domains greatly affect not only the morphology of the microstructure but also the local chemical phase equilibria. In Al-rich regions of the composition space we further observe the onset of the transformation of AlN-rich phases into their equilibrium wurtzite crystal structure. This work points to a wide palette of microstructures potentially accessible to these nitride systems and their tailoring is likely to result in significant improvements in the performance of transition metal nitride-based coating materials.   
    \end{abstract}


    \end{frontmatter}
    
    %}]

    %%%%%%%%%%%%%%%%%%%%%%%%%%%%%%%%%%%%%%%%%%%%%%%%%%%%%%%%%%%%%%%%%%%%%
    %%%%%%%%%%%%%%%%%%%%%%%%%%%%%%%%%%%%%%%%%%%%%%%%%%%%%%%%%%%%%%%%%%%%%
    %%%%%%%%%%%%%%%%%%%%%%%%%%%%%%%%%%%%%%%%%%%%%%%%%%%%%%%%%%%%%%%%%%%%%

    %\linenumbers
    \fontdimen2\font=0.5ex% inter word space
    %\small
    \justifying
    \section{Introduction}
    
    %% TMNs and Transformation from fcc to hcp and the reason...

Transition metal nitrides (TMNs) are generally characterized by having a unique set of properties---including very high hardness~\cite{abadias2010reactive}, relatively high melting points~\cite{abrikosov2011phase}, chemical inertness~\cite{ding2008corrosion}, oxidation resistance~\cite{donohue1997microstructure}---that make, many of them, suitable as coatings capable of protecting structural or functional materials from wear-related damage under harsh environments~\cite{mikula2016toughness}. Among TMNs, Ti$_{1-x}$Al$_x$N coatings with cubic NaCl (B1) structure have emerged as one of the most widely used materials for protective coatings for tooling~\cite{munz1986titanium} and other wear-resistance related applications \cite{paldey2003single}. Depending on their composition and operating conditions, cubic Ti$_{1-x}$Al$_x$N solid solutions often undergo phase separation between c-TiN- and c-AlN-rich nanoscale domains via \emph{coherent} spinodal decomposition~\cite{horling2002thermal,rachbauer2011decomposition}, which significantly enhances their hardness and makes them ideal for machining applications~\cite{horling2005mechanical}. 
These favourable properties are obtained for compositions that maintain the cubic symmetry. The nanostructures in phase-separated Ti$_{1-x}$Al$_x$N coatings are metastable, however, and under high temperature conditions (above $\sim$900$^o$C) prevalent during high-speed machining  ~\cite{knutsson2011machining}, thermally activated diffusion processes lead to the coarsening of the microstructure, loss of coherency and the ultimate transformation of the cubic c-AlN domains into their stable hexagonal wurtzite-type w-AlN form. This causes a dramatic loss of performance~\cite{rachbauer2011decomposition}, as this transformation is associated to a molar molar volume expansion of about 26 at.\%~\cite{bartosik2017fracture}. This lattice expansion tends to compromise the mechanical integrity of the coatings and negatively affects their performance.

Nowadays, many applications demand coatings capable of withstanding increasingly harsh operating conditions. The need to improve the performance of these coatings has lead to considerable efforts towards understanding the factors that control their microstructural evolution. Better understanding in these systems has produced a wide range of strategies to tailor their (nano)structrual features and thus optimize their performance. It is now known, for example, that multi-component alloying via addition of Cr or Zr in TMNs (Ti$_{1-x-y}$Al$_x$Cr$_y$N and Ti$_{1-x-y}$Al$_x$Zr$_y$N) changes the relative stability of the cubic and hexagonal phases and ultimately results in significant performance improvements~\cite{forsen2012mechanical,yang2013effect}. Recently, Lind $et$ $al.$, showed theoretically that the cubic-to-wurtzite phase transformation occurs at higher temperatures compared to TiAlN by addition of Cr~\cite{lind2013systematic}. Yang \etal~\cite{yang2013effect} investigated the effect of Zr additions on the structure and properties of TiAlN coatings and found that Zr resulted in an increased in hardness and improved oxidation resistance. On the other hand, Yalamanchi $et$ $al.$ \cite{yalamanchili2016growth} proposed interface energy minimization by growth of semi-coherent multilayer structures of TiN/ZrAlN that can offer both Koehler and coherency hardening mechanisms. Furthermore, recent studies report that the cutting performance of high-Al containing Zr-Al-N coatings is comparable to that of commercial grade Ti-Al-N coatings \cite{rogstrom2015wear}.
%Similarly, Chen $et$ $al.$ showed that increasing the Zr content from z=0 to 0.17, while keeping $x_{Al}\sim$0.5, results in a hardness increase from $\sim$33 to $\sim$37 GPa due to the promotion of cubic domains.% 

As interest in the use of Zr as a modifier of TiAlN coatings has increased, a number of theoretical and experimental investigations have focused on the phase stability and phase-decomposition behavior of Zr-containing TMN. Holec \etal~\cite{holec2011phase} investigated theoretically---and verified experimentally--- the phase stability of Zr$_{1-x}$Al$_x$N and Hf$_{1-x}$Al$_x$N alloys in comparison with the Ti$_{1-x}$Al$_x$N system and found that Zr (and Hf) contributed to a widening in the miscibility gap between the stable cubic (Ti/Zr/Hf-rich) and hexagonal (Al-rich) nitride phases. Significantly, this increases the chemical driving force for segregation in ZrAlN systems and widens the region where chemical separation-driven nanostructuring may be exploited for improved performance. Wang \etal~\cite{wang2015mechanical} found, via \emph{ab initio} calculations that the addition of Zr resulted in a transition in the Zener anisotropy, A$_z$, of Ti$_x$Al$_{1-x-y}$Zr$_y$N coatings from A$_z$ $>$ 1 to A$_z$ $<$ 1, affecting the preferred crystallographic direction for the elasto-chemical spinodal decomposition in these systems. Zhou \etal~\cite{zhou2017thermodynamic} investigated the thermodynamics of Ti--Al--Zr--N coatings within the CALPHAD (calculation of phase diagrams) framework by combining (limited) experimental data with first-principles calculations. They modelled the system as a pseudoternary and concluded that the system may undergo triple-spinodal decomposition. Their model constitutes a stepping stone towards the quantitative prediction of microstructural evolution via phase-field modelling, for example. 

As mentioned above, exploitation of the inherent tendency of (many) TMN systems to undergo phase separation and produce nanostructured microstructures has resulted in a marked improvement in the performance of ultra-hard coatings~\cite{wang2015mechanical,tasnadi2010significant,mikula2016toughness,munz1986titanium,rachbauer2011decomposition,aihua2012friction,horling2005mechanical,yang2013effect}. Exploration of the potential microstructural space by experiments alone is unrealistic and thus there has been an increased interest in developing and deploying quantitative simulations of the microstructure evolution in these systems. Gr{\"o}nhagen \etal~\cite{gronhagen2015phase} used a CALPHAD approach to describe the thermodynamics of Ti$_{1-x}$Al$_x$N systems and combined it with a Cahn-Hilliard model to describe the microstructure evolution in Ti--Al--N systems. Notably, they explicitly accounted for the presence of vacancies in the metal sublattice and concluded---through verification with indirect experimental evidence---that vacancies greatly affect the microstructural evolution in this system as their strong tendency to segregate to coherent AlN/TiN interfaces significantly enhances the kinetics of the phase separation. More recently, Zhou \etal~\cite{zhou2017effect} investigated the effect of Cr on the (metastable) phase equilibria and spinodal decomposition in c-TiAlN coatings. Similarly to Gr{\"o}nhagen \etal~\cite{gronhagen2015phase}, they coupled a CALPHAD model of the pseudoternary c-Ti$_{1-x-y}$Al$_x$Cr$_y$N system. Specifically, they investigated the phase separation behavior in the c-Ti$_{0.45}$Al$_{0.47}$Cr$_{0.08}$N system at 1000$^o$C and found that in this composition the system is located inside a two-phase miscibility gap, with Cr dissolving into TiN and AlN domains, although with a strong tendency for segregation at TiN/AlN interfaces.

While earlier works~\cite{gronhagen2015phase,zhou2017effect} have demonstrated the use of CALPHAD-informed phase field models to investigate the phase separation behavior in TMN systems, much remains to be done. Specifically, the significant difference between lattice parameters and elastic constants of different binary nitride systems means that elastic contributions to the microstructure evolution of (pseudo) ternary TMNs should be accounted for. Moreover, the elastic anisotropy of the different TMNs is expected to contribute significantly to the morphology of the microstructures undergoing phase separation. 

In this paper, we focus on the evolution of the c-Ti$_{1-x-y}$Al$_x$Zr$_y$N pseudoternary system over a wide range of compositions. The generalized Cahn-Hilliard formulation \cite{cahn1961spinodal} for chemical spinodal decomposition is extended to the elastochemical framework by coupling to microelasticity theory. The linear microelasticity problem resulting from compatibility conditions and consideration of the solution to the static (mechanical) equilibrium among microstructural constituents is employed to study the effect of elastic contributions on the microstructural evolution, inter-particle distance, and morphological change as well as the tendency towards the cubic-to-hexagonal transformation due to local strain fields. This paper is structured as follows: In section \ref{sec:model}, the modeling strategy is explained. The details of the utilized model parameters are provided in section \ref{sec:model_pars}. In Section \ref{sec:results}, the resultant microstructures under pure chemical and elastochemical simulations are discussed and compared. Furthermore, we investigate the onset of transformation of the metastable cubic AlN to wurtzite structure in section \ref{sec:Wurtzite-Rocksalt_onset} and present a summary on our findings and draw our conclusions in section \ref{sec:conculsions}.

%The sign and magnitude of the growth rates of misfitting particles can depend strongly on the crystallographic orientation of the particles and the degree of elastic anisotropy of the system. This is in contrast to a stress-free system wherein a change in the crystallographic orientation of the two particles in a cubic matrix has no effect on either the magnitude or sign of the particle growth rates \cite{} [Johnson-Abhinandanan-Vorhees 1990]. Our treatment is based on an optimized CALPHAD (CALculation of PHAse Diagram (CALPHAD) technique) \cite{} free-energy density function that couples composition, temperature, and the respective instabilities in composition to the strain space. We restrict this paper to the induced gradient in composition space and the established morphological changes in microstructure due to elastic inhomogeneities and crystallographic mismatches between the separating phases. The generalized Cahn-Hilliard formulation \cite{} for chemical spinodal decomposition is extended to the elastochemical framework by coupling to the microelasricity theory. The linear microelasticity problem resulting from compatibility and static equilibrium is employed to study the effect of field on the microstructural evolution, interparticular distance, and morphological change and the subsequent cubic-to-hexagonal transformation due to local strain fields. 

    %%%%%%%%%%%%%%%%%%%%%%%%%%%%%%%%%%%%%%%%%%%%%%%%%%%%%%%
    %%%%%%%%%%%%%%%%%%%%%%%%%%%%%%%%%%%%%%%%%%%%%%%%%%%%%%%
    %%%%%%%%%%%%%%%%%%%%%%%%%%%%%%%%%%%%%%%%%%%%%%%%%%%%%%%

    \section{Theoretical Background} \label{sec:model}
    \subsection{Kinetic Model}
    
    %\small
    \justifying
Here, we consider a pseudo-ternary alloy where its meso-scale structures are described by composition fields, $c_{\alpha} (\vec{r},t)$ defined at space ($\vec{r}$) for a given time ($t$), where $\alpha$ denotes the kind of species in this pseudo-ternary alloy. The thermodynamic energy functional or driving force for microstructural evolution is taken to be a function of chemical interactions due to compositional fluctuations and the local contractions/expansions associated with the composition dependence of elastic constants and lattice parameters. While gradients in the chemical potential of the constituent components are responsible for uphill diffusion (i.e. phase separation) within the spinodal region, strain energy resulting from local changes in lattice parameters usually suppresses or mitigates the phase-separating trends. Elastic interactions, however, may have more complex effects on the final microstructure of the system due to elastic anisotropy and the long-range elastic interactions between nanodomains. 

To treat this system, we write the total free energy function, $F^{tot}$ for a multi-component ($N$) inhomogeneous material as:
    
    %%%%%%%%%%%%%%%%%%
    \begin{align}\label{eqn:tot_en}
        F^{tot} \left( c,\nabla c,\varepsilon \right) = \int_\Omega \left( f^{interface} + f^{bulk} + f^{elastic} \right) d\Omega
    \end{align}
    %%%%%%%%%%%%%%%%%%
    
    \noindent
    where interface energy, $f^{interface}$, bulk energy, $f^{bulk}$, and strain energy, $f^{elastic}$ respectively are:

    %%%%%%%%%%%%%%%%%%
    \begin{equation}
       f^{interface} = \frac{1}{2} \sum_{\alpha=1}^N \sum_{\beta=1}^N \left[ \kappa_{\alpha} \delta_{\alpha\beta} + (1-\delta_{\alpha\beta})L_{\alpha\beta} \right] \nabla c_{\alpha} \nabla c_{\beta}   
    \end{equation}\label{eqn:int_en} \vspace{-1cm}
    %%%%%%%%%%%%%%%%%%
    
    %%%%%%%%%%%%%%%%%%%
    %\begin{align}\label{eqn:int_en}
    %    f^{chem} = \int_\Omega \left ( f^{interface} + f^{bulk} \right) d\Omega
    %\end{align}
    %%%%%%%%%%%%%%%%%%%
    
    %%%%%%%%%%%%%%%%%%
    \begin{align}\label{eqn:int_bulk}
        f^{bulk} =  f^{homogeneous}(c_1,c_2,...,c_N)
    \end{align}
    %%%%%%%%%%%%%%%%%%
    %\vspace{-1cm}
    %%%%%%%%%%%%%%%%%%
    \begin{align}\label{eqn:elas_en}
        f^{elas} = \frac{1}{2} \sigma_{ij}\varepsilon_{ij}^{el} 
    \end{align}
    %%%%%%%%%%%%%%%%%%  
    
    %\small
    \noindent
where $f^{interface}$ corresponds to the energy associated with gradients in non-homogeneous regions of the microstructure. $\kappa_{\alpha}$ and $L_{\alpha\beta}$ are symmetric tensors of composition-gradient energy coefficient for gradients in the $c_{\alpha}$ and $c_{\beta}$ composition fields. $\delta_{\alpha\beta}$ is the Kronecker's delta function. $f^{homogeneous}$ describes the energy of the homogeneous region of the microstructure, and is replaced by the CALPHAD free energy formalism for ternary system. $\sigma_{ij}$ and $\varepsilon_{ij}^{el}$ are the local stress and elastic strain in the material, respectively. In this paper, no gradients in the strain field are explicitly considered. Taking $f^{inhomogeneous}=f^{interface}+f^{bulk}$, $\kappa_{\alpha}$ and $L_{\alpha\beta}$ in a solution of uniform composition are given by: 
    
    %%%%%%%%%%%%%%%%%%
    \begin{align}\label{eqn:kappa_ii}
        \kappa_{\alpha} = \left[-2\frac{\partial^2f^{inhomogeneous}}{\partial c_{\alpha}\partial(\nabla^2c_{\alpha}) } + \frac{\partial^2f^{inhomogeneous}}{\partial(\nabla c_{\alpha})^2}  \right]_0
    \end{align}
    %%%%%%%%%%%%%%%%%% 
    %%%%%%%%%%%%%%%%%%
    \begin{align}\label{eqn:Kappa_ij}
        L_{\alpha\beta} = \left[\frac{\partial^2f^{inhomogeneous}}{\partial(\nabla c_{\alpha}).\partial(\nabla c_{\beta}) } \right]_0
    \end{align}
    %%%%%%%%%%%%%%%%%% 
    
    The evolution from the initially unstable mixed state to three stable phases is a highly nonlinear and complex process. The chemical part of the free energy that is composed of the interfacial and bulk energy contributions determines the compositions and volume fractions of the equilibrium phases. Additionally, the strain energy resulting from elastic heterogeneities affects the shapes and configurations of the domains. For a ternary alloy, the system of kinetic equations for the evolution of the composition of species ($\alpha= AlN, \beta=ZrN$) follow the two sets of kinetic equations \cite{cahn1961spinodal,ghosh2017particles}:
    
    %%%%%%%%%%%%%%%%%%
    \begin{equation}\label{eqn:kinetic_A}
        \frac{\partial c_{\alpha}}{\partial t} = \nabla. \Big(M_{\alpha} \nabla \frac{\delta F^{tot}}{\delta c_{\alpha}} \Big)
    \end{equation}
    %%%%%%%%%%%%%%%%%%
    %%%%%%%%%%%%%%%%%%
    \begin{equation}\label{eqn:kinetic_B}
        \frac{\partial c_{\beta}}{\partial t} = \nabla. \Big(M_{\beta} \nabla \frac{\delta F^{tot}}{\delta c_{\beta}} \Big)
    \end{equation}
    %%%%%%%%%%%%%%%%%%
    
    \noindent
where $M$ is the inherently positive effective rate of atomic diffusion for the species. $\frac{\delta F^{tot}}{\delta c_{\alpha}}$ is a generalized potential for the microstructural evolution and in this study it is a function of elastochemical interactions ($\mu^{tot}=\mu^{chemical}+\mu^{elastic}$). Equations \ref{eqn:kinetic_A} and \ref{eqn:kinetic_B} are nonlinear with respect to the composition field, although they are linear with respect to the driving force.%The diffusion mobility term M is related to the atomic mobilities of Ni and Al through M ¼ cAlcNiðcAlMNiþ cNiMAlÞ where MNi and MAl are obtained from the atomic mobility database of the fcc phase [37]. The determination of the parameter L in Eq. (11) is more difficult. Although L can be related to the interface mobility through a thin-interface analysis, the interface mobility is not known. However, it should be noted that an accurate value of L is not necessary in our particular study since the coarsening kinetics is diffusion-controlled. Therefore, we chose a value for L which ensures that the c0 precipitate evolution process is diffusion-controlled
    
    %%%%%%%%%%%%%%%%%%%%%%%%%%%%%%%%%%%%%%%%%%%%%%%%%%%%%%%
    %%%%%%%%%%%%%%%%%%%%%%%%%%%%%%%%%%%%%%%%%%%%%%%%%%%%%%%
    %%%%%%%%%%%%%%%%%%%%%%%%%%%%%%%%%%%%%%%%%%%%%%%%%%%%%%%

    \subsection{Thermodynamic Model}
    
The CALPHAD assessment of pseudoternary c-Ti$_{1-x-y}$Al$_x$Zr$_y$N system has been recently carried out by Zhou \etal~\cite{zhou2017thermodynamic} based on limited experimental data as well as free energies computed via \emph{ab initio} calculations based on Density Functional Theory (DFT). Although first-principles calculations predict that the final product of spinodal decomposition as c-TiN, c-AlN and c-TiZrN, the decomposition into c-AlN-, c-TiN- and c-ZrN-rich domains has been confirmed in Ti$_{0.30}$Al$_{0.46}$Zr$_{0.24}$N via STEM micrographs~\cite{lind2014high}. Consequently, the homogeneous c-Ti$_{1-x-y}$Al$_x$Zr$_y$N alloy is regarded to be a pseudo-ternary system where the product of decomposition consists of three species, i.e., c-TiN, c-ZrN and c-ZrN. Wang \etal~\cite{wang2012structural} calculated the enthalpy, entropy, and Gibbs energy of cubic nitrides over a wide range of temperatures by first-principles calculations. Considering the most common Redlich-Kister (R-K) polynomial expression for a ternary single-phase disordered substitutional solution having structure $\nu$ \cite{kaufman1970computer}, the free energy of c-Ti$_{1-x-y}$Al$_{x}$Zr$_{y}$N is expressed as a function of composition and temperature~\cite{zhou2017thermodynamic}: 
    
    %%%%%%%%%%%%%%%%%%
    %\footnotesize
    \begin{equation}\label{eq:cfenergy}
        f^\nu(c_{\alpha},T) = \sum_{\alpha}^N c_{\alpha} G_\alpha^o + RT\sum_{\alpha}^N{c_{\alpha} ln(c_{\alpha})} + \sum_{\alpha}^N \sum_{\beta>\alpha}^N c_\alpha c_\beta \sum_{\nu=0}^{\lambda} L_{\alpha\beta}^{\nu} (c_\alpha-c_\beta)^{\nu} + G^{xs}
    \end{equation}
    %%%%%%%%%%%%%%%%%%

    \noindent
    %\small
    where $G_{\alpha}^o$ is the reference Gibbs energy, $R$ is the ideal gas constant, and $L_{\alpha\beta}^{\nu}$ are the excess binary interaction parameters, and $\lambda=1$ represents a sub-regular system. $G^{xs}$ is a higher order interaction parameter defined as:

    %%%%%%%%%%%%%%%%%%
    \begin{equation}\label{eq:G_excess}
        G^{xs} = \sum_{\alpha}^{N-2} \sum_{\beta>\beta}^{N-1} \sum_{\gamma>\beta}^N \left( c_{\alpha}c_{\beta}c_{\gamma}L_{\alpha\beta\gamma} \right) %  c_i.L_{i} + c_j.L_{j} + c_k.L_{k})
    \end{equation}
    %%%%%%%%%%%%%%%%%%

    %\small
    \noindent
    where $L_{\alpha\beta\gamma}$ is the excess ternary interaction parameter taken to be $L_{\alpha\beta\gamma}=c_{\alpha}{^{\alpha}L_{\alpha\beta\gamma}}+c_{\beta}{^{\beta}L_{\alpha\beta\gamma}}+c_{\gamma}{^{\gamma}L_{\alpha\beta\gamma}}$. In this system, none of these interaction parameters are zero. In general, when one or more of the interaction parameters ($L_{\alpha\beta\gamma}$, $L_{\alpha\beta}^{\nu}$) are positive and large compared with $2RT$, repulsive interactions among species in the solution can be described. In this respect, when a binary miscibility gap on the two edges of a ternary alloy system exists, then the gap can extend across the ternary triangle between the two edges. The phase diagram of c-Ti$_{1-x-y}$Al$_x$Zr$_y$N \cite{zhou2017thermodynamic} indicates miscibility regions in all corners of the ternary phase diagram. The model optimized thermodynamic parameters are obtained from Refs.~\cite{zhou2017thermodynamic,wang2012structural} and summarized in Table \ref{tab:thermo_pars}. 
    
    %{ % begin box to localize effect of arraystretch change
    %\renewcommand{\arraystretch}{2.0}
    %%%%%%%%%%%%%%%%%%
    \begin{table*}[h!]
        \caption{The thermodynamic parameters used in the free energy model.}\label{tab:thermo_pars}
        %\scriptsize
        \centering
        \begin{adjustbox}{width=18.3cm}
        \begin{tabular}{lccccc} \toprule  
              Parameter                  &   Ref.                      & Unit                           & c-TiN              & c-AlN          & c-ZrN       \\ \midrule
            $G^0_{\alpha}$ at 298.15 $^{\circ}K$ & \multirow{2}{*}{\cite{wang2012structural}}    & \multirow{4}{*}{$J.mol^{-1}$}  & 9963    & 14868             & 6960                         \\
            $H_{\alpha}$ at 298.15 $^{\circ}K$   &    &                                & 19648   & 21381             & 18701                       \\
                $\sum_{\nu=0}^{\lambda} L_{\alpha\beta}^{\nu}(c_{\alpha}-c_{\beta})^{\nu}$           & \multirow{2}{*}{\cite{zhou2017thermodynamic}} &                            & 22129.97 + 5.25T + (-15.24+2.64T)($c_{TiN}-c_{AlN}$) & 42294.19-7.76T + (6024.33+0.72T)($c_{TiN}-c_{ZrN}$)   & 201137.88 - 91.53T + (143088.23 - 73.05T)($c_{AlN}-c_{ZrN}$)        \\ 
            $^{\alpha}L_{\alpha\beta\gamma}$ , $^{\beta}L_{\alpha\beta\gamma}$ and $^{\gamma}L_{\alpha\beta\gamma}$   &  &  & 9702.50+7.5T     & 67208.25 + 5.25T   & -18421.00 - 23.00T      \\ \bottomrule

        \end{tabular}
        \end{adjustbox}
    \end{table*}
    %%%%%%%%%%%%%%%%%%  
    %}
    
    %%%%%%%%%%%%%%%%%%%%%%%%%%%%%%%%%%%%%%%%%%%%%%%%%%%%%%%
    %%%%%%%%%%%%%%%%%%%%%%%%%%%%%%%%%%%%%%%%%%%%%%%%%%%%%%%
    %%%%%%%%%%%%%%%%%%%%%%%%%%%%%%%%%%%%%%%%%%%%%%%%%%%%%%%

    \subsection{Microelasticity: Periodic Strain Fields}\label{sec:microelasticity}
    %Solid-state phase transformations are accompanied by transformation strains, which may alter the nature of the transformation extent considerably. 
    Microelasticity is the regime in which a material undergoes a phase transition along with elastic interactions due to misfit strains and inhomogeneity in elastic properties. Crystallographic misfits result in a volume change when a phase transformation, and furthermore growth, happens. The stress-free transformation strains (SFTS) that would result from unconstrained crystallographic misfits are usually called eigenstrains, denoted by $\varepsilon_{ij}^{0}$. 
    %The lattice misfit is a rank-two tensor that corresponds to the eigenstrain associated with a misfit increment in different crystallographic orientations. 
    Therefore, the accommodation of these strains in elastically heterogeneous materials, e.g. multiphase and/or polycrystalline aggregates, determines the development of elastic strain $\varepsilon_{ij}^{el}$ and stress $\sigma_{ij}$ fields, and the favorable growth regimes.
    %Eigenstrain is, in general, an anisotropic material property. 
    
    \subsubsection{Kinematics and constitutive model}
    %%\small
    \justifying
     We assume that there is no change of macroscopic shape or volume during the phase transformation, so that the (compatible) total strain strain field $\varepsilon_{ij}^{tot}$  is linearly decomposed into two components, 
     %This infinitesimal strain tensor is the linear fraction of the Green-Lagrange strain tensor:
    
    %%%%%%%%%%%%%%%%%%
    \begin{equation}\label{eqn:strain_tensor}
        \varepsilon_{ij}^{tot} = \varepsilon_{ij}^{el}+ \varepsilon_{ij}^{0}
    \end{equation}
    %%%%%%%%%%%%%%%%%%
    where $\varepsilon_{ij}^{el}$ and $\varepsilon_{ij}^{0}$ are the position dependent elastic strain field and the dilatational eigenstrain derived from the lattice misfit, respectively. The eigenstrain $\varepsilon_{ij}^{0}$  for an isothermal state, over the composition domain is given by: 
    
    %%%%%%%%%%%%%%%%%%
    \begin{equation}
        \varepsilon_{ij}^{0} (c)=\sum_{\alpha=1}^N \varepsilon^T_{\alpha}\delta_{ij} h(c_{\alpha})
        \label{e:varepsilon_0}
    \end{equation}
    %%%%%%%%%%%%%%%%%%
    
    \noindent
    where ${\varepsilon^T_{\alpha}}$ is the lattice misfit, and $h(c_{\alpha})$ is an interpolation function taken based on Ref. \cite{moelans2008introduction} that interpolates the lattice misfit over the domain. We assume that the material is governed by Hooke's constitutive law so that the position dependent stress field $\sigma_{ij}$ is then obtained according to:
    
    %%%%%%%%%%%%%%%%%%
    \begin{equation}
        \sigma_{ij}= C_{ijkl}\varepsilon_{kl}^{el}= C_{ijkl} \left( \varepsilon_{kl}^{tot} - \varepsilon_{kl}^0 \right)
        \label{e:sigma}
    \end{equation}
    %%%%%%%%%%%%%%%%%%
    
    Here $C_{ijkl}$ is the composition dependent elastic stiffness tensor, and is obtained by interpolating across the unit cell through the interpolation function $g(c_{\alpha})$ taken from Ref. \cite{moelans2008introduction}:
        
        %%%%%%%%%%%%%%%%%%
        %\begin{equation}
        %    C_{ijkl} = \left( \sum_{i=1}^n \alpha \left( c_i \right) C^{-1}_i \right)^{-1}
        %\end{equation}
        %%%%%%%%%%%%%%%%%%
        
        %%%%%%%%%%%%%%%%%%
        \begin{equation}
            C_{ijkl} (c)= C^{eff}_{ijkl} + \sum_{\alpha=1}^{N-1} g(c_{\alpha}) \left( C_{ijkl}^{\alpha} (T,c_{\alpha}) - C_{ijkl}^0 (T,c_0) \right)
        \label{e:C}    
        \end{equation}
        %%%%%%%%%%%%%%%%%%    
        
    \noindent
    where $C^{eff}_{ijkl}$ represents the effective elastic stiffness tensor and is selected to be the average of elastic tensors of the product phases of the spinodal decomposition. $C^0_{ijkl}$ and $C^{\alpha}_{ijkl}$, which are function of temperature and composition, represent the elastic tensor of the reference phase and the elastic tensors of the other phases, respectively.  %In some cases, it can also be replaced by voight \cite{}, Mori-Tanaka \cite{} or self-consistant approximation \cite{} of the domain to correct the effective elastic tensor based on the overall volume fraction of phases. % (matrix phase depending on the overall composition of the alloy)
    
    As the composition field is periodic, the elastic moduli and the eigenstrains are thus periodic on the material domain $\Omega$. Therefore, the total strain field $\varepsilon_{ij}^{tot}$ is periodic  on $\Omega$ and can be expressed as the sum of its average homogeneous strain tensor ${E}_{ij}$  in the domain $\Omega$ and the periodic fluctuation strain field $\varepsilon^{\ast}_{ij}$:
    
        \begin{equation}
            \varepsilon_{ij}^{tot}=E_{ij}+\varepsilon^{\ast}_{ij}
            \label{e:epsilont}
        \end{equation}

     \noindent   
      where ${E}_{ij}=\left\langle \{\varepsilon_{ij}^{tot}\}\right\rangle$, being  $\langle\{\cdot\}\rangle=\frac{1}{\Omega}\int_{\Omega }\cdot \ \mathrm{d}\Omega$ the average quantity. The position dependent compatible periodic strain field $\varepsilon^{\ast}_{ij}$ is now given by:
        %The periodicity of eigen displacement fields ($u^0$) implies that the average of $\varepsilon^0$ on the domain vanishes and therefore that the average of $\varepsilon$ is $\varepsilon_{ij}^{tot}$.
            %%%%%%%%%%%%%%%%%%
        \begin{equation}\label{eqn:periodic_strain_tensor}
            \varepsilon_{ij}^{\star} = \frac{1}{2} \Bigg( \frac{\partial u^{\star}_i}{\partial r_j}+\frac{\partial u^{\star}_j}{\partial r_i} \Bigg)
        \end{equation}
        %%%%%%%%%%%%%%%%%%

        \subsubsection{Static Equilibrium}

    Adding equation  \ref{e:epsilont} in equation \ref{e:sigma}  we restate the mechanical equilibrium as: % (Refer to for instance Malvern, 1969; Mura, 1987 )
        
        %%%%%%%%%%%%%%%%%%
        %\fontdimen2\font=0.25ex% inter word space
        \begin{equation}\label{eqn:mechanical_eqn_set}
          \left\{ 
            \arraycolsep=1.4pt\def\arraystretch{2.0}
            \thickmuskip=0.05\thickmuskip
            \begin{array}{lcl}
                \frac{\partial }{\partial r_j} \left\{C_{ijkl} \left( E_{kl} + \varepsilon^{\star}_{kl} - \varepsilon^{0}_{kl} \right) \right\} = 0 & \mbox{on} & \Omega \\ 
                \varepsilon^{\star}_{kl} & \mbox{periodic on} & \Omega
            \end{array}\right. 
        \end{equation}
        %%%%%%%%%%%%%%%%%%
    Substituting now Eq. \ref{e:C} and Eq. \ref{e:varepsilon_0}, and $\varepsilon_{ij}^{\star}$ in terms of displacement in the mechanical equilibrium equation (Eq. \ref{eqn:mechanical_eqn_set}), and considering the symmetry properties of both the elastic stiffness matrices and strains, we leads into the following equation:
    
    % \begin{equation}
    \begin{multline}
       \left[C^{eff}_{ijkl} \frac{\partial^2 }{\partial r_j\partial r_k} \right]u^{\star}_l(\vec{r})=C^{eff}_{ijkl} \sum_{\alpha=1}^N \varepsilon^T_{\alpha}\delta_{kl} \frac{\partial h(c_{\alpha})}{\partial r_j}- \sum_{\alpha=1}^{N-1} \left( C_{ijkl}^{\alpha} - C_{ijkl}^0 \right)E_{kl}\frac{\partial g(c_{\alpha})}{\partial r_j}+ \\ \sum_{\alpha=1}^{N-1} \left( C_{ijkl}^{\alpha} - C_{ijkl}^0 \right)\varepsilon^T_{\alpha}\delta_{kl}\frac{\partial \{g(c_{\alpha})h(c_{\alpha})\}}{\partial r_j} -  \left[\sum_{\alpha=1}^{N-1} \left( C_{ijkl}^{\alpha} - C_{ijkl}^0 \right) \frac{\partial }{\partial r_j}\left( g(c_{\alpha}) \frac{\partial }{\partial r_k}\right) \right]u^{\star}_l(\vec{r})
        \label{e:equi_pde}  
     \end{multline}
    
    %\end{equation}
    
    The implicit equation shown in Eq.\ref{e:equi_pde} can be efficiently solved transforming it into the $Fourier$ space and solving it through the standard FFT based iterative solvers described by Khachaturyan \cite{khachaturyan2013theory}, Mura \cite{mura2013micromechanics}, Moulinec-Suquet \cite{moulinec1998numerical}, Gururajan \etal~\cite{gururajan2007phase}, and Lebensohn \etal~\cite{lebensohn2004macroscopic}. In the FFT-based method the domain $\Omega$ needs to be first discretized by means of a regular grid in real space $\{\boldsymbol{r}\}$. In turn, this partition of Cartesian space determines a corresponding grid of the same (element-wise) size in Fourier space $\{\boldsymbol{\xi}\}$. We can now write Eq.\ref{e:equi_pde} for the iteration $n$ as:
    
    % \begin{equation}
    \begin{multline}
       \hat{u}^{\star(n)}_l(\boldsymbol{\xi})=-{\rm i}{\xi}_j\left[C^{eff}_{ijkl} {\xi}_j {\xi}_k \right]^{-1}[C^{eff}_{ijkl} \sum_{\alpha=1}^N \varepsilon^T_{\alpha}\delta_{kl} \hat{h}(c_{\alpha}(\boldsymbol{\xi}))- \sum_{\alpha=1}^{N-1} \left( C_{ijkl}^{\alpha} - C_{ijkl}^0 \right)E^{n-1}_{kl}\hat{g}(c_{\alpha}(\boldsymbol{\xi}))+\\
       \sum_{\alpha=1}^{N-1} \left( C_{ijkl}^{\alpha} - C_{ijkl}^0 \right)\varepsilon^T_{\alpha}\delta_{kl}\hat{g}(c_{\alpha}(\boldsymbol{\xi}))\hat{h}(c_{\alpha}(\boldsymbol{\xi})) -  \sum_{\alpha=1}^{N-1} \left( C_{ijkl}^{\alpha} - C_{ijkl}^0 \right)  \hat{g}(c_{\alpha}(\boldsymbol{\xi})) {\xi}_k \hat{u}^{\star(n-1)}_l(\boldsymbol{\xi})]
        \label{e:FFTequi_pde}  
     \end{multline}
    
    %\end{equation}
    \noindent
    where the symbol ($\hat{.}$) denotes the Discrete Fourier transform of the corresponding quantity and $E^{n-1}_{kl}$ is the homogeneous strain field obtained in the iteration $n-1$. Note that FFT based algorithms are strain-control based methods and require an imposed homogeneous strain. Under imposed macroscopic stress $\Sigma_{ij}$ or imposed eigenstrain problems (i.e. $\Sigma_{ij}=\left\langle\{\sigma_{ij} \}\right\rangle=0$), the prescribed strain needs to be controlled based on prescribed overall stress as indicated in \cite{michel1999effective,gururajan2007phase}, and thus has to be corrected every iteration. In this work we follow the update in the overall strain proposed by Gururajan \etal~\cite{gururajan2007phase}, so that for a given iteration $n$, the homogeneous strain $E_{ij}^{(n)}$ is prescribed as: 
    
       \begin{equation}\label{eqn:stress}
        E_{ij}^{(n)} = S_{ijkl} \left[\Sigma_{kl} + \left\langle\{\sigma_{kl}^{0} \}\right\rangle - \left\langle\{\sigma_{kl}^{\star(n)} \}\right\rangle \right]
        \end{equation}
     \noindent
      where $S_{ijkl}=[\left\langle\{C_{ijkl} \}\right\rangle]^{-1}$ is the compliance matrix corresponding to the averaged stiffness matrices of the cell and  $\sigma_{kl}^{0}=[\left\langle\{C_{ijkl} \varepsilon_{kl}^{0}\}\right\rangle]$  and    $\sigma_{kl}^{\star(n)}=[\left\langle\{C_{ijkl} \varepsilon_{kl}^{\star(n)}\}\right\rangle]$.
            
    The described approach is similar to the one proposed by Gururajan \etal~\cite {gururajan2007phase}, which in this work has been extended to multiphase problems. As in that work we assume during the first iteration $\sum_{\alpha=1}^{N-1} ( C_{ijkl}^{\alpha} - C_{ijkl}^0)=0$. The convergence of the problem is achieved when the $L^2$ norm of $||\boldsymbol{u}^{n+1}-\boldsymbol{u}^{n}||<10^{-8}$.
        \noindent

    %%%%%%%%%%%%%%%%%%
    %\begin{equation}\label{eqn:}
    %    \thickmuskip=0.05\thickmuskip
    %    E_{ij} = S_{ijkl} \big[\sigma^A_{kl} + <\{\sigma_{kl}^0 \}> - <\{\sigma_{kl}^{\star} \}> \big]
    %\end{equation}
    %%%%%%%%%%%%%%%%%% 
    
    %Substituting for $C_{ijkl}$ and $\varepsilon^{0}_{kl}$ in terms of composition, and $\varepsilon^{\star}_{kl}$ in terms of the displacement field in the first row of Eq. \ref{eqn:mechanical_eqn_set}, and using the symmetry properties of the elastic constants and strains, we obtain:

    %%%%%%%%%%%%%%%%%%
    %\begin{equation}
    %\begin{align*}
    %    \left[ C^{eff}_{ijkl}  \frac{\partial^2 }{\partial r_j \partial r_k} + \Delta C_{ijkl} \frac{\partial }{\partial r_j} \big( \alpha(c) \frac{\partial}{\partial r_k} \big) \right] u^{\star}_{l}(r) = &\\
    %    C^{eff}_{ijkl} \varepsilon^T \delta_{kl} \frac{\partial \beta(c)}{\partial r_j} - \Delta C_{ijkl} E_{kl} \frac{\partial \alpha(c)}{\partial r_j} &\\
    %    + \Delta C_{ijkl} \varepsilon^T \delta_{kl} \frac{\partial \{\alpha(c)\beta(c)\}}{\partial r_j}  \\
    %\end{align*}
    %\end{equation}
    %%%%%%%%%%%%%%%%%% 
    
\subsection{Numerical Integration}

    %The iterative algorithm is based on the use of a homogeneous reference medium of stiffness $C_{ijkl}$. The periodicity of the solution is inherent to the spectral method due to the $Fourier$ approximation of the composition and elastic fields.
    In order to solve the phase field microstructural evolution problem, a semi-implicit $Fourier$ spectral approach \cite{chen1998applications} is used for the system of Cahn-Hilliard equations. For each iteration (i.e. at each time step), we solve the microelasticity problem for a given composition at a given time, as described in section \ref{sec:microelasticity}.In all simulations, we start with 2\% initial random perturbation in alloy composition in the simulation cell with periodic boundary conditions. The composition of the alloys that are studied, and the corresponding metastable region in the ternary phase diagram is indicated in Table \ref{tab:sim_alloy_comp}. 
    
    The simulations are performed in (2D) unit cells with spatial dimension of 50$\times$50 $nm$ and 160$\times$160 $nm$. The models are discretized with 512$\times$512 mesh points. A sensitive mesh analysis of the influence of the mesh size in the local micromechanical response is included in \ref{sec:appendix}. Additionally, 3D simulations for two alloy compositions are provided in the main body of the paper---note that he have also included further 3D simulations in \textbf{Supplementary Material}. The 3D simulation unit cells represent a size of $50\times50\times16.6$ $nm$, and are discretized with 128$\times$128$\times$30 mesh points.
    
    The time increment in our numerical solution have been chosen as $\delta t$=$2\times10^{-4}$ seconds. As the simulations proceed, we track the volume fraction of phases by approximating the number of unit cells that contain the appropriate phase, and terminate the simulations when the change in the calculated volume fraction is less than 1\%. 
    
    %%%%%%%%%%%%%%%%%%
    %%%%%%%%%%%%%%%%%%
    \begin{table}[htpb!]
        \caption{Compositions, and the respective conditions of the model}\label{tab:sim_alloy_comp}
        \scriptsize
        \centering
        %\begin{adjustbox}{width=18.1cm}
        %\begin{adjustbox}{width=15cm}
        \begin{tabular}{clccccccc} \toprule
        
            Group & Case & Alloy Composition                            & Temp. ($^\circ$C) & Region & Spinodal region type          & $\varepsilon^T_{AlN}$ & $\varepsilon^T_{ZrN}$ & $\varepsilon^T_{TiN}$  \\ \midrule
            \multirow{3}{*}{One} & 1    & c-Ti$_{0.71}$Al$_{0.05}$Zr$_{0.24}$N         & \multirow{13}{*}{1200} & I      & \multirow{13}{*}{Positive indefinite region}     & Negative & \multirow{13}{*}{Positive} & 0  \\ 
            &     2    & c-Ti$_{0.30}$Al$_{0.46}$Zr$_{0.24}$N         &       & II     &     & 0 &  & Positive \\
            &     3    & c-Ti$_{0.06}$Al$_{0.70}$Zr$_{0.24}$N         &       & III    &     & 0 &  & Positive \\ \cmidrule(lr{1em}){2-3}

            \multirow{10}{*}{two} &1    & c-Ti$_{0.70}$Al$_{0.25}$Zr$_{0.05}$N         &                 & \multirow{9}{*}{IV}      &     & Negative &  & 0 \\ 
            &2    & c-Ti$_{0.65}$Al$_{0.30}$Zr$_{0.05}$N         &       &       &     & Negative        &   & 0 \\
            &3    & c-Ti$_{0.60}$Al$_{0.35}$Zr$_{0.05}$N         &       &       &     & Negative        &   & 0 \\ 
            &4    & c-Ti$_{0.55}$Al$_{0.40}$Zr$_{0.05}$N         &       &       &     & Negative        &   & 0 \\ 
            &5    & c-Ti$_{0.50}$Al$_{0.45}$Zr$_{0.05}$N         &       &       &     & Negative        &   & 0 \\ 
            &6    & c-Ti$_{0.45}$Al$_{0.50}$Zr$_{0.05}$N         &       &       &     & 0 &   & Positive \\
            &7    & c-Ti$_{0.40}$Al$_{0.55}$Zr$_{0.05}$N         &       &       &     & 0 &   & Positive \\ 
            &8    & c-Ti$_{0.35}$Al$_{0.60}$Zr$_{0.05}$N         &       &       &     & 0 &   & Positive \\ 
            &9    & c-Ti$_{0.30}$Al$_{0.65}$Zr$_{0.05}$N         &       &       &     & 0 &   & Positive \\  \cmidrule(lr{1em}){2-3}
            &10   & c-Ti$_{0.25}$Al$_{0.70}$Zr$_{0.05}$N         &       & III   &     & 0 &   & Positive \\ \bottomrule

            \end{tabular}
        %\end{adjustbox}
    \end{table}
    %%%%%%%%%%%%%%%%%%
    %%%%%%%%%%%%%%%%%%

    %%%%%%%%%%%%%%%%%%%%%%%%%%%%%%%%%%%%%%%%%%%%%%%%%%%%%%%
    %%%%%%%%%%%%% Interface Conditions %%%%%%%%%%%%%%%%%%%%
    %%%%%%%%%%%%%%%%%%%%%%%%%%%%%%%%%%%%%%%%%%%%%%%%%%%%%%%
    %\subsubsection{Interface Conditions: Coherent Solid Interfaces versus Incoherent Solid Interfaces} 
    
    %Along each interface, there are conditions for mechanical equilibrium, and a condition for phase change equilibrium. They both depend on the nature of this interface. Along an incoherent solid-solid boundary, the equilibrium equations are:
    
    %%%%%%%%%%%%%%%%%%
    %\begin{equation}
    %\begin{equation}
    %    T_{ij}^{\alpha} n^{\alpha}_j = \omega^{\alpha}n_i^{\alpha}  
    %\end{equation}\label{eqn:}
    %\end{equation}
    %%%%%%%%%%%%%%%%%% 
        %%%%%%%%%%%%%%%%%%
    %\begin{equation}
    %\begin{equation}
    %    T_{ij}^{\beta} n^{\beta}_j   = \omega^{\beta}n_i^{\beta}
    %\end{equation}\label{eqn:}
    %\end{equation}
    %%%%%%%%%%%%%%%%%% 
    %%%%%%%%%%%%%%%%%%
    %\begin{equation}
    %\begin{equation}
    %    \omega^{\alpha} = \omega^{\beta}
    %\end{equation}\label{eqn:}
    %\end{equation}
    %%%%%%%%%%%%%%%%%%     
    
    %where $n_i^{\alpha}$ (resp. $n_i^{\beta}$) are the components of the normal to the interface oriented from $\alpha$ to $\beta$ (resp. $\beta$ to $\alpha$). They all contain o) and hence the Lagrange multiplier MaK

    %%%%%%%%%%%%%%%%%%%%%%%%%%%%%%%%%%%%%%%%%%%%%%%%%%%%%%%
    %%%%%%%%%%%%%%%%%%%%%%%%%%%%%%%%%%%%%%%%%%%%%%%%%%%%%%%
    %%%%%%%%%%%%%%%%%%%%%%%%%%%%%%%%%%%%%%%%%%%%%%%%%%%%%%%

    \section{Determination of elastic parameters}\label{sec:model_pars}

    %%%%%%%%%%%%%%%%%%
    %%%%%%%%%%%%%%%%%% Stress-free transformation strain
    %%%%%%%%%%%%%%%%%% Stress-free transformation strain
    %%%%%%%%%%%%%%%%%% Stress-free transformation strain
    %%%%%%%%%%%%%%%%%%
    
\subsection{Stress-free transformation strain}
    
\begin{figure}[h!]
        \centering
        \subfloat[]{\includegraphics[angle=0,origin=c,scale=0.18]{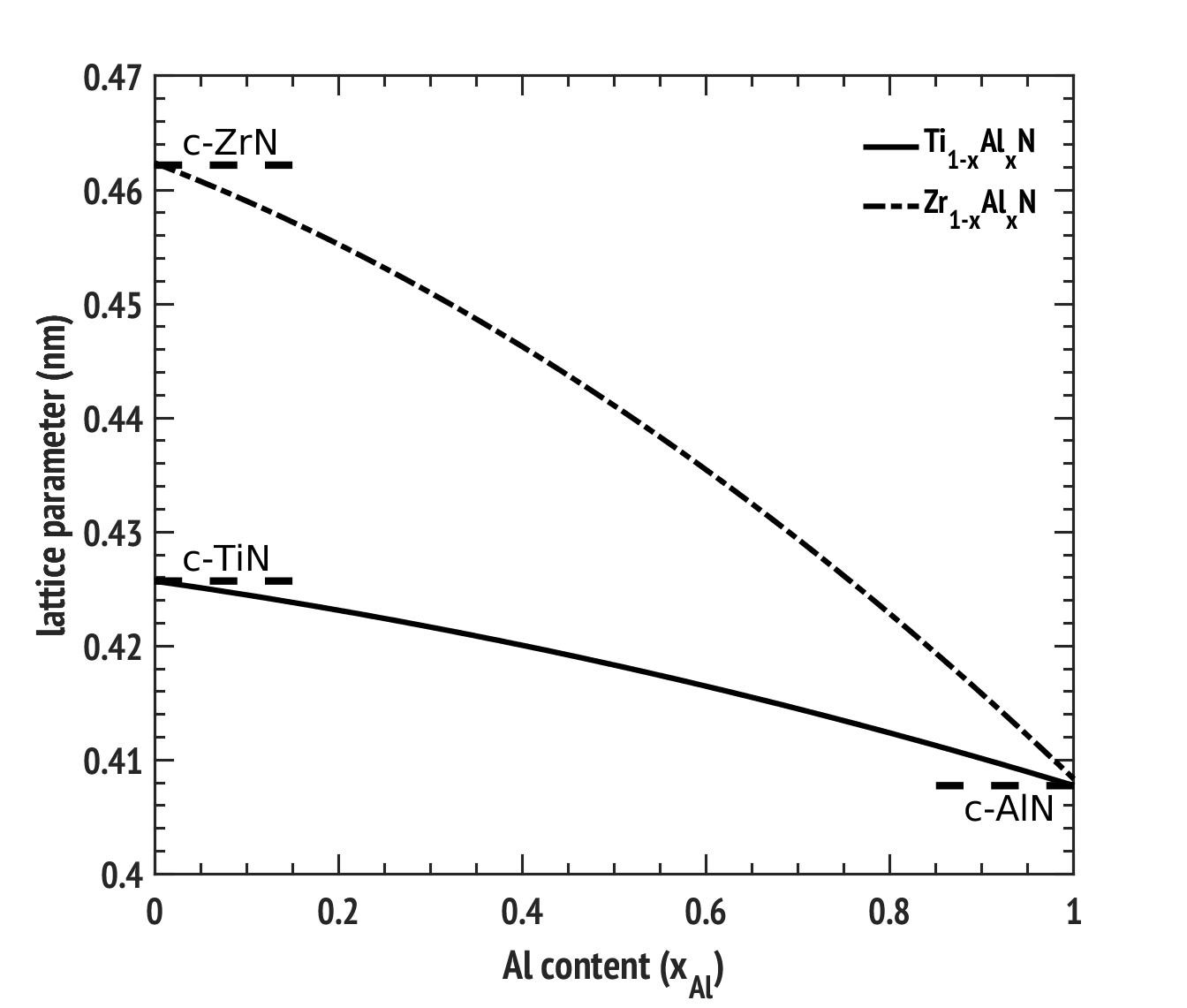}}%\\ \vspace{-0.35cm}
        \subfloat[]{\includegraphics[angle=0,origin=c,scale=0.196]{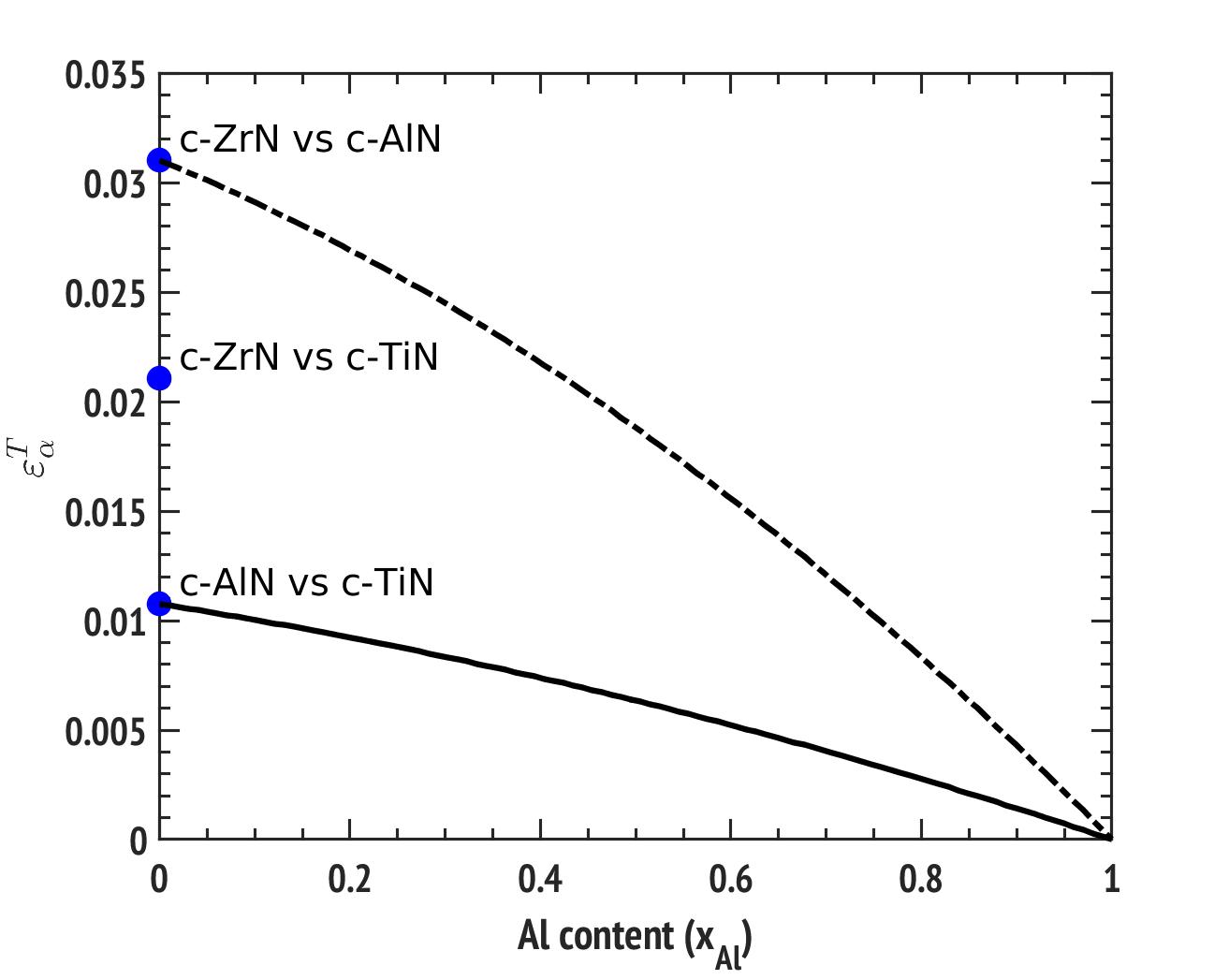}} \vspace{-0.25cm}
        \caption{ a) Lattice parameters of Ti$_{1-x}$Al$_{x}$N \cite{alling2007mixing} and Zr$_{1-x}$Al$_{x}$N \cite{sheng2008phase}, as a function of Al content. b) Estimated SFTS values between different phases as a function of composition.}
        \label{fig:lattice_par2}
    \end{figure}

The quadratic fit on \emph{ab-initio}-based calculated lattice parameters for c-Ti$_{1-x}$Al$_{x}$N and c-Zr$_{1-x}$Al$_{x}$N phases as a function of Al content are shown in Fig.~\ref{fig:lattice_par2}a. The figure illustrates that c-Ti$_{1-x}$Al$_{x}$N exhibits the most linear behaviour. This is, however, a consequence of a smaller difference between the lattice constants of c-AlN and c-TiN as compared with c-AlN and c-ZrN. The lattice parameters of binary c-AlN, c-TiN and c-ZrN phases are also indicated in this diagram by the dashed lines. These data are the corresponding values in room temperature.
    
    The SFTS for the binary nitride orderings was calculated and is approximately linear in c-Ti$_{1-x}$Al$_{x}$N, as seen in Fig.~\ref{fig:lattice_par2}b. The resulting fit for $\varepsilon^{T}_{\alpha}$ for c-Ti$_{1-x}$Al$_{x}$N and c-Zr$_{1-x}$Al$_{x}$N mixtures is shown in Fig.~\ref{fig:lattice_par2}b. The sign of the SFTS changes depending on the selection of the reference phase. The phase with the higher composition resides in the matrix, and is selected to be the reference phase. While our phase field calculations are performed at 1200 $^\circ$C, we assumed that the misfit at this temperature is the same as in room temperature. 
    
    %The pseudo-ternary alloy system in our study has the property that the lattice misfit depending on the type of the phase or Al content changes in such a way that negative, positive or zero misfit is possible.}
    
    %%%%%%%%%%%%%%%%%%
    %%%%%%%%%%%%%%%%%%

    %%%%%%%%%%%%%%%%%%
    %%%%%%%%%%%%%%%%%% Elastic stiffness of binary phases
    %%%%%%%%%%%%%%%%%% Elastic stiffness of binary phases
    %%%%%%%%%%%%%%%%%% Elastic stiffness of binary phases
    %%%%%%%%%%%%%%%%%%

\subsection{Elastic stiffness of binary phases} 
    
The spinodal decomposition may be influenced by elastic anisotropy \cite{fratzl1999modeling}, and the hardness enhancement observed upon age hardening relies on a shear modulus difference between the formed domains \cite{koehler1970attempt} as well as their coherency strain \cite{abrikosov2011phase}. Thus, it is of primary interest to determine the elastic properties of  (c)-Ti$_{1-x-y}$Al$_{x}$Zr$_{y}$N as a function of Al content ($x_{Al}$), and/or Zr content ($x_{Zr}$). We approximate this with the elastic properties of c-Ti$_{1-x}$Al$_{x}$N , (c)-Zr$_{1-x}$Al$_{x}$N, and (c)-Ti$_{1-x}$Zr$_{x}$N alloys as a function of composition and temperature.  %
    
\begin{figure*}[h!]
        \centering
         \subfloat[]{\includegraphics[angle=0,origin=c,scale=0.1355]{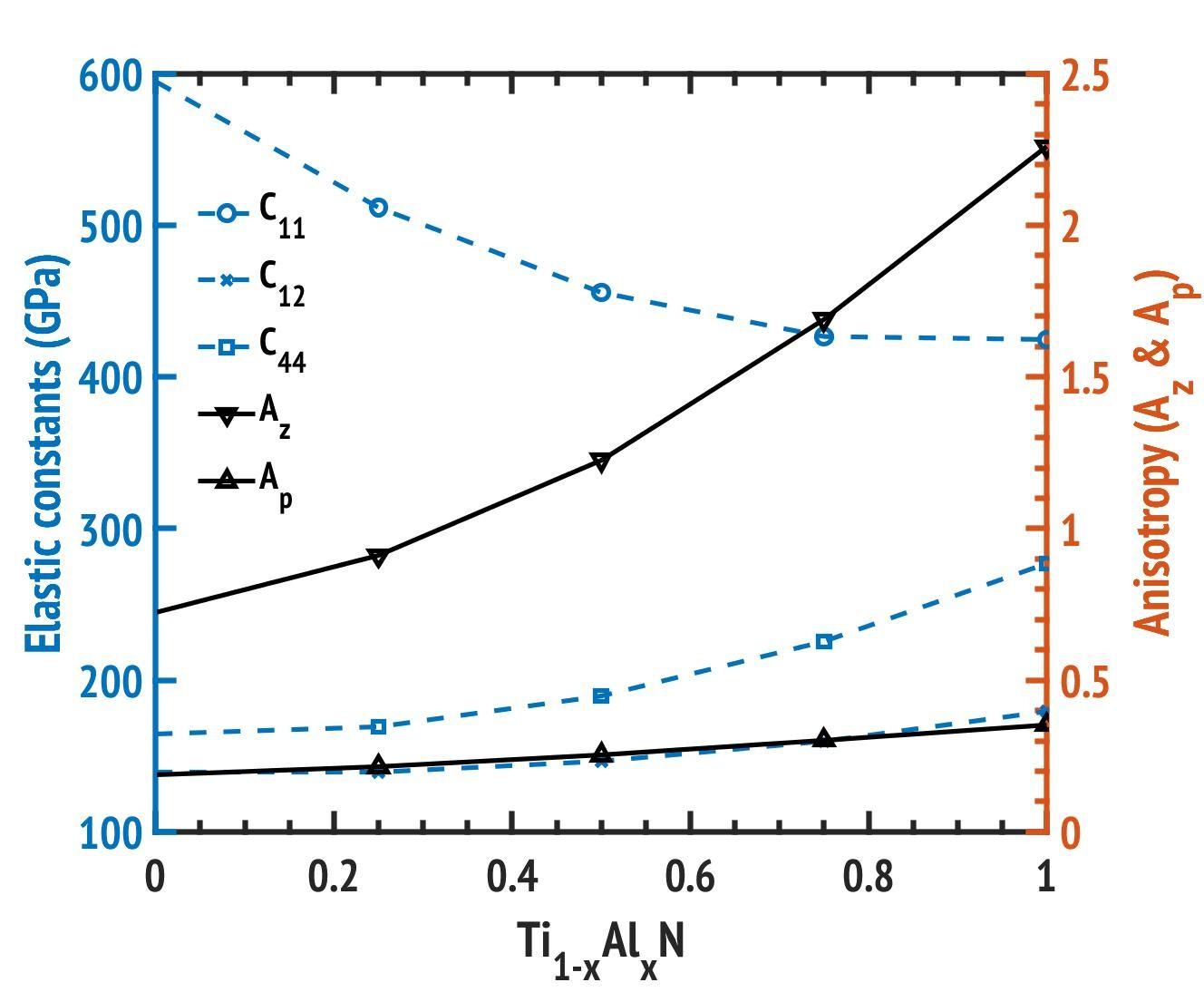}}
         \subfloat[]{\includegraphics[angle=0,origin=c,scale=0.13]{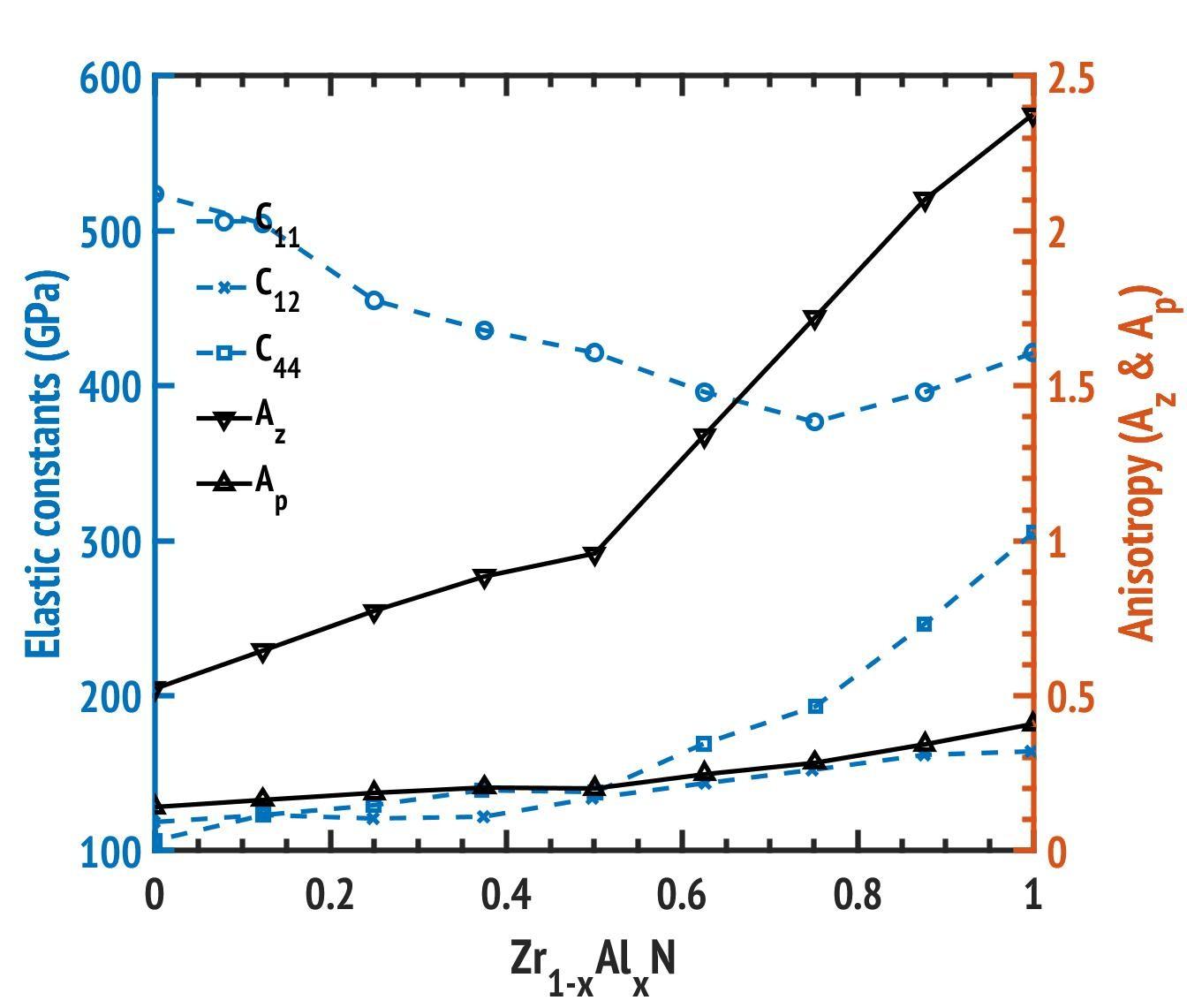}}
         \subfloat[]{\includegraphics[angle=0,origin=c,scale=0.13]{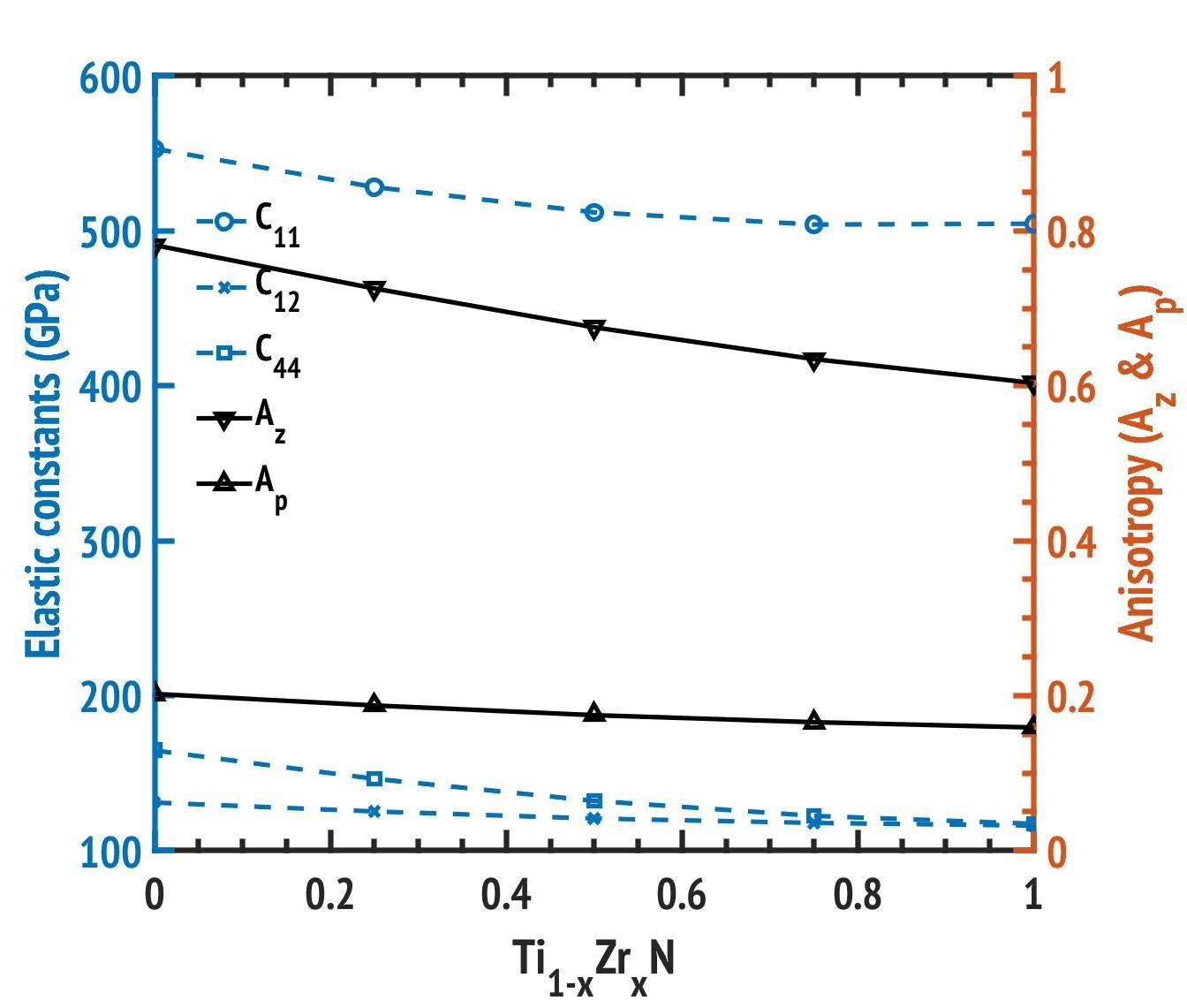}}
        \caption{ Single-crystal elastic constants (C$_{11}$, C$_{12}$, and C$_{44}$) of cubic (a) Ti$_{1-x}$Al$_{x}$N \cite{tasnadi2010significant}, (b) Zr$_{1-x}$Al$_{x}$N \cite{holec2014macroscopic}, and (c) Ti$_{1-x}$Zr$_{x}$N \cite{abadias2012structure} ordered structures, and the Zener$’$s anisotropy ratio A$_z$ as functions of composition and temperature in the three corner of the phase diagram.}
        \label{fig:elastic_cst}
    \end{figure*}

    The \emph{ab-initio} based elastic constants (C$_{11}$, C$_{12}$, and C$_{44}$), and mechanical behaviour of (c)-Ti$_{1-x}$Al$_{x}$N, (c)-Zr$_{1-x}$Al$_{x}$N, and (c)-Zr$_{1-x}$Ti$_{x}$N are addressed by many \cite{tasnadi2010significant,holec2014macroscopic,abadias2012structure,balasubramanian2018energetics,sangiovanni2018inherent,wang2017systematic,yang2017multiaxial,mikula2016toughness}, among which the utilized ones are depicted in Fig.~\ref{fig:elastic_cst}. Also, the deformation and micromechanical response of Ti(C,N) and Zr(C,N) micro pillars have been shown to be linear elastic with a high yield strength of 14 GPa \cite{el2018investigations}. Therefore, Hook's law should be sufficient to study the response of the nitrite systems. The elastic constants in (c)-Ti$_{1-x}$Al$_{x}$N and (c)-Zr$_{1-x}$Al$_{x}$N solutions are based on the Al content, and in (c)-Zr$_{1-x}$Ti$_{x}$N is based on the Zr content. Figure~\ref{fig:elastic_cst}.a illustrates that (c)-Ti$_{1-x}$Al$_{x}$N alloys behave smoothly along the fitted line that connects the parent binary compounds, c-AlN and c-TiN. C$_{12}$ and C$_{44}$ increase with the amount of Al concentration, while C$_{11}$ shows a pronounced decrease. 
    
The elastic constants (C$_{12}$ and C$_{44}$) of (c)-Zr$_{1-x}$Al$_{x}$N behave in a similar manner to (c)-Ti$_{1-x}$Al$_{x}$N (i.e. they increase with Al content.). This increase is not very smooth for the case of C$_{44}$ as it is for (c)-Ti$_{1-x}$Al$_{x}$N, however, as it sharply increases when $x_{Al}>0.45$. On the other hand, C$_{11}$ behaves differently, and it first decreases up to $x_{Al}\approx0.7$, and then increases.  There are only a few experimental reports on the mechanical properties of (c)-Zr$_{1-x}$Al$_{x}$N monolithic films in the literature. In (c)-Ti$_{1-x}$Zr$_{x}$N, the behaviour of elastic constants (C$_{11}$, C$_{12}$, and C$_{44}$) are provided as a function of Zr content and they show somewhat a different behaviour. In this case, all elastic constants decrease smoothly as a function of Zr content which is different than both of the other solutions, where at least one constant behaves in the reverse direction of the other constants. The Zener anisotropy $A_z = 2C_{44}/(C_{11} - C_{12})$, and normalized linear compressibility $A_p = C_{44}/(C_{11} + 2C_{12})$ are also provided in Fig.~\ref{fig:elastic_cst} for each system. 

    %%%%%%%%%%%%%%%%%%
    %%%%%%%%%%%%%%%%%%
    
    \section{Results and Discussions}\label{sec:results}
    %%%%%%%%%%%%%%%%%%
    %%%%%%%%%%%%%%%%%% Phase Stability Analysis
    %%%%%%%%%%%%%%%%%% Phase Stability Analysis
    %%%%%%%%%%%%%%%%%% Phase Stability Analysis
    %%%%%%%%%%%%%%%%%% Phase Stability Analysis
    %%%%%%%%%%%%%%%%%%
\subsection{Structural Properties and Phase Stabilities from a Thermodynamic Perspective}
    
The important concept of intrinsic stability and the thermodynamics of systems under externally controlled thermodynamic variables have been investigated by, among many, Gibbs, \cite{gibbs1961scientific} Tisza, \cite{tisza1961thermodynamics} Cahn, \cite{cahn1984simple} Huh-Johnson \cite{huh1995intrinsic}, Yi \etal~\cite{yi2018strain}. De Fontaine \cite{de1979configurational} has shown that the stability properties can be deduced from examination of the determinant of the well-known $Hessian$ matrix ($H$) formed from second derivatives of the free energy with respect to the concentration degrees of freedom. In a similar fashion, Kikuchi \cite{kikuchi1987second} proposed a criterion based on $Hessian$ evaluation to determine first and second order transitions. In ternary systems, this $Hessian$ matrix which is shown in Eq. \ref{eqn:eigenvalues} has three eigenvalues. 
    
    \begin{equation}\label{eqn:eigenvalues}
        \begin{aligned}  
        H =   
        \begin{bmatrix}   \vspace{8pt}
        \frac{\partial^2 G^{tot}_{\alpha,\beta,\gamma}}{\partial^2 \alpha}          & \frac{\partial^2 G^{tot}_{\alpha,\beta,\gamma}}{\partial \alpha \partial \beta} & \frac{\partial^2 G^{tot}_{\alpha,\beta,\gamma}}{\partial \alpha \partial \gamma}  \\ \vspace{8pt}
        \frac{\partial^2 G^{tot}_{\alpha,\beta,\gamma}}{\partial \beta \partial \alpha} & \frac{\partial^2 G^{tot}_{\alpha,\beta,\gamma}}{\partial \beta}            & \frac{\partial^2 G^{tot}_{\alpha,\beta,\gamma}}{\partial \beta \partial \gamma}  \\ \vspace{8pt}
        \frac{\partial^2 G^{tot}_{\alpha,\beta,\gamma}}{\partial \gamma \partial \alpha} & \frac{\partial^2 G^{tot}_{\alpha,\beta,\gamma}}{\partial \gamma \partial \beta} & \frac{\partial^2 G^{tot}_{\alpha,\beta,\gamma}}{\partial^2 \gamma}           \\
        \end{bmatrix}  
        \end{aligned}
    \end{equation}
    
It is well established that to have an intrinsic instability it is necessary to have at least one negative eigenvalue with respect to infinitesimally small perturbations in the relevant degrees of freedom. If two of the eigenvalues of $H$ are negative for a given average concentration (i.e. ($H$) is negative definite) then the ternary solution is unstable with respect to fluctuations along any direction in composition space. The case of both eigenvalues being positive results in a stable solution. For $H$ being indefinite, fluctuations in some directions in composition space are unstable while the remaining directions are stable. In all selected regions (regions I, II, III and IV) of Ti$_{1-x-y}$Al$_{x}$Zr$_{y}$N, only one of the three eigenvalues is negative and alloys within these regions are unstable only along one direction in the ternary composition triangle. The negative eigenvalues give the direction in the composition space that corresponds to the maximal decrease of the Gibbs free energy.

    %Fig. \ref{fig:TiAlZN_mic} illustrate the microstructure of the (c)-TiAlZrN alloy for selected compositions in these regions at T=1200$^\circ$C.  
    
    %- Point 1 is located in the metastable region I where a phase separation to ZrN and TiN phases is expected. In this region, two out of three eigenvalues of the Hessian matrix are negative and these eigenvalues $\lambda$ determine unstable fluctuation modes. 
    
\begin{figure*}[h!]
        \centering \vspace{-0.4cm}
        \subfloat[]{\includegraphics[angle=0,origin=c,scale=0.093]{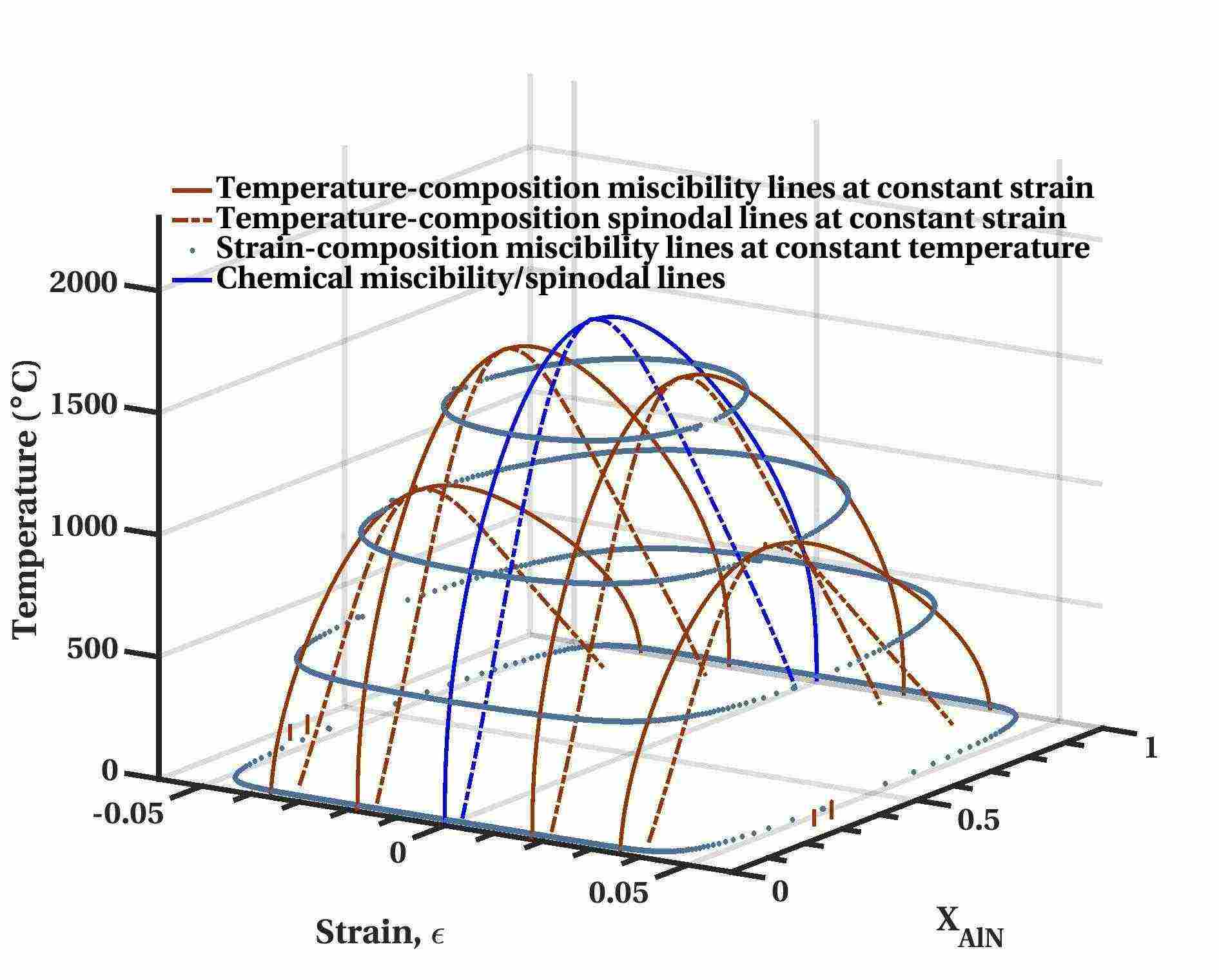}}   \vspace{-0.15cm} \hspace{-0.35cm}
        \subfloat[]{\includegraphics[angle=0,origin=c,scale=0.093]{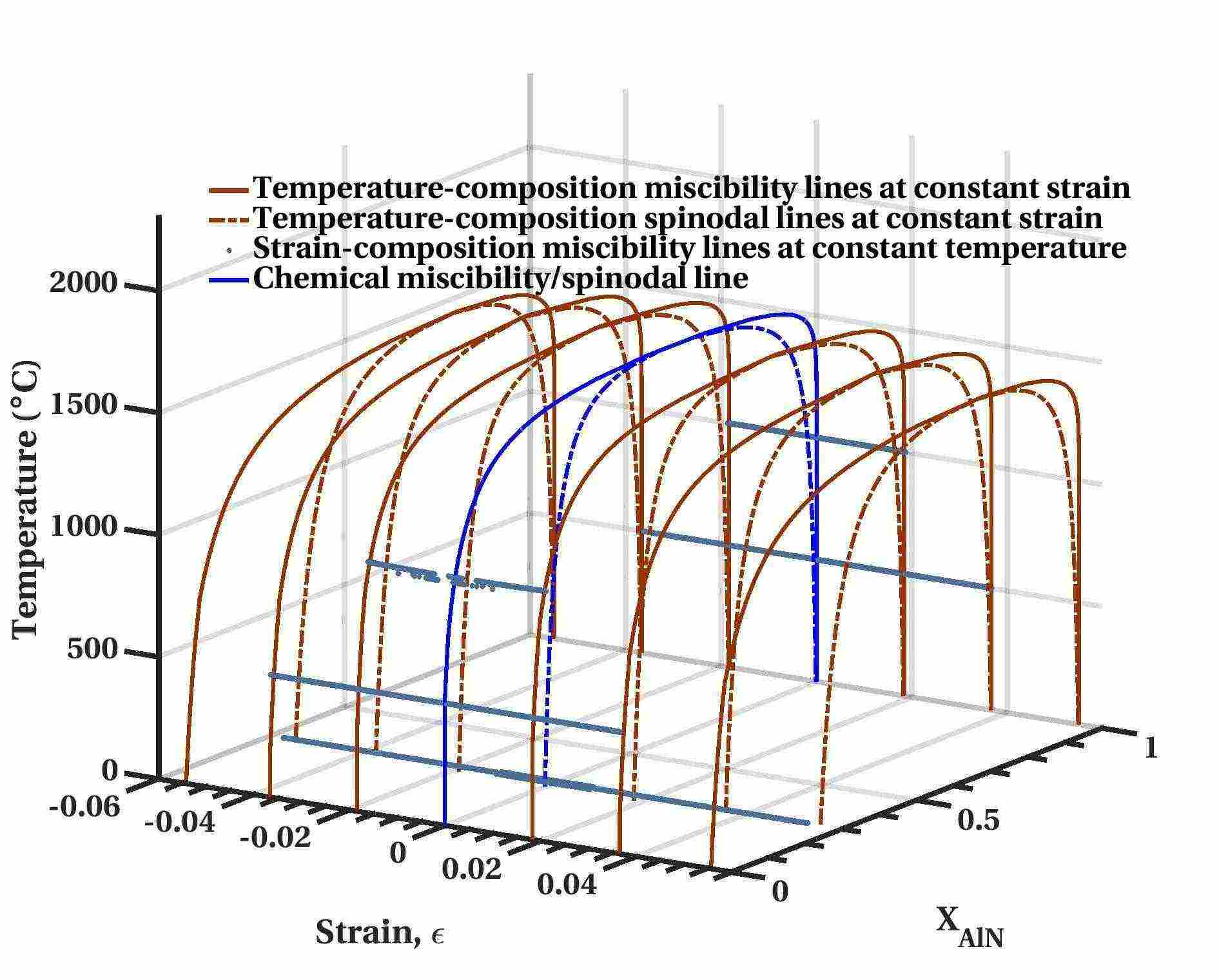}}   \vspace{-0.15cm} \hspace{-0.35cm}       \subfloat[]{\includegraphics[angle=0,origin=c,scale=0.093]{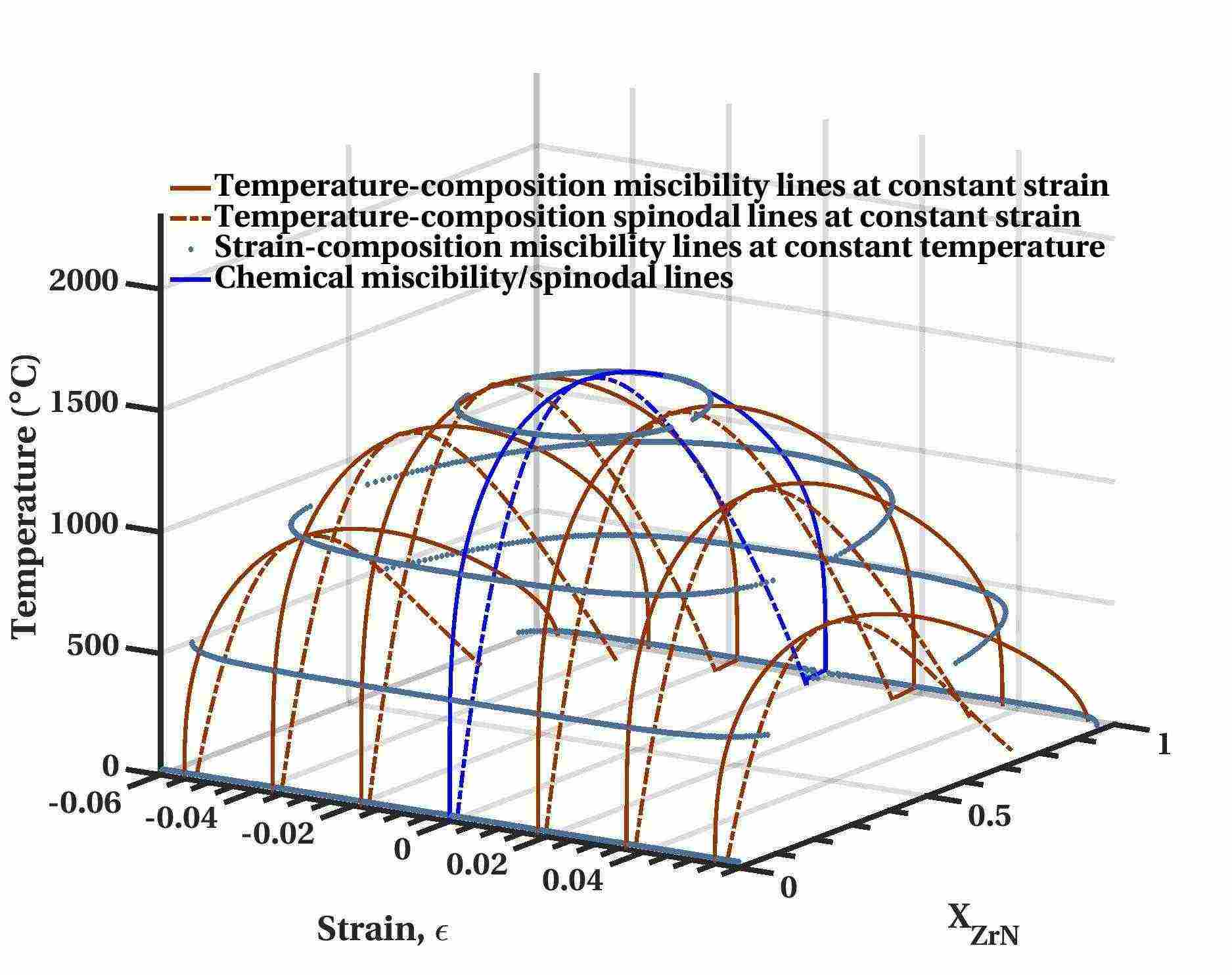}}\\ \vspace{-0.15cm} 
        \subfloat[]{\includegraphics[angle=0,origin=c,scale=0.093]{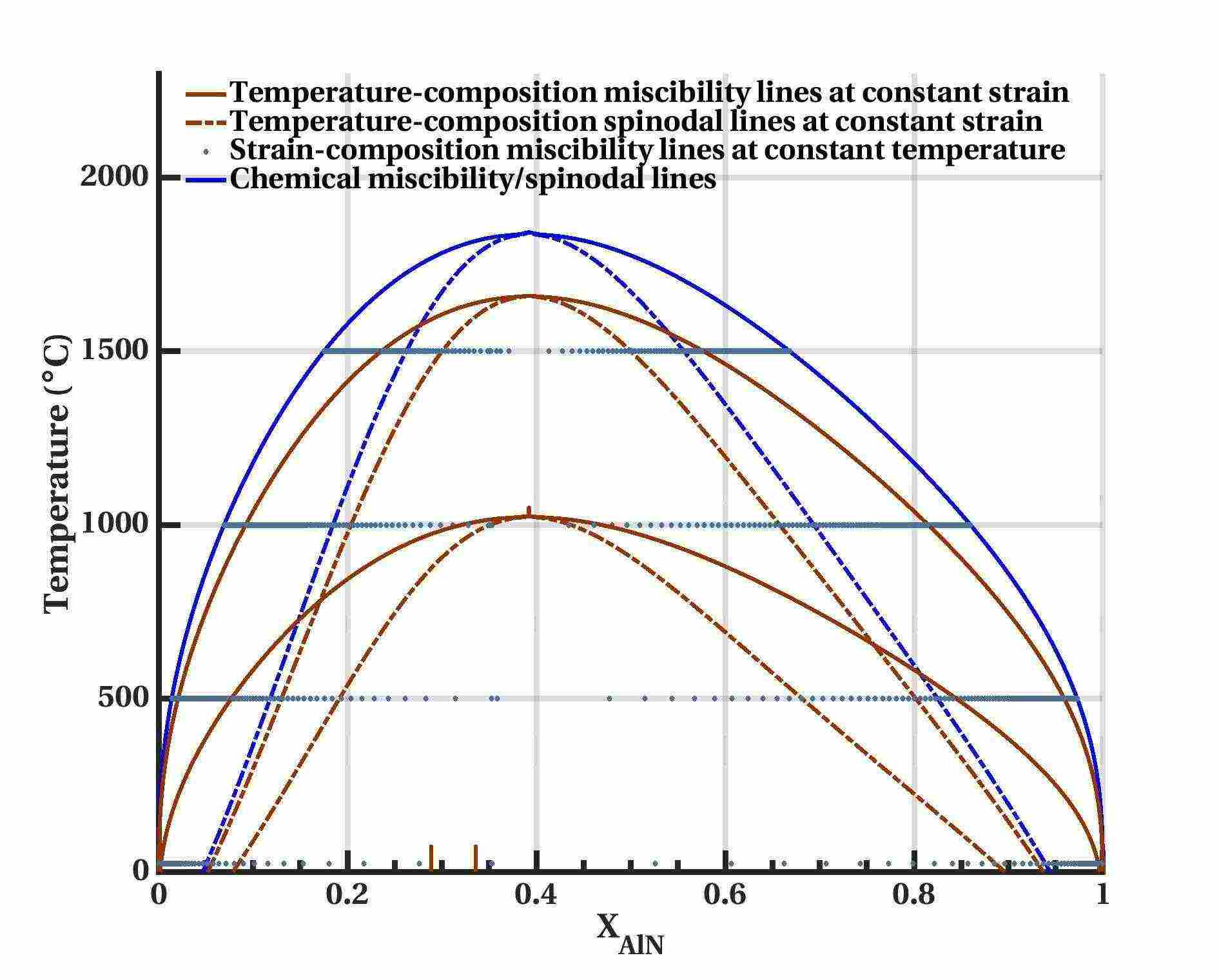}}   \vspace{-0.15cm} \hspace{-0.35cm}
        \subfloat[]{\includegraphics[angle=0,origin=c,scale=0.093]{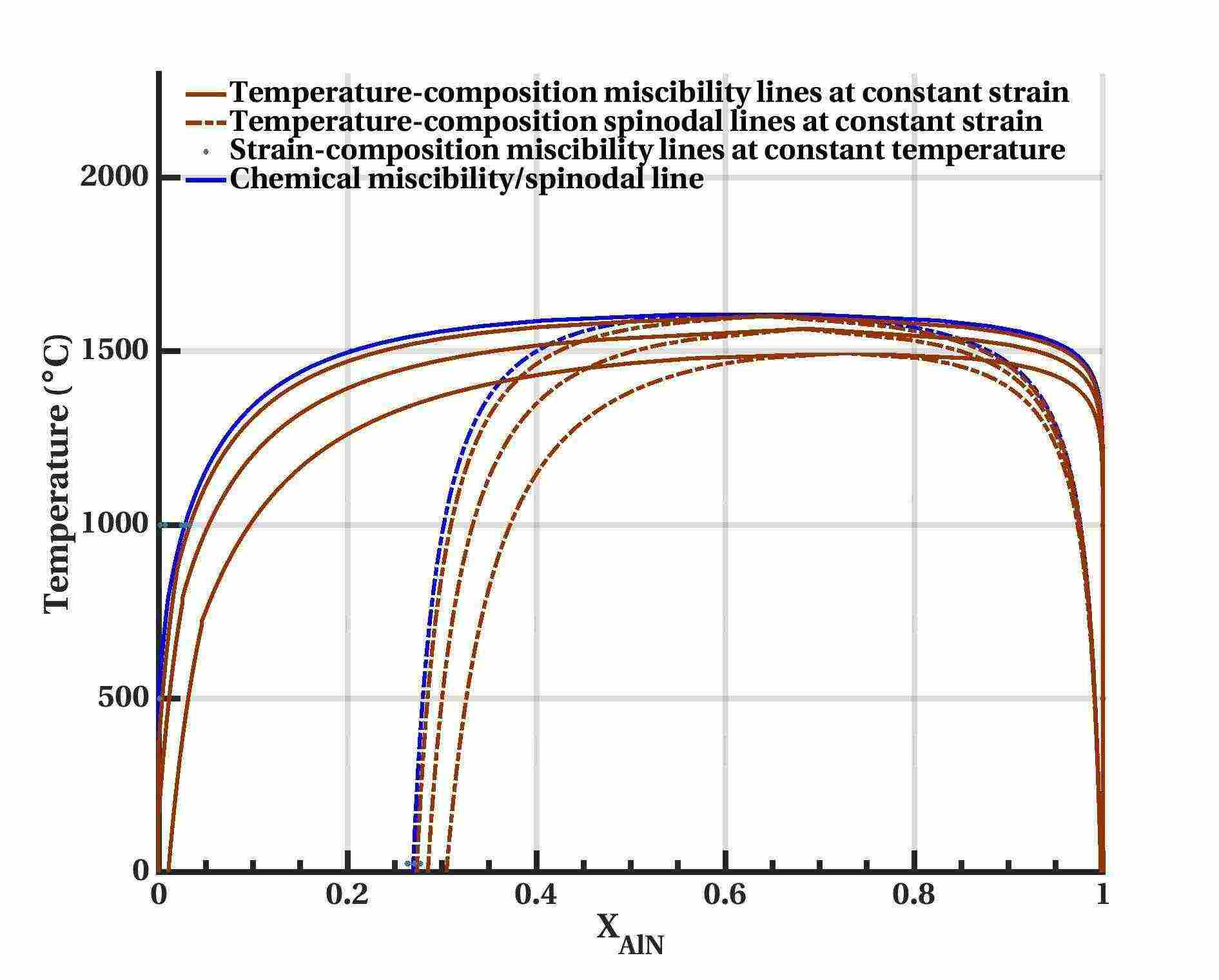}}   \vspace{-0.15cm} \hspace{-0.35cm}
        \subfloat[]{\includegraphics[angle=0,origin=c,scale=0.093]{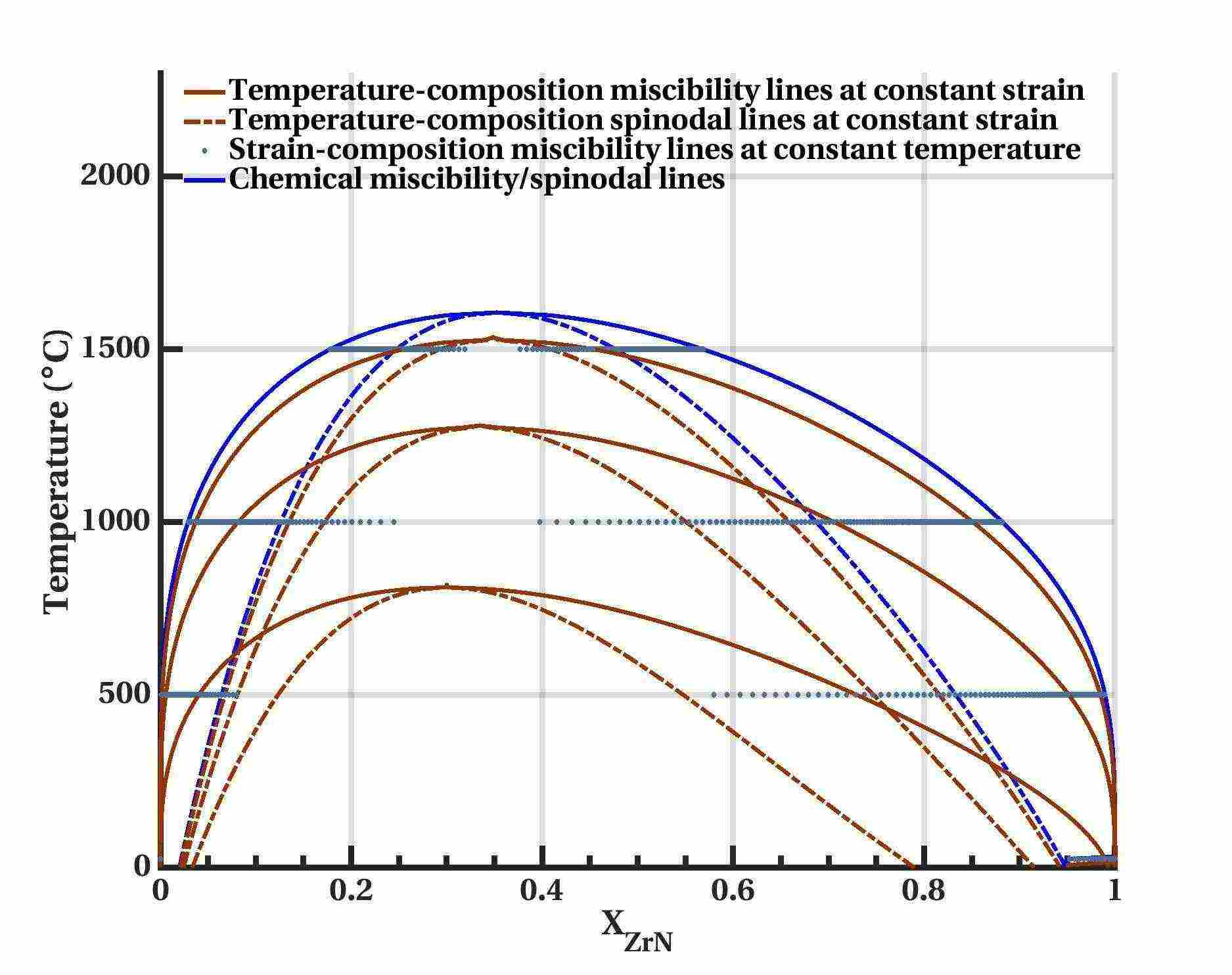}}   \vspace{-0.10cm}        
        \caption{ Pseudo-binary phase diagrams on the sides of Ti$_{1-x-y}$Al$_{1-x}$Zr$_{y}$N pseudo-ternary triangle. a,c) TiN-AlN b,e) AlN-ZrN c,f) ZrN-TiN. The chemical lines shown in blue (online) are same as Ref. \cite{zhou2017thermodynamic}. The elastochemical lines indicated by the brown and green indicate the effect of elastic strain on the suppression of the miscibility gaps in strain and temperature space. Brown lines are for $\varepsilon^T_{\alpha}=-0.054,-0.036,-0.018,0,0.018,0.036,0.054$. The miscibility gap in TiN-AlN system under $\varepsilon^T_{\alpha}=-0.054,0.054$ disappeared entirely. }
        \label{fig:TiAlZN_PD}
    \end{figure*}

Following the limit of metastability, we have spanned the chemical miscibility line around the strain, composition and temperature space. Figure~\ref{fig:TiAlZN_PD} shows the obtained three-dimensional phase diagram in strain-composition-temperature space for the three miscibility gaps that are present in the corners of the ternary phase diagram. In this figure, three subsets of lines are present for constant strains and constant temperatures indicated by vertical and horizontal lines, respectively. The blue (online) lines in each sub-figure indicate the chemical-only miscibility, and brown lines indicate the suppressed miscibility gaps for constant strains. The brown lines are symmetric with respect to the blue line (zero strain miscibility gap). A close look at the brown lines in Figures~ \ref{fig:TiAlZN_PD}d, \ref{fig:TiAlZN_PD}e, and \ref{fig:TiAlZN_PD}f reveals that a dilatational SFTS of $\pm$0.54 suppresses the miscibility from $\sim$1600$^\circ$C to below 1000$^\circ$C in Ti$_{1-x}$Al$_{x}$N and Ti$_{1-x}$Zr$_{x}$N systems. However, it doesn't have any noticeable effect on the metastable Zr$_{1-x}$Al$_{x}$N system and it only suppresses the miscibility gap less than 100$^\circ$C. This suggests that the strength of the elastic driving force is considerably less than the chemical one, and it can't compete with the chemical reactions to mitigate it. This is consistent with observations that show that the metastable Zr$_{1-x}$Al$_{x}$N system decomposes to its equilibrium phases of c-ZrN and w-AlN \cite{sheng2008phase} above 900$^\circ$C. However, the predicted Zr$_{1-x}$Al$_{x}$N phase diagram shown in Fig.~\ref{fig:TiAlZN_PD}c illustrates a wide miscibility gap, and predicts the decomposition pathway to c-ZrN and c-AlN phases as an intermediate structure. This has been attributed to a large misfit strain between c-ZrN and c-AlN hindering the coherent isostructural decomposition pathway. The large difference in lattice parameters of (c)-AlN, and (c)-ZrN is depicted in Fig.~\ref{fig:lattice_par2}. We note that Sheng \etal~\cite{sheng2008phase} calculated temperature-composition phase diagram for (c)-Zr$_{1-x}$Al$_{x}$N predicts a slightly skewed miscibility gap toward c-AlN comparing to the phase diagram in this study. 
    
    %%%%%%%%%%%%%%%%%%
    %%%%%%%%%%%%%%%%%%   
    
\begin{figure}[htb!]
        \centering

        \subfloat[]{\includegraphics[width=0.200\columnwidth]{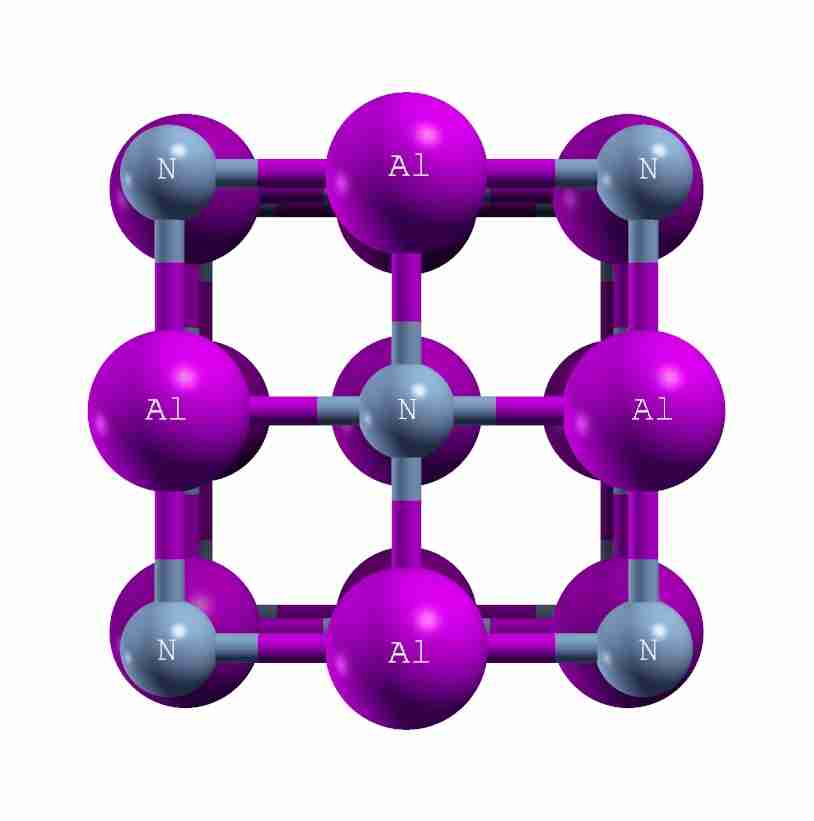}
        \includegraphics[width=0.200\columnwidth]{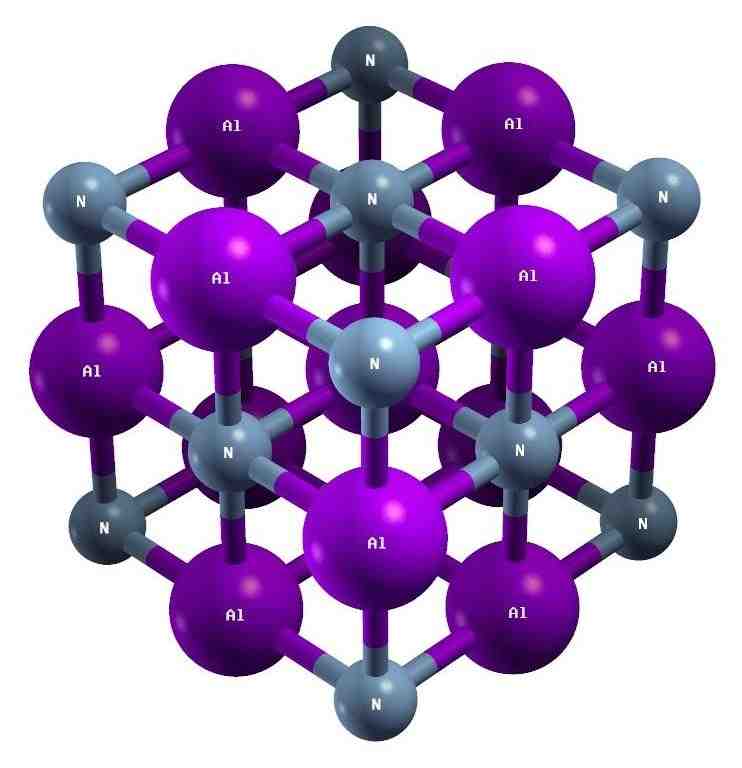}}
        \subfloat[]{\raisebox{0.4cm}{\includegraphics[width=0.250\columnwidth]{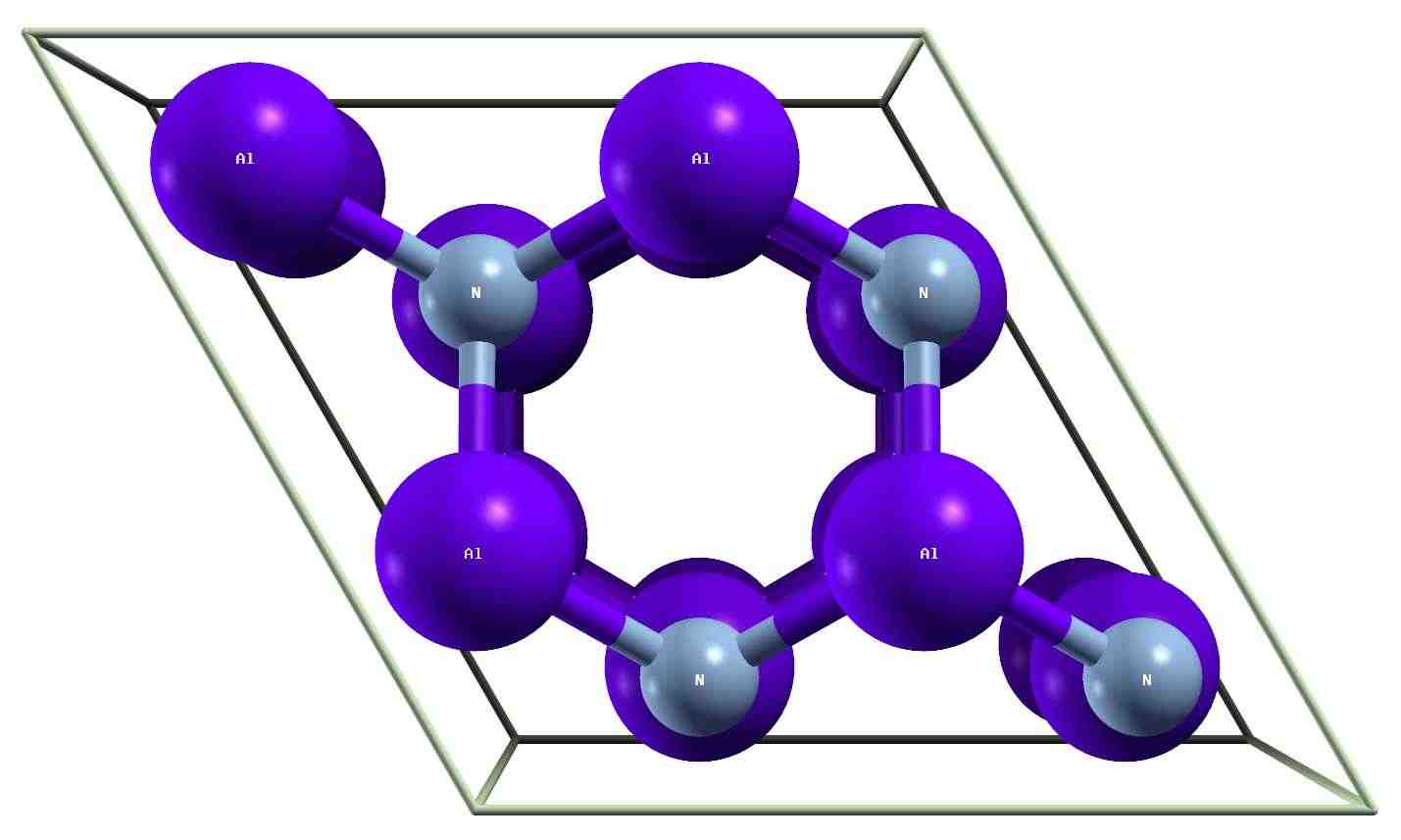}
        \includegraphics[width=0.270\columnwidth]{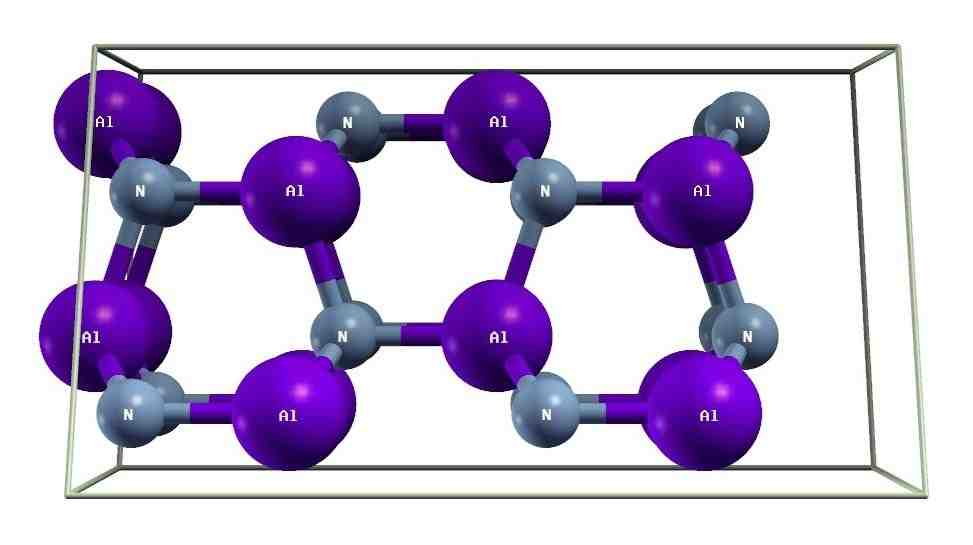}}}\\
        
        \caption{ Unit cells of the (a) cubic ($Fm\bar{3}m$) TiN, AlN, and ZrN versus (b) wurtzite ($P6_3/mc$) AlN ordered structures. The larger purple atoms represent Ti, Al, or Zr, and the smaller gray atoms represent N in the structure.}
        \label{fig:crystal_struct}
    \end{figure}

\begin{figure}[h!]
        \centering
        \subfloat[]{\includegraphics[scale=0.45]{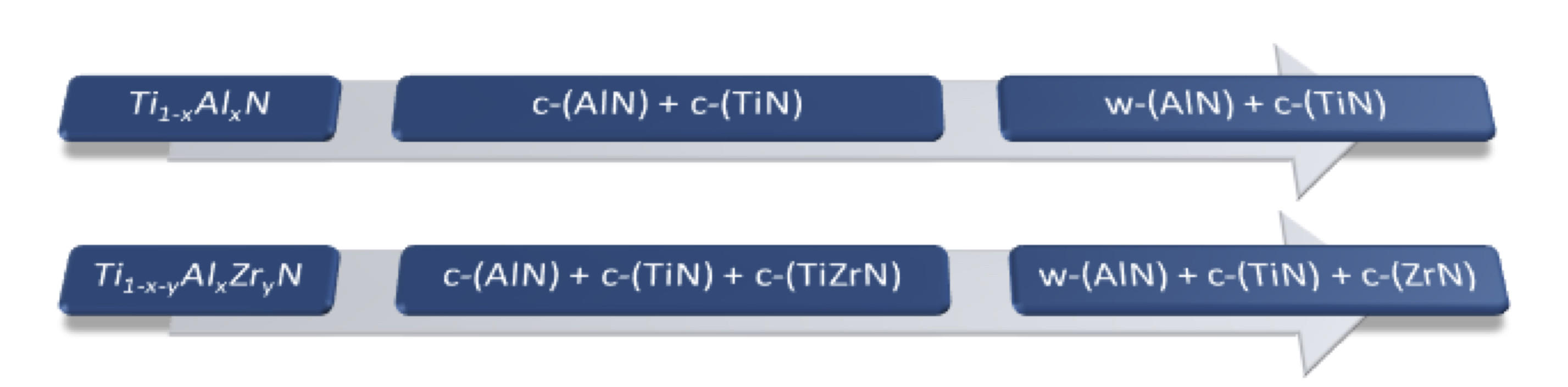}}\\ \vspace{-0.15cm}
        \subfloat[B1 structure]{\includegraphics[width=0.17\columnwidth]{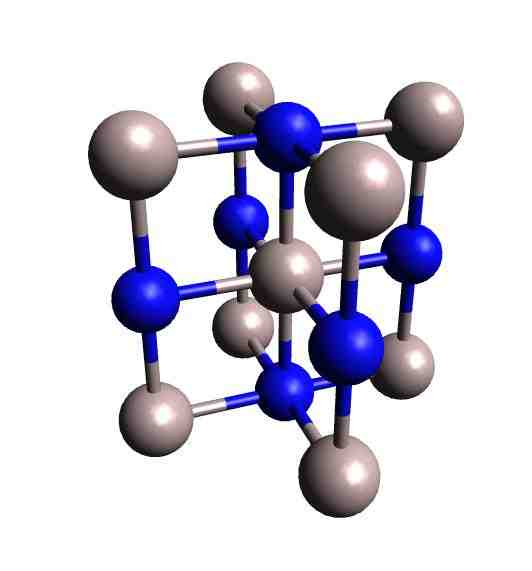}}  \raisebox{3.5 em}{\LARGE$\xrightarrow{}$} \vspace{-0.05cm}
        \subfloat[\emph{bct} versus \emph{hcp} hypothetical intermediate structures]{
        \begin{longfbox}[
   border-radius=15pt,
   border-width=2pt,
   border-top-left-radius={15pt,15pt},
   border-right-style=dashed,
   border-left-style=dashed,
   %border-left-color=dashed,
   padding={0.1em,15pt},
   width=0.4\linewidth,
]
        {\includegraphics[width=0.39\columnwidth]{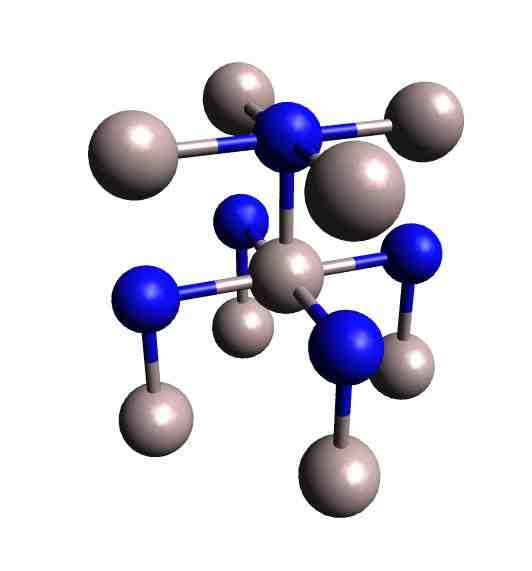} \raisebox{3.5 em}{\large{OR}}
        \includegraphics[width=0.39\columnwidth]{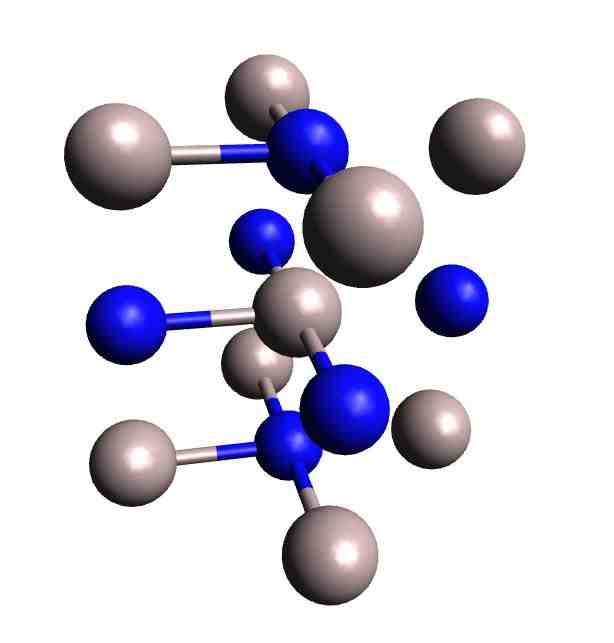} 
        }\end{longfbox}
        } 
        \raisebox{3.5 em}{\LARGE$\xrightarrow{}$} \vspace{-0.05cm} 
        \subfloat[B4 structure]{\includegraphics[width=0.17\columnwidth]{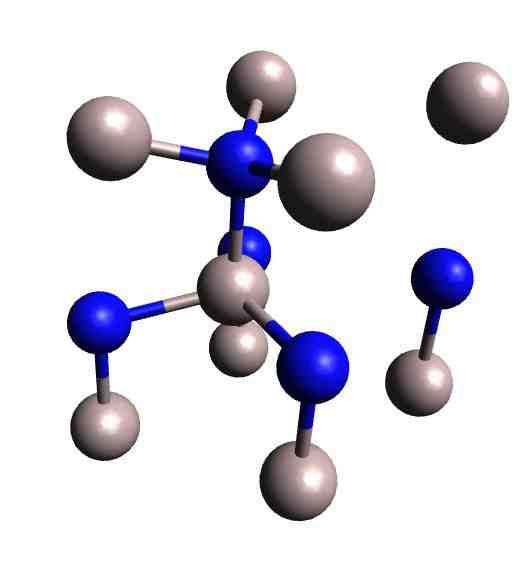}} \vspace{-0.05cm}         
        \caption{a) The sequence of phase transition in Ti$_{0.5}$Al$_{0.5}$N, and Ti$_{0.34}$Al$_{0.66}$N as it is shown by XRD measurements in \cite{mayrhofer2003self} above $\sim$850$^\circ$C, comparing with the predicted sequence of transition in Ti$_{1-x-y}$Al$_{x}$Zr$_y$N above $\sim$1100$^\circ$C. b, c, and d) The notion of structural transition of cubic (B1) AlN to wurtzite (B4) AlN through hypothetical \emph{bct} or \emph{hcp} intermediate stages. Only one of the intermediate structures (\emph{hcp}) is energetically favorable in AlN.}
        \label{fig:stability_sequence}
    \end{figure}

A cubic ($Fm\bar{3}m$) unit cell for TiN, AlN, and ZrN phases is depicted in Fig. \ref{fig:crystal_struct}. The lattice parameter of (c)-TiN, (c)-AlN, and (c)-ZrN at room temperature is 4.26, 4.069 \cite{alling2007mixing}, and 4.61 nm \cite{sheng2008phase}, respectively. A wurtzite ($P6_3/mc$) structure for w-AlN is also depicted in this figure. It is known that (c)-AlN often times transforms to (w)-AlN which is energetically more stable. Such a transition occurs after the alloy initially undergo spinodal decomposition into coherent cubic nanometer-size domains. Figure \ref{fig:stability_sequence}a demonstrates the sequence of phase transitions in Ti$_{1-x}$Al$_{x}$N and Ti$_{1-x-y}$Al$_{x}$Zr$_{y}$N. This transition is experimentally validated in Ti$_{0.5}$Al$_{0.5}$N, and Ti$_{0.34}$Al$_{0.66}$N for various ranges of compositions and temperatures. Experimental analysis by Mayrhofer \etal~\cite{mayrhofer2003self} confirms that the (c)-Ti$_{1-x}$Al$_{x}$N becomes self-organized during thermal annealing treatments at elevated temperatures. On the other hand, using \emph{ab initio}-reinforced predictions, Holec \etal~\cite{holec2011phase} show that the rock salt cubic structure, and wurtzite AlN structures in Ti$_{1-x}$Al$_{x}$N are stable for $x_{Al}<0.70$, and $x_{Al}>0.70$, respectively. Experimental studies show that the rock salt type AlN is stable for $x_{Al}<0.43$ in Zr$_{1-x}$Al$_{x}$N, and further increase of Al causes change of structure to (w)-AlN. In this respect, Zr addition in Ti$_{1-x-y}$Al$_{x}$Zr$_{y}$N system is shown to reduce or prohibit the formation of the (w)-AlN phase largely. Chen \etal~\cite{chen2011influence} verifies the cubic AlN structure for $x<0.6-0.7$ and $y\leq0.4$ composition ranges by magnetron sputtering method. Furthermore, Fig.~\ref{fig:stability_sequence}b, c, and d illustrate that the Minimal Energy Pathway (MEP) for this structural transition may cross from two hypothetical routes where tetragonal ($I\bar{4}mm$) and hexagonal ($P6_3/mmc$) are possible intermediate structures. We will discuss this transition in more detail in section \ref{sec:Wurtzite-Rocksalt_onset}.
    
    %\floatsetup[figure]{style=plain,subcapbesideposition=top}
    %%%%%%%%%%%%%%%%%%
    %%%%%%%%%%%%%%%%%%
    
    %\subsection{Microstructure, and Morphological Evolution}
    %%%%%%%%%%%%%%%%%%
    %%%%%%%%%%%%%%%%%% Simulations versus experiments
    %%%%%%%%%%%%%%%%%% Simulations versus experiments
    %%%%%%%%%%%%%%%%%% Simulations versus experiments
    %%%%%%%%%%%%%%%%%% Simulations versus experiments
    %%%%%%%%%%%%%%%%%%    
    
\subsubsection{Simulations versus experiments: Estimating the Kinetic Parameters} 
    
We modeled the microstructural evolution of thirteen (initially) homogeneous (c)-Ti$_{1-x-y}$Al$_{x}$Zr$_{y}$N pseudo-ternary alloy compositions shown in Table \ref{tab:sim_alloy_comp} with cubic structure during isothermal annealing at 1200$^\circ$C. These choices correspond to different thermodynamic conditions in different unstable regions of the ternary phase diagram of this system. The first set of three compositions reside in regions I (TiN-ZrN mixture), II (TiN-AlN-ZrN mixture), and III (AlN-ZrN mixture) of the phase-diagram, respectively over a straight line with constant Zr concentration of 0.24. The second set of compositions are located in region IV (TiN-AlN mixture) of the phase diagram. We compare the phase-field microstructure with the experimental Scanning Electron Transmission Microscopy (STEM) results first, and then study the elasto-chemical regime of microstructure evolution for a range of Al content in region IV of the phase diagram which seems to be a suitable region for alloy design due to lower Zr content \cite{yalamanchili2016growth,sheng2008phase,chen2011influence}. We also compared the elasto-chemical microstructure evolution regime with the conventional chemical-only microstructural evolution. The obtained microstructures were visualized by integrating the three composition fields on one layer while the elemental map of a selected rectangular region is indicated in the corner of each microstructure for each microstructure.

\begin{figure}[h!]
        \centering
        \subfloat[]{\includegraphics[width=0.375\columnwidth]{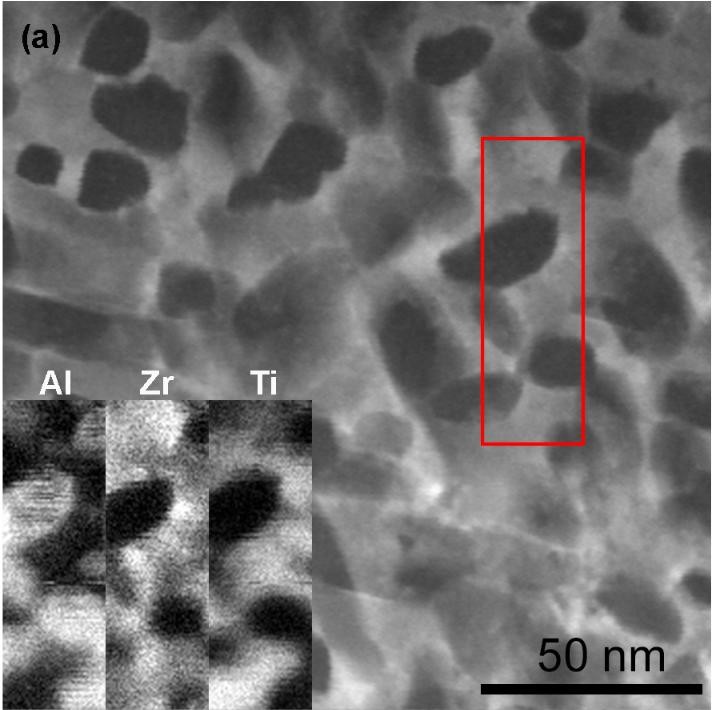} }
        \subfloat[]{\scalebarbackgroundsabc{\includegraphics[width=0.37\columnwidth]{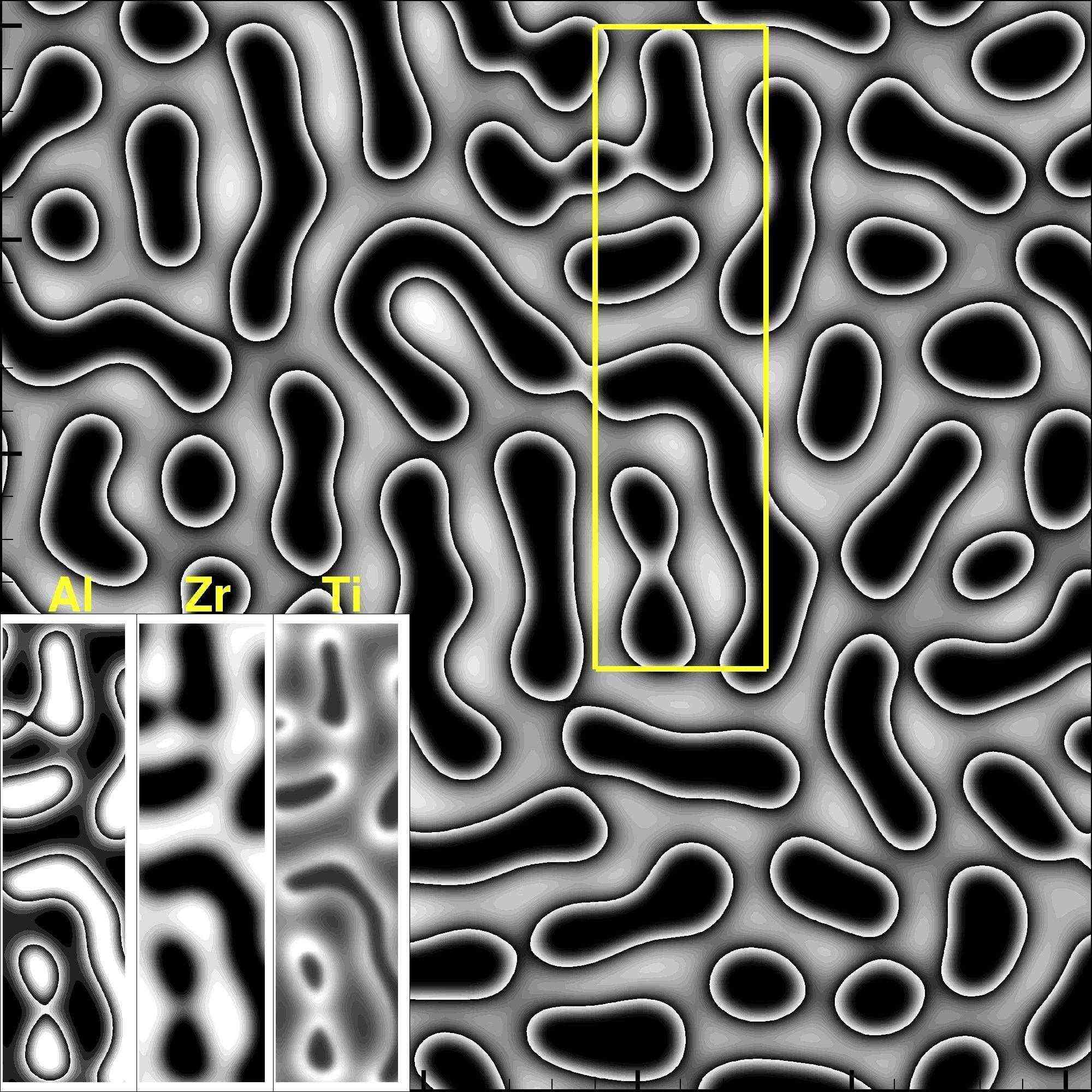}}{950}{5.3}{50} }
        \caption{(a) Elemental contrast STEM micrographs of Ti$_{0.30}$Al$_{0.46}$Zr$_{0.24}$N annealed for 2 hours at 1100 $^\circ$C reproduced from \cite{lind2014high} with permission, (b) computational result in this work annealed for 2 hours at 1200 $^\circ$C. The bright regions in the elemental maps in the left corner of the overview images indicate the corresponding elemental map.}
        \label{fig:exps_vs_sims}
    \end{figure}

We start by comparing the implemented coupled multi-physics model with experimental microstructural information. Lind $et$ $al.$ \cite{lind2014high} shows the process of spinodal decomposition in Ti$_{0.30}$Al$_{0.46}$Zr$_{0.24}$N samples that were prepared by the cathodic arc evaporation approach, and deposited to obtain a coating thickness of 2 $\mu$m. The obtained substrates underwent spinodal decomposition during annealing treatment in a vacuum chamber at an isothermal temperature of 1100$^\circ$C for 2 hours. The high angle annular dark field STEM image, and EDX elemental maps of this coating is illustrated in Fig.~\ref{fig:exps_vs_sims}a and confirm the phase separation. The darker contrast in the overview STEM image shows presence of (w)-AlN-rich phases. Most of the bright contrast (high Z-contrast elements) regions are dominantly enriched in either Ti or Zr corresponding to the two cubic (Ti,Zr)N phases. Lind \etal~\cite{lind2014high} suggests this observation is due to the incomplete sequence of phase separation into the stable phases. This is confirmed by the elemental map that is demonstrated in the EDX elemental maps recorded for the red rectangle area of 25$\times$75 $nm^2$. The study also confirms that slower rate of decomposition in Ti$_{0.13}$Al$_{0.69}$Zr$_{0.19}$N (Zr-rich sample) comparing to that of Ti$_{0.30}$Al$_{0.46}$Zr$_{0.24}$N (Zr-poor sample) is the direct consequence of the lower driving force for decomposition.    
     
    %%%%%%%%%%%%%%%%%%
    %%%%%%%%%%%%%%%%%%
    %%%%%%%%%%%%%%%%%%
    %%%%%%%%%%%%%%%%%%
        
The predicted microstructure using the phase-field modeling is shown to the right of the experimental microstructure in Fig.~\ref{fig:exps_vs_sims}b. The elemental map of the selected rectangular area indicated by the yellow box is extracted from the overview microstructure similar to EDX map in the experimental micrograph, and is represented in the corner of the phase-field overview microstructure. Similar to STEM image, the darker contrast in the overview microstructure shows AlN rich regions, and the bright regions indicate either TiN or ZrN. The elemental map shown in the corner of this image represents AlN, ZrN, and TiN maps indicated by white color from left to right, respectively. The bright regions in the extracted rectangular EDX maps and phase-field elemental maps agree well. Although, phase-field elemental map provides more insight on how TiN and ZrN form and/or segregate in the microstructure. 
    
In this respect, the sequence of phase separation is consistent with the \emph{ab-initio} results reported in Ref. \cite{lind2014high}. The phase-field modeling reveal that the process of separation starts with the initial decomposition of ZrN and AlN phases, and then TiN precipitates later. Both experimental, and computational studies clearly reveal that AlN phases precipitate as particles in the form of semi-ellipsoidal or worm-like morphologies. During initial stages of the separation, AlN appear to form small worm-like domains that turn to elongated structures later at the coarsening stages. It is difficult to clearly interpret the matrix phase from the experimental micrograph. On the other hand, phase-field modeling reveals that ZrN is the matrix phase that hosts AlN as the reinforcing precipitates. Furthermore, TiN tends to first segregate or form a very narrow region around the AlN precipitates. 
    
    The interface mobility and gradient energy coefficient ($\kappa_{\alpha}$) are estimated by matching the overall morphology of the obtained microstructure with the experimental STEM, and evolution time up to a reasonable coarsening, and particle size distibution. The values are summarized in Table \ref{tab:kinetic_pars}. The interfacial mobility is taken as a function of diffusivity ($M_{\alpha}=D_{\alpha}/{RT}$) where $D_{\alpha}$ is the diffusivity, and $R$ is the gas constant. AlN diffusivity is estimated based on $D_{AlN}=1.4\times10^{-5}exp({{Q}/{RT}})$ proposed by Knutsson~\etal~\cite{knutsson2013microstructure} for Ti$_{1-x}$Al$_x$N system where $Q$ varies between 2.4 eV and 3.2 eV. Different self-diffusion data for N and Zr diffusion in $ZrN_x$ exist in the literature \cite{hultman2000thermal}, and the range of these data is disperse. Due to lack of sufficient diffusion data in Zr based nitirites, we estimate the kinetic input for ZrN based on the general characteristics of the experimental micrograph in Fig.~\ref{fig:exps_vs_sims}a. Although, M and $\kappa_{\alpha}$ are usually a function of composition, and temperature, we used a constant parameter in the respective interfaces. 
    
    %classifying the precipitate phase and matrix phase based on the amount of elements in the sides of pseudo-ternary phase-diagram (regions I,II, and III) is valid. While experiments show that Zr, and Ti are spread all over the micrograph outside the AlN particles, 
    
    %%%%%%%%%%%%%%%%%%
    %%%%%%%%%%%%%%%%%%
    \begin{table}[h!]
        \caption{The estimated kinetic parameters in this study by means of comparing with experimental microstructure.} \label{tab:kinetic_pars} %\vspace{-0.25cm}
        \scriptsize
        \centering
        %\begin{adjustbox}{width=8.1cm,height=0.60cm}
        \begin{tabular}{ccc} \toprule
        
            Mobility matrix (M$_{\alpha\beta}$) $m^2.J^{-1}s^{-1}$         &  $\kappa_{\alpha}$  ($J.m^{-1}$)   & $L_{\alpha\beta}$  ($J.m^{-1}$)  \\ \midrule 
            $\begin{bmatrix} 2.230\times10^{-21} & 0 \\ 0 & 2.230\times10^{-25} \\ \end{bmatrix}$  & $\begin{bmatrix} 6.329\times10^{-16}\\ 6.513\times10^{-16} \end{bmatrix}$   &  $\begin{bmatrix} 2.550\times10^{-16} & 0 \\  0 & 2.550\times10^{-16}  \\  \end{bmatrix}$  \\ \bottomrule

            \end{tabular}
        %\end{adjustbox}
    \end{table}
    %%%%%%%%%%%%%%%%%%
    %%%%%%%%%%%%%%%%%%
    
    %Figure~\ref{fig:3D_mic} compares the 3D microstrcutures of Ti$_{0.70}$Al$_{0.25}$Zr$_{0.05}$N alloy in the case of chemical and elastochemical modeling. AlN particles are dispersed with similar particular distances in both cases. Although, elastic effects slow down the kinetics of reaction. The dimension of the simulation cell is $50\times50\times16.6$ $nm$. These results for Ti$_{1-x-0.05}$Al$_{x}$Zr$_{0.05}$N alloys will be discussed thoroughly in the upcoming sections. 
    
\begin{figure}[h!]
    \centering
    \raisebox{0.25cm}{a)}
    \includegraphics[scale=0.12]{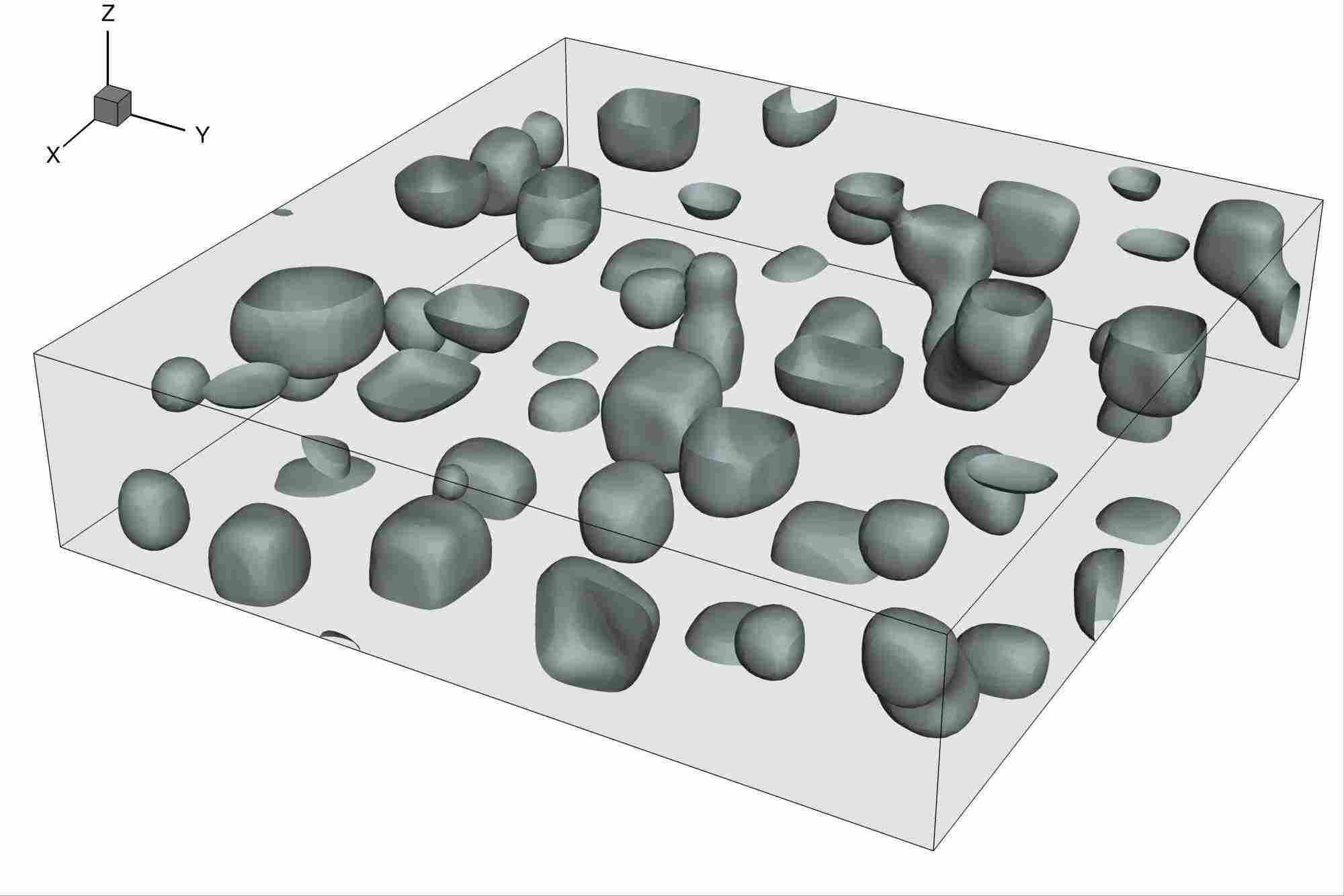} \vspace{-0.10cm}    
    \raisebox{0.25cm}{b)}
    \includegraphics[scale=0.12]{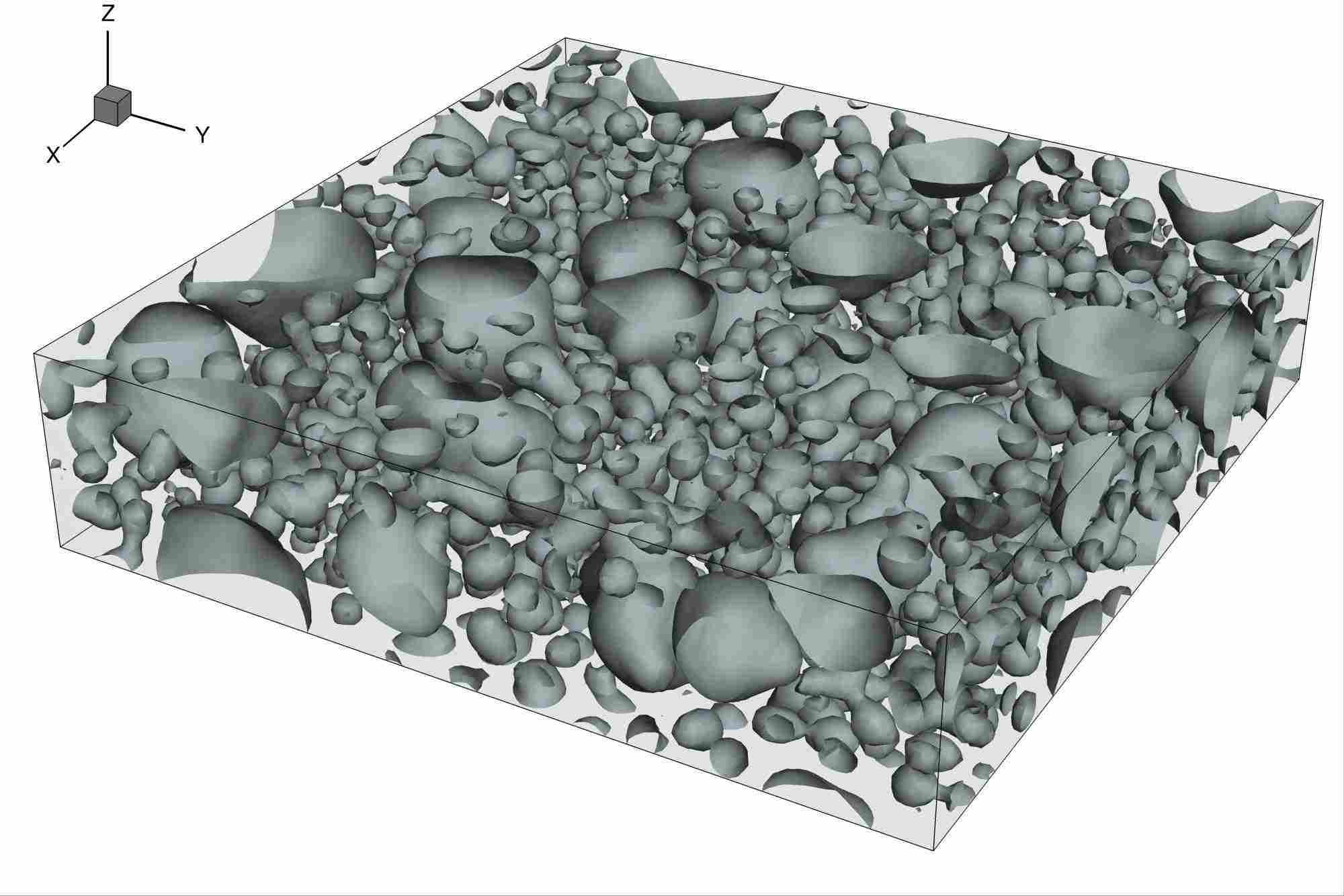} \vspace{-0.25cm}
    %\caption{3D microstructure of Ti$_{0.70}$Al$_{0.25}$Zr$_{0.05}$N for a) chemical, and b) elastochemical growth regime showing AlN spheres versus AlN cuboids in TiN matrix. ZrN has not yet segregate entirely, and is not depicted in these images.}
    \caption{3D microstructures of a) Ti$_{0.70}$Al$_{0.25}$Zr$_{0.05}$N and b) Ti$_{0.30}$Al$_{0.65}$Zr$_{0.05}$N alloys. The Ti$_{0.70}$Al$_{0.25}$Zr$_{0.05}$N contains AlN cuboid particles, and Ti$_{0.30}$Al$_{0.65}$Zr$_{0.05}$N represents a bimodal microstructure with large TiN semi-spheres in AlN matrix, along with the secondary smaller ZrN particles.}
    
    \label{fig:3D_mic}
    \end{figure}

    Figure~\ref{fig:3D_mic}a~and~\ref{fig:3D_mic}b compare the 3D microstrcutures of Ti$_{0.70}$Al$_{0.25}$Zr$_{0.05}$N, and Ti$_{0.30}$Al$_{0.65}$Zr$_{0.05}$N alloys in the case of elastochemical modeling, respectively. The figure illustrates the unimodal and bimodal distribution of the particles in the first and second alloy, respectively. AlN cuboid precipitates reside in TiN matrix in the case of unimodal microstructure, and TiN large semi-spherical particles along with secondary small ZrN precipitates in the AlN matrix are illustrated in this figure. The microstructural characteristics of Ti$_{1-x-0.05}$Al$_{x}$Zr$_{0.05}$N alloys will be discussed thoroughly in the upcoming sections. 

    %%%%%%%%%%%%%%%%%%
    %%%%%%%%%%%%%%%%%%
    %%%%%%%%%%%%%%%%%% Chemical Phase Separation
    %%%%%%%%%%%%%%%%%% Chemical Phase Separation
    %%%%%%%%%%%%%%%%%% Chemical Phase Separation
    %%%%%%%%%%%%%%%%%%
    %%%%%%%%%%%%%%%%%%
    
    \subsection{Phase Separation during Chemical Growth Regime}
    
\begin{figure*}[h!]
        \centering
        
        %% 0.25
        \subfloat[x\textsubscript{Al}=0.25]{\scalebarbackground{\includegraphics[scale=0.048]{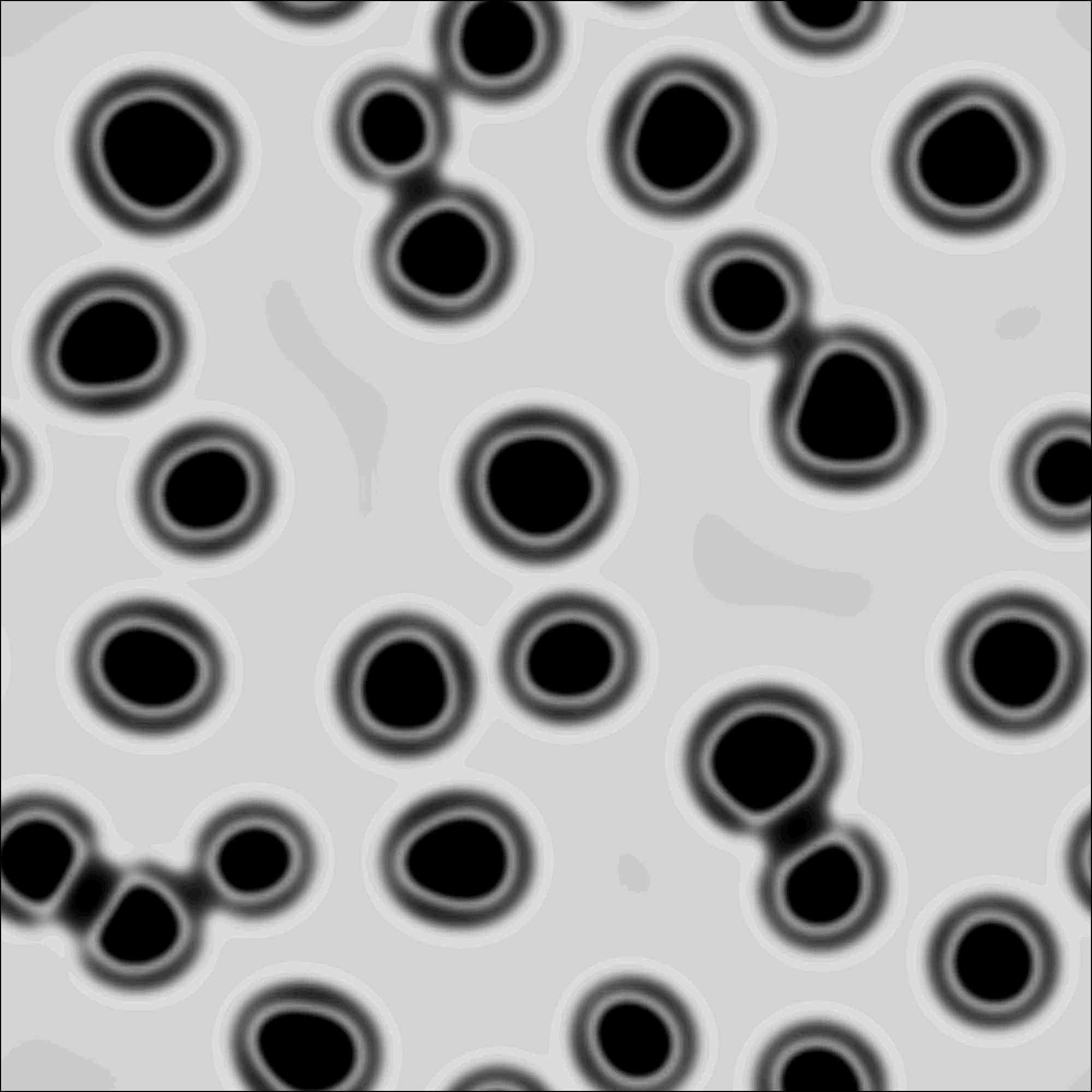}}{495}{9.887}{10} }
        %% 0.30
        \subfloat[x\textsubscript{Al}=0.30]{\scalebarbackground{\includegraphics[scale=0.048]{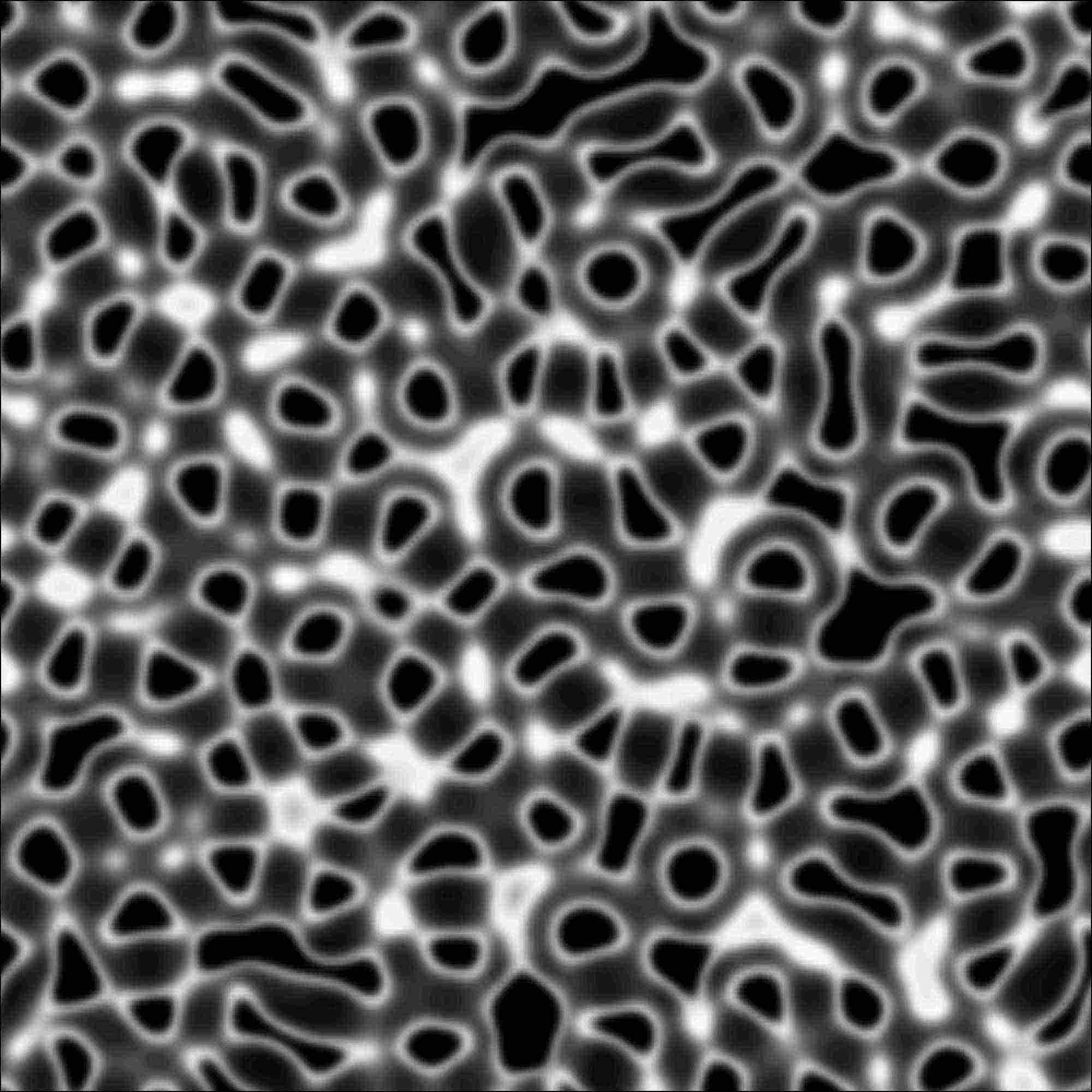}}{495}{9.887}{10} } 
        %% 0.35
        \subfloat[x\textsubscript{Al}=0.35]{\scalebarbackground{\includegraphics[scale=0.048]{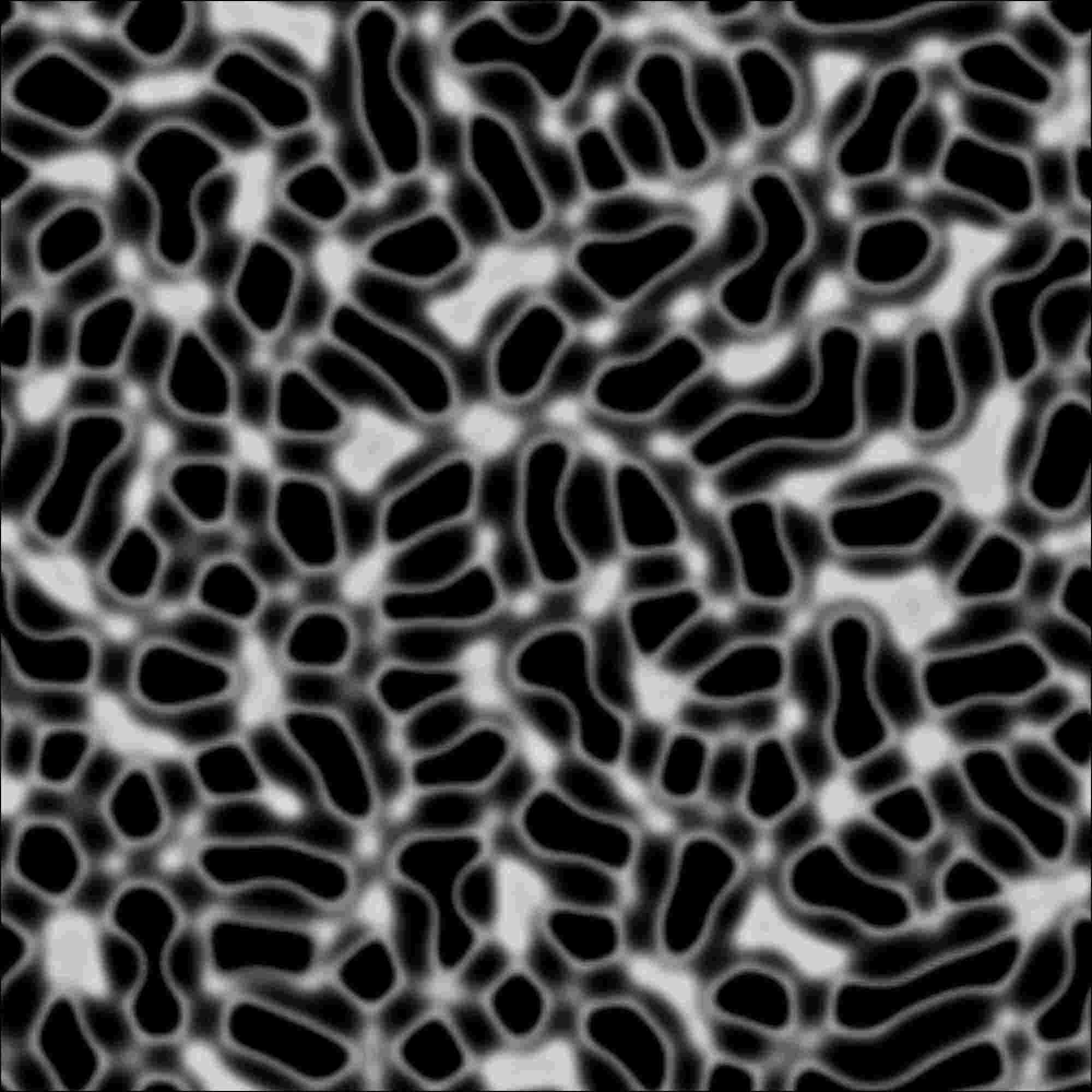}}{495}{9.887}{10} }
        %% 0.40
        \subfloat[x\textsubscript{Al}=0.40]{\scalebarbackground{\includegraphics[scale=0.048]{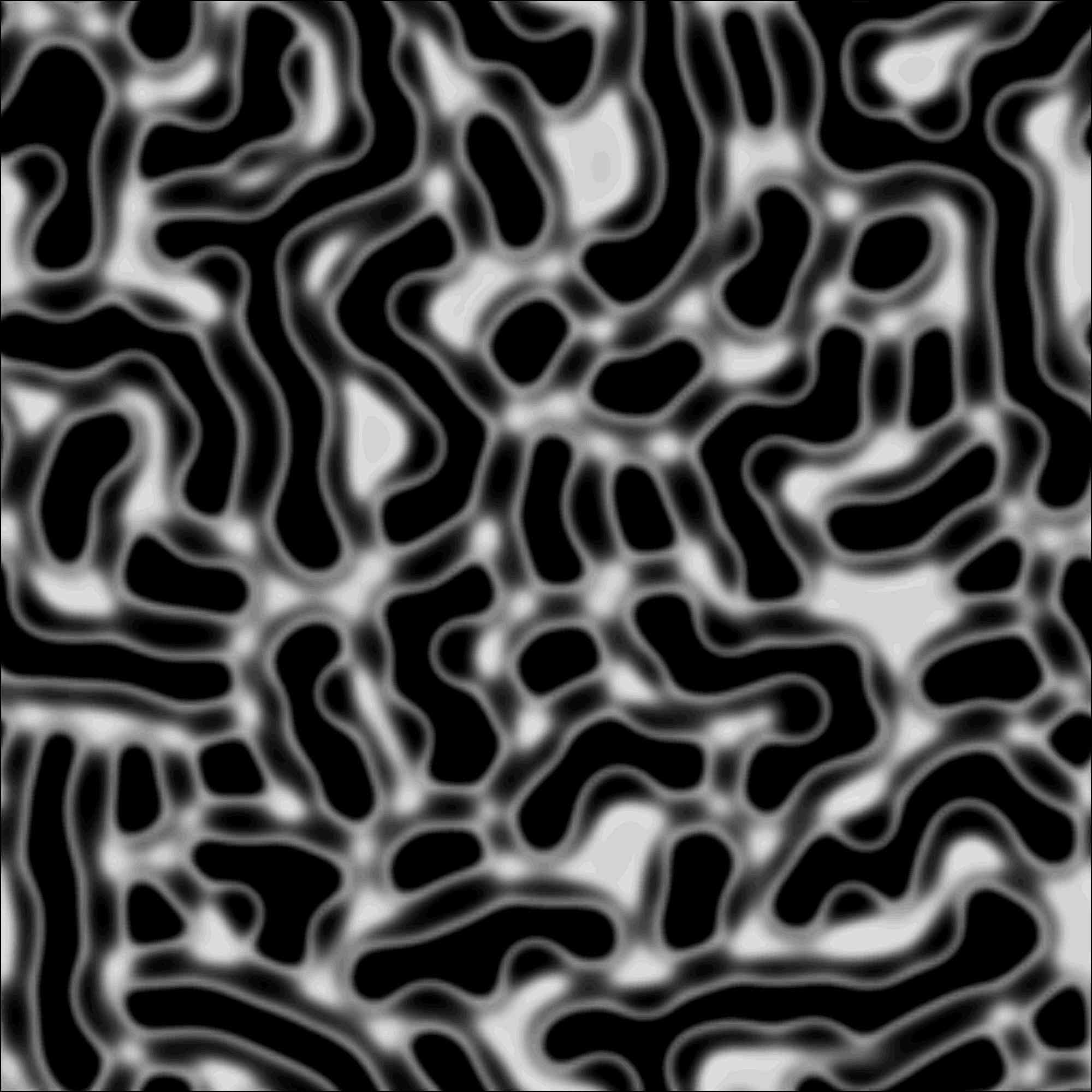}}{495}{9.887}{10} }
        %% 0.45
        \subfloat[x\textsubscript{Al}=0.45]{\scalebarbackground{\includegraphics[scale=0.048]{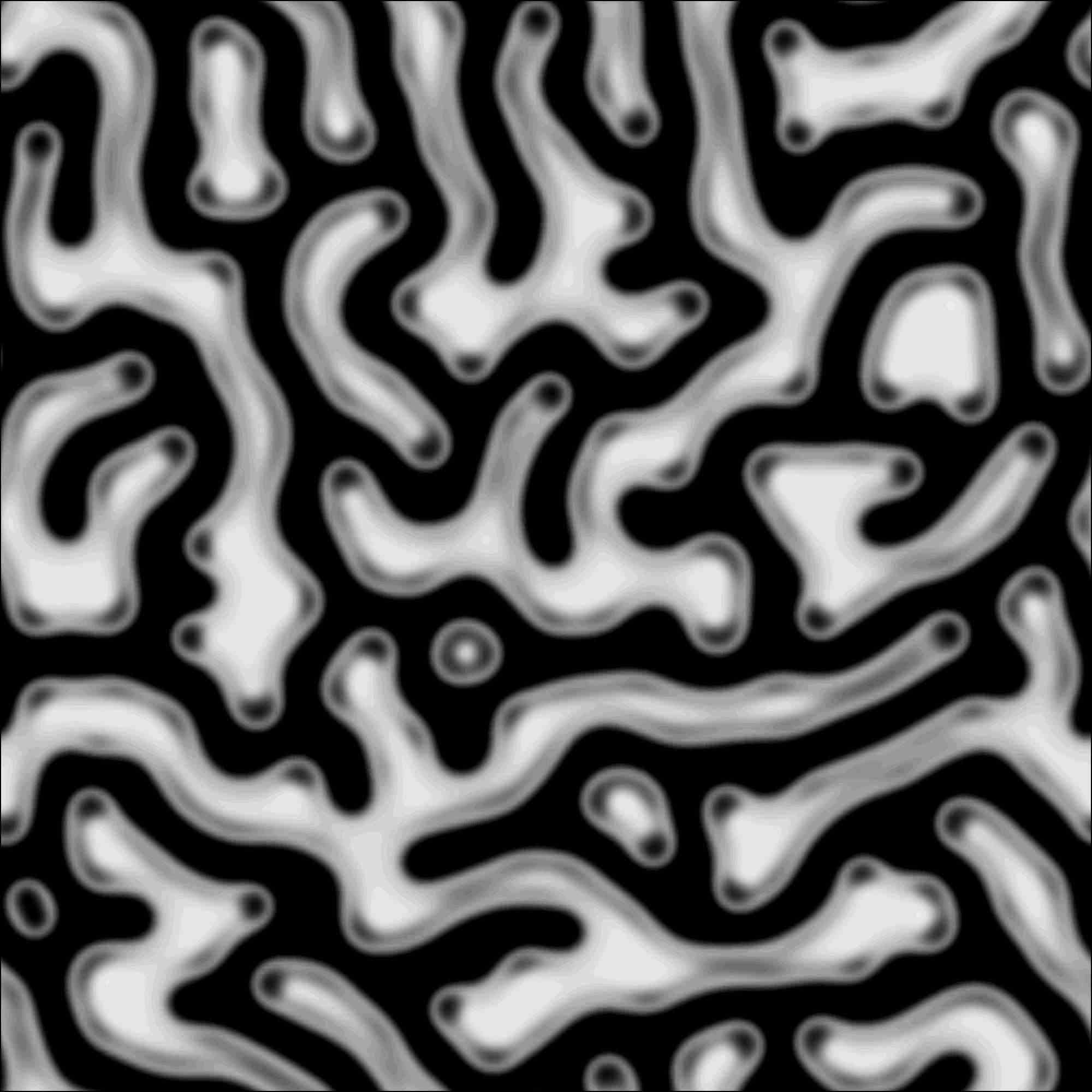}}{495}{9.887}{10} }\\ \vspace{-0.35cm}
        %% 0.50
        \subfloat[x\textsubscript{Al}=0.50]{\scalebarbackground{\includegraphics[scale=0.048]{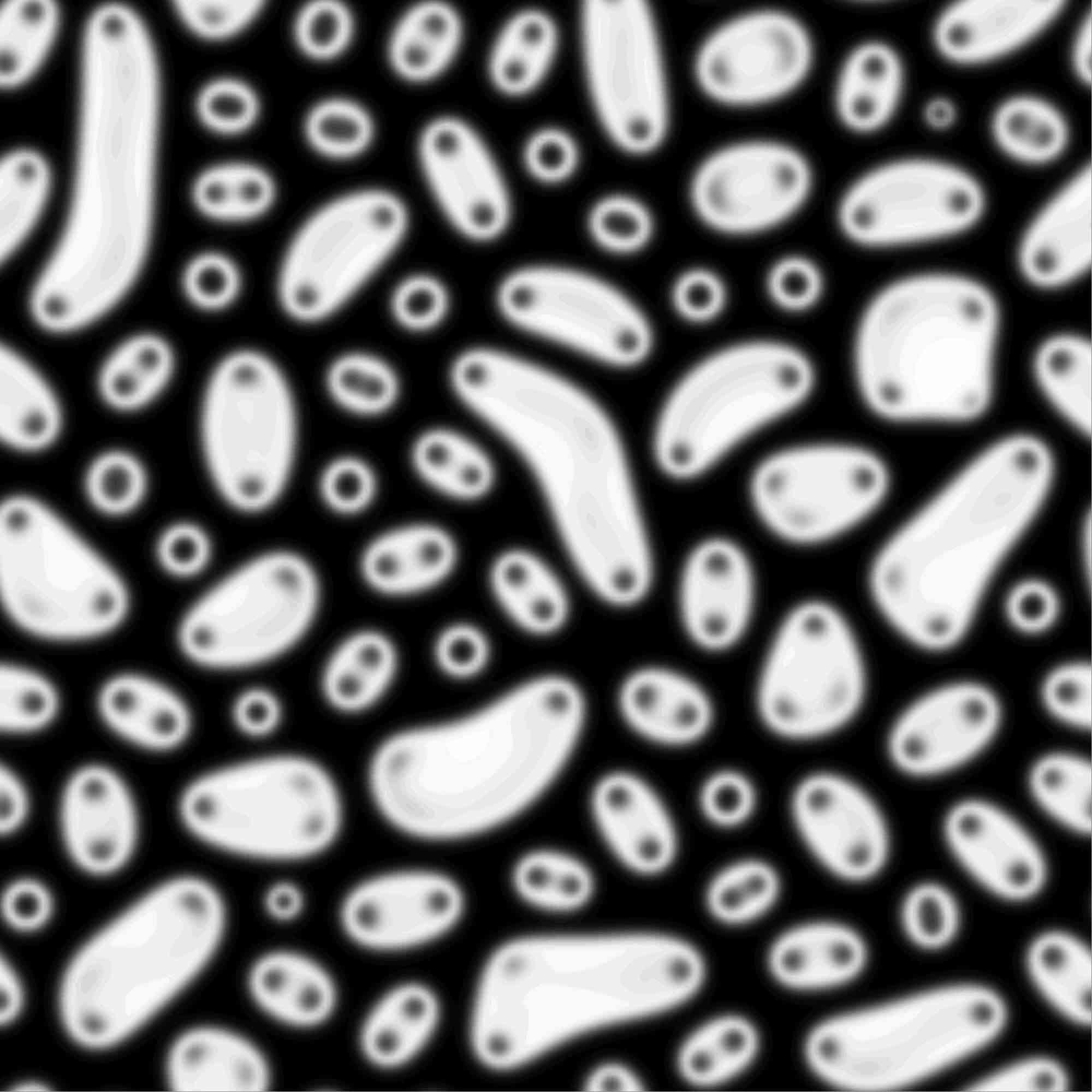}}{495}{9.887}{10} }
        %% 0.55
        \subfloat[x\textsubscript{Al}=0.55]{\scalebarbackground{\includegraphics[scale=0.048]{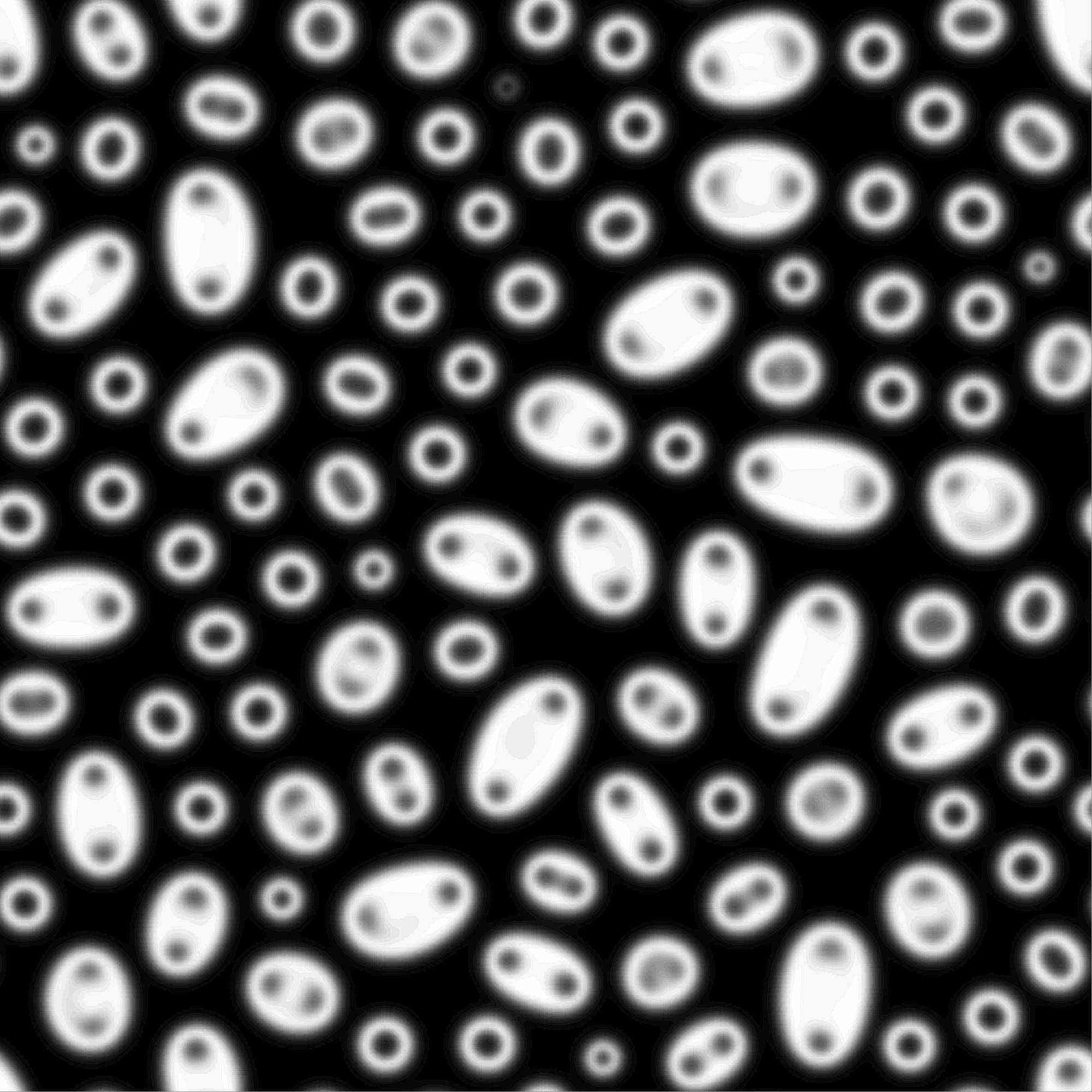}}{495}{9.887}{10} }
        %% 0.60
        \subfloat[x\textsubscript{Al}=0.60]{\scalebarbackground{\includegraphics[scale=0.048]{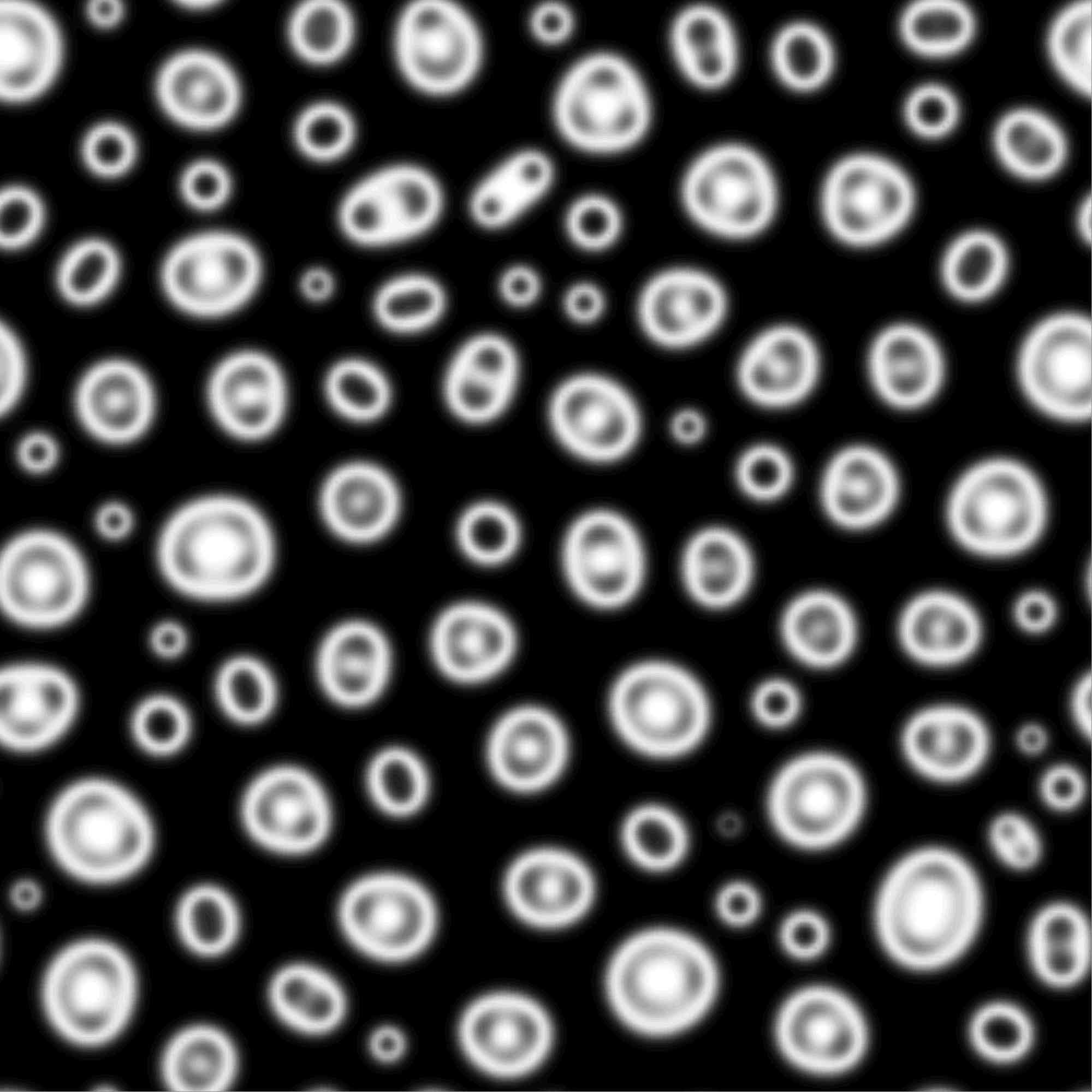}}{495}{9.887}{10} }
        %% 0.65
        \subfloat[x\textsubscript{Al}=0.65]{\scalebarbackground{\includegraphics[scale=0.048]{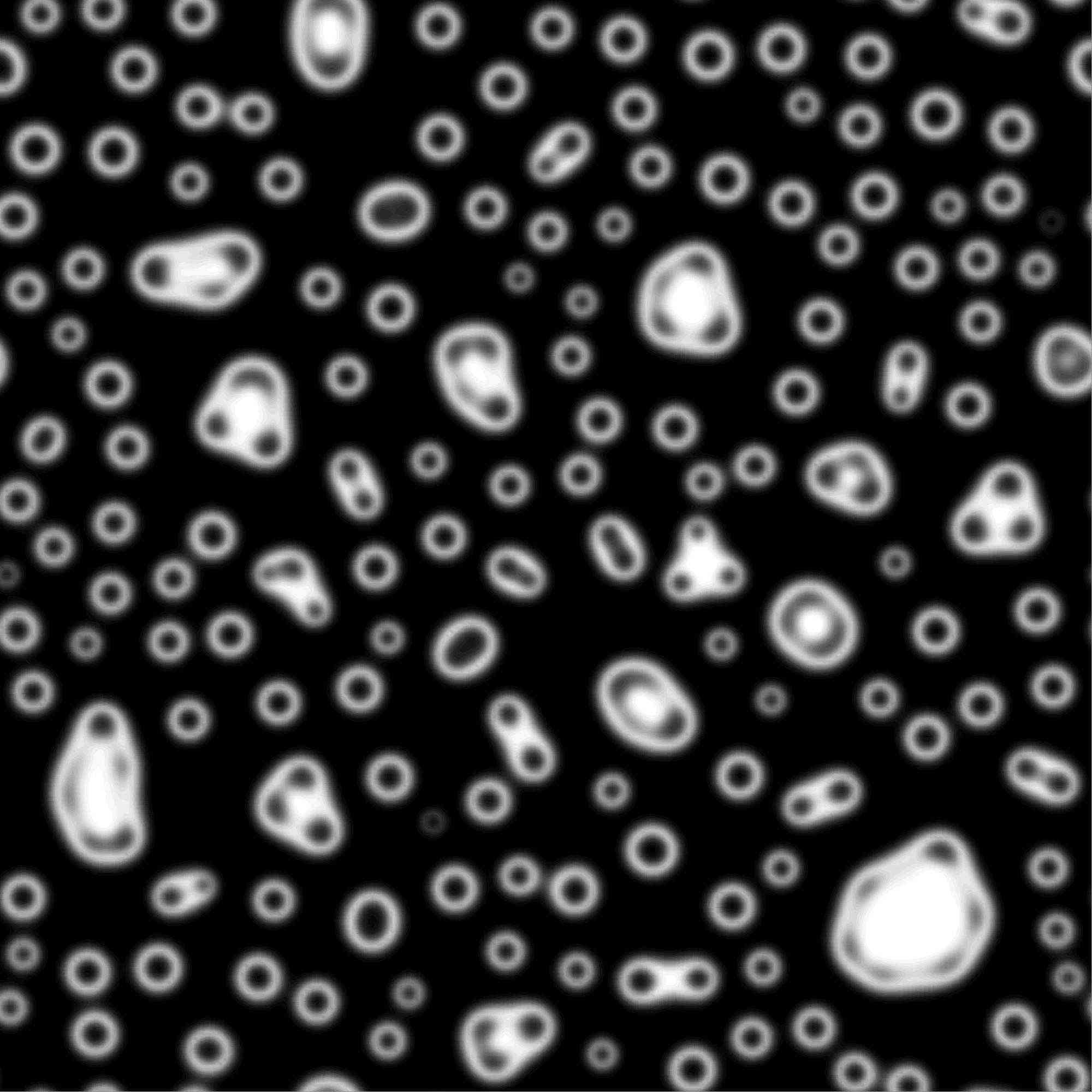}}{495}{9.887}{10} }
        %% 0.70
        \subfloat[x\textsubscript{Al}=0.70]{\scalebarbackground{\includegraphics[scale=0.048]{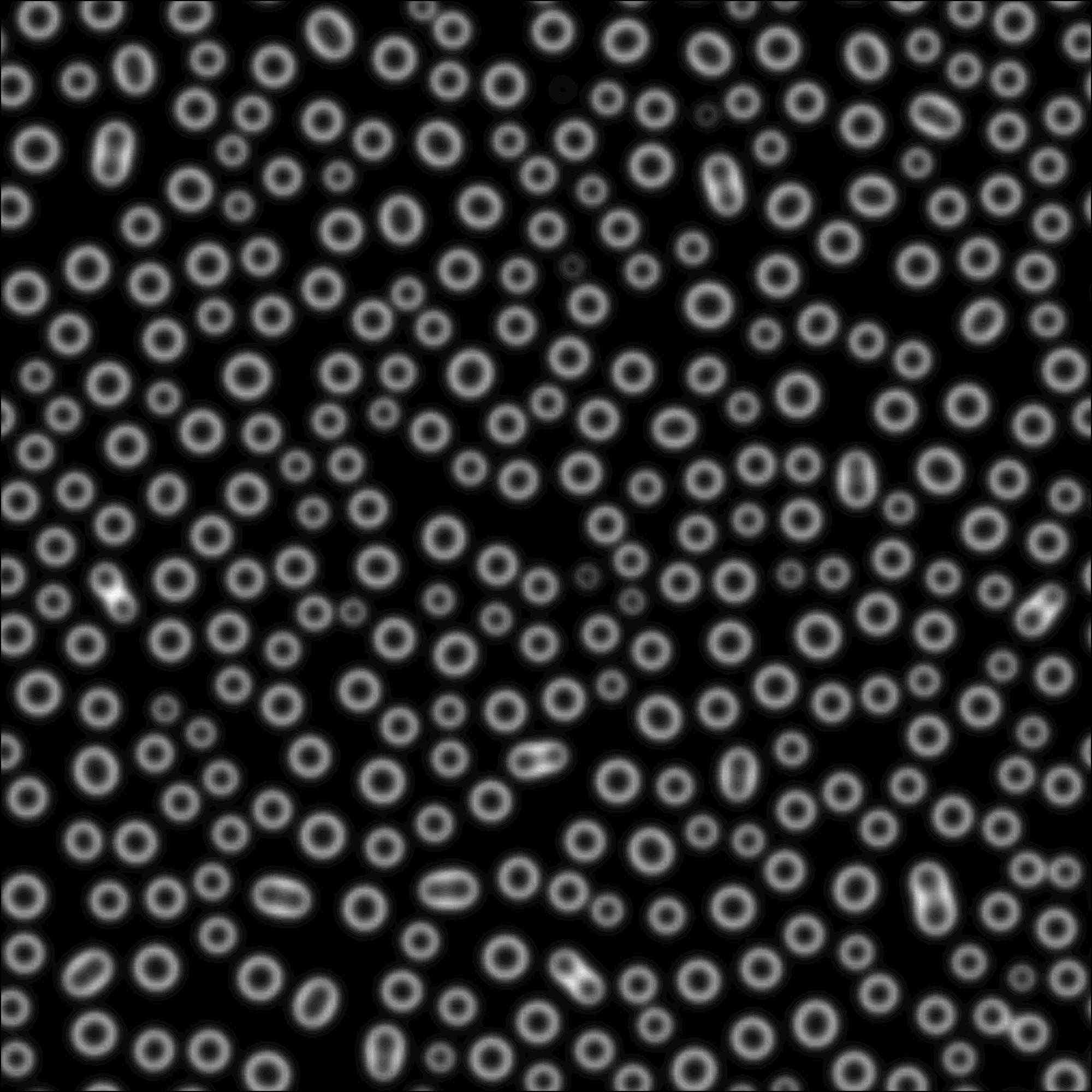}}{495}{9.887}{10} }
        
        \caption{ Comparing the microstructure of Ti\textsubscript{1-x-0.05}Al\textsubscript{x}Zr\textsubscript{0.05}N alloy during annealing at T=1200$^\circ$C for chemical-only simulations for $0.25\leq x_{Al}\leq0.70$ at the same frozen time. Color map is \raisebox{1mm}{\fcolorbox{black!90!gray}{black}{\null}} (c)-AlN, \raisebox{1mm}{\fcolorbox{black}{white}{\null}} (c)-TiN, \raisebox{1mm}{\fcolorbox{black!90!gray}{gray}{\null}} (c)-ZrN. }
        \label{fig:chem-only-sims}
    \end{figure*}

In this section, we investigate the microstructure of the pseudo-ternary Ti$_{1-x-0.05}$Al$_{x}$Zr$_{0.05}$N alloy in the TiN-AlN corner (region IV) of the phase-diagram for a wide range of Al compositions. The two-dimensional solute distributions during chemical growth regime at the isothermal state of 1200$^\circ$C is depicted in Fig.~\ref{fig:chem-only-sims}. The respective change in microstructure for fixed $x_{Al}$ intervals, and the local composition profile is represented by gray-scale levels. During the early stages of decomposition, the solid solution with small composition fluctuation transforms into two-phase regions of AlN and TiN. Each phase interchanges between precipitate or matrix depending on the value of $x_{Al}$, similar to the decomposition process in binary systems. ZrN only appears as the third phase after full decomposition of AlN and TiN phases, segregating in different locations (grain boundaries, particle, or matrix phase) depending on the value of $x_{Al}$. 
    
    \fboxsep=1.3mm \fboxrule=0.25mm
    %%%%%%%%%%%%%%%%%%
    %%%%%%%%%%%%%%%%%%

Figure \ref{fig:chem-only-sims} shows that when $x_{Al}$ is as low as 0.25, the microstructure is composed of semi-circular (droplet) AlN precipitates with ZrN surrounding them and TiN residing in the matrix. A gradual increase of $x_{Al}$ to 0.30 changes the morphology from semi-circular phases to worm-like precipitates. These worm-like domains turn into to bi-continuous structures around $x_{Al}=0.45$. Further increase of Al composition up to $x_{Al}=0.70$ abandons the bi-continuous morphology and retains the droplet morphology again. However, this time TiN phases precipitate in the AlN matrix. 
    
In the Al-poor side of the miscibility gap, depending on the average AlN particular distance, and/or volume fraction, ZrN either segregates in the grain boundaries of AlN particles (Fig.~\ref{fig:chem-only-sims}.a) or acts as the interconnecting phase between the AlN precipitates by forming a worm-like morphology (Fig.~\ref{fig:chem-only-sims}b, c, and d). When $0.30\leq x_{Al}\leq0.40$, the structure is dominant by interconnected network of Al- and Zr-rich phases. When $x_{Al}=0.45$ the structure appears to be bi-continuous, and ZrN only segregates in the dead end corner of this structure. Understanding the degree of bi-countinuity requires 3D modeling to fully interpret the structure of these domains as it is recently proposed by Kwon \etal, and others \cite{kwon2007coarsening,kwon2009topology,chen2010morphological}. However, we are not interested in characterization of the topology of these structures in this study. We note that an ideal bi-continuous structure may potentially be observed in alloys with composition $x_{Al}=0.45$. 
    
When $x_{Al}\geq 0.50$, TiN and AlN phases switch the role in the microstructure. When $x_{Al}=0.50$, the structure is again a disrupted worm-like morphology TiN large precipitates are distributed everywhere, and ZrN segregates inside the these precipitates rather than forming the interconnected network of precipitates as it is for $0.30\leq x_{Al}\leq0.40$. In addition, limited number of small ZrN circular precipitates also start to form in between the larger TiN particles in this composition. For $x_{Al}\geq 0.65$, the combinations of TiN and ZrN particles show a bimodal distribution in the microstructure. In general, the circular precipitates tend to get smaller in the Al-rich corner of the miscibility gap in comparison with the the Al-poor corner. This indicates that the kinetics of evolution gets slower by increasing the Al content, due to a decrease in the chemical driving force driving the microstructure evolution. Furthermore, even though the Zr content is constant in the selected alloy, ZrN segregates faster in Al rich side of the miscibility gap. 
    
These computational insights provide a rigorous understanding on how the morphology changes by a slight change in $x_{Al}$, and could pave the way for proper alloy design to achieve the desired mechanical responses based on the microstructure design. The compositional boundaries indicated here for the observed morphology may change slightly in 3D modeling. The 3D simulations for the same conditions are provided in the supplementary document.

    %Chemical-only simulations also suggest that cubic ZrN phase is stable for wide range of aluminum content, though the role in the microstructure change from segregating in the grain boundaries for very low Al content to the interconnected network of ZrN and AlN phases where 0.30$<x_{Al}<$0.45, and precipitation inside TiN phases. 

    %First we present the chemical-only simulations and then compare the results with elasto-chemical results in the next section. The two-dimensional solute distributions, and the change in the microstructure by increasing the Al content at the isothermal state of 1200$^\circ$C is shown in Fig.~\ref{fig:chem-only-sims} for the pseudo-ternary Ti$_{1-x-0.05}$Al$_{x}$Zr$_{0.05}$N. 
    
    %As can be seen, the solid solution with small composition fluctuation transforms into a highly interconnected three-phase structure at the intermediate, and later stages of the growth. During the early stages, the initial single-phase matrix with slight composition fluctuation starts to decompose into a circular or interconnected network consisting of Ti- and Al-rich domains, while ZrN phases appear later. At the early stages, decomposition follows similar to the binary systems. However, at the later stages, all three phases decompose entirely.
    
    %%%%%%%%%%%%%%%%%%
    %%%%%%%%%%%%%%%%%%
    %%%%%%%%%%%%%%%%%% Elasto-chemical Phase Separation
    %%%%%%%%%%%%%%%%%% Elasto-chemical Phase Separation
    %%%%%%%%%%%%%%%%%% Elasto-chemical Phase Separation
    %%%%%%%%%%%%%%%%%%
    %%%%%%%%%%%%%%%%%%
    
    \subsection{Phase Separation in the Elasto-chemical Regime}
    
The process of chemical phase separation in the previous section demonstrated the alignment of AlN and TiN precipitates, like strings of pearls, along the chemically enforced modulations. On the other hand, internal strain is an inevitable consequence of the phase transformations in the solid state(they are usually accompanied by finite lattice deformations), and stresses have to be considered, which may cause large deviations from chemical equilibrium \cite{fratzl1999modeling}. Elastic interactions arising from the difference of lattice spacing, and elastic constants between the coherent phases can have a strong influence on the process of phase separation (coarsening) in alloys. If the elastic moduli are different in the two phases, the elastic  interactions may accelerate, slow down or even stop the phase separation process. A brief look to the elastic properties of this alloy indicated in Fig.~\ref{fig:elastic_cst} demonstrate an stiff system with very high elastic constants inferring the necessity to take into account the elastic interactions upon phase transformations in this system. 

    %%%%%%%%%%%%%%%%%%
    \begin{figure*}[!ht]
        \centering
        
        %% 0.25
        \subfloat[x\textsubscript{Al}=0.25]{\scalebarbackground{\includegraphics[scale=0.048]{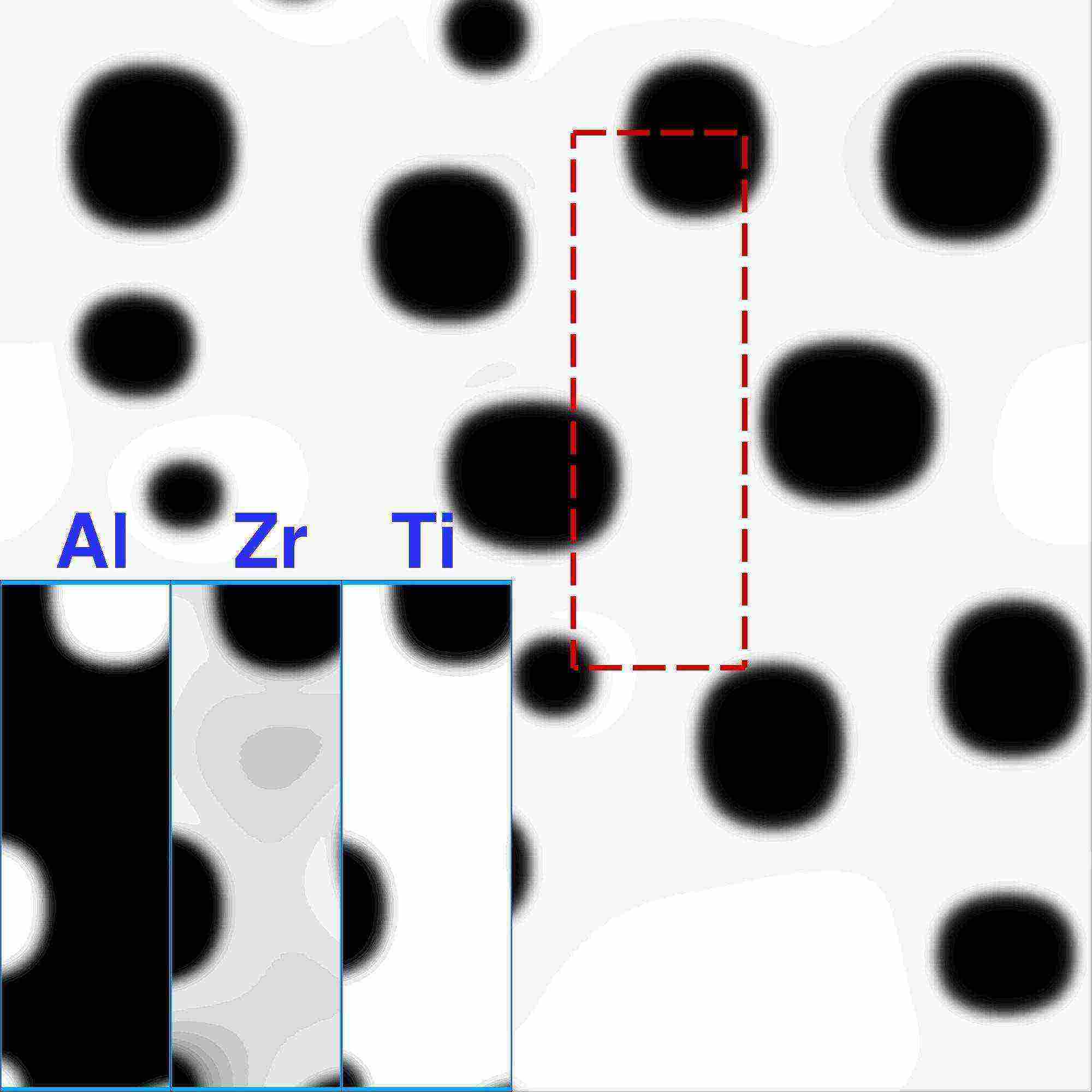}}{495}{9.887}{10} } 
        %% 0.30
        \subfloat[x\textsubscript{Al}=0.30]{\scalebarbackground{\includegraphics[scale=0.048]{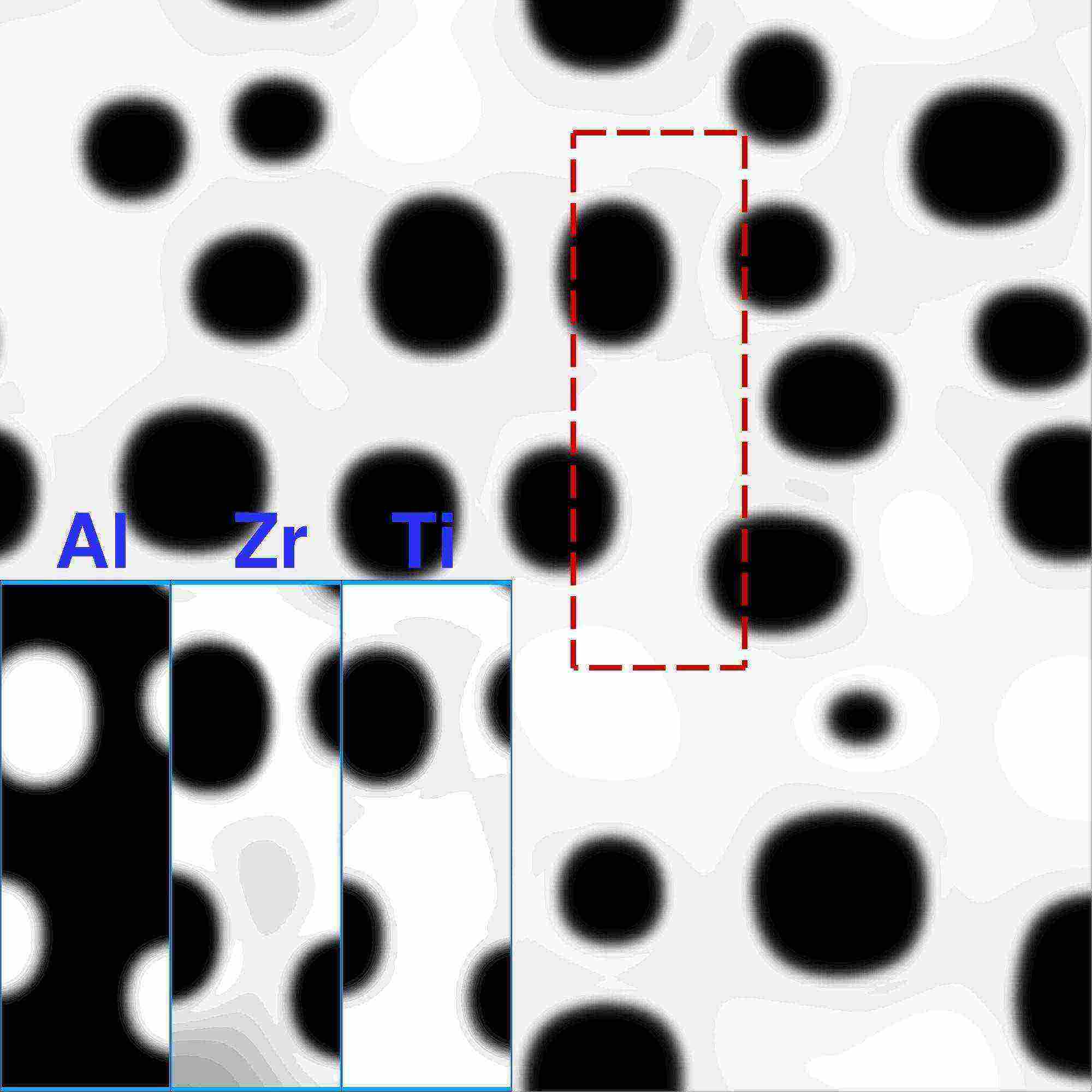}}{495}{9.887}{10} } 
        %% 0.35
        \subfloat[x\textsubscript{Al}=0.35]{\scalebarbackground{\includegraphics[scale=0.048]{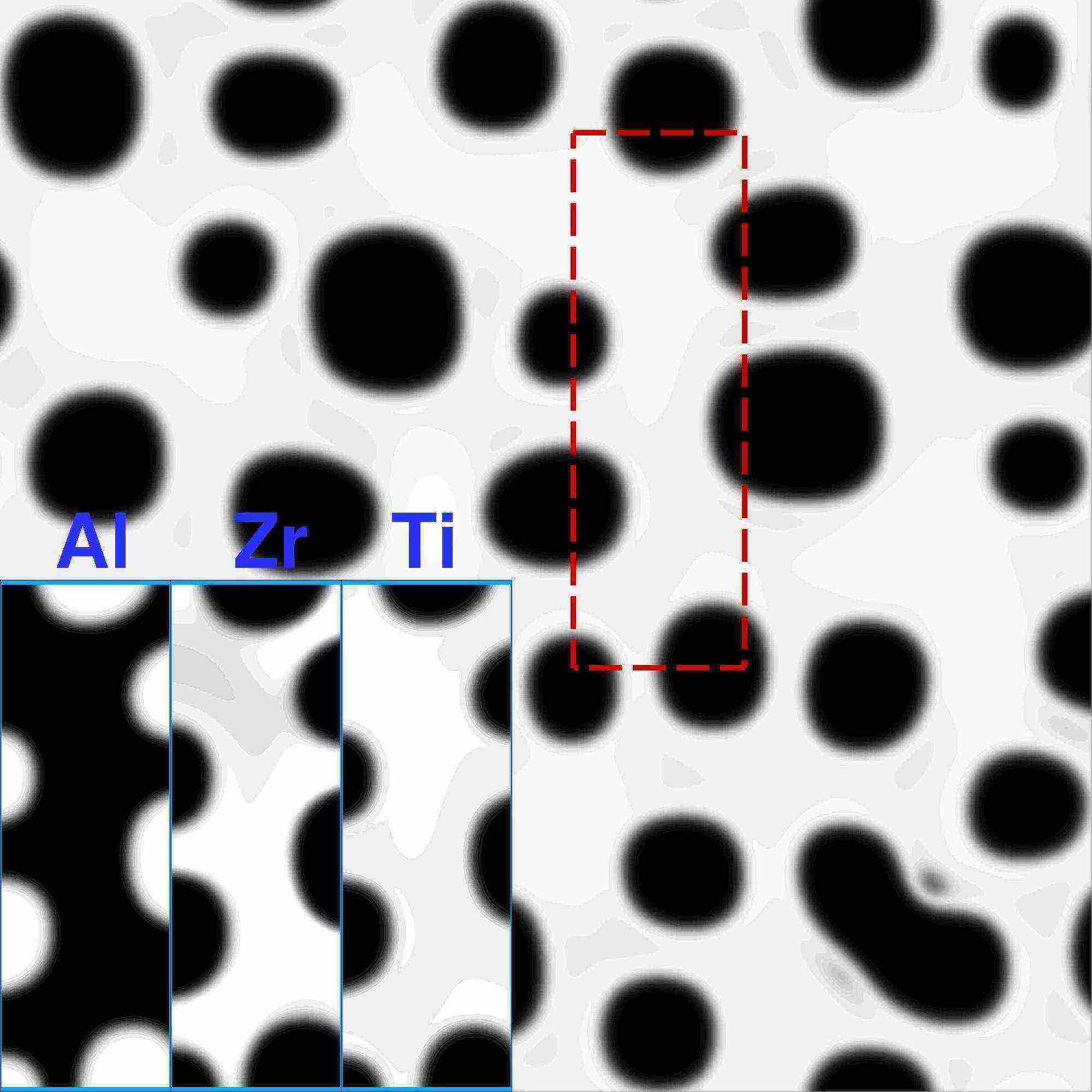}}{495}{9.887}{10} } 
        %% 0.40
        \subfloat[x\textsubscript{Al}=0.40]{\scalebarbackground{\includegraphics[scale=0.048]{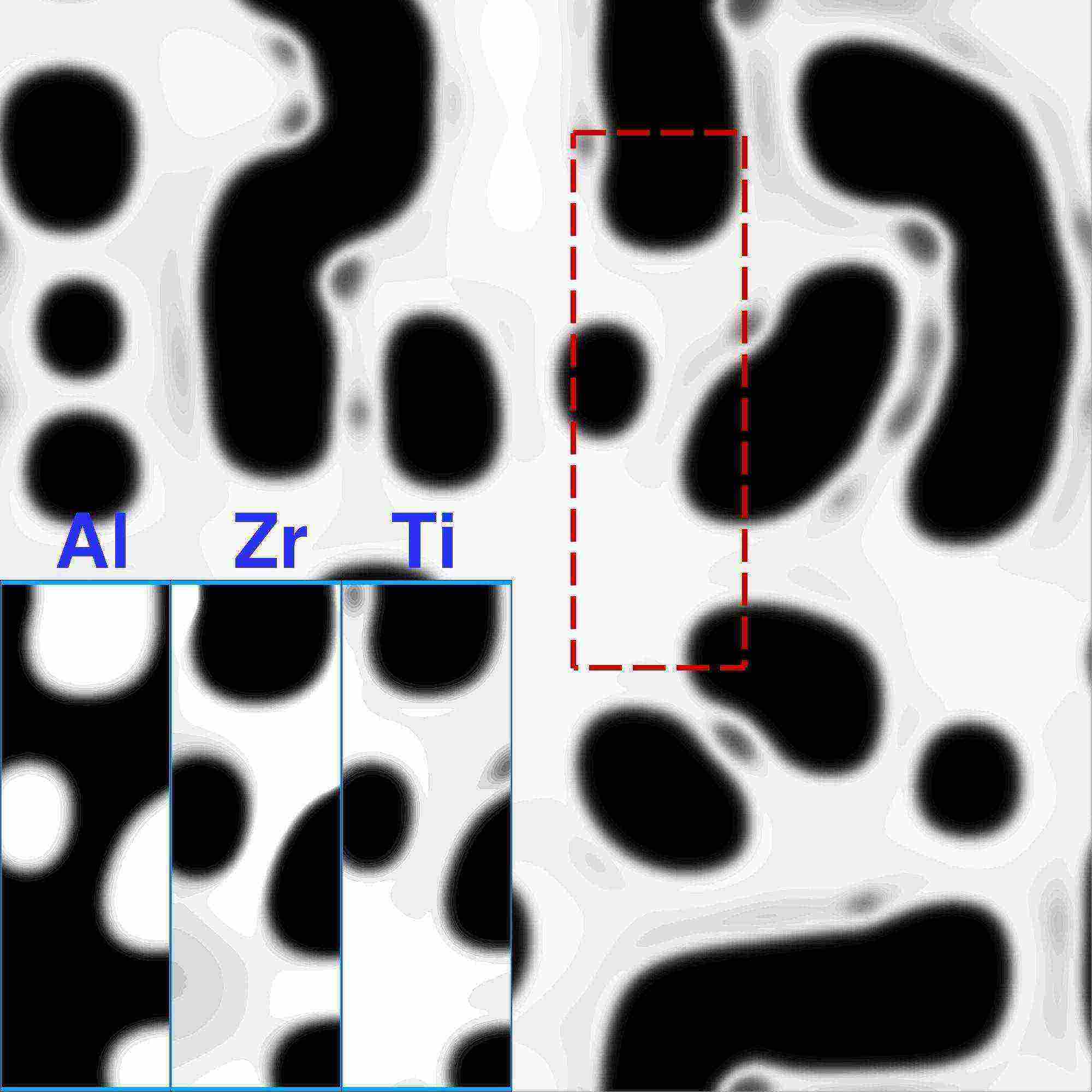}}{495}{9.887}{10} } 
        %% 0.45
        \subfloat[x\textsubscript{Al}=0.45]{\scalebarbackground{\includegraphics[scale=0.048]{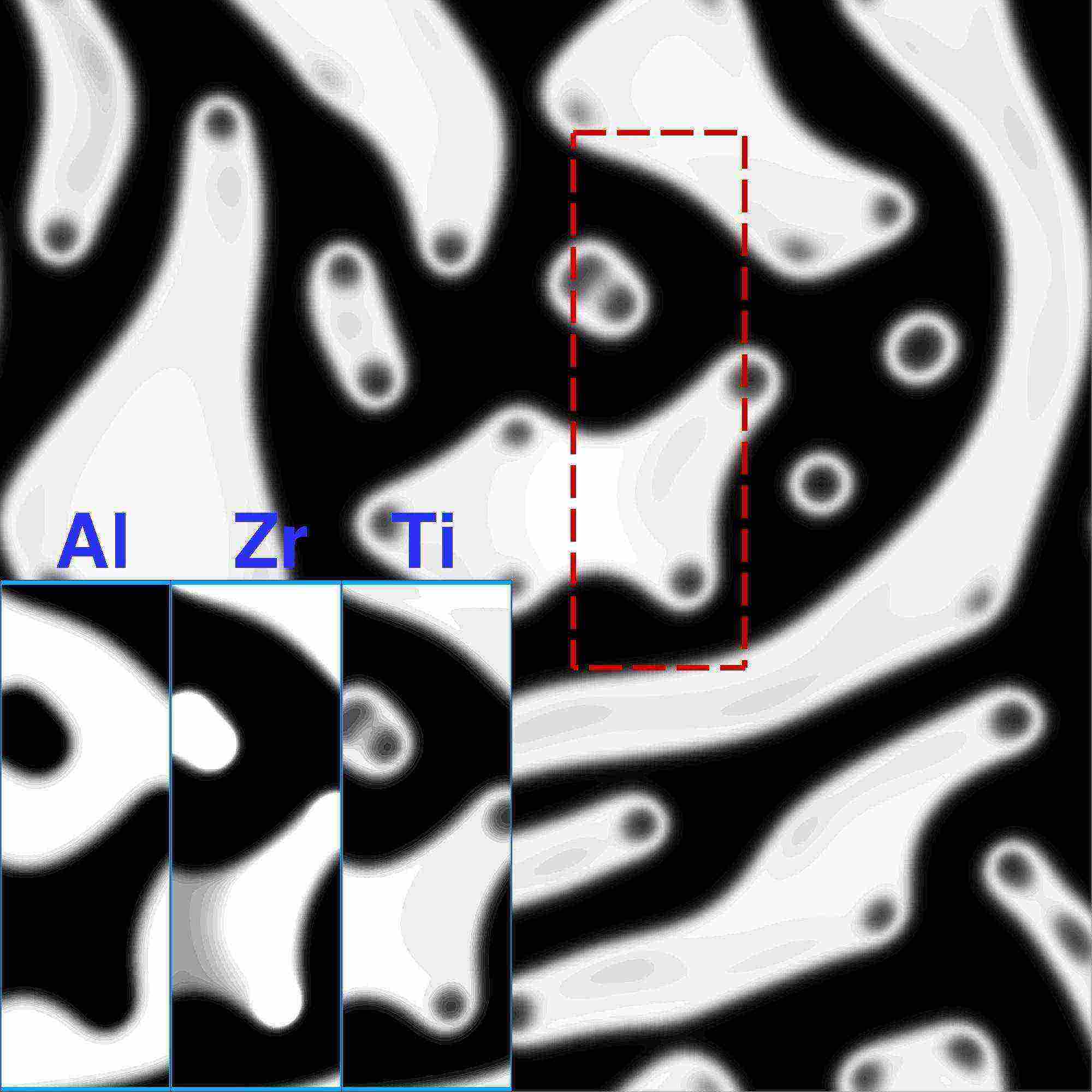}}{495}{9.887}{10} } \\ \vspace{-0.35cm} 
        %% 0.50
        \subfloat[x\textsubscript{Al}=0.50]{\scalebarbackground{\includegraphics[scale=0.048]{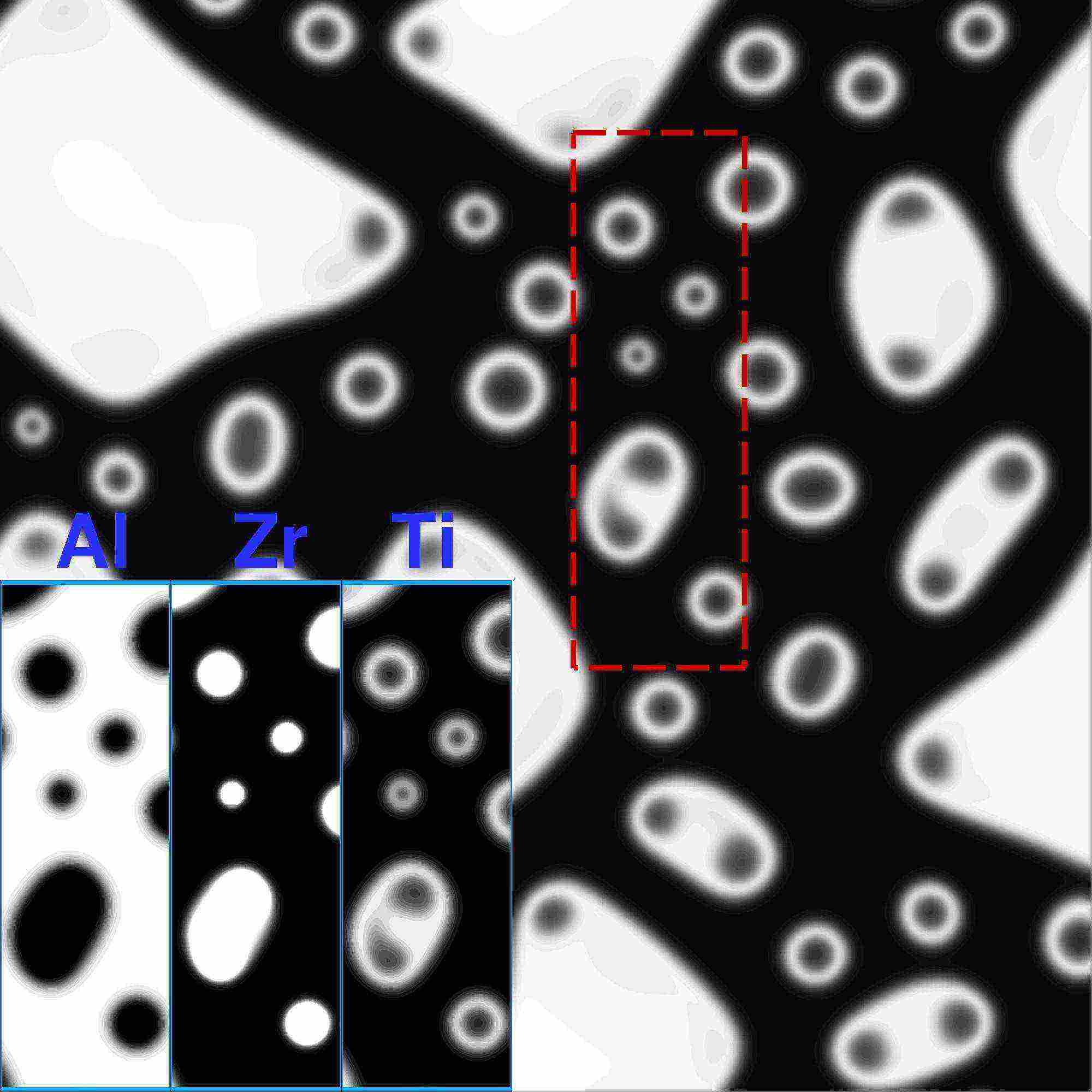}}{495}{9.887}{10} } 
        %% 0.55
        \subfloat[x\textsubscript{Al}=0.55]{\scalebarbackground{\includegraphics[scale=0.048]{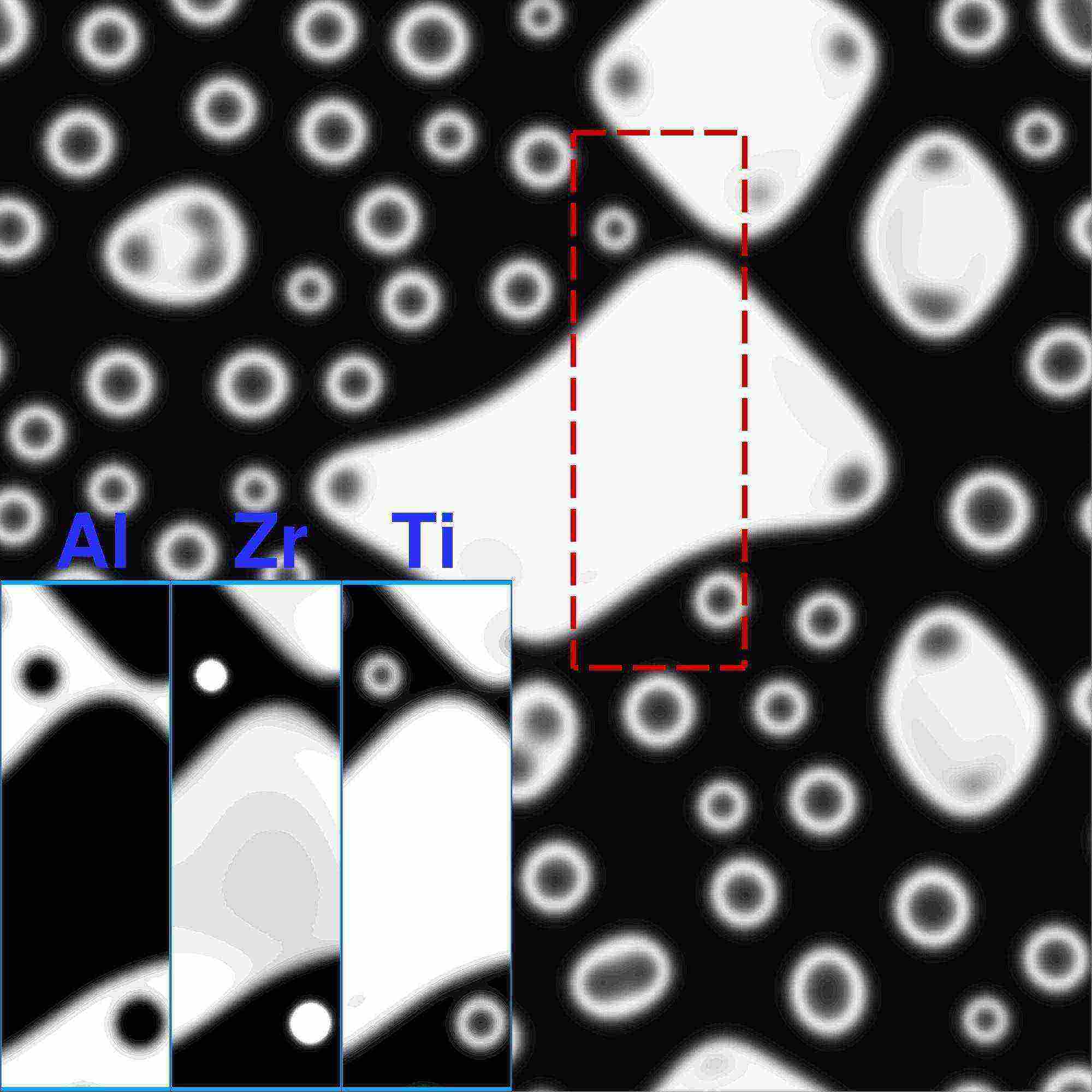}}{495}{9.887}{10} } 
        %% 0.60
        \subfloat[x\textsubscript{Al}=0.60]{\scalebarbackground{\includegraphics[scale=0.048]{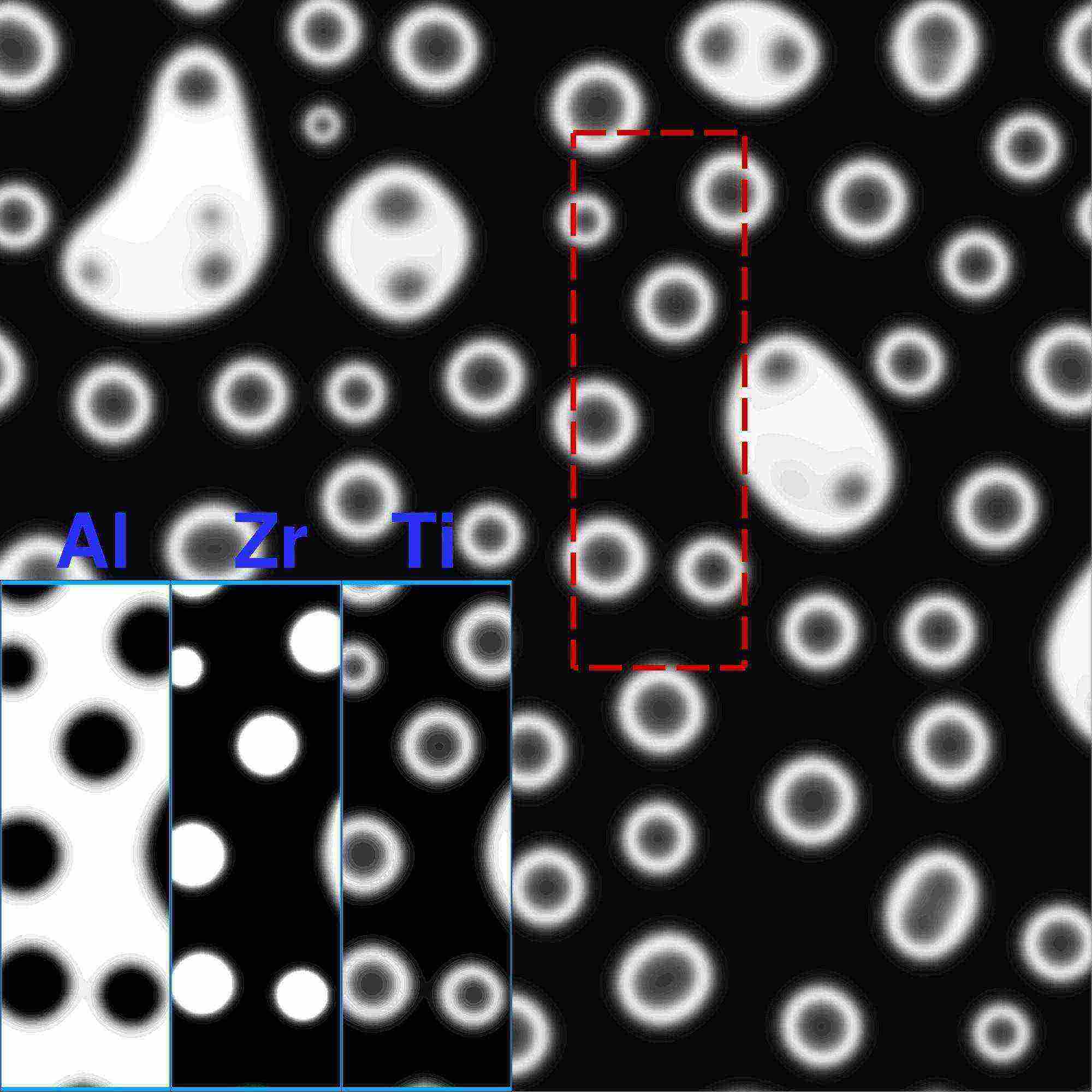}}{495}{9.887}{10} } 
        %% 0.65
        \subfloat[x\textsubscript{Al}=0.65]{\scalebarbackground{\includegraphics[scale=0.048]{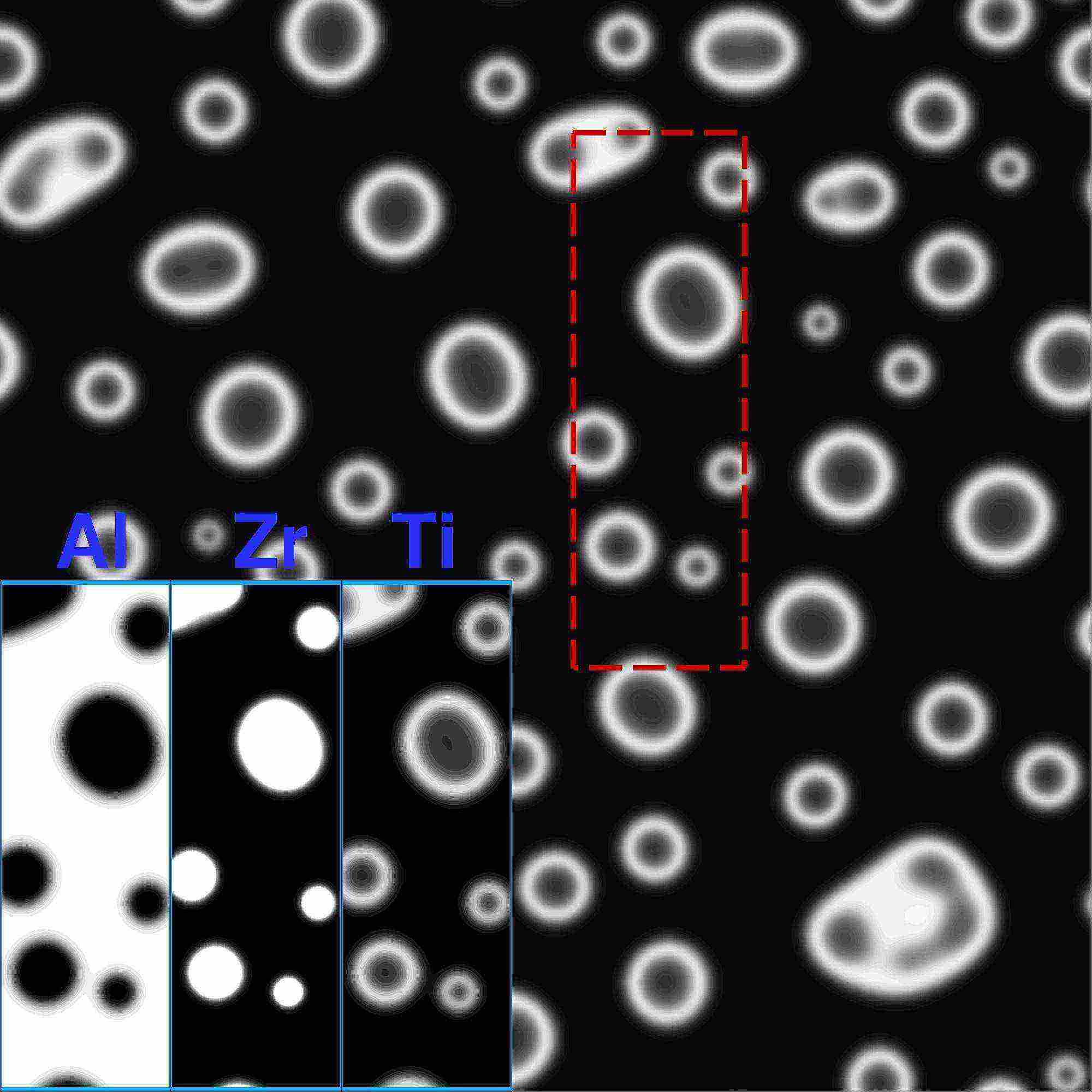}}{495}{9.887}{10} }         
        %% 0.70
        \subfloat[x\textsubscript{Al}=0.70]{\scalebarbackground{\includegraphics[scale=0.048]{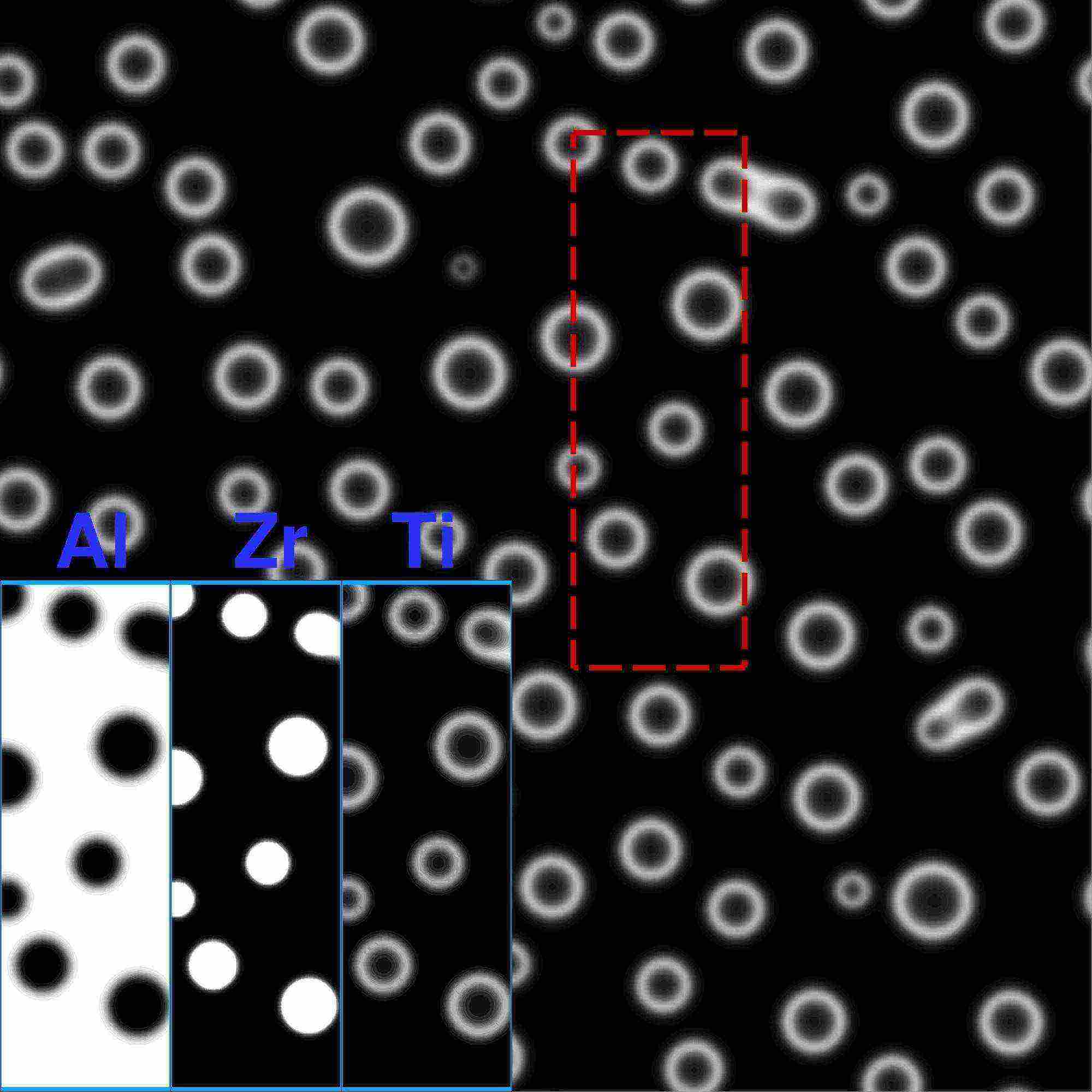}}{495}{9.887}{10} } 
        
        \caption{ Comparing the microstructure of Ti\textsubscript{1-x-0.05}Al\textsubscript{x}Zr\textsubscript{0.05}N alloy during annealing at T=1200$^\circ$C in elasto-chemical regime for $0.25\leq x_{Al}\leq0.70$ at the same frozen time. Color map in the overview image is \raisebox{1mm}{\fcolorbox{black!90!gray}{black}{\null}} (c)-AlN, \raisebox{1mm}{\fcolorbox{black}{white}{\null}} (c)-TiN, \raisebox{1mm}{\fcolorbox{black!90!gray}{gray}{\null}} (c)-ZrN. The bright regions in the elemental maps in the left corner of the overview images indicate the corresponding elemental map. }
        \label{fig:elasto-chem-sims}
    \end{figure*}
    %%%%%%%%%%%%%%%%%%
    %soft directions.
    
Figure~\ref{fig:elasto-chem-sims} illustrates the frozen two-dimensional microstructures for the same conditions investigated in the previous section, with the only difference being that elastic interactions are considered here. The most evident observation is that the precipitates turn to form only in round, cube-like, or worm-like morphologies, depending on the Al content of the alloy, and ZrN does not segregate in any form in Al-poor side of the miscibility gap, due to large SFTS between ZrN and AlN, and lower chemical driving force for ZrN separation. The initial stage of phase separation is similar to the previous section. However, the growth trend changes entirely as we observed the decomposition of AlN and TiN. This behaviour changes for the $x_{Al}\geq0.45$ range. 
    
For $x_{Al}=0.25$, cuboid AlN phases are observed instead of the round precipitates as illustrated in the chemical-only growth regime in Fig.~\ref{fig:chem-only-sims}a. Elemental maps indicate that Zr and Ti are dispersed in the matrix together, forming a (Ti,Zr)N mixture/solid solution. A similar microstructure is observed for the $0.30\leq x_{Al}\leq0.35$ range. The interconnected structure in the chemical-only growth regime turns into cuboid AlN phases in a (Ti,Zr)N matrix for this range of composition, too. The generally accepted reason for cube-like shapes of coherent secondary phases inside a matrix is the effect of anisotropic elasticity. For $x_{Al}=0.40$, the AlN particles elongate along the elastically soft direction in the (Ti,Zr)N matrix, and form worm-like domains instead of cuboids. The bi-countionus structure is disrupted entirely in the case of $x_{Al}=0.45$, and elongated structures happen to be present together with small circular particles during the coarsening stages as opposed to the case of chemical-only evolution. In this case, the elemental maps in the left corner of the microstructure reveal that the matrix phase only consists of the AlN phase. Small secondary particles mainly contain ZrN while TiN segregates along the grain boundaries of these phases, while the larger particles are (Ti,Zr)N mixture phases. 
    
In the Al-poor region in the composition space the precipitates appear always as AlN domains. On the other hand, in the Al-rich region AlN becomes the matrix phase and one can observe TiN, ZrN and/or (Ti,Zr)N precipitates, depending on the stage of the coarsening process. In this range of compositions, during the initial transient, the fine TiN phases appear with large amounts of ZrN present along the matrix/particle interfaces. Later, ZrN migrates into the interior region of the larger TiN particles and together form larger (Ti,Zr)N, leading to a rapid increase in the mean radius of larger particles that exist as a (Ti,Zr)N mixture. Smaller ZrN particles are also present in this stage where TiN is segregated in the grain boundaries. This is entirely different than what is observed in the chemical-only phase-field modeling in the previous section. The continuous increase of $x_{Al}$ shown in Fig.~\ref{fig:elasto-chem-sims}f to \ref{fig:elasto-chem-sims}j depicts a fortifying effect in the size of the particles due to higher cooperative elastic energy which assists elemental diffusion rate. Accordingly, the elastic energy tends to make the size of inclusions uniform. However, over the transitional competition between capillary and elastic interactions, the capillary interactions often become dominant. The effects of such rival/cooperative driving forces are more evident in Al rich side of the miscibility gap. Also, the particles retain their near spherical morphology in this composition range. 

In this work, we didn't examine the stabilizing effect of elastic interactions opposing the tendency of surface energy to coarsen inclusions at the expense of the small ones (inverse coarsening), and it is not clear whether the phenomenon originally coined by \cite{johnson1984elastic} occurs in this system at all. Preliminary observations of the present simulations, however, suggest that specific particle size distributions, and/or inter-particle distances, and coherency strain can favour locally the inverse coarsening in this multi-particle system after decomposition finishes for a short time during growth/coarsening stage. A follow up study will be provided considering the point that the notion of inverse coarsening \cite{su1996dynamics} has been put into question in two-phase solids by the work of Onuki and Nishimori \cite{onuki1991eshelby}.

    %%%%%%%%%%%%%%%%%%
    %%%%%%%%%%%%%%%%%%
    %%%%%%%%%%%%%%%%%% Particle Size Distribution
    %%%%%%%%%%%%%%%%%% Particle Size Distribution
    %%%%%%%%%%%%%%%%%% Particle Size Distribution
    %%%%%%%%%%%%%%%%%%
    %%%%%%%%%%%%%%%%%%    
    
    \subsection{Particle Size Distribution}
    
During chemical-only phase evolution, an interesting deviation from unimodal phase distribution is observed. This is evident in Fig.~\ref{fig:chem-only-sims}h where a bimodal distribution of particles in the microstructure is demonstrated. To be more precise, TiN precipitates with a relatively large size, irregular shape and low number density; and secondary ZrN particles with a small size, spherical shape and high number density are clearly seen during isothermal evolution in the Ti$_{0.30}$Al$_{0.65}$Zr$_{0.05}$N composition. The large TiN particles are evenly distributed in the AlN matrix with the secondary ZrN precipitates separating them. The bimodal growth behaviour reported in this study is similar to the observed phenomena in Ni-based super alloys \cite{coakley2010coarsening,li2018influence,singh2011influence}. Similar to the experimental observations in \cite{coakley2010coarsening} that suggest there is a transition point in unimodal to bimodal growth, the phase-field studies show that the multimodal growth in this nitride coating starts after full decomposition of TiN and AlN particles which happens during the early stages. Accordingly, the multimodal distribution is only observed after the early stages of decomposition, although it continues throughout the coarsening stages. 
    
\begin{figure}[h!]
        \centering
        %{\scalebarbackground{\includegraphics[width=0.47\columnwidth]{images/case5_elastic_2/055_005_00755.jpeg}}{495}{9.887}{10} }
        %{\scalebarbackground{\includegraphics[width=0.47\columnwidth]{images/multimodal_mics/phi1_00,200,000.jpeg}}{495}{9.887}{10} }
        %\includegraphics[width=0.47\columnwidth]{images/multimodal_mics/NiAl_gamma_prime.jpg}
        %\includegraphics[width=0.4\columnwidth]{images/case5_elastic_2/055_005_00755.jpeg}
%        {\scalebarbackground{\includegraphics[width=0.40\columnwidth]{images/multimodal_mics/phi1_00,200,000.jpeg}}{495}{9.887}{10} }
        \subfloat[]{\scalebarbackground{\includegraphics[width=0.309\columnwidth]{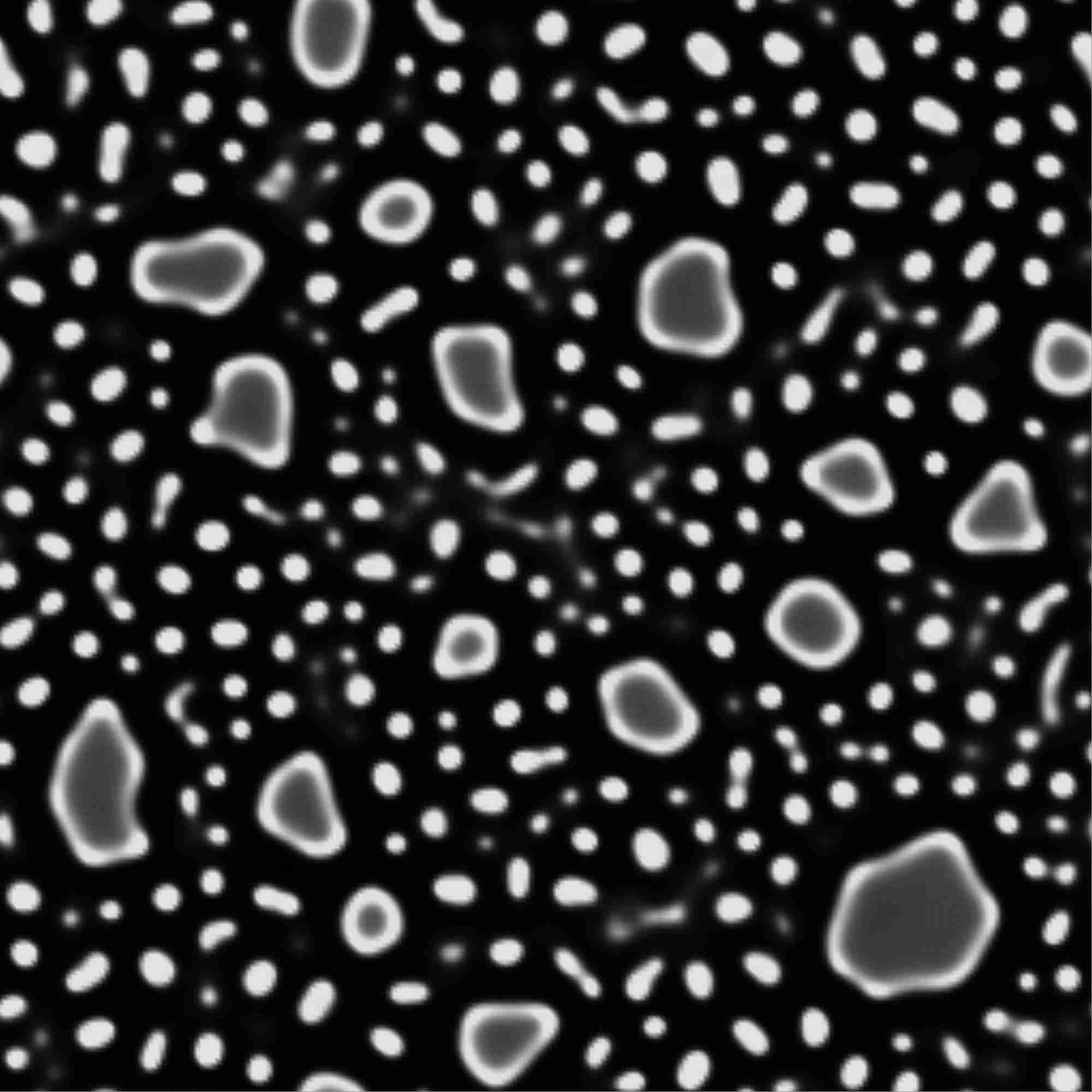}}{495}{9.887}{10} } \vspace{-0.18cm}
        \subfloat[]{\includegraphics[width=0.31\columnwidth]{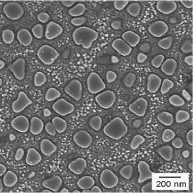}} \vspace{-0.18cm} \\
        \subfloat[]{\scalebarbackground{\includegraphics[width=0.309\columnwidth]{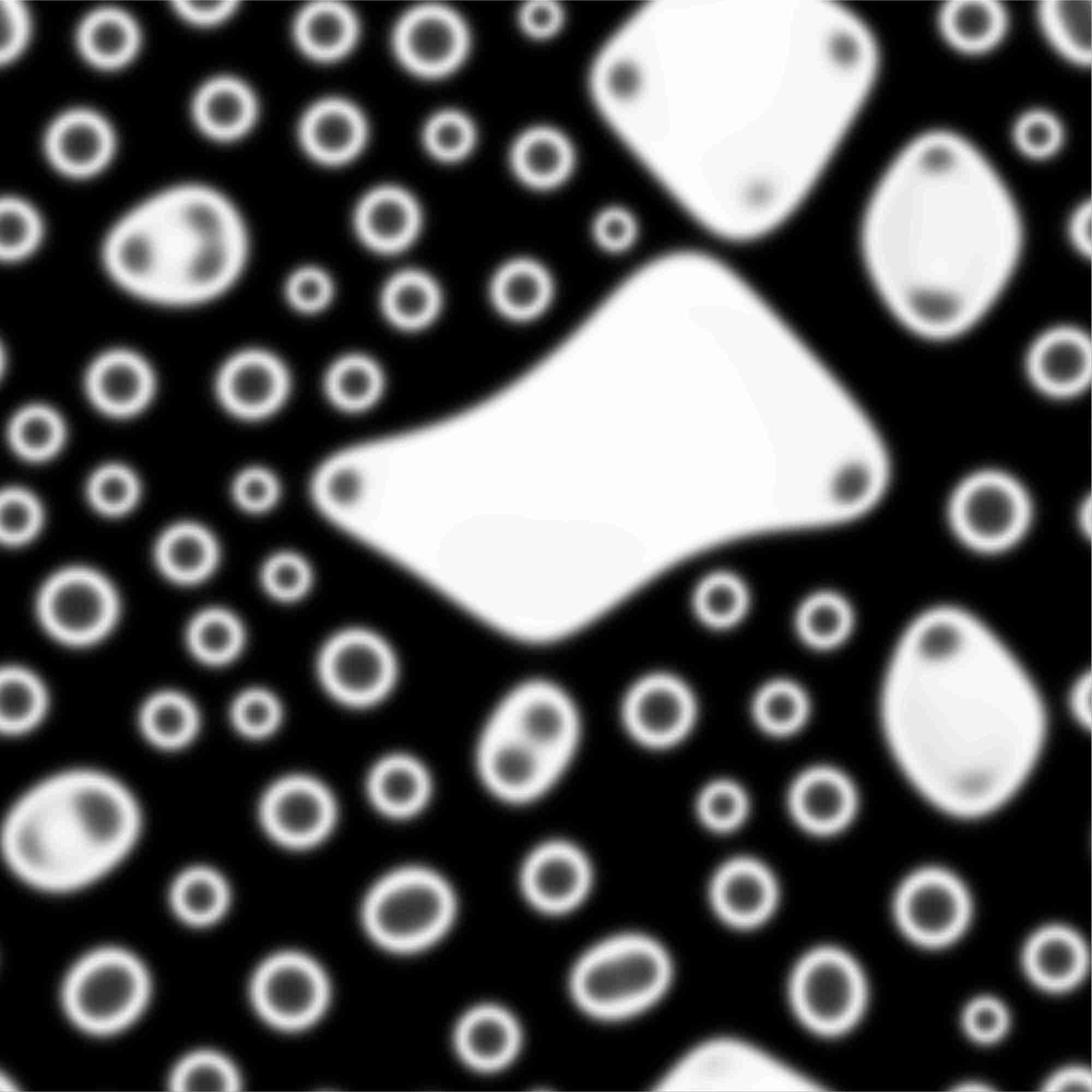}}{495}{9.887}{10} }\vspace{-0.18cm}
        \subfloat[]{\includegraphics[width=0.31\columnwidth]{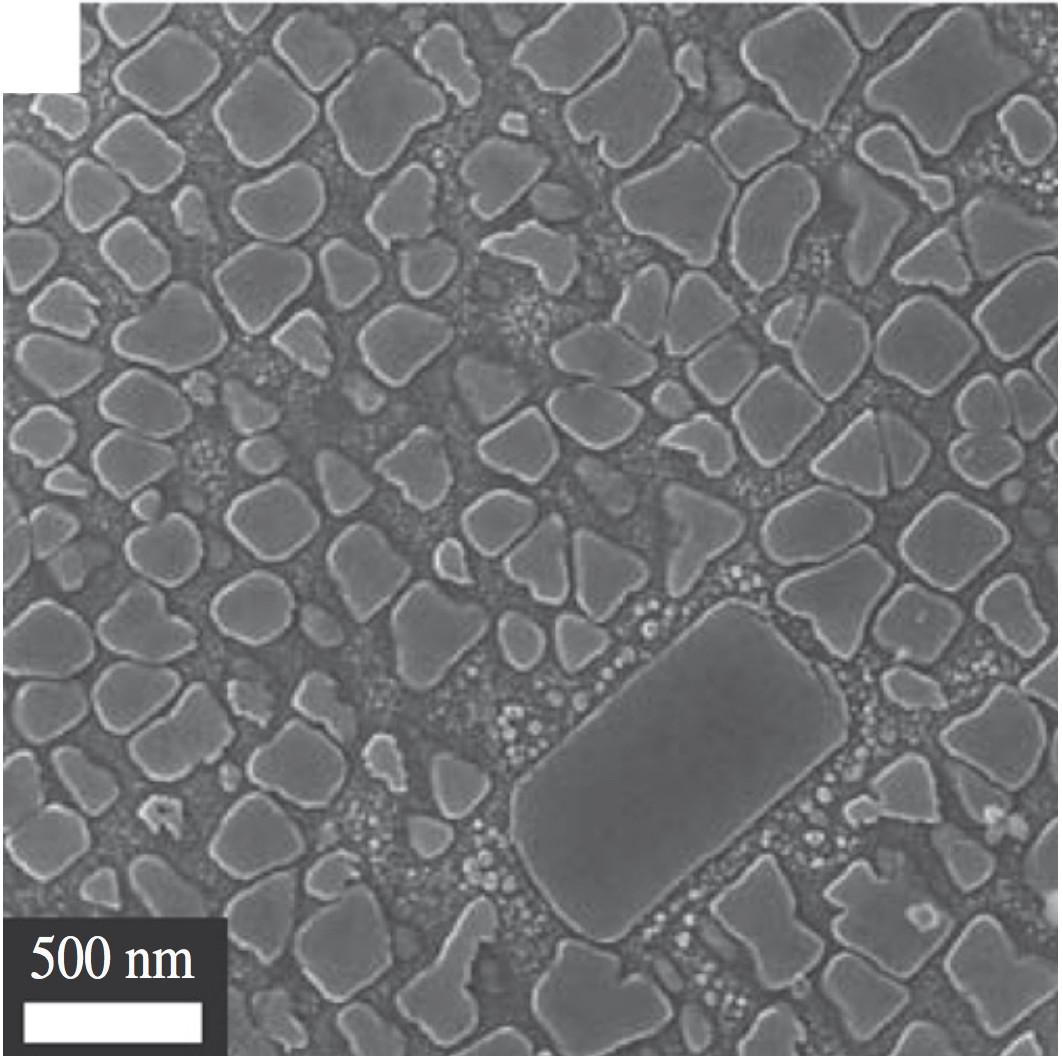}} \vspace{-0.18cm} 
        \caption{ a,c) Phase-field microstructures, b,d) actual experimental microstructures. (a and b) Qualitative comparison of bimodal phase distribution in a) Ti$_{0.30}$Al$_{0.65}$Zr$_{0.05}$N system showing large TiN particles \raisebox{1mm}{\fcolorbox{black}{gray}{\null}} in AlN matrix \raisebox{1mm}{\fcolorbox{black}{black!90!gray}{\null}}, surrounded by small secondary ZrN phases \raisebox{1mm}{\fcolorbox{black}{white}{\null}}, similar to b) $L1_2$ precipitates ($\gamma^{'}$) in FGH96 polycrystalline Ni superalloy ({reproduced from \cite{li2018influence}, with permission from Elsevier}). (c and d) Qualitative comparison of trimodal phase distribution in c) Ti$_{0.40}$Al$_{0.55}$Zr$_{0.05}$N system showing large TiN particles \raisebox{1mm}{\fcolorbox{black}{white}{\null}} in AlN matrix \raisebox{1mm}{\fcolorbox{black}{black!90!gray}{\null}}, surrounded by small secondary ZrN phases \raisebox{1mm}{\fcolorbox{black}{gray}{\null}}, similar to d)  $L1_2$ precipitates ($\gamma^{'}$) in a Ni115 alloy ({reproduced from \cite{coakley2010coarsening}, with permission from Elsevier}).}
        \label{fig:compare_bimodal_trimodal}
    \end{figure}

Figure~\ref{fig:compare_bimodal_trimodal}a, and b compares an example of such bimodal microstructure in the Zr poor (y=0.05), Al rich corner (x=0.65) of Ti$_{1-x-y}$Al$_{x}$Zr$_{y}$N system with an experimental micrograph (of a completely different chemical system). The obtained phase-field micrograph with bimodal particle distribution qualitatively matches with the bimodal distribution of $L1_2$ ($\gamma^{'}$) precipitates in FGH96 polycrystalline Ni-based super alloys used in turbine applications. Furthermore, a trimodal distribution of particles is observed during elasto-chemical evolution (Refer to Fig.~\ref{fig:elasto-chem-sims}) in the Al-rich Ti$_{1-x-0.05}$Al$_{x}$Zr$_{0.05}$N composition. The microstructure in Fig.~\ref{fig:elasto-chem-sims}g illustrates a single large TiN particle where its size is 251.9 $nm^2$, secondary TiN particles with average size of $\sim$25.49 $nm^2$, and fine tertiary ZrN particles with average particle size of $\sim$4.37 $nm^2$. This trimodal phase distribution can be compared with experimental trimodal microstructure of Ni115 alloy shown in Fig.~\ref{fig:compare_bimodal_trimodal}d. Trimodal structures have potential to achieve combinations of physical and mechanical properties that are unattainable with the individual phases, such as strength, ductility, and high-strain-rate deformation.
    
    %%%%%%%%%%%%%%%%%%  

    %%%%%%%%%%%%%%%%%%        
%    \vspace{-0.20cm}
    %%%%%%%%%%%%%%%%%%  
    %\begin{figure}
    %    \centering
    %    \caption{a) Phase-field microstructure, b) actual experimental microstructure. Qualitative comparison of trimodal phase distribution in a) Ti$_{0.40}$Al$_{0.55}$Zr$_{0.05}$N system showing large TiN particles \raisebox{1mm}{\fcolorbox{black}{gray}{\null}} in AlN matrix \raisebox{1mm}{\fcolorbox{black}{black!90!gray}{\null}}, surrounded by small secondary ZrN phases \raisebox{1mm}{\fcolorbox{black}{white}{\null}}, similar to b)  $L1_2$ precipitates ($\gamma^{'}$) in a Ni superalloy sample ({reproduced from \cite{li2018influence}, with permission from Elsevier}).}
    %    \label{fig:compare_trimodal}
    %\end{figure}    
    %%%%%%%%%%%%%%%%%% 
    
    %%%%%%%%%%%%%%%%%%
    %%%%%%%%%%%%%%%%%%
    %%%%%%%%%%%%%%%%%% Origin of wurtzite - rocksalt phase transition
    %%%%%%%%%%%%%%%%%% Origin of wurtzite - rocksalt phase transition
    %%%%%%%%%%%%%%%%%% Origin of wurtzite - rocksalt phase transition
    %%%%%%%%%%%%%%%%%%
    %%%%%%%%%%%%%%%%%%
     
    \subsection{Origin of Wurtzite-Rocksalt Phase Transition in AlN} \label{sec:Wurtzite-Rocksalt_onset}
    
The transition of c-AlN (B1 structure) into w-AlN (B4 structure) is detrimental to hardness in TM-Al-N coatings, and is associated with different phenomena such as the transition of coherent interfaces to incoherent ones. The transformation peak is located at 1,280$^\circ$C for Ti\textsubscript{0.34}Al\textsubscript{0.37}Zr\textsubscript{0.29}N, which is higher than that of the Ti\textsubscript{0.48}Al\textsubscript{0.52}N by 70$^\circ$ C. The recent study by Saha \etal~\cite{saha2016understanding} enlightened the hypothesis that the transformation is triggered by defects at $c-\{0\bar{1}1\}$ growth fronts that offer a nearly invariant plane with respect to the parallel $w-\{2\bar{1}\bar{1}0\}$ planes. The incoherent interfaces pave the road for strain localization, in the form of the lattice misfit between wurtzite and rock salt phases. 

    Even though the mechanism by which this transformation takes place is not precisely known, the MEP for it might pass from tetragonal phase (I4mm) ($O_h \rightarrow D_{4h}$) with five-fold coordinated atoms or a hexagonal structure ($P6_3/mmc$). These transition sequences are illustrated in Fig.~\ref{fig:stability_sequence}b, c, and d. Despite GaN \cite{yao2013b,qian2013variable} and lnN \cite{duan2016different} cases which tetragonal route is preferred, \emph{ab-initio} calculations suggest that the hexagonal transition route due to shear deformation is more favorable in the case of AlN \cite{zhang2009deformation,duan2016different}. Here, we look at this transition as a mechano-chemical phase transition \cite{rudraraju2016mechanochemical}, where the high symmetry phase decomposes martensitically into another phase with different structural variants. This is a typical solid-state displacive (diffusionless) phase transition (martensitic transformation) that breaks crystal symmetry through the development of spontaneous anisotropic lattice strain \cite{falk1990three}. Hence, we assumed that the AlN structural phase transition in the nitrite coatings is a deformation-induced transformation and is driven by the elastic instability (($C_{11}-C_{22})/2\rightarrow0$) or by lattice mode softening.

%Even though the mechanism by which this transformation takes place is not precisely known, the minimal energy pathway (MEP) for it is accompanied with a transition from high symmetry cubic phase (B1) to intermediate lower symmetry tetragonal phase (I4mm) ($O_h \rightarrow D_{4h}$) with five-fold coordinated atoms, and then the B4 structure. This transition sequence is illustrated in Fig.~\ref{fig:stability_sequence}b, c, and d. Here, we look at this transition as a mechano-chemical phase transition \cite{rudraraju2016mechanochemical}, where the high symmetry phase decomposes martensitically into another phase with different structural variants. This is a typical solid-state displacive (diffusionless) phase transition (martensitic transformation) that breaks crystal symmetry through the development of spontaneous anisotropic lattice strain \cite{falk1990three}. Hence, we assumed that the AlN structural phase transition in the nitrite coatings is a deformation-induced transformation and is driven by the elastic instability (($C_{11}-C_{22})/2\rightarrow0$) or by lattice mode softening.   

    %%%%%%%%%%%%%%%%%%
    %%%%%%%%%%%%%%%%%%  
    \subsubsection{Adoption of an Energy Model}
    %et al. refers the mechanochemical decomposition is a phenomenon that is likely present in many multi-component materials but has to date been overlooked as a mechanism by which a high symmetry phase can decompose martensitically sort of cubic to tetragonal transiotion.
    
    %Pre-martensitic phenomena, also called martensite precursor effects, are long standing critical issues of martensitic phase transformation that have not been fully understood. 
To remedy the onset of transformation from the thermodynamics standpoint, we consider an elastic continuum, representing a cubic prototype phase which may deform hyperelastically into other structural forms (e.g. square to rectangular lattice (SR), rectangle to oblique (RO), and/or square to oblique (SO) transition, etc.). A thermodynamic driving force exists for segregation by strain to induce these structural transformations in the strain space to any of the tetragonal variants, oblique structures, and/or successive transitions. This potential can also be coupled with composition space to cover the phase transitions in this alloy. Following Refs.~\cite{rudraraju2016mechanochemical,steinbach2011phase,falk1990three,rasmussen2001three,nittono1982phenomenological,barsch1984twin,saxena1993pretransformation}, we express a thermodynamic potential ($\Phi$), in terms of symmetry adopted order parameters ($e_i$) appropriate for cubic symmetry for a structural transition:
    
    %%%%%%%%%%%%%%%%%%%%%%%%%%%
    \begin{equation} \label{eq:phi}
        %\Phi = Ae_2^2 + B(-3e_2^2) + C(e_2^4)
        %\Phi = A(e_2^2+e_3^2) + Be_3(e_3^2-3e_2^2) + C(e_2^2+e_3^2) 
        \Phi = A(e_2^2+e_3^2) + Be_3(e_3^2-3e_2^2) + C(e_1-E(e_2^2+e_3^2))^2 + D(e_2^2+e_3^2)^2 + E(e_4^2 + e_5^2 + e_6^2) 
    \end{equation}    
    %%%%%%%%%%%%%%%%%%%%%%%%%%%
    
    \noindent
    where coefficients A, B, C, D, and E are generally function of temperature and pressure. For a fourth order free energy expansion 21 material parameters are required to interpret an energy function in the strain space \cite{falk1990three}, and this number can be reduced by results motivated with experiments. The energy landscape for the phase transitions in this alloy system including the structural transitions will be addressed in the future study. Here, we investigate the order parameters defined in Table~ \ref{tab:strain_order_pars} in terms of the linear elastic strain, $\varepsilon_{ij}$ obtained from Eq. \ref{eqn:strain_tensor}: 
      
    \begin{table}[h!]
        \centering
        %\scriptsize
        \caption{Strain order parameters}\label{tab:strain_order_pars}
        \begin{tabular}{lcc} \toprule
            3D & 2D & Definition \\ \midrule
            $e_1 = (\varepsilon_{11} + \varepsilon_{22} + \varepsilon_{33})/ \sqrt{3}$ &  $e_1 = (\varepsilon_{11} + \varepsilon_{22})/ \sqrt{2}$                   & \multirow{1}{*}{dillation strain}  \\ 
            $e_2 = (\varepsilon_{11} - \varepsilon_{22})/\sqrt{2}$                  &  \multirow{2}{*}{$e_2 = (\varepsilon_{11} - \varepsilon_{22})/\sqrt{2}$}   & \multirow{2}{*}{deviatoric strain} \\
            $e_3 = (\varepsilon_{11} + \varepsilon_{22} - 2\varepsilon_{33})/\sqrt{6}$ &                                                     &                                    \\
            $e_4 = 2\varepsilon_{23}$                                             &   \multirow{3}{*}{$e_6 = 2\varepsilon_{12}$}             & \multirow{3}{*}{shear strain}      \\
            $e_5 = 2\varepsilon_{13}$                                             &                                                          &                                    \\
            $e_6 = 2\varepsilon_{12}$                                             &                                                          &                                    \\ \bottomrule
        \end{tabular}
    \end{table}      

The measures $e_2$ and $e_3$ are especially suited as order parameters to describe cubic to tetragonal distortions. In a 2D structure, $e_1$, $e_2$, and $e_6$ reduce to the dilatation, deviatoric, and shear strain, respectively, where $e_2$ uniquely maps the square lattice into the two rectangular variants (SR). The total free energy functional for such a phase transition has four minima in $\{e_2$,$e_3\}$ space, one at the origin (zero strain) corresponding to the cubic phase, and three at other strain positions representing the three variants of tetragonal distortions. This energy model specifically gains importance when the strain energy is an important part of the total energy for a phase transition. This model is also useful in coupling the composition and strain space $\left(F^{tot}(c_i,e_i) = f^{chem}(c_i) + \Phi(e_i)\right)$, and understanding the structural transitions such as the ones that happen in many martensitic phenomena. It will be further developed to study the evolution of weakly to moderately first-order transitions using phase-field modeling.

    \subsubsection{The Tendency to Transform Based on the Strain Order Parameters}
    
Phase transformation is not only influenced by chemical driving forces, though surface energy differences (Gibbs-Thomson effect), crystallite size and subsequent elastic strains can favor the formation of a particular phases in metastable systems. In this section, we first study the value, and distribution of the obtained strain order parameters, and then re-investigate the Zener anisotropy ($A_z$) in the alloy.
    
\begin{figure*}[h!]
        \centering
        %% 0.25
        \subfloat[x\textsubscript{Al}=0.25]{\includegraphics[scale=0.060]{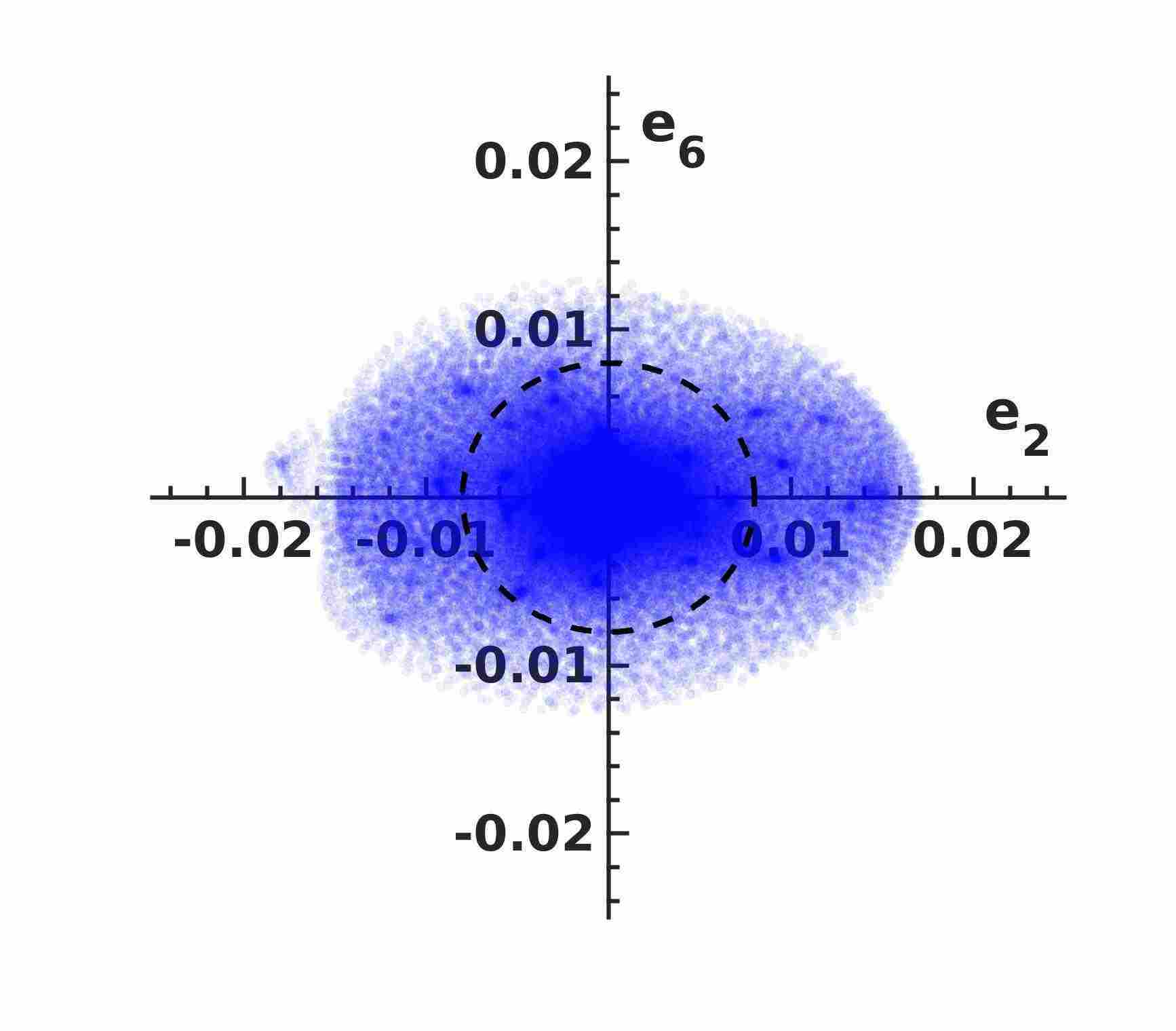}} 
        %% 0.30
        \subfloat[x\textsubscript{Al}=0.30]{\includegraphics[scale=0.060]{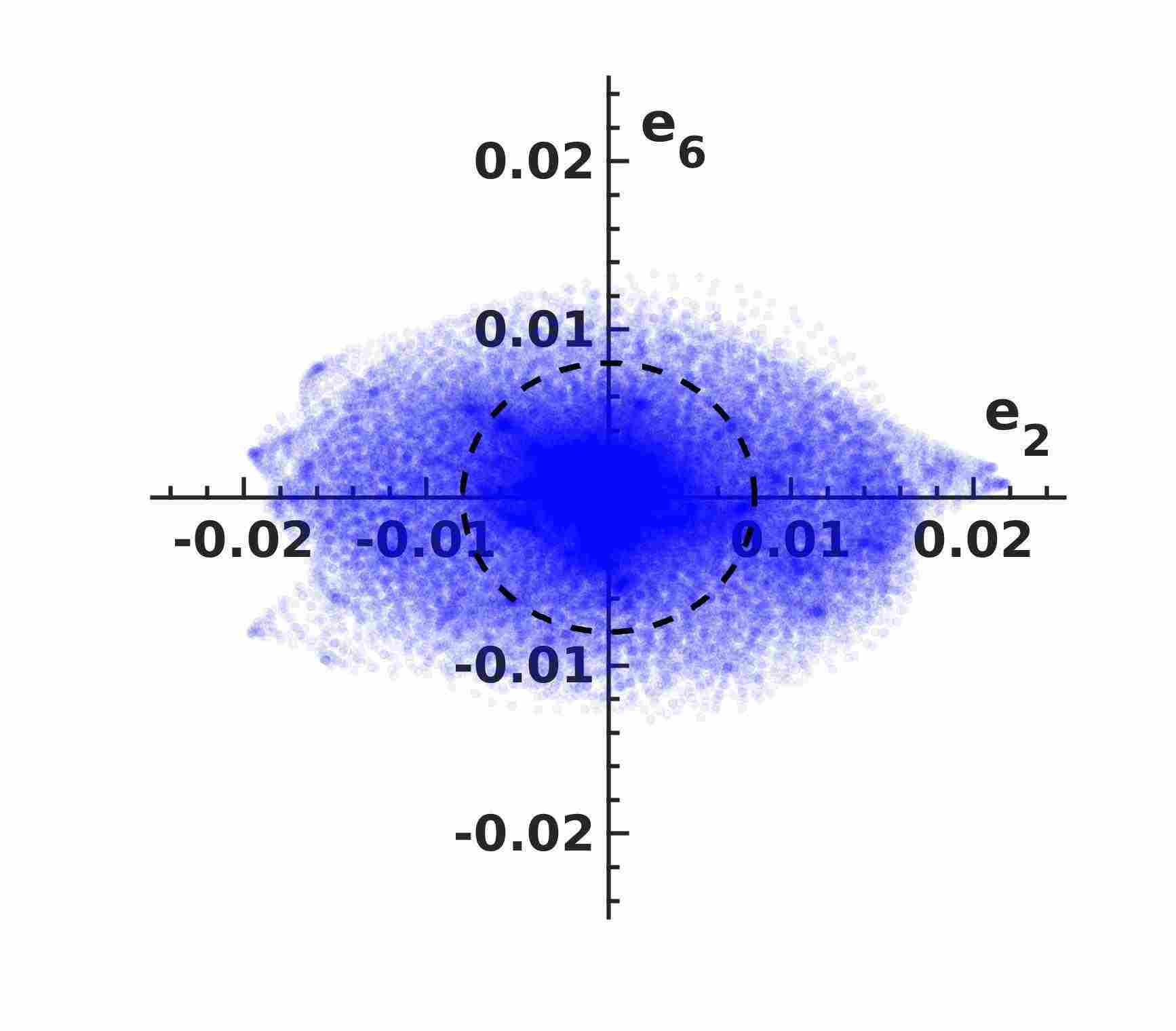}} 
        %% 0.35
        \subfloat[x\textsubscript{Al}=0.35]{\includegraphics[scale=0.060]{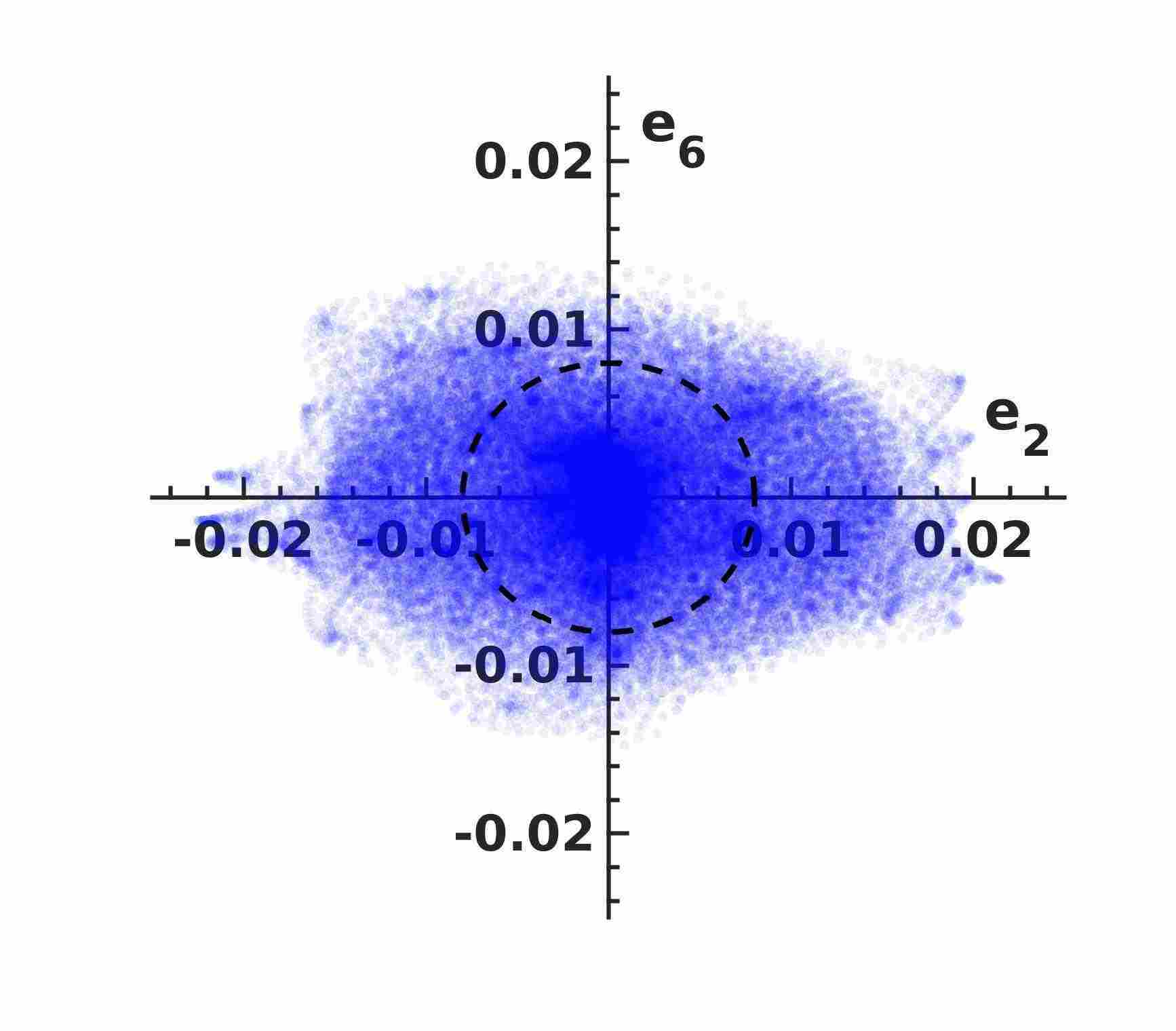}} 
        %% 0.40
        \subfloat[x\textsubscript{Al}=0.40]{\includegraphics[scale=0.060]{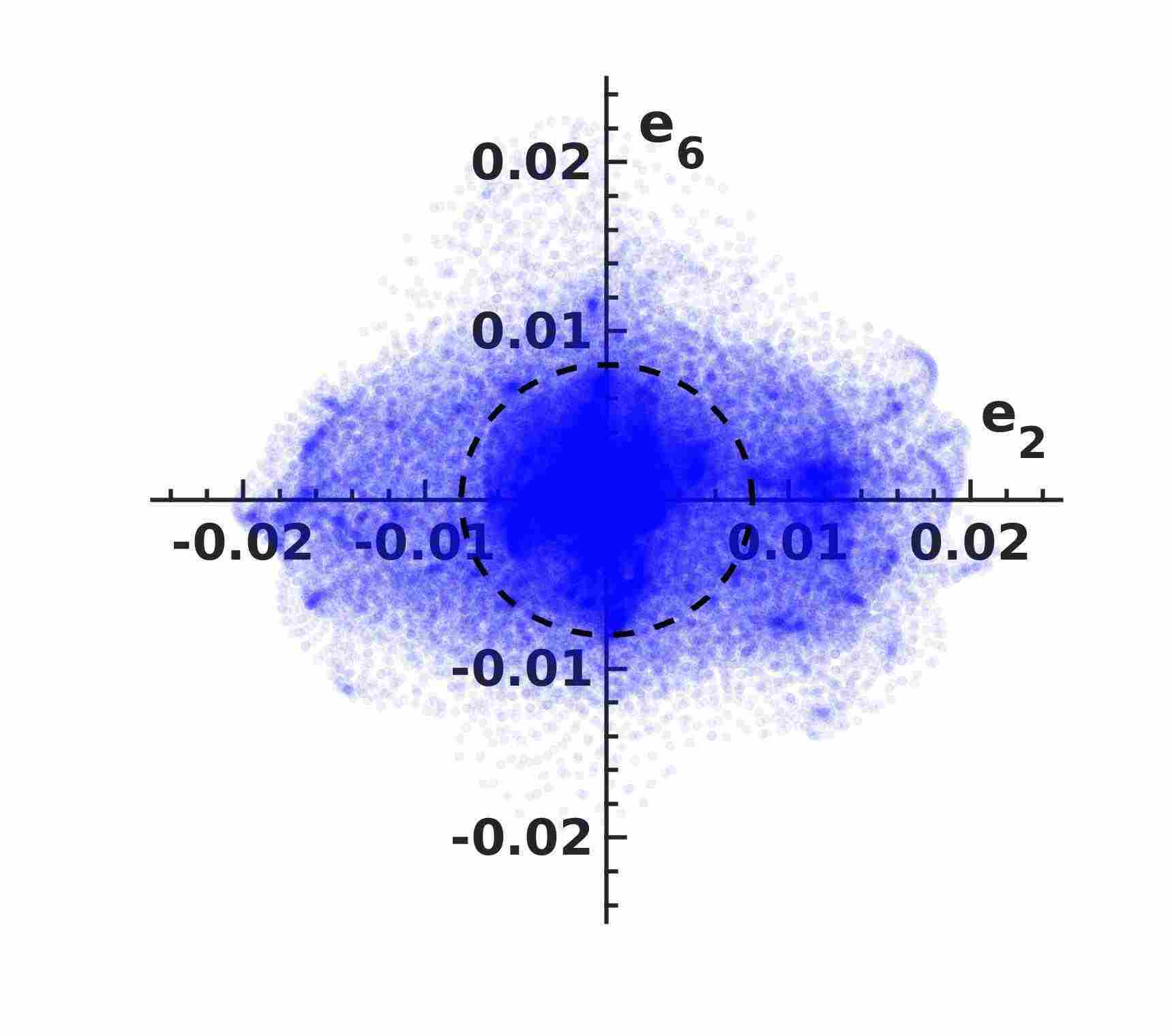}}
        %% 0.45
        \subfloat[x\textsubscript{Al}=0.45]{\includegraphics[scale=0.060]{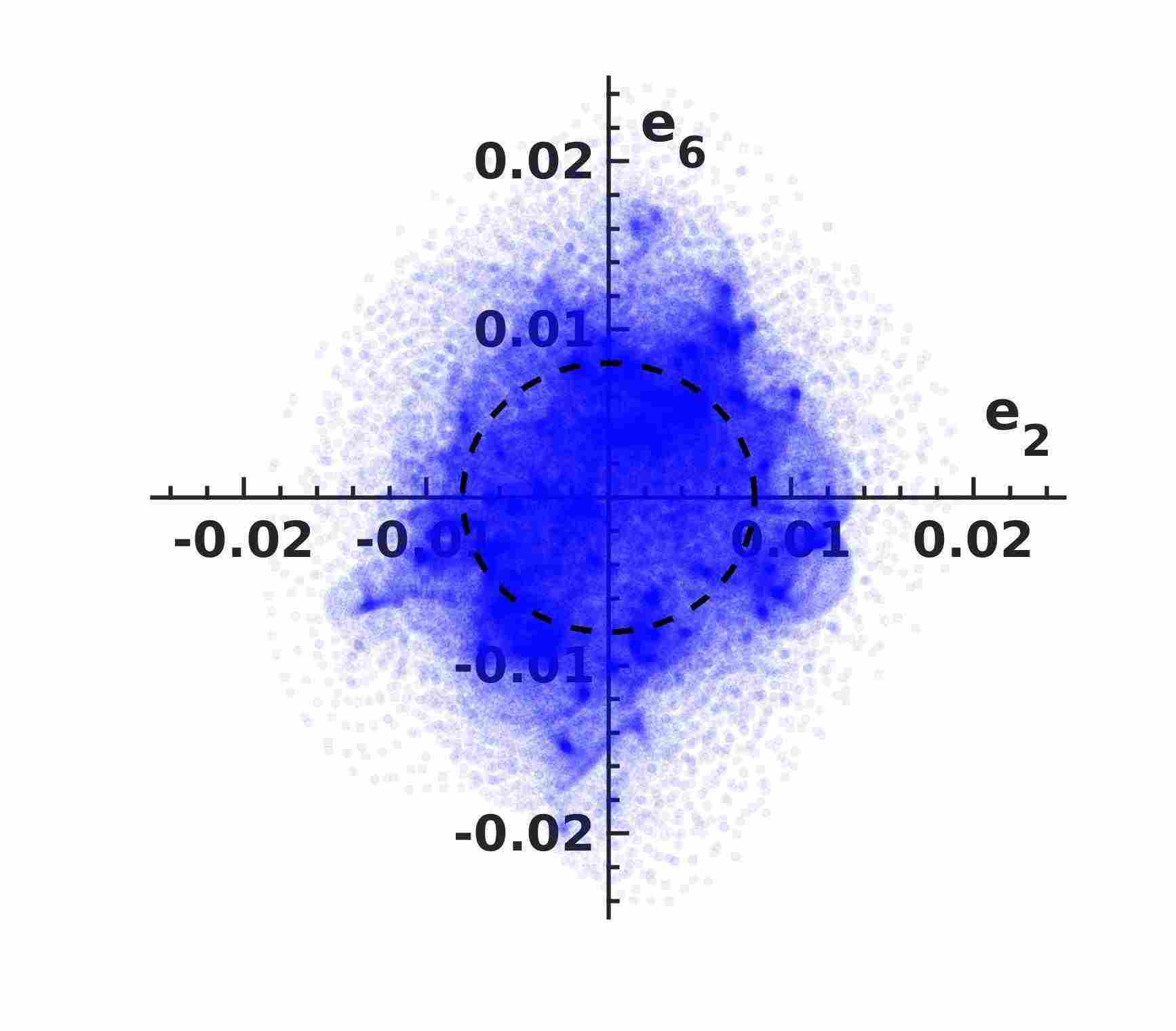}}\\
        %% 0.50
        \subfloat[x\textsubscript{Al}=0.50]{\includegraphics[scale=0.060]{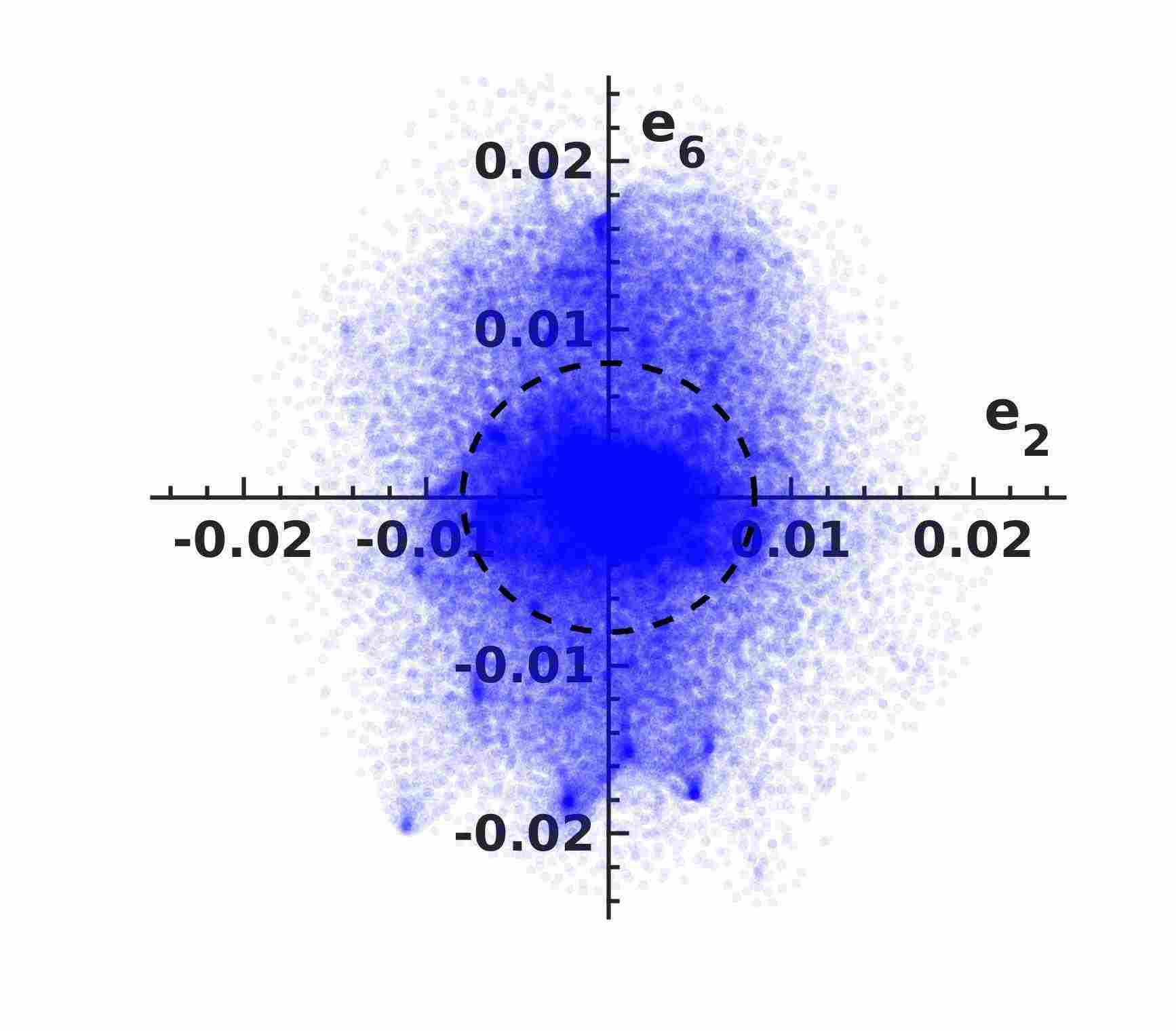}}
        %% 0.55
        \subfloat[x\textsubscript{Al}=0.55]{\includegraphics[scale=0.060]{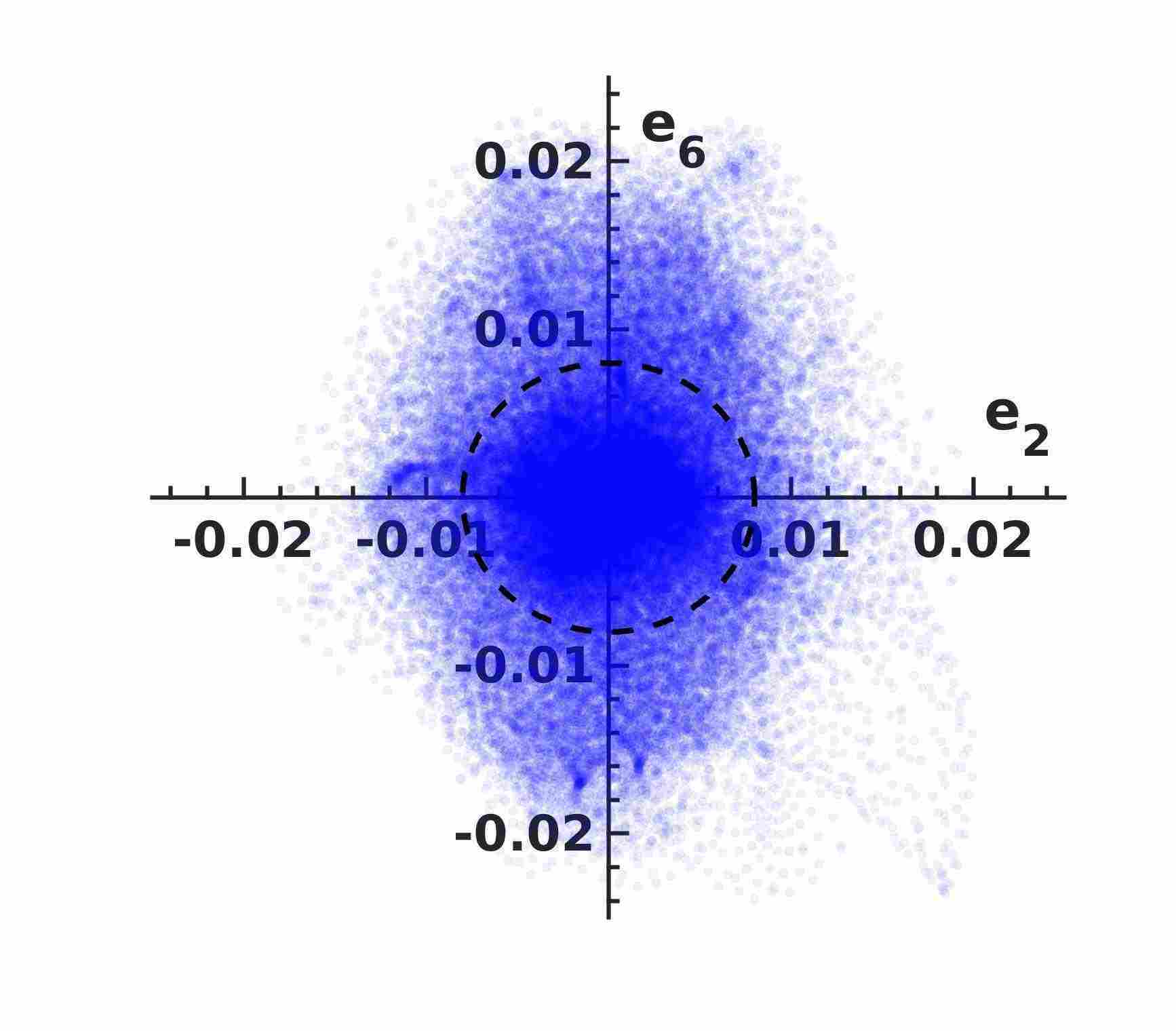}}
        %% 0.60
        \subfloat[x\textsubscript{Al}=0.60]{\includegraphics[scale=0.060]{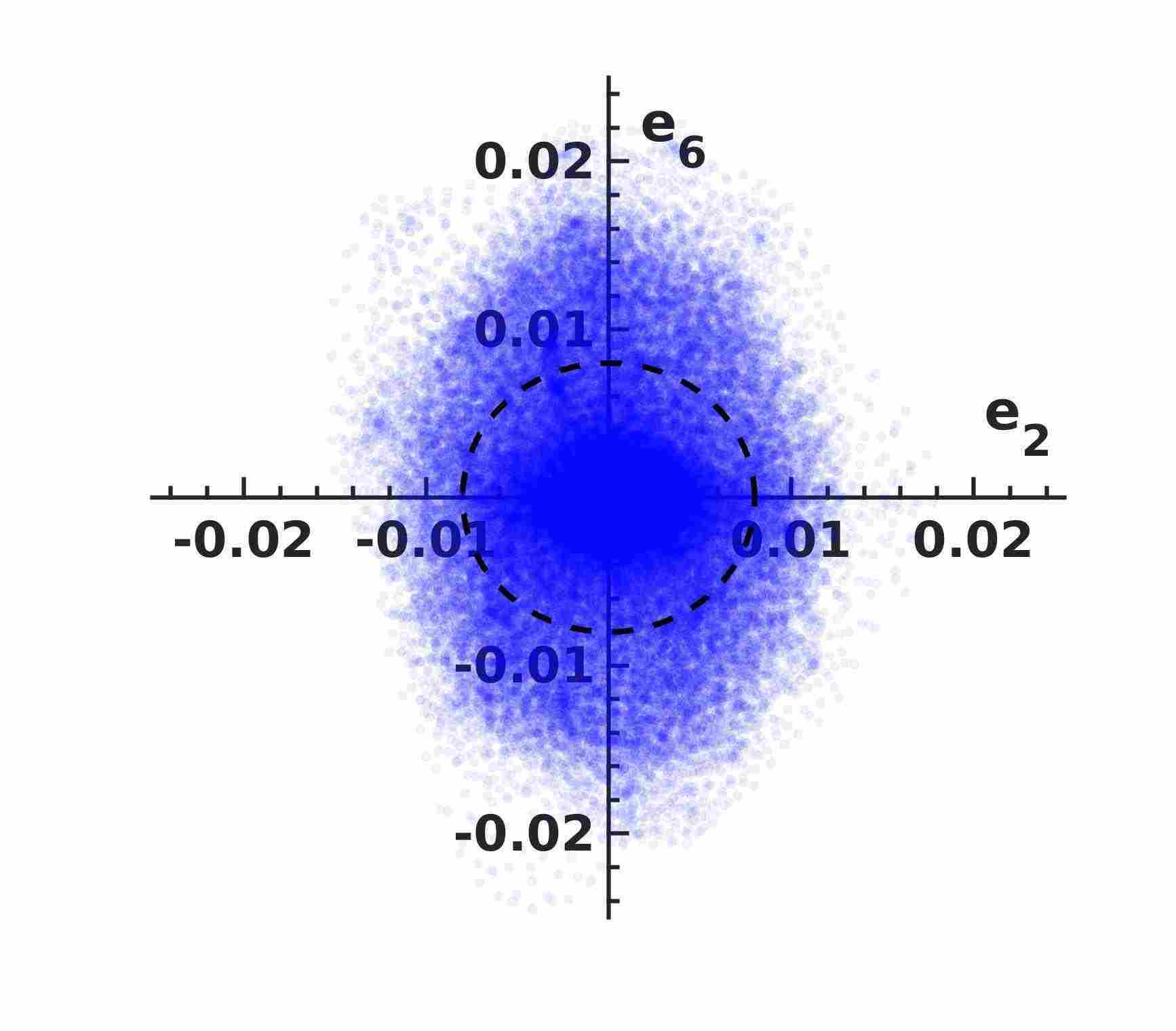}}
        %% 0.65
        \subfloat[x\textsubscript{Al}=0.65]{\includegraphics[scale=0.060]{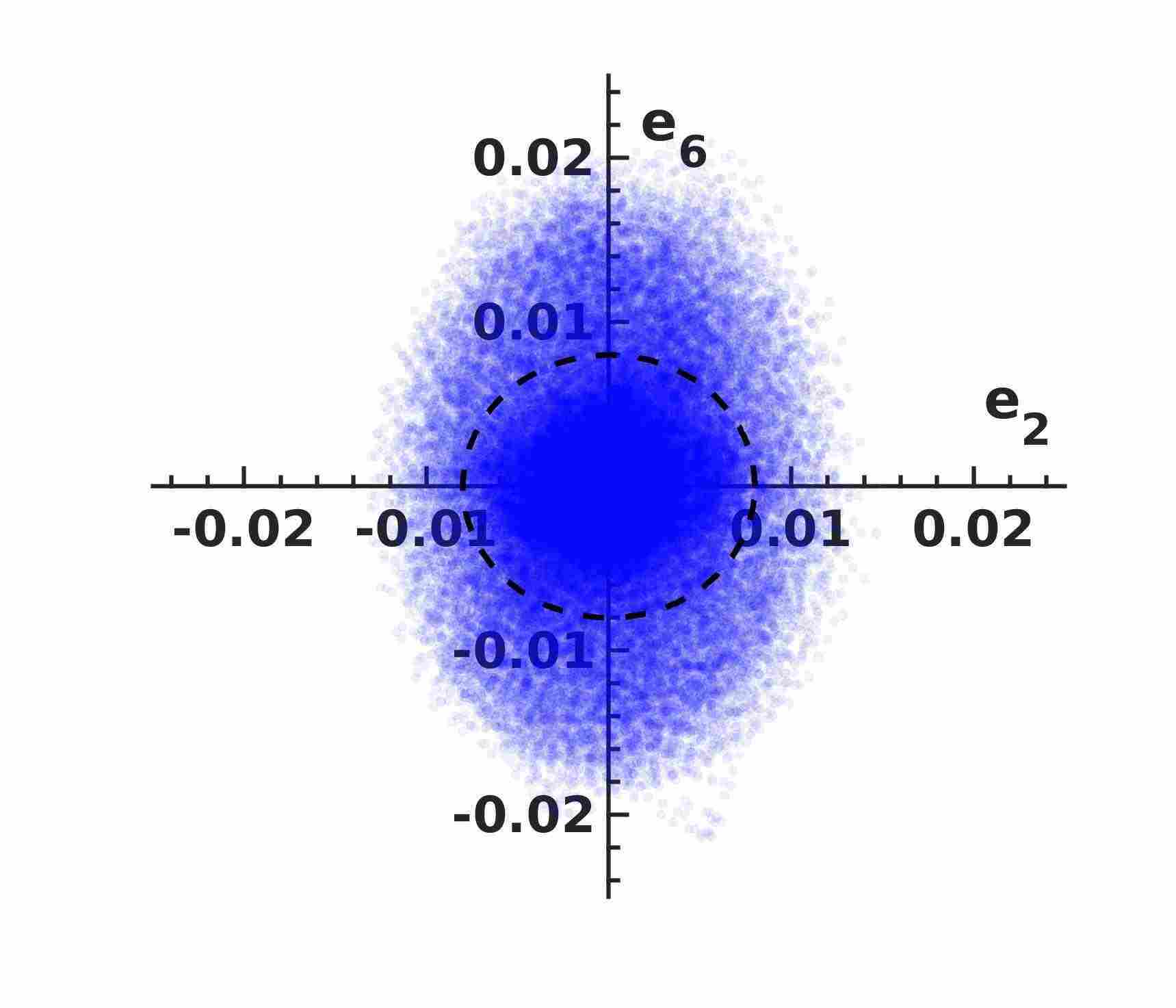}}
        %% 0.70
        \subfloat[x\textsubscript{Al}=0.70]{\includegraphics[scale=0.060]{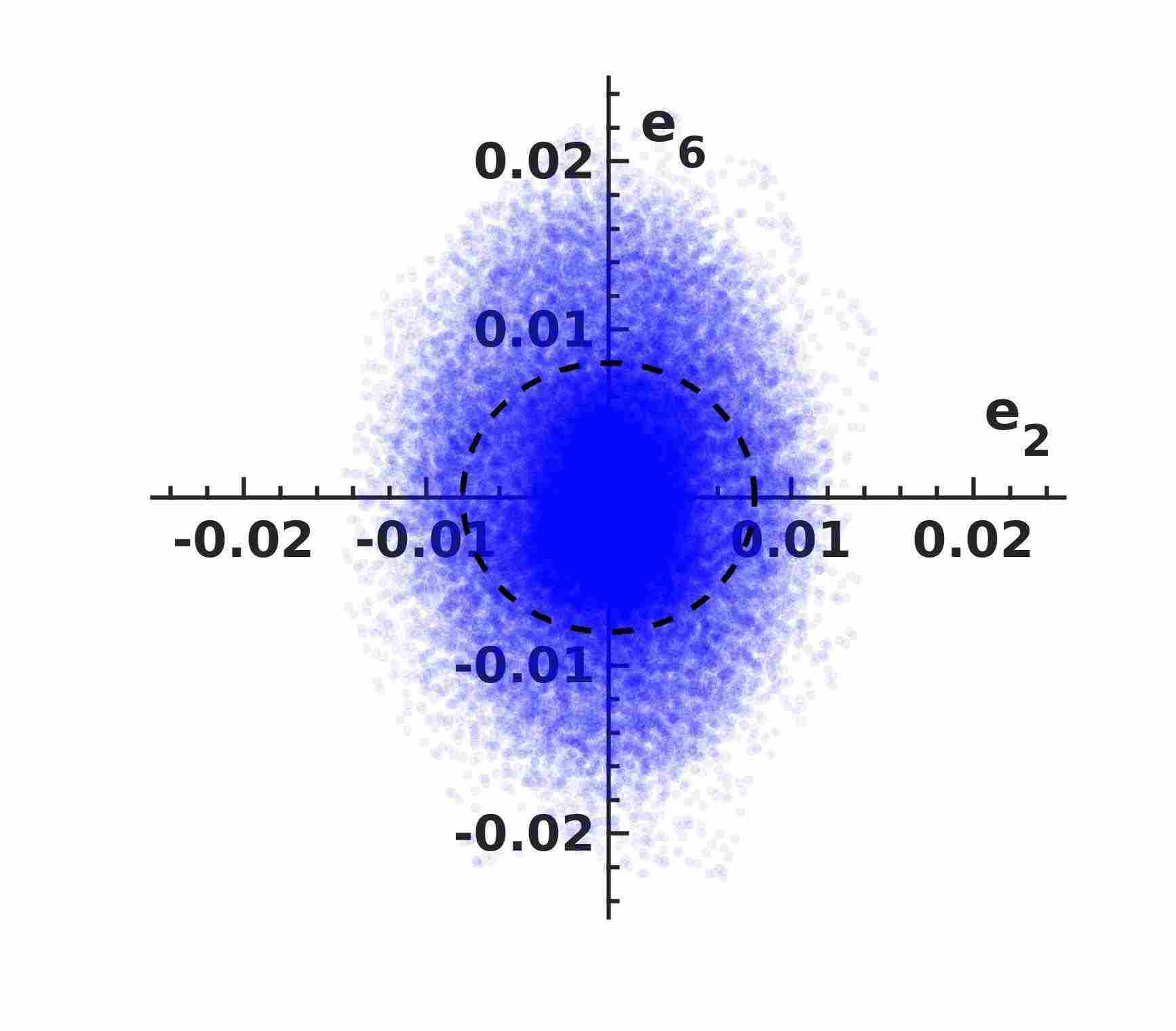}}

        \caption{ The distribution of ($e_2$,$e_6$) points for the given microstructures in Fig.~\ref{fig:elasto-chem-sims} indicating the tendency of deviatoric and/or shear strains to transform the cubic structure to other structural varients. }
        \label{fig:scatter_e2_e6}
    \end{figure*}

Figure~\ref{fig:scatter_e2_e6} illustrates the distribution of ($e_2$,$e_6$) data points in the form of blue (online) clouds extracted from the microstructures that are demonstrated in Fig.~\ref{fig:elasto-chem-sims}. The spatial distribution of the values of $e_2$ and $e_6$ in the domain depicts an overall shear to deviatoric ratio ($\sfrac{e_6}{e_2}$) for a given microstructure which can be further utilized to interpret the tendency of a given domain to structurally deform to other crystallographic structures. Accordingly, the strain cloud in Fig.~\ref{fig:scatter_e2_e6} infers that the $\sfrac{e_6}{e_2}$ ratio increases by increasing the Al content in the microstructure.

   %%%%%%%%%%%%%%%%%%
    %%%%%%%%%%%%%%%%%% 
    
On the other hand, a strain threshold in $\{e_2,e_6\}$ space is taken to distinguish the cubic region in the center of the plot from the rest of the strain space. The black circles drawn in the center of the clouds in Fig.~\ref{fig:scatter_e2_e6} determine the border of this region. In this area, the fluctuations in the strain space are assumed not to cause any structural phase transition, and the cubic structure would preserve its integrity. Outside this region, however, the strains tend to deform the cubic phase into other structural variants such as \emph{bct}, and/or wurtzite. 
    
\begin{figure}[h!]
        \centering
        \includegraphics[scale=0.50]{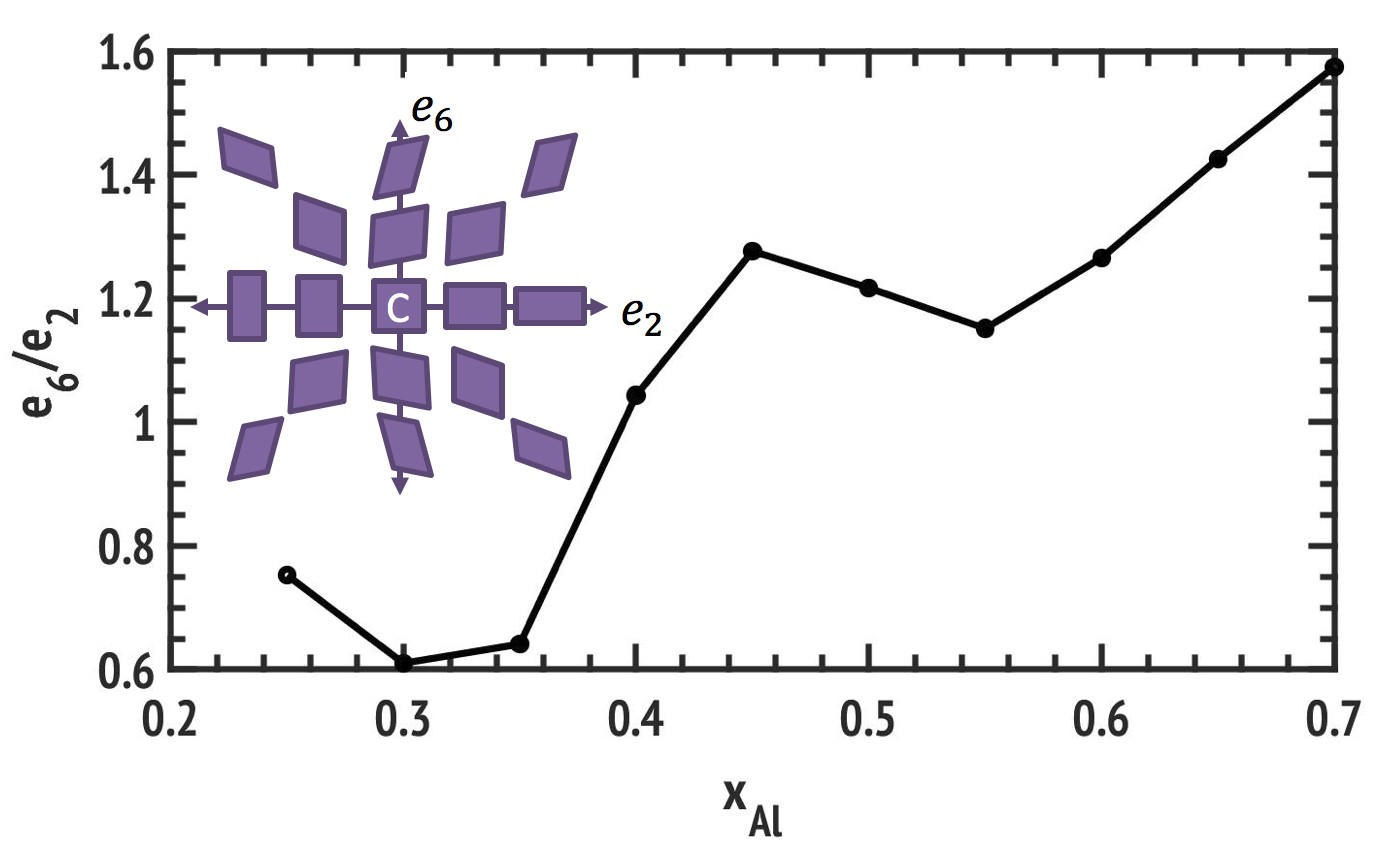}   \vspace{-0.15cm}
        \caption{ $\sfrac{e_6}{e_2}$ ratio versus the change in Al content in Ti$_{1-x-0.05}$Al$_x$Zr$_{0.05}$ alloy for determining the Cubic $\rightarrow$ wurtzite transformation tendancy based on Al content, and the respective change in the microstructure.}
        \label{fig:e6/e2_ratio}
    \end{figure}

An increase in the $\sfrac{e_6}{e_2}$ ratio outside the defined cubic region infers that the microstructrual redistribution of phases, particular distances, elastic properties, and/or other factors which affect the strain cloud tend to increase the chance of structural transition from cubic phase to wurtzite phase. While the ratio is 0.7526 for $x_{Al}=0.25$, it increases to 1.5746 for $x_{Al}=0.70$ in these plots, suggesting an increase in the shear strain order parameter as Al increases in the microstructures. We interpret this with the tendency for structural deformation, which is in accordance with the experimental \cite{chen2011influence} observations, and the \emph{ab-initio} calculations \cite{holec2014macroscopic,zhang2009deformation,duan2016different}. The schematic plot in the interior left corner of Fig.~\ref{fig:e6/e2_ratio} depicts the interplay between the strain order parameters in $\{e_2,e_6\}$ space, and their role in SR, SO, and RO transitions. The increase in $\sfrac{e_6}{e_2}$ ratio is not smooth, and it declines slightly for $0.45\leq x_{Al}\leq0.55$ range and it increases again as indicated in Fig.~\ref{fig:e6/e2_ratio}. This later observation is also consistent with the experimental study of Chen \etal \cite{chen2011influence} that suggested a high hardness values of $\sim$40 GPa over the elevated temperature range of 700-1100 $^\circ$C in Ti$_{0.40}$Al$_{0.55}$Zr$_{0.05}$N alloy. They attribute this prominent hardness results to the capability of the alloy in maintaining more cubic AlN phases, and retarding the formation of wurtzite AlN. 
    
As discussed above, the elastic constants soften non-uniformly with temperature, and both Zener anisotropy ($A_z$), and compressiblity ($A_p$) factors increases non-linearly with Al content, and temperature; though Zr addition reduces this effect to some extent. Elastic isotropy ($A_z=1$) only occurs at $x_{Al}=0.25$, and $T\approx300 ^\circ$C in the Ti$_{1-x}$Al$_{x}$N system. Similarly, it occurs at $x_{Al}=0.50$, and $T\approx260 ^\circ$C in Zr$_{1-x}$Al$_{x}$N system. The slope of $A_z$ factor is significantly higher than the slope of $A_p$. $A_z$/$A_p$ ratio is higher than 2 for Al content over 0.75 indicating a severe transformation tendency. Due to high elastic mismatch between ZrN and TiN/AlN phases, large deviations from the chemical-only regime is viable for ZrN phases. Also, upon confronting sufficient strain at the c-AlN interfaces, when the strain corresponds to the Bain strain, the necessary condition to trigger the cubic $\rightarrow$ wurtzite cell transition occurs. 

%The maps of the strain order parameters $e_2$, and $e_6$ in the simulation cell for different Al compositions in the Ti$_{1-x-0.05}$Al$_{x}$Zr$_{0.05}$N are provided in the supplementary document. Due to high elastic mismatch between ZrN and TiN/AlN phases, large deviations from the chemical-only regime occurs. Upon providing sufficient strain at the c-AlN interfaces, when the strain corresponds to the Bain strain, the necessary condition to trigger the cubic $\rightarrow$ \emph{bct} $\rightarrow$ wurtizite cell transition occurs. %Later, an appropriate compression along the z-axis together with a uniform expansion along the x- and y-axes would be enough to convert the \emph{bct} cell of ALN into the body-centred cubic or body-centred tetragonal cells of martensite, bainite or Widmanstatten ferrite. 
    % the elastic moduli of the system in Voigt notation is defined $C_{KL}=\sfrac{\partial^2 F}{\partial e_K\partial e_L}$. 
    
    %%%%%%%%%%%%%%%%%%

\section{Conclusions}\label{sec:conculsions}
    
While material property predictions based on multi-phase-field microstructural modeling and density functional theory (DFT) calculations have been shown to be valuable in guiding materials design efforts, an honest appraisal of the predictive capabilities regarding metastable phase formation is a sobering experience, but useful to try. In this respect, here we have studied the microstructural aspects and high temperature annealing properties of quaternary (pseudoternary) metastable (c)-Ti$_{1-x-y}$Al$_{x}$Zr$_{y}$N system with NaCl cubic (c) structure at 1,200$^\circ$C with a phase-field model where the free energy functions were given by CALPHAD expressions, and the atomic mobilities were parameterized using the available experimental observations. Both free energy functions and atomic mobilities were considered experimentally validated. In this work, we studied the high temperature elasto-chemical behaviour of TiAlZrN system by the application of microelasticity theory extended to study diffusional growth problems of elastically misfitting phases. We showed that the morphology, and growth depends on (i) chemical driving force, (ii) misfit strain, (iii) the elastic modulus of constituent phases, and subsequent (iv) stresses. 
    
Shape instabilities are observed for growing precipitates in different compositions. Particle shape transitions with symmetry reduction, sphere, and cuboids are observed for a wide range of Al compositions. In general, by decreasing Al content in region IV of the ternary phase diagram by means of appreciable lattice mismatch and coherency strain, the particles may evolve from randomly distributed spheres to sharply aligned cubes and rods as composition changes. For high Al systems that form mixed TiN, and ZrN precipitates, the matrix is sheared but the precipitate are not, so the sphere is a good (low-energy) configuration when the matrix is more flexible. Bimodal phase growth is observed in high Al ($x_{Al}\sim0.65$) region of Ti$_{1-x-0.05}$Al$_{x}$Zr$_{0.05}$N system which can enhance the mechanical properties significantly. Trimodal growth is observed when elasto-chemical interactions get more important in high Al coating system. These observations are similar to the experimental observations in Ni super alloys. 
    
Theory and simulation are combined to describe the effect of elastic strains, and the role of microstructure on the onset of cubic AlN transition to wurtzite structure in TMNs. Assuming strains have a significant role in this transition, strain order parameters are utilized to study the tendency of obtained microstructures in inducing this phase transition. It is shown that the shear strain gains more importance in the microstrurue as Al content increases paving the road for cubic $\rightarrow$ wurtzite transition. Age hardening effect in Ti$_{1-x-y}$Al$_{x}$Zr$_{y}$N coatings has a direct relationship with Al and Zr content. As a guideline, a better age-hardening property may be achieved around $0.40\leq x_{Al}\leq0.55$ where the higher anisotropy due to the high Al content is balanced by Zr addition, and resulted in more elongated Al- and Ti-enriched domains.

    \bigskip
    \section*{Acknowledgments}
    The authors would like to thank Dr. Peter Voorhees for the useful discussion on the results obtained thorugh phase-field modeling. Vahid Attari is also grateful for the useful insights provided by Dr. Amine Benzerga, and Mr. Lei Xu. The authors acknowledge Terra supercomputing facility of the Texas A\&M University, for providing computing resources useful in conducting the research reported in this paper. This research was supported by the National Science Foundation under NSF Grant No. CMMI-1462255. RA and AC also acknowledge support from NSF Grant No. CMMI-1534534.
    
    %%%%%%%%%%%%%%%%%%%%%%%%%%%%%%%%%%%%%%%%%%%%%%%%%%%%%%%
    %%%%%%%%%%%%%%%%%%%%%%%%%%%%%%%%%%%%%%%%%%%%%%%%%%%%%%%
    %%%%%%%%%%%%%%%%%%%%%%%%%%%%%%%%%%%%%%%%%%%%%%%%%%%%%%%
    \bigskip
    \bibliography{Paper_nitrites} 

\begin{thebibliography}{10}
\expandafter\ifx\csname url\endcsname\relax
  \def\url#1{\texttt{#1}}\fi
\expandafter\ifx\csname urlprefix\endcsname\relax\def\urlprefix{URL }\fi
\expandafter\ifx\csname href\endcsname\relax
  \def\href#1#2{#2} \def\path#1{#1}\fi

\bibitem{abadias2010reactive}
G.~Abadias, L.~Koutsokeras, S.~Dub, G.~Tolmachova, A.~Debelle, T.~Sauvage,
  P.~Villechaise, {Reactive Magnetron Co-sputtering of Hard and Conductive
  Ternary Nitride Thin Films: Ti--Zr--N and Ti--Ta--N}, Journal of Vacuum
  Science \& Technology A: Vacuum, Surfaces, and Films 28~(4) (2010) 541--551.

\bibitem{abrikosov2011phase}
I.~A. Abrikosov, A.~Knutsson, B.~Alling, F.~Tasn{\'a}di, H.~Lind, L.~Hultman,
  M.~Od{\'e}n, Phase stability and elasticity of tialn, Materials 4~(9) (2011)
  1599--1618.

\bibitem{ding2008corrosion}
X.-z. Ding, A.~Tan, X.~Zeng, C.~Wang, T.~Yue, C.~Sun, {Corrosion Resistance of
  CrAlN and TiAlN Coatings Deposited by Lateral Rotating Cathode Arc}, Thin
  Solid Films 516~(16) (2008) 5716--5720.

\bibitem{donohue1997microstructure}
L.~Donohue, I.~Smith, W.-D. M{\"u}nz, I.~Petrov, J.~Greene, {Microstructure and
  Oxidation-resistance of {Ti$_{1-x-y-z}$Al$_x$Cr$_y$Y$_z$N} Layers Grown by
  Combined Steered-arc/Unbalanced Magnetron Sputter Deposition}, Surface and
  Coatings Technology 94 (1997) 226--231.

\bibitem{mikula2016toughness}
M.~Mikula, D.~Pla{\v{s}}ienka, D.~G. Sangiovanni, M.~Sahul, T.~Roch,
  M.~Truchl{\`y}, M.~Gregor, L.~{\v{C}}aplovi{\v{c}}, A.~Plecenik,
  P.~K{\'u}{\v{s}}, {Toughness Enhancement in Highly NbN-alloyed Ti--Al--N Hard
  Coatings}, Acta Materialia 121 (2016) 59--67.

\bibitem{munz1986titanium}
W.-D. M{\"u}nz, {Titanium Aluminum Nitride Films: A New Alternative to TiN
  Coatings}, Journal of Vacuum Science \& Technology A: Vacuum, Surfaces, and
  Films 4~(6) (1986) 2717--2725.

\bibitem{paldey2003single}
S.~PalDey, S.~Deevi, {Single Layer and Multilayer Wear Resistant Coatings of
  (Ti, Al) N: A Review}, Materials Science and Engineering: A 342~(1-2) (2003)
  58--79.

\bibitem{horling2002thermal}
A.~H{\"o}rling, L.~Hultman, M.~Od{\'e}n, J.~Sj{\"o}l{\'e}n, L.~Karlsson,
  {Thermal Stability of Arc Evaporated High Aluminum-content Ti$_{1-x}$Al$_x$N
  Thin Films}, Journal of Vacuum Science \& Technology A: Vacuum, Surfaces, and
  Films 20~(5) (2002) 1815--1823.

\bibitem{rachbauer2011decomposition}
R.~Rachbauer, S.~Massl, E.~Stergar, D.~Holec, D.~Kiener, J.~Keckes,
  J.~Patscheider, M.~Stiefel, H.~Leitner, P.~Mayrhofer, {Decomposition Pathways
  in Age Hardening of Ti-Al-N Films}, Journal of Applied Physics 110~(2) (2011)
  023515.

\bibitem{horling2005mechanical}
A.~H{\"o}rling, L.~Hultman, M.~Od{\'e}n, J.~Sj{\"o}l{\'e}n, L.~Karlsson,
  {Mechanical Properties and Machining Performance of Ti$_{1-x}$Al$_x$N-coated
  Cutting Tools}, Surface and Coatings Technology 191~(2-3) (2005) 384--392.

\bibitem{knutsson2011machining}
A.~Knutsson, M.~Johansson, L.~Karlsson, M.~Od{\'e}n, {Machining Performance and
  Decomposition of TiAlN/TiN Multilayer Coated Metal Cutting Inserts}, Surface
  and Coatings Technology 205~(16) (2011) 4005--4010.

\bibitem{bartosik2017fracture}
M.~Bartosik, C.~Rumeau, R.~Hahn, Z.~Zhang, P.~Mayrhofer, {Fracture Toughness
  and Structural Evolution in the TiAlN System upon Annealing}, Scientific
  reports 7~(1) (2017) 16476.

\bibitem{forsen2012mechanical}
R.~Fors{\'e}n, {Mechanical Properties and Thermal Stability of Reactive Arc
  Evaporated Ti-Cr-Al-N Coatings}, Ph.D. thesis, Link{\"o}ping University
  Electronic Press (2012).

\bibitem{yang2013effect}
B.~Yang, L.~Chen, Y.~X. Xu, Y.~B. Peng, J.~C. Fen, Y.~Du, M.~J. Wu, {Effect of
  Zr on Structure and Properties of Ti--Al--N coatings with Varied Bias},
  International Journal of Refractory Metals and Hard Materials 38 (2013)
  81--86.

\bibitem{lind2013systematic}
H.~Lind, F.~Tasnadi, I.~Abrikosov, {Systematic Theoretical search for Alloys
  with Increased Thermal Stability for Advanced Hard Coatings Applications},
  New Journal of Physics 15~(9) (2013) 095010.

\bibitem{yalamanchili2016growth}
K.~Yalamanchili, F.~Wang, H.~Aboulfadl, J.~Barrirero, L.~Rogstr{\"o}m,
  E.~Jim{\'e}nez-Pique, F.~M{\"u}cklich, F.~Tasnadi, M.~Od{\'e}n, N.~Ghafoor,
  {Growth and Thermal Stability of TiN/ZrAlN: Effect of Internal Interfaces},
  Acta Materialia 121 (2016) 396--406.

\bibitem{rogstrom2015wear}
L.~Rogstr{\"o}m, M.~Johansson-J{\~o}esaar, L.~Land{\"a}lv, M.~Ahlgren,
  M.~Od{\'e}n, {Wear Behavior of ZrAlN Coated Cutting Tools during Turning},
  Surface and Coatings Technology 282 (2015) 180--187.

\bibitem{holec2011phase}
D.~Holec, R.~Rachbauer, L.~Chen, L.~Wang, D.~Luef, P.~H. Mayrhofer, {Phase
  Stability and Alloy-related Trends in Ti--Al--N, Zr--Al--N and Hf--Al--N
  Systems from First Principles}, Surface and Coatings Technology 206~(7)
  (2011) 1698--1704.

\bibitem{wang2015mechanical}
A.~Wang, M.~He, R.~Zhang, Y.~Du, D.~Chen, B.~Fan, S.-L. Shang, Z.-K. Liu,
  {Mechanical Properties and Spinodal Decomposition of
  Ti$_x$Al$_{1-x-y}$Zr$_y$N Coatings}, Physics Letters A 379~(36) (2015)
  2037--2040.

\bibitem{zhou2017thermodynamic}
J.~Zhou, L.~Zhang, L.~Chen, Y.~Du, Z.~Liu, {A Thermodynamic Description of
  Metastable c-TiAlZrN Coatings with Triple Spinodally Decomposed Domains},
  Journal of Mining and Metallurgy B: Metallurgy 53~(2) (2017) 85--93.

\bibitem{tasnadi2010significant}
F.~Tasn{\'a}di, I.~A. Abrikosov, L.~Rogstr{\"o}m, J.~Almer, M.~P. Johansson,
  M.~Od{\'e}n, {Significant Elastic Anisotropy in Ti$_{1-x}$Al$_x$N Alloys},
  Applied Physics Letters 97~(23) (2010) 231902.

\bibitem{aihua2012friction}
L.~Aihua, D.~Jianxin, C.~Haibing, C.~Yangyang, Z.~Jun, {Friction and Wear
  Properties of TiN, TiAlN, AlTiN and CrAlN PVD Nitride Coatings},
  International Journal of Refractory Metals and Hard Materials 31 (2012)
  82--88.

\bibitem{gronhagen2015phase}
K.~Gr{\"o}nhagen, J.~{\AA}gren, M.~Od{\'e}n, Phase-field modelling of spinodal
  decomposition in tialn including the effect of metal vacancies, Scripta
  Materialia 95 (2015) 42--45.

\bibitem{zhou2017effect}
J.~Zhou, L.~Zhang, L.~Chen, {Effect of Cr on Metastable Phase Equilibria and
  Spinodal Decomposition in c-TiAlN Coatings: A CALPHAD and Cahn-Hilliard
  study}, Surface and Coatings Technology 311 (2017) 231--237.

\bibitem{cahn1961spinodal}
J.~W. Cahn, On spinodal decomposition, Acta metallurgica 9~(9) (1961) 795--801.

\bibitem{ghosh2017particles}
S.~Ghosh, A.~Mukherjee, T.~Abinandanan, S.~Bose, Particles with selective
  wetting affect spinodal decomposition microstructures, Physical Chemistry
  Chemical Physics 19~(23) (2017) 15424--15432.

\bibitem{lind2014high}
H.~Lind, R.~Pilemalm, L.~Rogstr{\"o}m, F.~Tasnadi, N.~Ghafoor, R.~Fors{\'e}n,
  L.~J. S.~Johnson, M.~P. Johansson-J{\"o}esaar, M.~Od{\'e}n, I.~A. Abrikosov,
  {High Temperature Phase Decomposition in Ti$_x$Zr$_y$Al$_z$N}, AIP Advances
  4~(12) (2014) 127147.

\bibitem{wang2012structural}
A.~Wang, S.~Shang, D.~Zhao, J.~Wang, L.~Chen, Y.~Du, Z.-K. Liu, T.~Xu, S.~Wang,
  {Structural, Phonon and Thermodynamic Properties of fcc-based Metal Nitrides
  from First-principles Calculations}, Calphad 37 (2012) 126--131.

\bibitem{kaufman1970computer}
L.~Kaufman, H.~Bernstein, {Computer Calculation of Phase Diagrams. With Special
  Reference to Refractory Metals}, Academic Press Inc, New York, 1970.

\bibitem{moelans2008introduction}
N.~Moelans, B.~Blanpain, P.~Wollants, An introduction to phase-field modeling
  of microstructure evolution, Calphad 32~(2) (2008) 268--294.

\bibitem{khachaturyan2013theory}
A.~G. Khachaturyan, Theory of structural transformations in solids, Courier
  Corporation, 2013.

\bibitem{mura2013micromechanics}
T.~Mura, Micromechanics of defects in solids, Springer Science \& Business
  Media, 2013.

\bibitem{moulinec1998numerical}
H.~Moulinec, P.~Suquet, A numerical method for computing the overall response
  of nonlinear composites with complex microstructure, Computer methods in
  applied mechanics and engineering 157~(1-2) (1998) 69--94.

\bibitem{gururajan2007phase}
M.~Gururajan, T.~Abinandanan, Phase field study of precipitate rafting under a
  uniaxial stress, Acta Materialia 55~(15) (2007) 5015--5026.

\bibitem{lebensohn2004macroscopic}
R.~Lebensohn, Y.~Liu, P.~P. Castaneda, Macroscopic properties and field
  fluctuations in model power-law polycrystals: full-field solutions versus
  self-consistent estimates, in: Proceedings of the Royal Society of London A:
  Mathematical, Physical and Engineering Sciences, Vol. 460, The Royal Society,
  2004, pp. 1381--1405.

\bibitem{michel1999effective}
J.-C. Michel, H.~Moulinec, P.~Suquet, Effective properties of composite
  materials with periodic microstructure: a computational approach, Computer
  methods in applied mechanics and engineering 172~(1-4) (1999) 109--143.

\bibitem{chen1998applications}
L.~Q. Chen, J.~Shen, Applications of semi-implicit fourier-spectral method to
  phase field equations, Computer Physics Communications 108~(2-3) (1998)
  147--158.

\bibitem{alling2007mixing}
B.~Alling, A.~V. Ruban, A.~Karimi, O.~E. Peil, S.~Simak, L.~Hultman,
  I.~Abrikosov, Mixing and decomposition thermodynamics of c- ti 1- x al x n
  from first-principles calculations, Physical Review B 75~(4) (2007) 045123.

\bibitem{sheng2008phase}
S.~Sheng, R.~Zhang, S.~Veprek, Phase stabilities and thermal decomposition in
  the zr1- xalxn system studied by ab initio calculation and thermodynamic
  modeling, Acta Materialia 56~(5) (2008) 968--976.

\bibitem{fratzl1999modeling}
P.~Fratzl, O.~Penrose, J.~L. Lebowitz, Modeling of phase separation in alloys
  with coherent elastic misfit, Journal of Statistical Physics 95~(5-6) (1999)
  1429--1503.

\bibitem{koehler1970attempt}
J.~Koehler, Attempt to design a strong solid, Physical review B 2~(2) (1970)
  547.

\bibitem{holec2014macroscopic}
D.~Holec, F.~Tasnadi, P.~Wagner, M.~Fri{\'a}k, J.~Neugebauer, P.~H. Mayrhofer,
  J.~Keckes, {Macroscopic Elastic Properties of Textured ZrN-AlN
  Polycrystalline Aggregates: From \emph{ab initio} Calculations to Grain-scale
  Interactions}, Physical Review B 90~(18) (2014) 184106.

\bibitem{abadias2012structure}
G.~Abadias, V.~Ivashchenko, L.~Belliard, P.~Djemia, Structure, phase stability
  and elastic properties in the ti1--xzrxn thin-film system: Experimental and
  computational studies, Acta Materialia 60~(15) (2012) 5601--5614.

\bibitem{balasubramanian2018energetics}
K.~Balasubramanian, S.~V. Khare, D.~Gall, Energetics of point defects in
  rocksalt structure transition metal nitrides: Thermodynamic reasons for
  deviations from stoichiometry, Acta Materialia 159 (2018) 77--88.

\bibitem{sangiovanni2018inherent}
D.~Sangiovanni, Inherent toughness and fracture mechanisms of refractory
  transition-metal nitrides via density-functional molecular dynamics, Acta
  Materialia 151 (2018) 11--20.

\bibitem{wang2017systematic}
F.~Wang, D.~Holec, M.~Od{\'e}n, F.~Muecklich, I.~A. Abrikosov, F.~Tasnadi,
  Systematic ab initio investigation of the elastic modulus in quaternary
  transition metal nitride alloys and their coherent multilayers, Acta
  Materialia 127 (2017) 124--132.

\bibitem{yang2017multiaxial}
W.~Yang, G.~Ayoub, I.~Salehinia, B.~Mansoor, H.~Zbib, Multiaxial
  tension/compression asymmetry of ti/tin nano laminates: Md investigation,
  Acta Materialia 135 (2017) 348--360.

\bibitem{el2018investigations}
I.~El~Azhari, J.~Garcia, M.~Zamanzade, F.~Soldera, C.~Pauly, L.~Llanes,
  F.~M{\"u}cklich, Investigations on micro-mechanical properties of
  polycrystalline ti (c, n) and zr (c, n) coatings, Acta Materialia 149 (2018)
  364--376.

\bibitem{gibbs1961scientific}
J.~W. Gibbs, Scientific Papers: Thermodynamics, Vol.~1, Dover Publications,
  1961.

\bibitem{tisza1961thermodynamics}
L.~Tisza, The thermodynamics of phase equilibrium, Annals of Physics 13~(1)
  (1961) 1--92.

\bibitem{cahn1984simple}
J.~Cahn, F.~Larch{\'e}, A simple model for coherent equilibrium, Acta
  metallurgica 32~(11) (1984) 1915--1923.

\bibitem{huh1995intrinsic}
J.-Y. Huh, W.~Johnson, Intrinsic thermodynamic stability of stressed coherent
  systems, Acta metallurgica et materialia 43~(4) (1995) 1631--1642.

\bibitem{yi2018strain}
S.-i. Yi, V.~Attari, M.~Jeong, J.~Jian, S.~Xue, H.~Wang, R.~Arroyave, C.~Yu,
  Strain-induced suppression of the miscibility gap in nanostructured mg 2
  si--mg 2 sn solid solutions, Journal of Materials Chemistry A 6~(36) (2018)
  17559--17570.

\bibitem{de1979configurational}
D.~De~Fontaine, Configurational thermodynamics of solid solutions, in: Solid
  state physics, Vol.~34, Elsevier, 1979, pp. 73--274.

\bibitem{kikuchi1987second}
R.~Kikuchi, Second hessian determinant as the criterion for order (first or
  second) of phase transition, Physica A: Statistical Mechanics and its
  Applications 142~(1-3) (1987) 321--341.

\bibitem{mayrhofer2003self}
P.~H. Mayrhofer, A.~H{\"o}rling, L.~Karlsson, J.~Sj{\"o}l{\'e}n, T.~Larsson,
  C.~Mitterer, L.~Hultman, Self-organized nanostructures in the ti--al--n
  system, Applied Physics Letters 83~(10) (2003) 2049--2051.

\bibitem{chen2011influence}
L.~Chen, D.~Holec, Y.~Du, P.~H. Mayrhofer, Influence of zr on structure,
  mechanical and thermal properties of ti--al--n, Thin Solid Films 519~(16)
  (2011) 5503--5510.

\bibitem{knutsson2013microstructure}
A.~Knutsson, J.~Ullbrand, L.~Rogstr{\"o}m, N.~Norrby, L.~Johnson, L.~Hultman,
  J.~Almer, M.~Johansson~J{\"o}esaar, B.~Jansson, M.~Od{\'e}n, Microstructure
  evolution during the isostructural decomposition of tialn$-$a combined
  in-situ small angle x-ray scattering and phase field study, Journal of
  Applied Physics 113~(21) (2013) 213518.

\bibitem{hultman2000thermal}
L.~Hultman, Thermal stability of nitride thin films, Vacuum 57~(1) (2000)
  1--30.

\bibitem{kwon2007coarsening}
Y.~Kwon, K.~Thornton, P.~W. Voorhees, {Coarsening of Bicontinuous Structures
  via Nonconserved and Conserved Dynamics}, Physical Review E 75~(2) (2007)
  021120.

\bibitem{kwon2009topology}
Y.~Kwon, K.~Thornton, P.~Voorhees, {The Topology and Morphology of Bicontinuous
  Interfaces During Coarsening}, EPL (Europhysics Letters) 86~(4) (2009) 46005.

\bibitem{chen2010morphological}
Y.-c.~K. Chen, Y.~S. Chu, J.~Yi, I.~McNulty, Q.~Shen, P.~W. Voorhees, D.~C.
  Dunand, {Morphological and Topological Analysis of Coarsened Nanoporous Gold
  by x-ray Nanotomography}, Applied Physics Letters 96~(4) (2010) 043122.

\bibitem{johnson1984elastic}
W.~Johnson, On the elastic stabilization of precipitates against coarsening
  under applied load, Acta Metallurgica 32~(3) (1984) 465--475.

\bibitem{su1996dynamics}
C.-H. Su, P.~Voorhees, The dynamics of precipitate evolution in elastically
  stressed solids${-}${I}. inverse coarsening, Acta materialia 44~(5) (1996)
  1987--1999.

\bibitem{onuki1991eshelby}
A.~Onuki, H.~Nishimori, {On Eshelby's Elastic Interaction in Two-phase Solids},
  Journal of the Physical Society of Japan 60~(1) (1991) 1--4.

\bibitem{coakley2010coarsening}
J.~Coakley, H.~Basoalto, D.~Dye, {Coarsening of a Multimodal Nickel-base
  Superalloy}, Acta Materialia 58~(11) (2010) 4019--4028.

\bibitem{li2018influence}
M.~Li, J.~Coakley, D.~Isheim, G.~Tian, B.~Shollock, {Influence of the Initial
  Cooling Rate from ${\gamma}^{'}$ Supersolvus Temperatures on Microstructure
  and Phase Compositions in a Nickel Superalloy}, Journal of Alloys and
  Compounds 732 (2018) 765--776.

\bibitem{singh2011influence}
A.~Singh, S.~Nag, J.~Hwang, G.~Viswanathan, J.~Tiley, R.~Srinivasan, H.~Fraser,
  R.~Banerjee, {Influence of Cooling Rate on the Development of Multiple
  Generations of ${\gamma}^{'}$ Precipitates in a Commercial Nickel base
  Superalloy}, Materials characterization 62~(9) (2011) 878--886.

\bibitem{saha2016understanding}
B.~Saha, S.~Saber, E.~A. Stach, E.~P. Kvam, T.~D. Sands, {Understanding the
  Rocksalt-to-Wurtzite Phase Transformation through Microstructural Analysis of
  (Al,Sc)N Epitaxial Thin Films}, Applied Physics Letters 109~(17) (2016)
  172102.

\bibitem{yao2013b}
Y.~Yao, D.~D. Klug, B 4- b 1 phase transition of gan under isotropic and
  uniaxial compression, Physical Review B 88~(1) (2013) 014113.

\bibitem{qian2013variable}
G.-R. Qian, X.~Dong, X.-F. Zhou, Y.~Tian, A.~R. Oganov, H.-T. Wang, Variable
  cell nudged elastic band method for studying solid--solid structural phase
  transitions, Computer Physics Communications 184~(9) (2013) 2111--2118.

\bibitem{duan2016different}
Y.~Duan, L.~Qin, H.~Liu, Different evolutionary pathways from b4 to b1 phase in
  aln and inn: metadynamics investigations, Journal of Physics: Condensed
  Matter 28~(20) (2016) 205403.

\bibitem{zhang2009deformation}
R.~Zhang, S.~Veprek, Deformation paths and atomistic mechanism of b4→ b1
  phase transformation in aluminium nitride, Acta Materialia 57~(7) (2009)
  2259--2265.

\bibitem{rudraraju2016mechanochemical}
S.~Rudraraju, A.~Van~der Ven, K.~Garikipati, Mechanochemical spinodal
  decomposition: a phenomenological theory of phase transformations in
  multi-component, crystalline solids, npj Computational Materials 2 (2016)
  16012.

\bibitem{falk1990three}
F.~Falk, P.~Konopka, Three-dimensional landau theory describing the martensitic
  phase transformation of shape-memory alloys, Journal of Physics: Condensed
  Matter 2~(1) (1990) 61.

\bibitem{steinbach2011phase}
I.~Steinbach, O.~Shchyglo, {Phase-field Modelling of Microstructure Evolution
  in Solids: Perspectives and Challenges}, Current opinion in solid state and
  materials science 15~(3) (2011) 87--92.

\bibitem{rasmussen2001three}
K.~Rasmussen, T.~Lookman, A.~Saxena, A.~Bishop, R.~Albers, S.~Shenoy,
  {Three-dimensional Elastic Compatibility and Varieties of Twins in
  Martensites}, Physical review letters 87~(5) (2001) 055704.

\bibitem{nittono1982phenomenological}
O.~Nittono, Y.~Koyama, {Phenomenological Considerations of Phase
  Transformations in Indium-rich Alloys}, Japanese Journal of Applied Physics
  21~(5R) (1982) 680.

\bibitem{barsch1984twin}
G.~Barsch, J.~Krumhansl, {Twin Boundaries in Ferroelastic Media without
  Interface Dislocations}, Physical Review Letters 53~(11) (1984) 1069.

\bibitem{saxena1993pretransformation}
A.~Saxena, G.~Barsch, Pretransformation strain modulations in proper
  ferroelastics, Physica D: Nonlinear Phenomena 66~(1-2) (1993) 195--204.

\end{thebibliography}
    %\bibliographystyle{elsarticle-num}
    %\bibliographystyle{model1a-num-names}

    %%%%%%%%%%%%%%%%%%%%%%%%%%%%%%%%%%%%%%%%%%%%%%%%%%%%%%%
    %%%%%%%%%%%%%%%%%%%%%%%%%%%%%%%%%%%%%%%%%%%%%%%%%%%%%%%
    %%%%%%%%%%%%%%%%%%%%%%%%%%%%%%%%%%%%%%%%%%%%%%%%%%%%%%%
    %\renewcommand\thefigure{\thesection.\arabic{figure}}    
    %\setcounter{figure}{0}
    \appendix 
    \counterwithin*{figure}{section}
    \counterwithin*{table}{section}
    
    \section{Benchmark microelasticity: Numerical vs analytical solution}\label{sec:appendix}
    
    The numerical test of the $Fourier$ spectral solver for the microelasticity problem, and comparison with the analytical solution is provided here. The analytical solution for only specific particle shapes are available in the literature (when the interface is sharp) \cite{mura2013micromechanics}. Here, we compare the numerical solutions for the normalized principal stress components as a function of normalized distance from the centre of a circular (cylindrical inhomogeneous precipitate) inclusion. The effect of the domain size and discretization is also investigated for further optimization of the accuracy of the solver with computation cost.
    
\begin{figure}[!htb]
        \centering 
        %\subfloat[]{\includegraphics[scale=0.050]{images/Inclusion.jpeg}}
        \subfloat[]{\includegraphics[width=0.50\columnwidth]{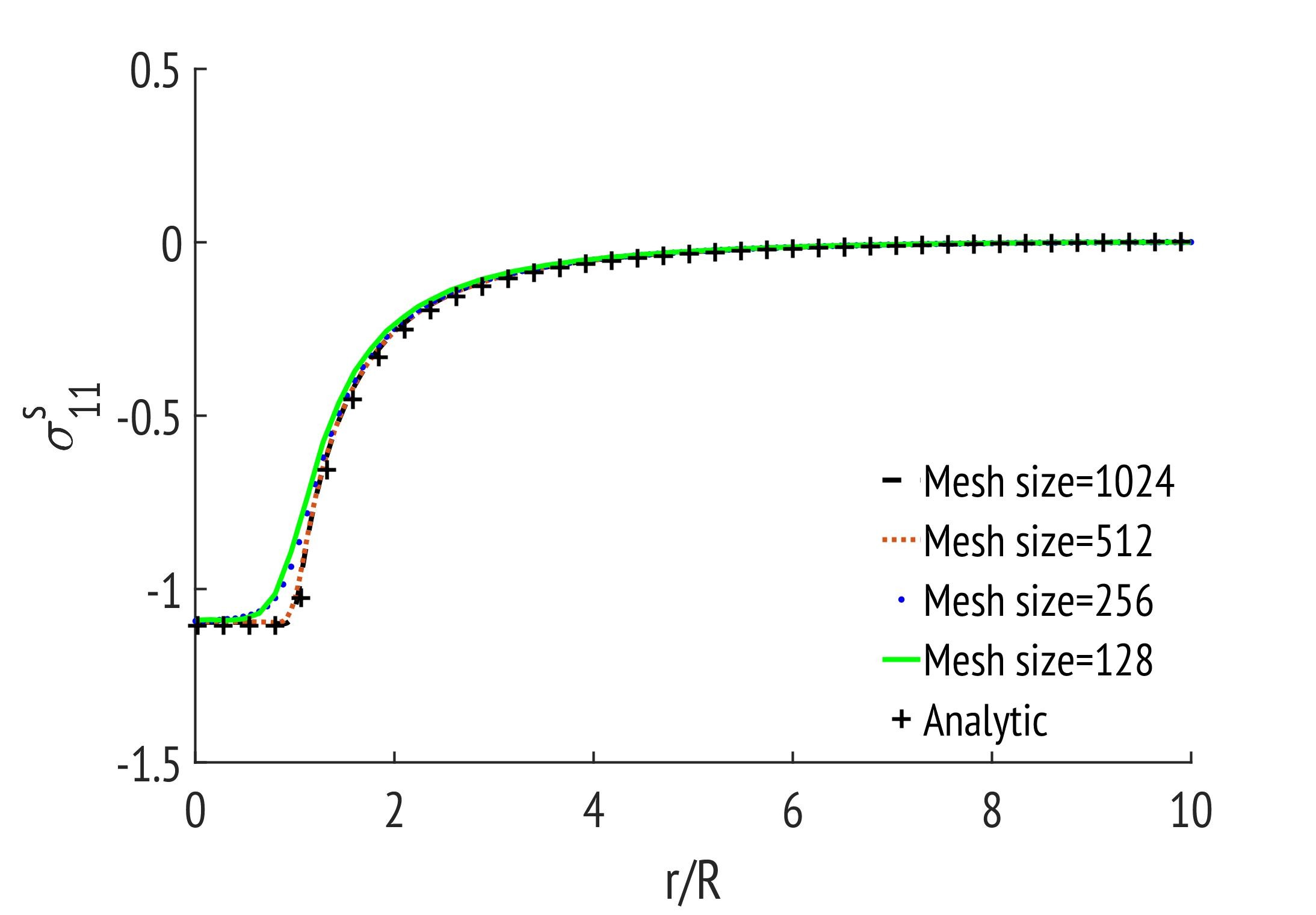}}
        \subfloat[]{\includegraphics[width=0.50\columnwidth]{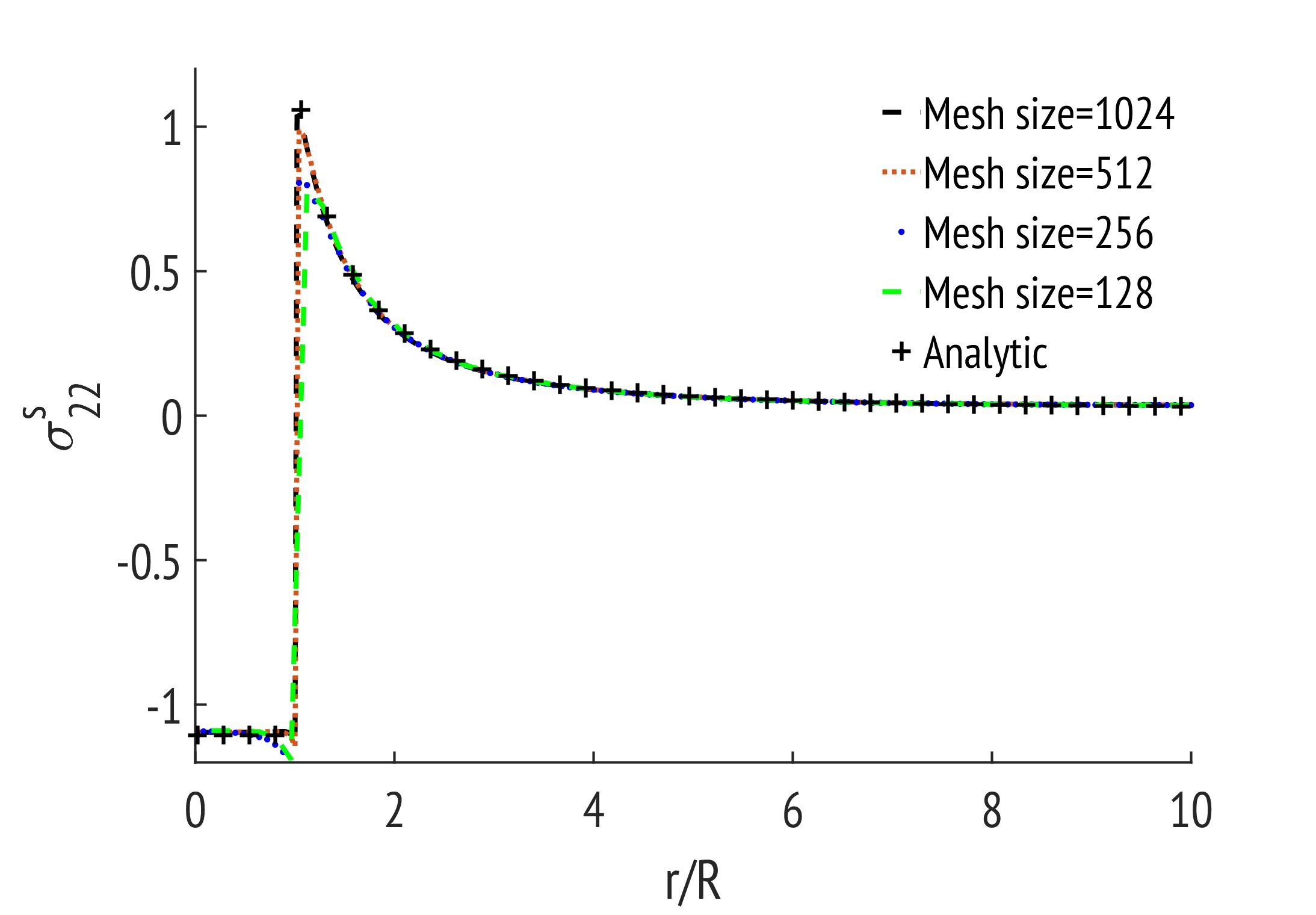}}\\
        \caption{Benchmark result for a circular inclusion in a square domain for different mesh sizes. The figures illustrates the normalized analytical \cite{mura2013micromechanics} and numerical solutions of stress components $\sigma^s_{11}$ and $\sigma^s_{22}$ as a function of normalized distance (r/R) along the x-axis. The length is normalized by R, the precipitate radius, while the stress is normalized by the characteristic stress $C_{44}\varepsilon^T$. Different domain discritization (1024, 512, 256, and 128) are carried out to test the accuracy of the solver.}
        \label{fig:analytic_test}
    \end{figure}

    The principal stress components (normalized by $C_{44}\varepsilon^T$) as a function of normalized distance (r/R) from the centre of a circular precipitate along the x-axis is shown in Fig.~\ref{fig:analytic_test}a and \ref{fig:analytic_test}b. The precipitate is softer than the matrix phase and the materials are elastically isotropic ($A_z=1$) where the elastic constant contrast between the inclusion and matrix is 50\%. This is a sufficiently large difference comparing to typical two-phase systems. The eigenstrain ($\varepsilon^T$) value in the particle and matrix is taken to be 0.01 and 0, respectively. The radius of the particle in the center is taken to be 0.1 of the width of the square domain. The corresponding analytical solutions are also shown (using plus markers) for direct comparison. $\sigma^s_{11}$ is continuous throughout the particle-matrix interface, and $\sigma^s_{22}$ is discontinues at this interface. In addition, the mean absolute relative error between the mesh size of 1024 and 512 is 0.0083 for $\sigma^s_{11}$ values and 0.0047 for $\sigma^s_{22}$ values. Table~\ref{tab:error_mesh_size} summarizes the relative errors between different mesh sizes that are used in benchmarking with the reference mesh size (1024).
    
    %%%%%%%%%%%%%%%%%%
    %%%%%%%%%%%%%%%%%%  
    
    \begin{table}[h!]
        \centering
        %\scriptsize
        \begin{tabular}{lccc} \toprule
             Stress                          & Mesh size & Ref. mesh size          & Mean absolute relative error\\ \midrule
             \multirow{3}{*}{$\sigma^s_{11}$}  &  512      & \multirow{6}{*}{1024}   & 0.0083 \\
                                             &  256      &                         & 0.0380 \\
                                             &  128      &                         & 0.1254 \\
             \multirow{3}{*}{$\sigma^s_{22}$}  &  512      &                         & 0.0047 \\
                                             &  256      &                         & 0.0408 \\
                                             &  128      &                         & 0.0707 \\       \bottomrule        
        \end{tabular}
        \caption{The mean absolute relative error between different discretizations and the reference mesh size (1024).}
        \label{tab:error_mesh_size}
    \end{table}

   %%%%%%%%%%%%%%%%%%%%%%%%%%%%%%%%%%%%%%%%%%%%%%%%%%%%%%%
    %%%%%%%%%%%%%%%%%%%%%%%%%%%%%%%%%%%%%%%%%%%%%%%%%%%%%%%
    %%%%%%%%%%%%%%%%%%%%%%%%%%%%%%%%%%%%%%%%%%%%%%%%%%%%%%%
    %%%%%%%%%% Merge with supplemental materials %%%%%%%%%%
    \renewcommand\thefigure{\thesection.\arabic{figure}}    
    \setcounter{figure}{0} 
    \newcommand{\beginsupplement}{%
        \setcounter{table}{0}
        \renewcommand{\thetable}{S\arabic{table}}%
        \setcounter{figure}{0}
        \renewcommand{\thefigure}{S\arabic{figure}}%
     }

    \clearpage
    \newpage
    \onecolumn
    
    %\singlespacing
    \setstretch{0.1}
    \section*{Supplementary Results} 
    \beginsupplement
    
\begin{figure*}[!ht]
        \centering
        %% 0.25
        \raisebox{1.6cm}{$x_{Al} = 0.25$}
        \subfloat[]{\scalebarbackground{\includegraphics[scale=0.045]{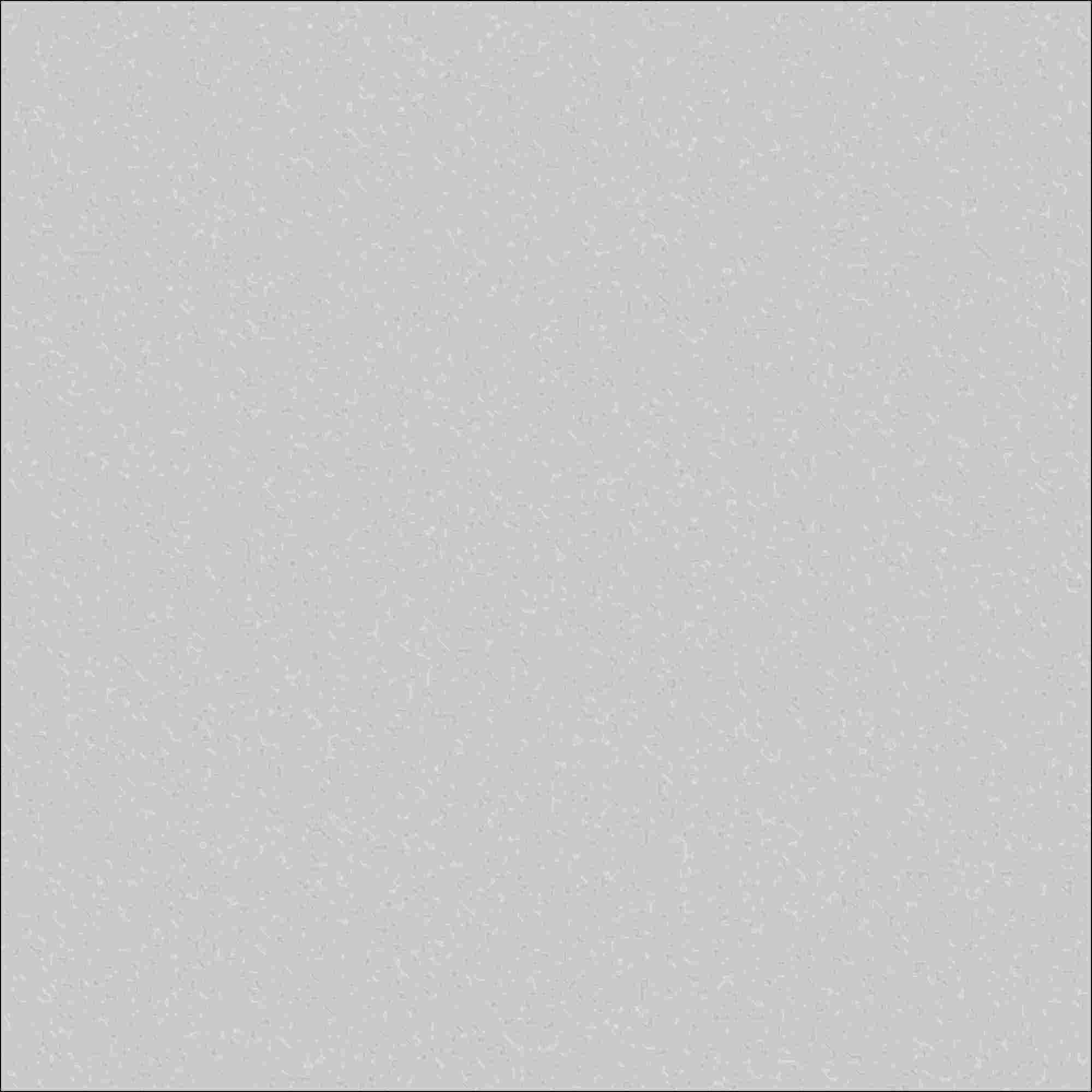}}{495}{9.887}{10} } \vspace{-0.1cm}
        \subfloat[]{\scalebarbackground{\includegraphics[scale=0.045]{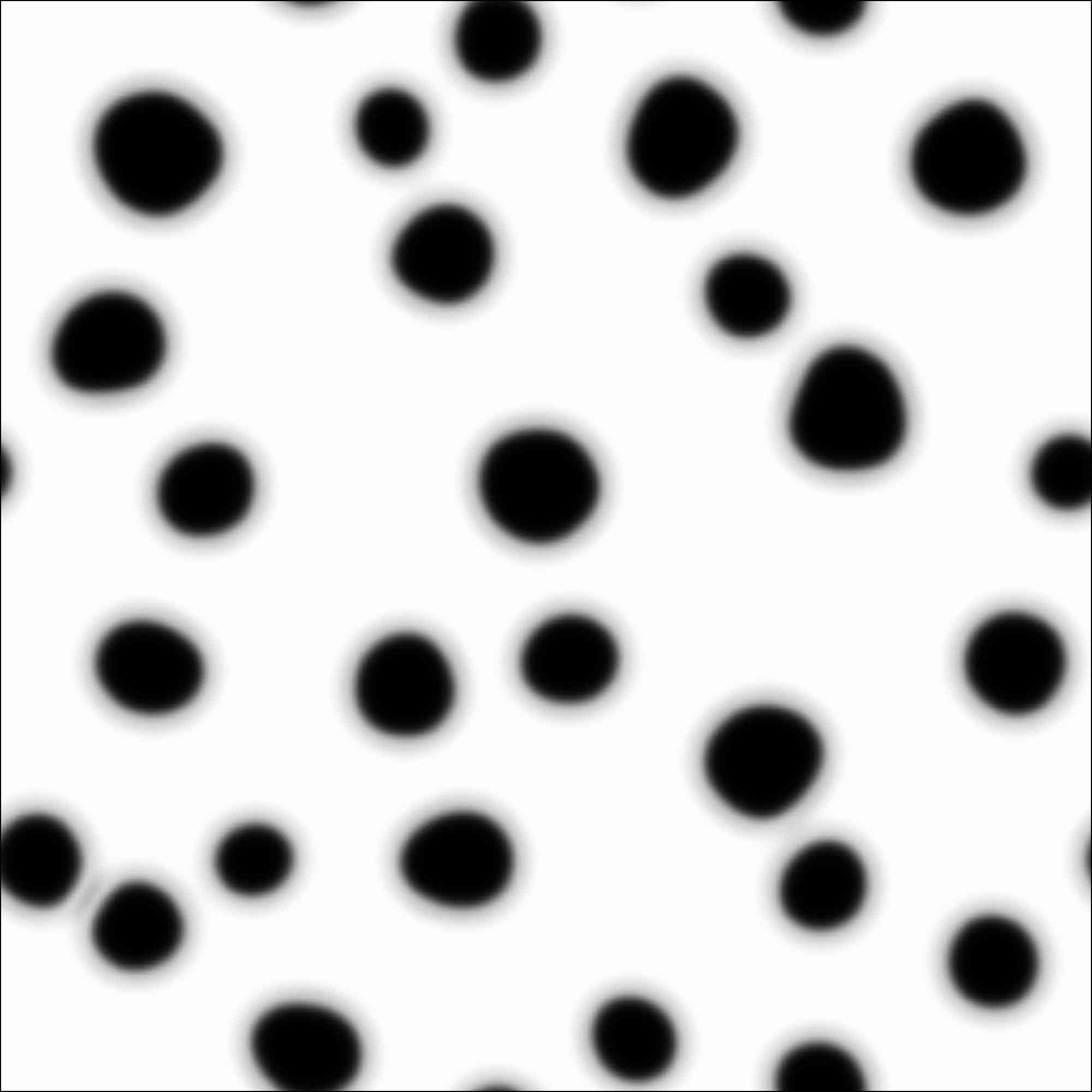}}{495}{9.887}{10} } \vspace{-0.1cm}
        \subfloat[]{\scalebarbackground{\includegraphics[scale=0.045]{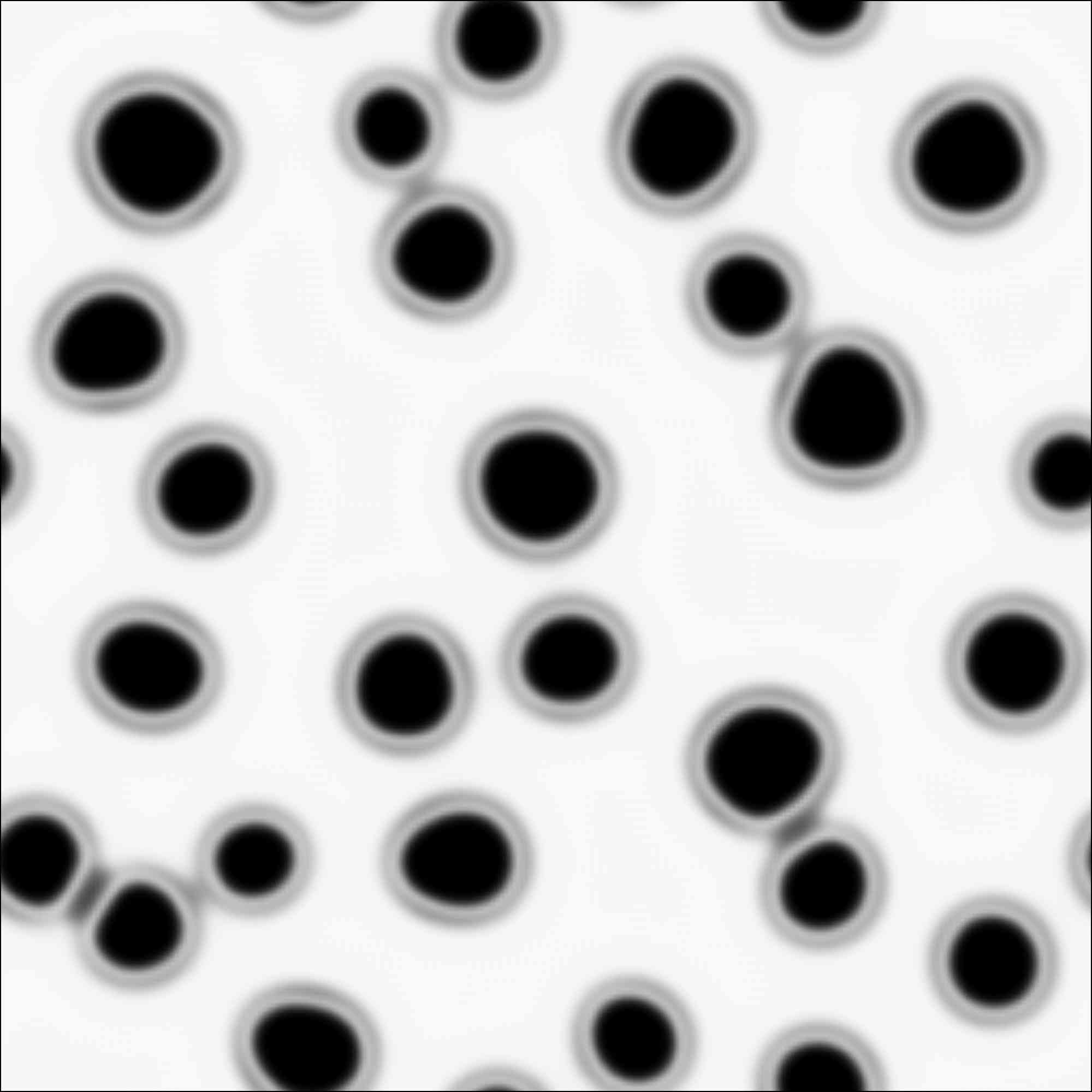}}{495}{9.887}{10} } \vspace{-0.1cm}
        \subfloat[]{\scalebarbackground{\includegraphics[scale=0.045]{images/case5_chem/0.25_0.05/phi3_00500.jpeg}}{495}{9.887}{10} }\\ \vspace{-0.1cm}
        %% 0.30
        \raisebox{1.6cm}{$x_{Al} = 0.30$}
        \subfloat[]{\scalebarbackground{\includegraphics[scale=0.045]{images/case5_chem/0.30_0.05/all_merged_phi3_1.jpeg}}{495}{9.887}{10} } \vspace{-0.1cm}
        \subfloat[]{\scalebarbackground{\includegraphics[scale=0.045]{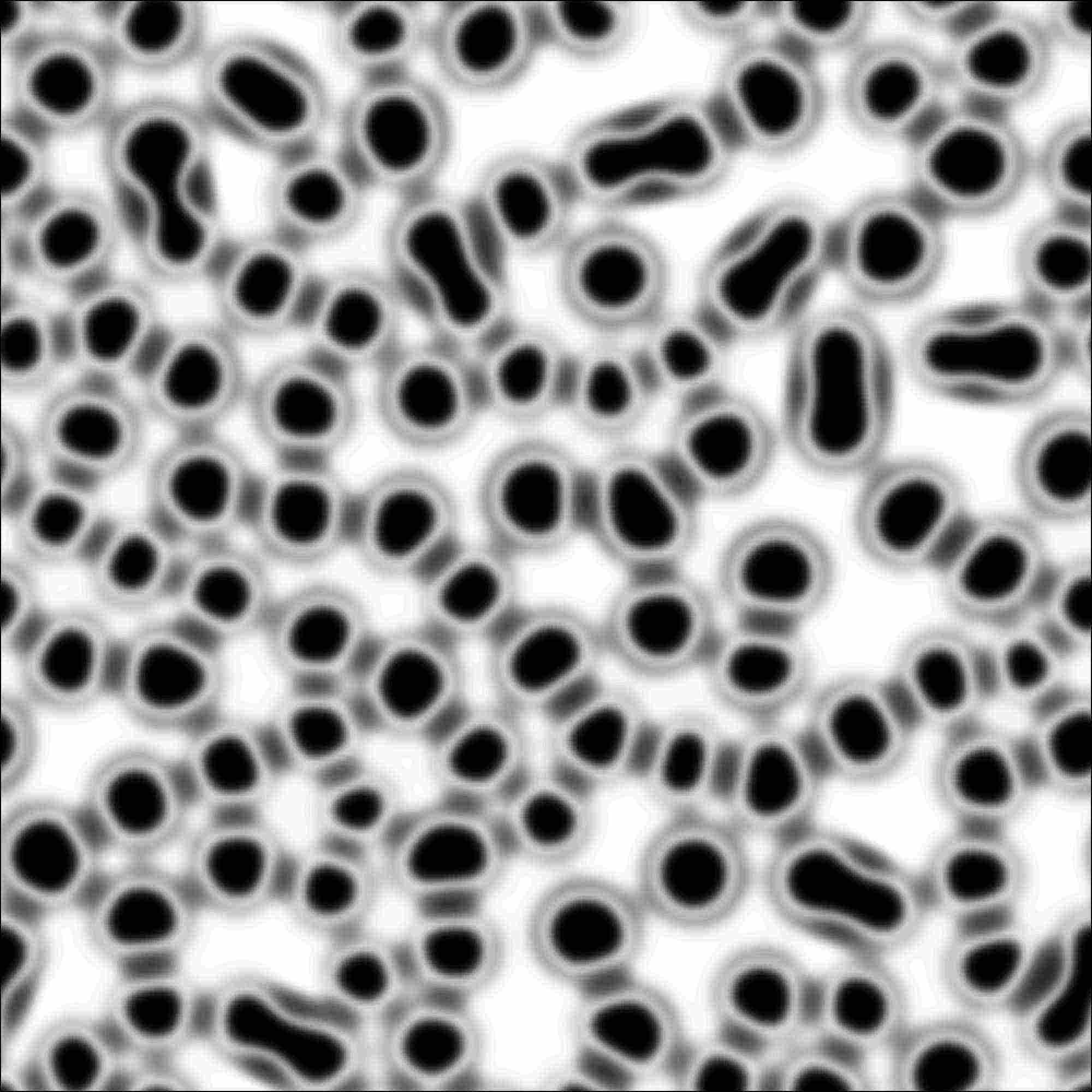}}{495}{9.887}{10} } \vspace{-0.1cm}
        \subfloat[]{\scalebarbackground{\includegraphics[scale=0.045]{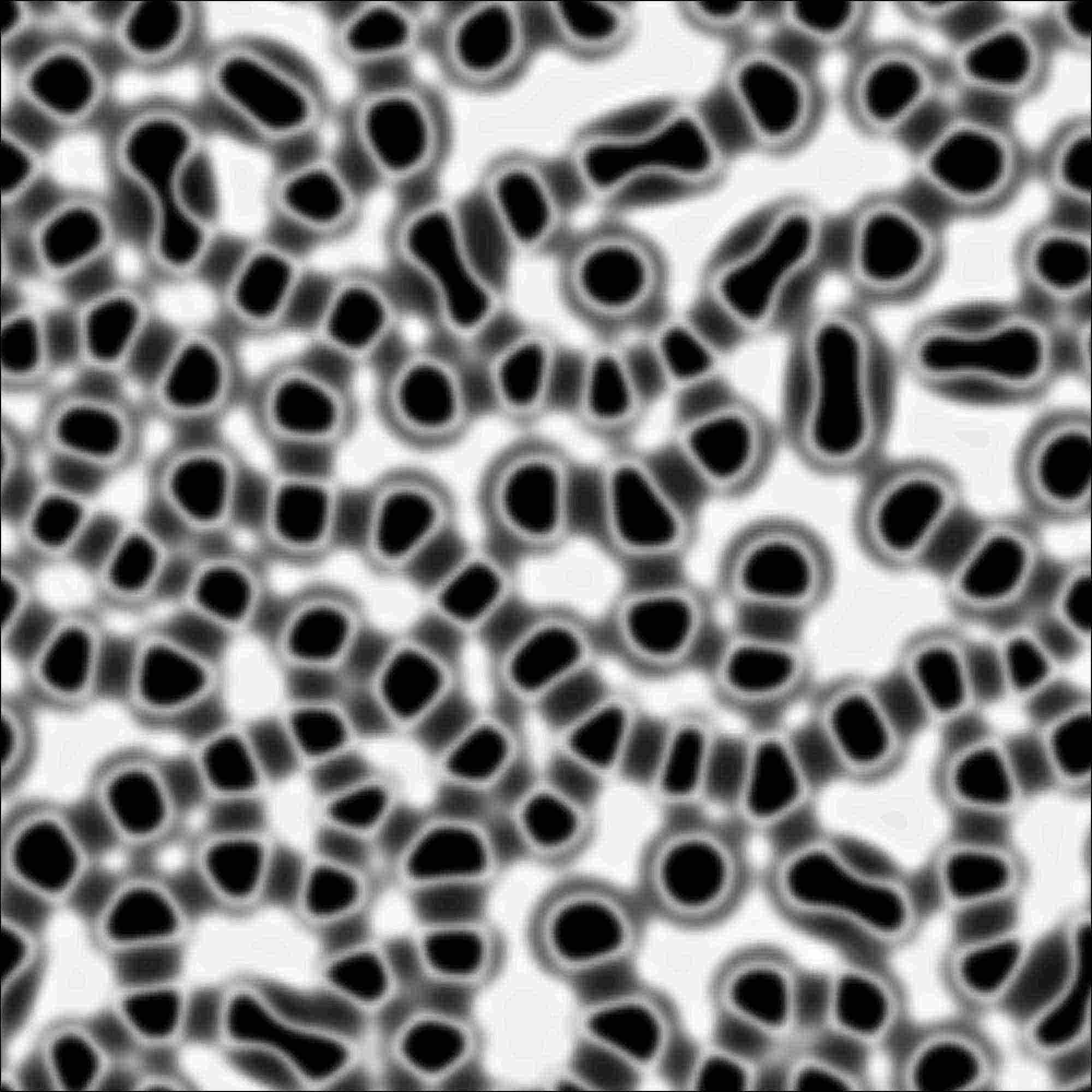}}{495}{9.887}{10} } \vspace{-0.1cm}
        \subfloat[]{\scalebarbackground{\includegraphics[scale=0.045]{images/case5_chem/0.30_0.05/all_merged_phi3_15.jpeg}}{495}{9.887}{10} } \\ \vspace{-0.1cm}
        %% 0.35
        \raisebox{1.6cm}{$x_{Al} = 0.35$}
        \subfloat[]{\scalebarbackground{\includegraphics[scale=0.045]{images/case5_chem/0.30_0.05/all_merged_phi3_1.jpeg}}{495}{9.887}{10} } \vspace{-0.1cm}
        \subfloat[]{\scalebarbackground{\includegraphics[scale=0.045]{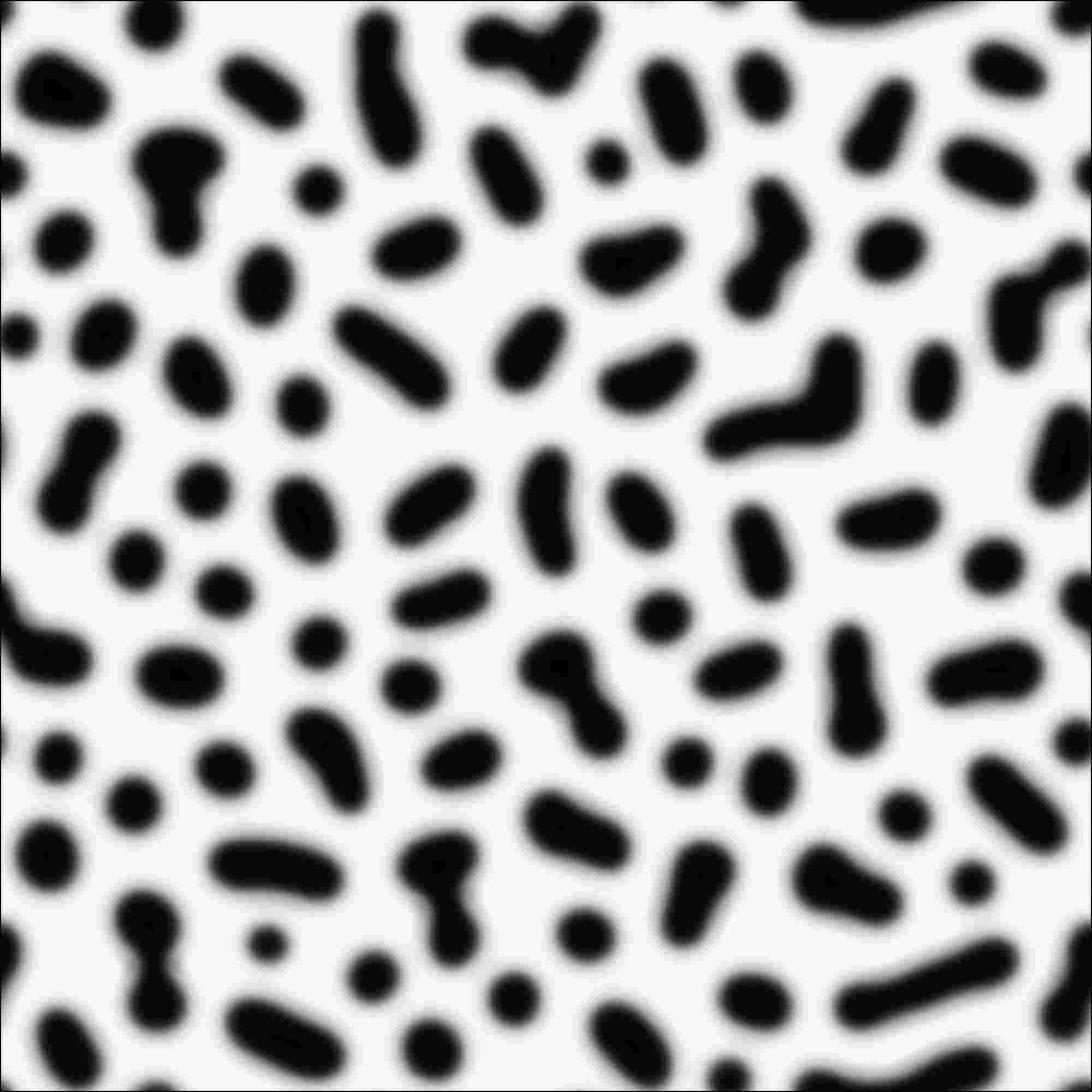}}{495}{9.887}{10} } \vspace{-0.1cm}
        \subfloat[]{\scalebarbackground{\includegraphics[scale=0.045]{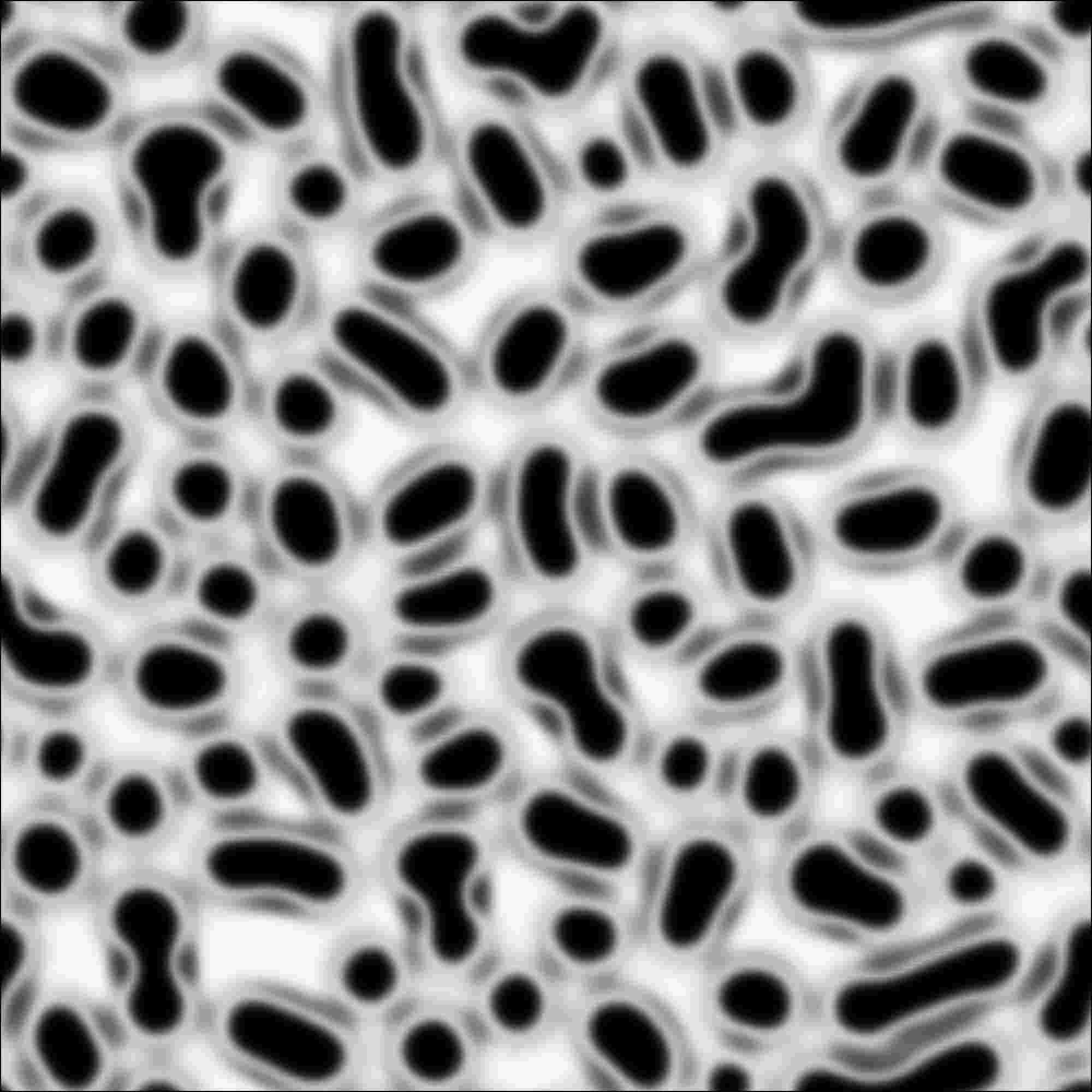}}{495}{9.887}{10} } \vspace{-0.1cm}
        \subfloat[]{\scalebarbackground{\includegraphics[scale=0.045]{images/case5_chem/0.35_0.05/phi3_00515.jpeg}}{495}{9.887}{10} } \\ \vspace{-0.1cm}
        %% 0.40
        \raisebox{1.6cm}{$x_{Al} = 0.40$}
        \subfloat[]{\scalebarbackground{\includegraphics[scale=0.045]{images/case5_chem/0.30_0.05/all_merged_phi3_1.jpeg}}{495}{9.887}{10} } \vspace{-0.1cm}
        \subfloat[]{\scalebarbackground{\includegraphics[scale=0.045]{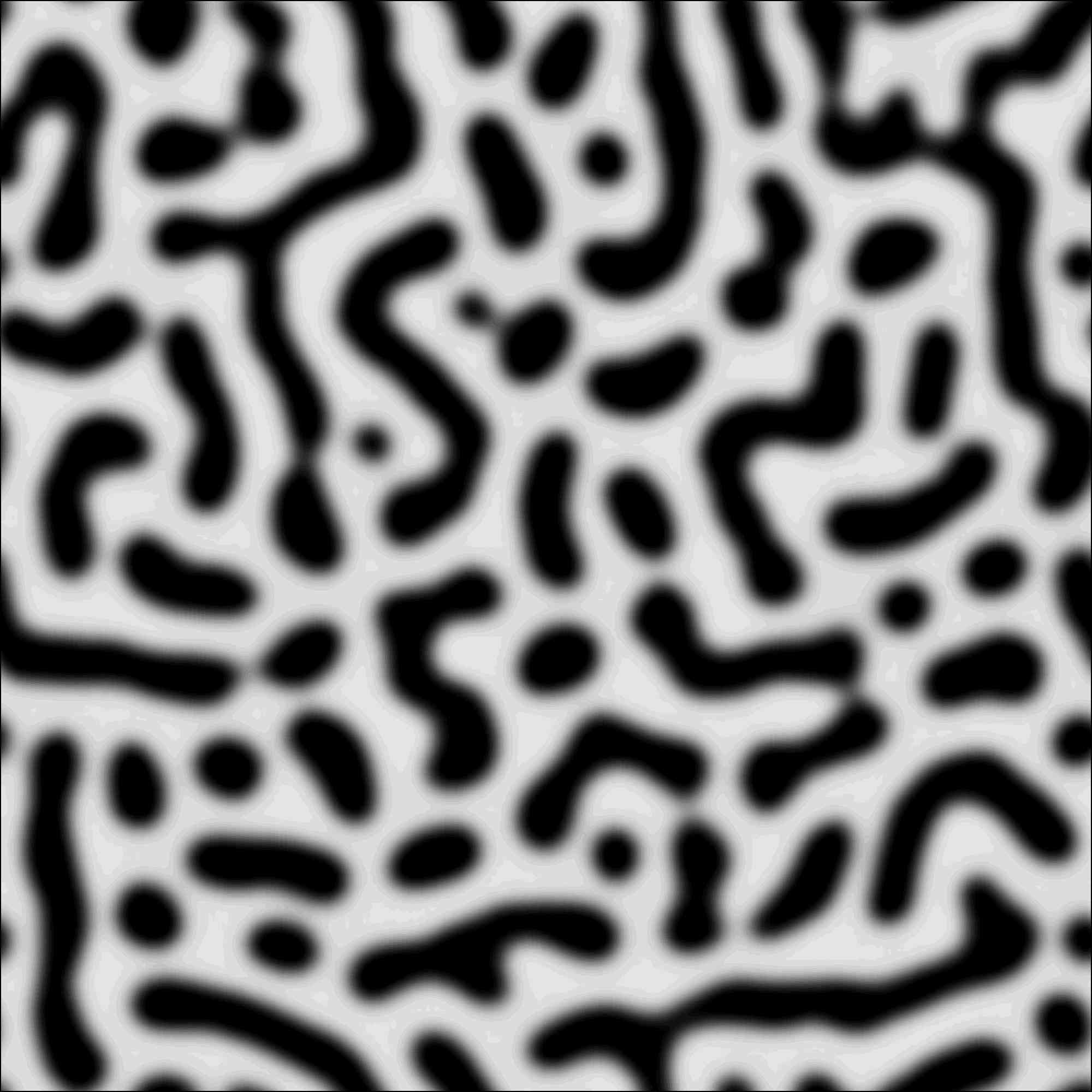}}{495}{9.887}{10} } \vspace{-0.1cm}
        \subfloat[]{\scalebarbackground{\includegraphics[scale=0.045]{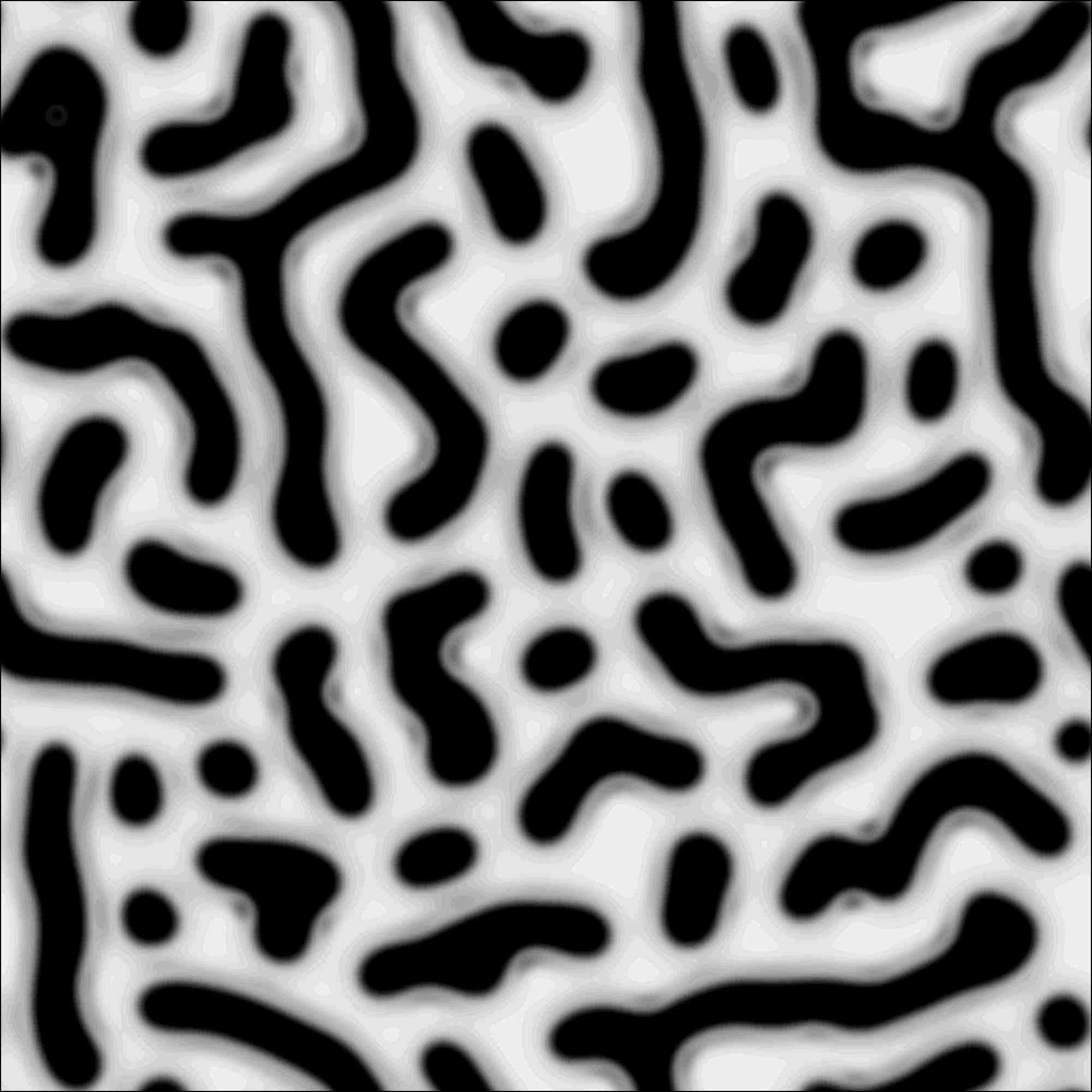}}{495}{9.887}{10} } \vspace{-0.1cm}
        \subfloat[]{\scalebarbackground{\includegraphics[scale=0.045]{images/case5_chem/0.40_0.05/phi3_00251.jpeg}}{495}{9.887}{10} } \\ \vspace{-0.1cm}
        %% 0.45
        \raisebox{1.6cm}{$x_{Al} = 0.45$}
        \subfloat[]{\scalebarbackground{\includegraphics[scale=0.045]{images/case5_chem/0.30_0.05/all_merged_phi3_1.jpeg}}{495}{9.887}{10} } \vspace{-0.1cm}
        \subfloat[]{\scalebarbackground{\includegraphics[scale=0.045]{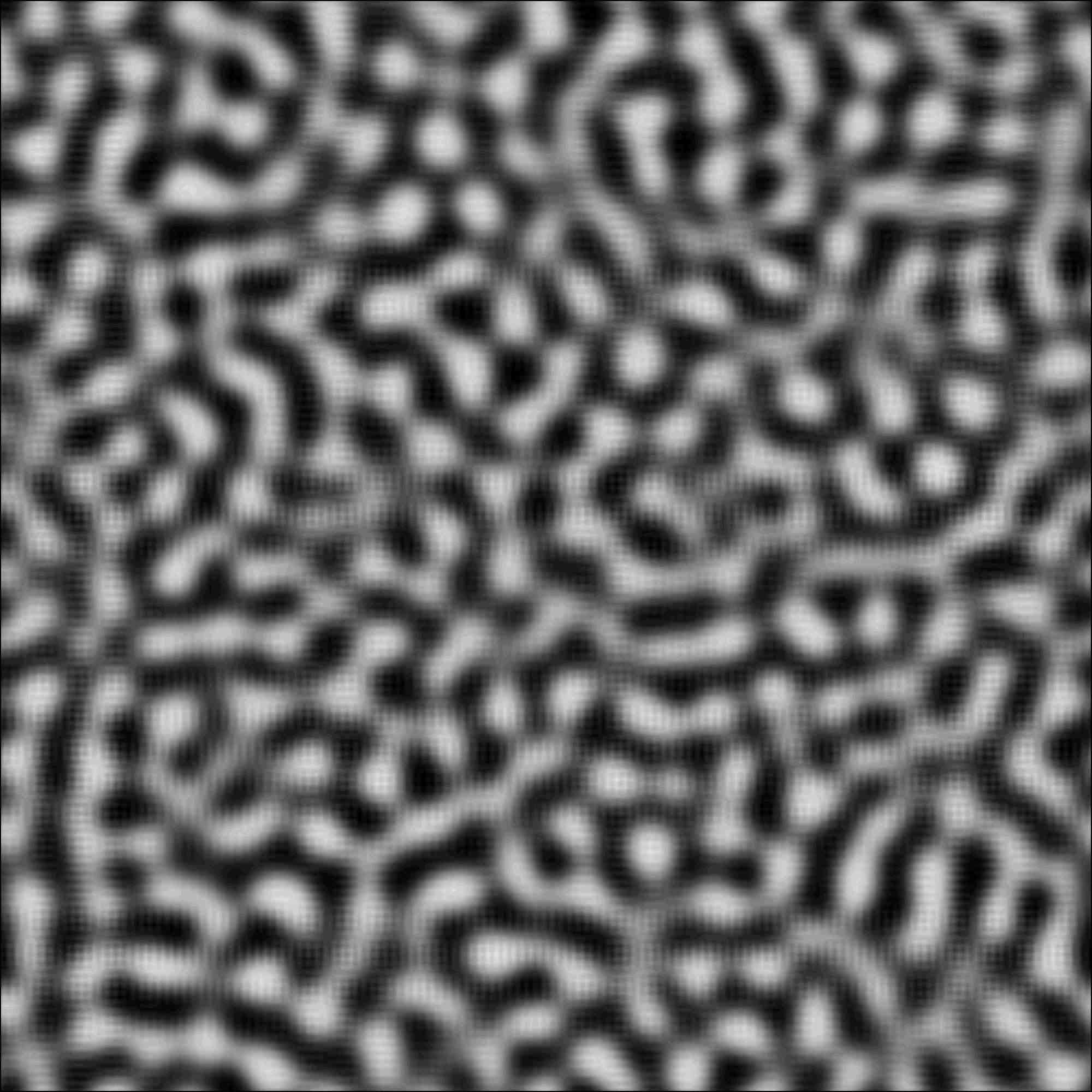}}{495}{9.887}{10} } \vspace{-0.1cm}
        \subfloat[]{\scalebarbackground{\includegraphics[scale=0.045]{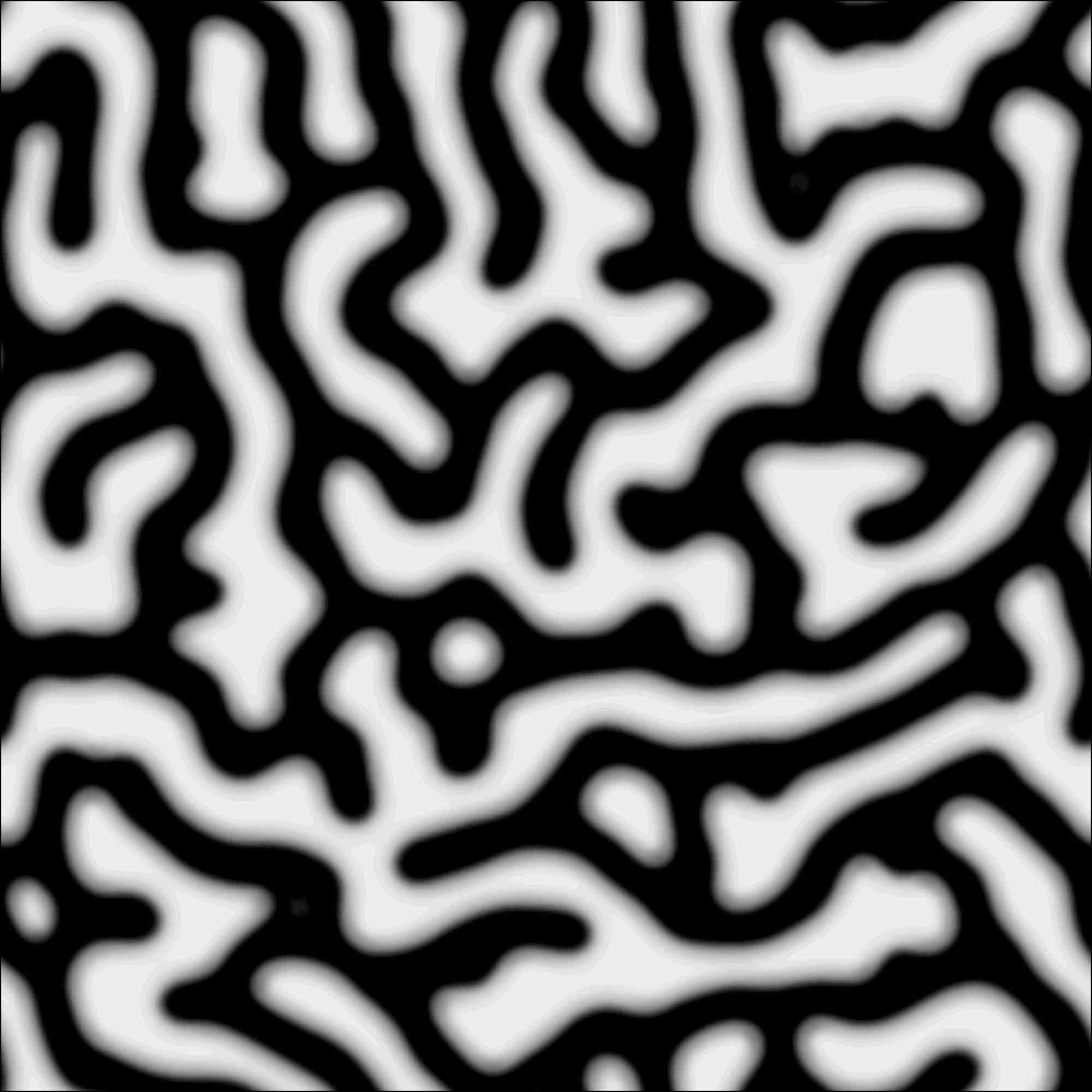}}{495}{9.887}{10} } \vspace{-0.1cm}
        \subfloat[]{\scalebarbackground{\includegraphics[scale=0.045]{images/case5_chem/0.45_0.05/phi3_00500.jpeg}}{495}{9.887}{10} } \\ \vspace{-0.1cm}
        
        \caption{Evolution of the microstructure of Ti$_{1-x-0.05}$Al$_x$Zr$_{0.05}$N system for different Al content during annealing at T=1200$^\circ$C. Micrographs in each column do not correspond to the same time. Each row correspond to an specifc Al composition shown to the left of each set. The evolution of the microstructures over time demonstrates that AlN and TiN phases segregate first, and ZrN appears later.} 
        \label{fig:time_evolution_1}
    \end{figure*}

\begin{figure*}[h!]
        \centering
        
        %% 0.50
        \raisebox{1.5cm}{$x_{Al} = 0.50$}
        \subfloat[]{\scalebarbackground{\includegraphics[scale=0.050]{images/case5_chem/0.30_0.05/all_merged_phi3_1.jpeg}}{495}{9.887}{10} } \vspace{-0.1cm}
        \subfloat[]{\scalebarbackground{\includegraphics[scale=0.050]{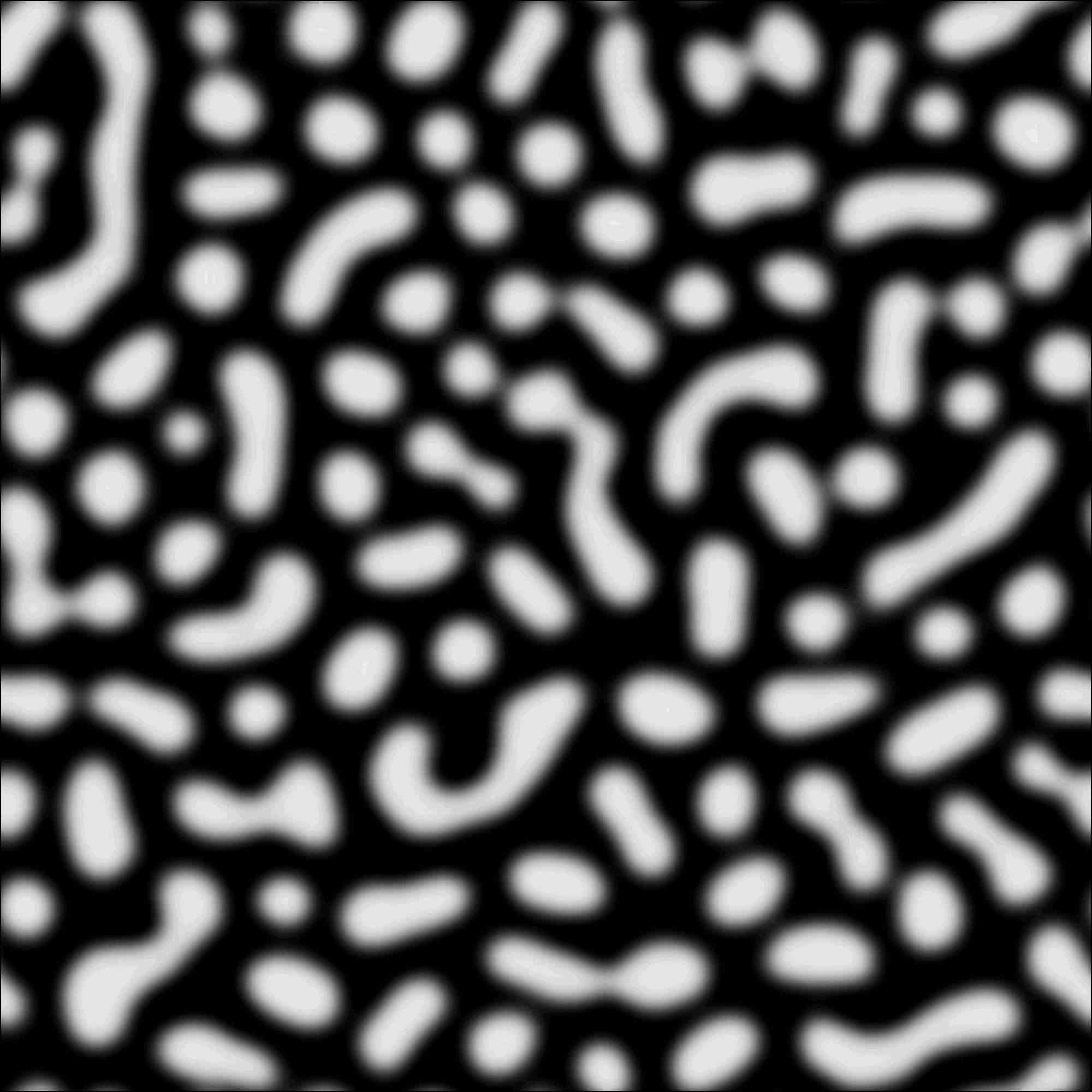}}{495}{9.887}{10} } \vspace{-0.1cm}
        \subfloat[]{\scalebarbackground{\includegraphics[scale=0.050]{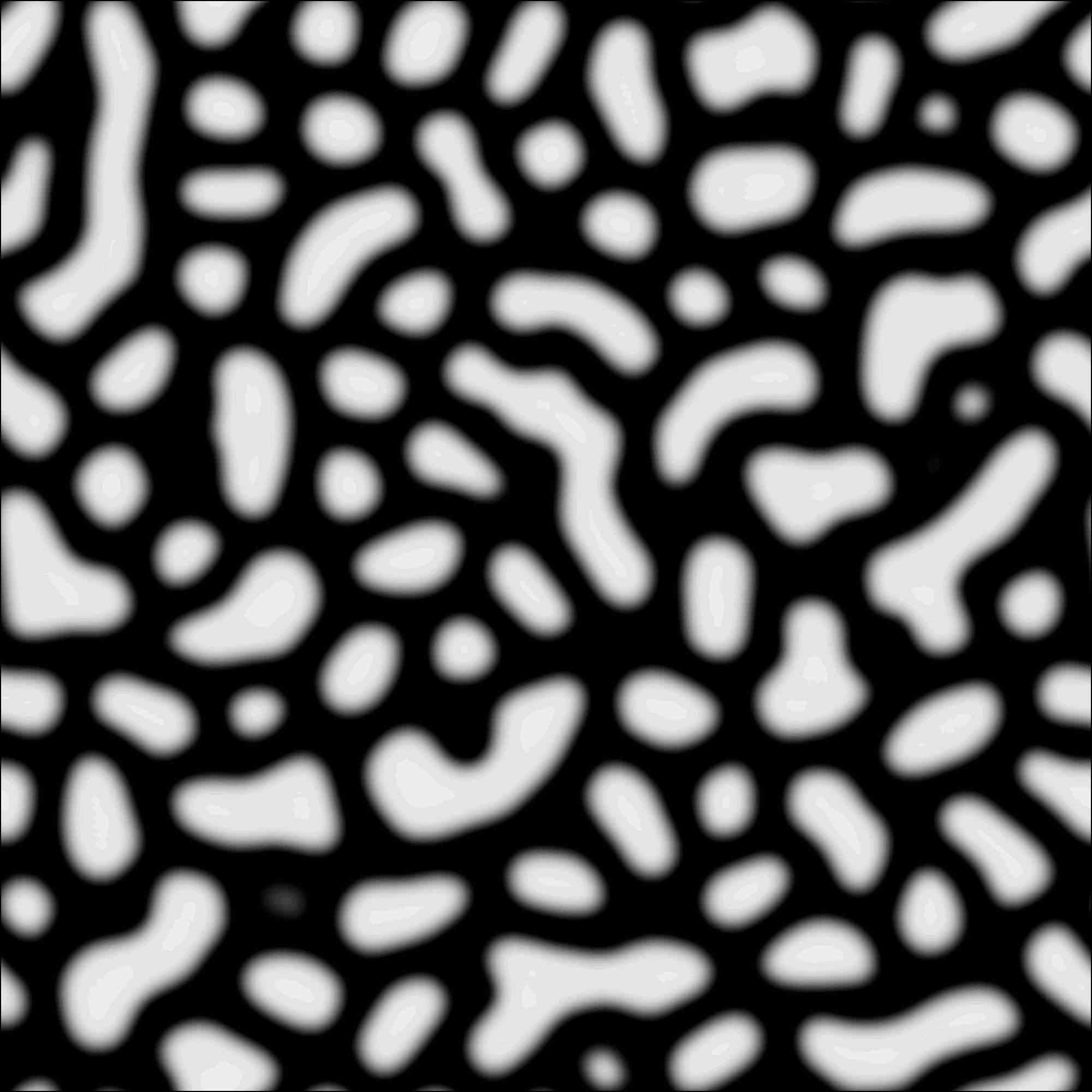}}{495}{9.887}{10} } \vspace{-0.1cm}
        \subfloat[]{\scalebarbackground{\includegraphics[scale=0.050]{images/case5_chem/0.50_0.05/phi3_00100_dark.jpeg}}{495}{9.887}{10} } \\ \vspace{-0.1cm}
        %% 0.55
        \raisebox{1.5cm}{$x_{Al} = 0.55$}
        \subfloat[]{\scalebarbackground{\includegraphics[scale=0.050]{images/case5_chem/0.30_0.05/all_merged_phi3_1.jpeg}}{495}{9.887}{10} } \vspace{-0.1cm}
        \subfloat[]{\scalebarbackground{\includegraphics[scale=0.050]{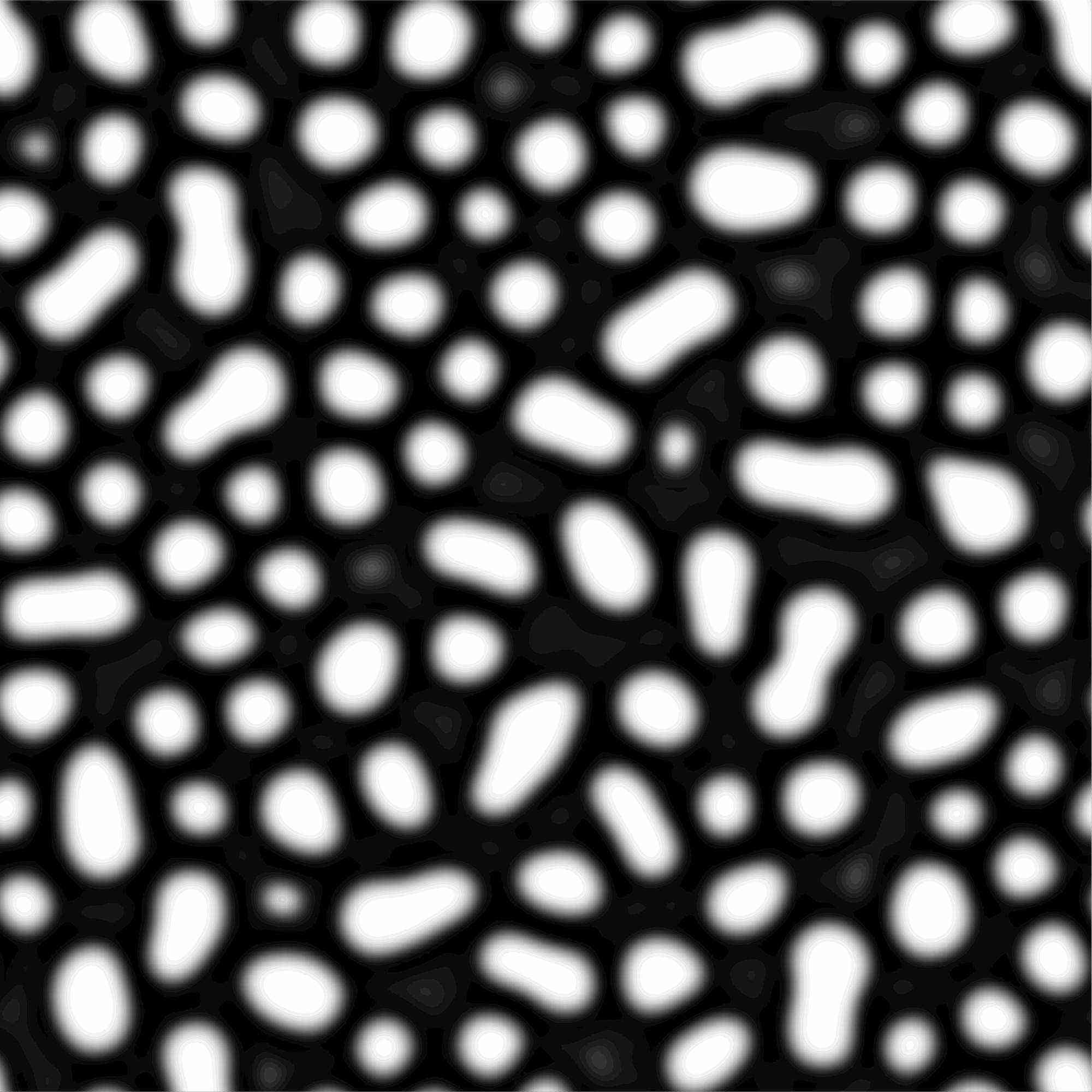}}{495}{9.887}{10} } \vspace{-0.1cm}
        \subfloat[]{\scalebarbackground{\includegraphics[scale=0.050]{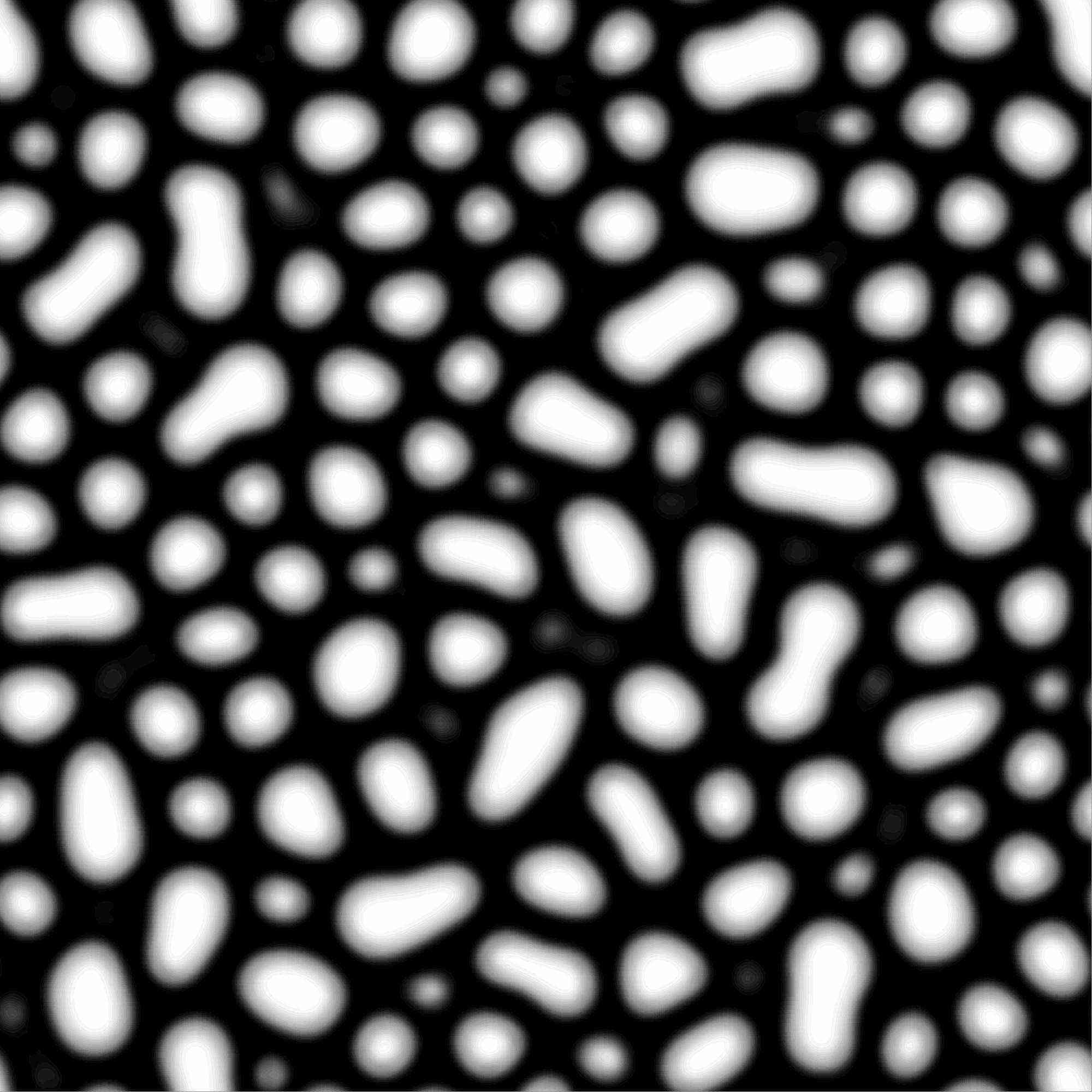}}{495}{9.887}{10} } \vspace{-0.1cm}
        \subfloat[]{\scalebarbackground{\includegraphics[scale=0.050]{images/case5_chem/0.55_0.05/phi3_00100_dark.jpeg}}{495}{9.887}{10} } \\ \vspace{-0.1cm}
        %% 0.60
        \raisebox{1.5cm}{$x_{Al} = 0.60$}
        \subfloat[]{\scalebarbackground{\includegraphics[scale=0.050]{images/case5_chem/0.30_0.05/all_merged_phi3_1.jpeg}}{495}{9.887}{10} } \vspace{-0.1cm}
        \subfloat[]{\scalebarbackground{\includegraphics[scale=0.050]{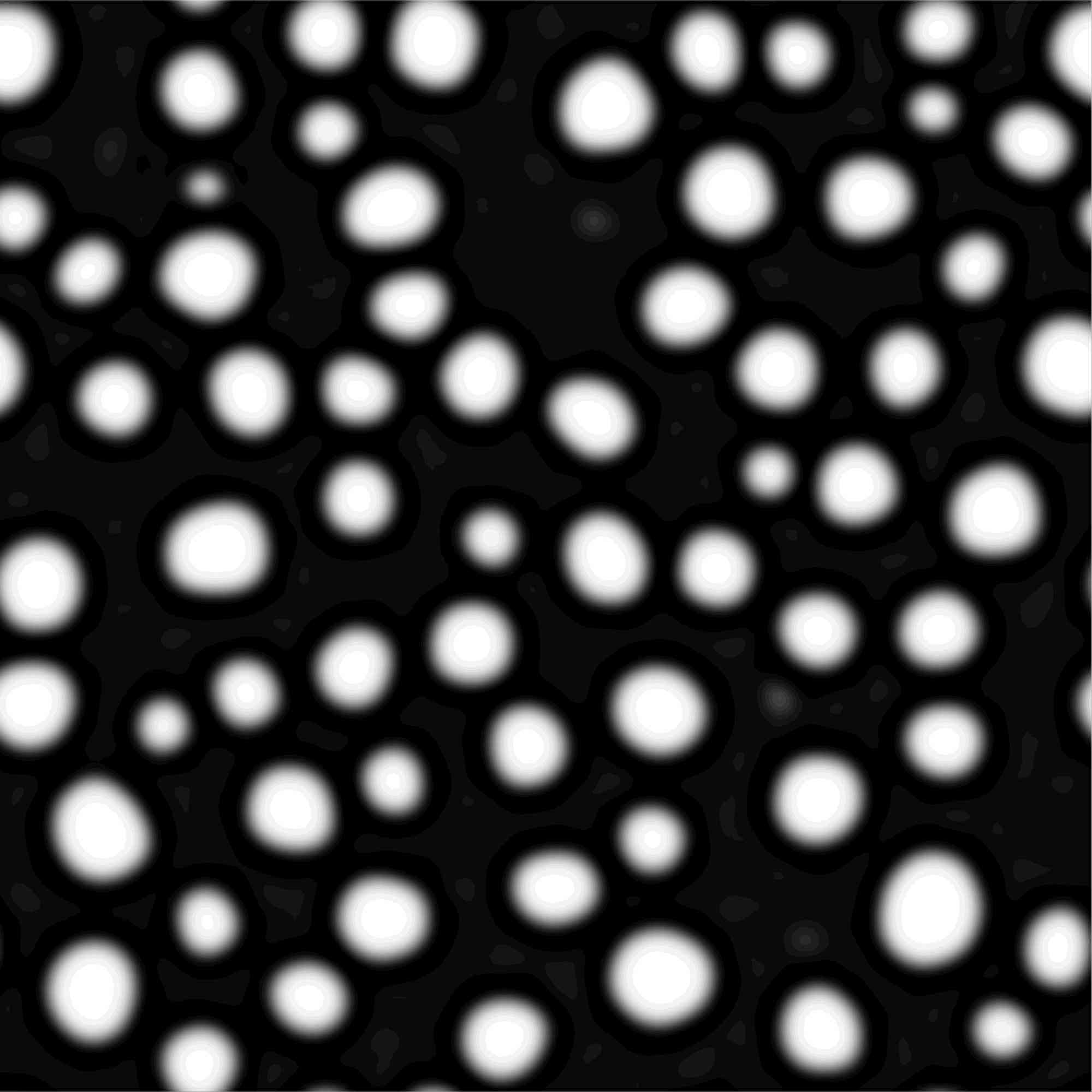}}{495}{9.887}{10} } \vspace{-0.1cm}
        \subfloat[]{\scalebarbackground{\includegraphics[scale=0.050]{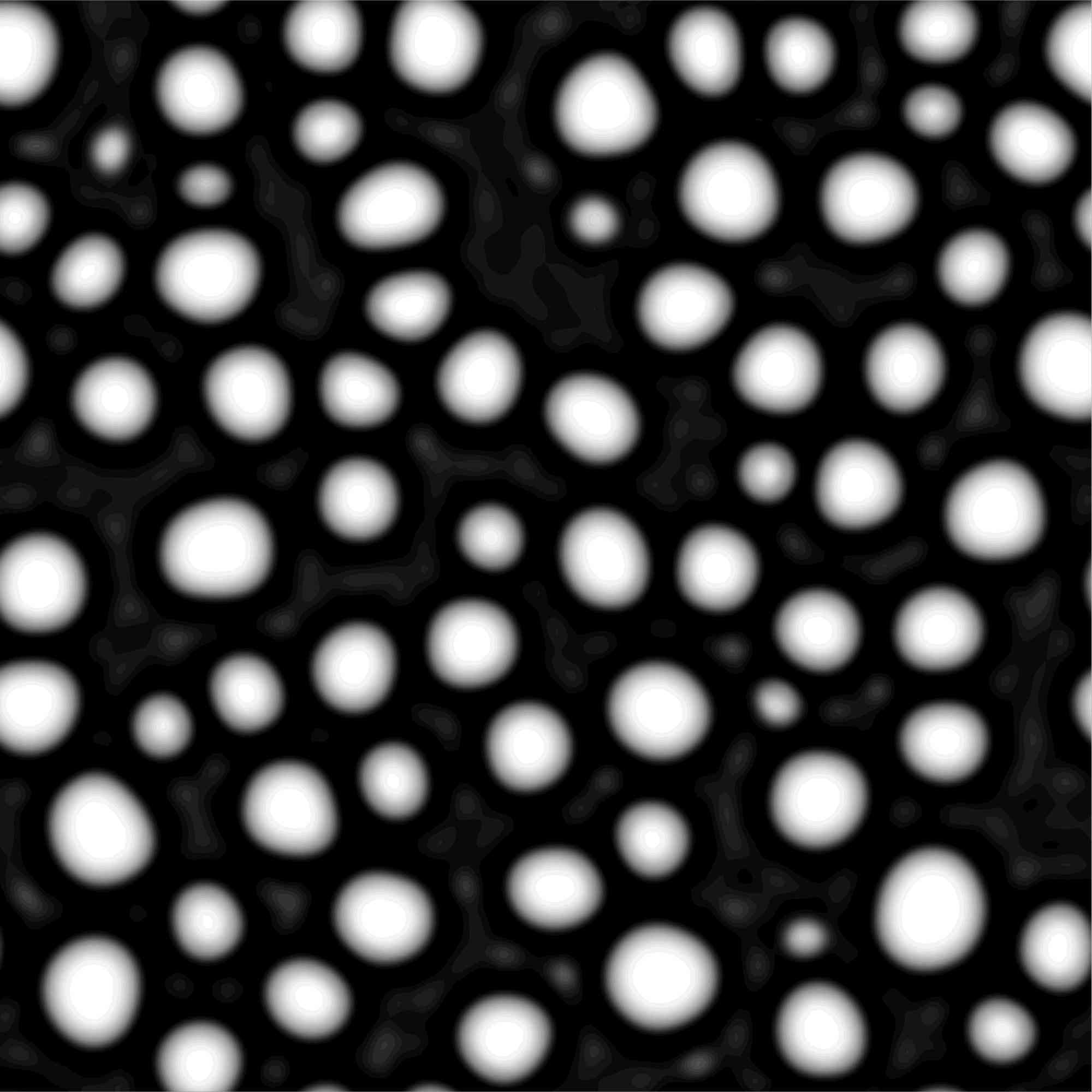}}{495}{9.887}{10} } \vspace{-0.1cm}
        \subfloat[]{\scalebarbackground{\includegraphics[scale=0.050]{images/case5_chem/0.60_0.05/phi3_00100_dark.jpeg}}{495}{9.887}{10} } \\ \vspace{-0.1cm}
        %% 0.65
        \raisebox{1.5cm}{$x_{Al} = 0.65$}
        \subfloat[]{\scalebarbackground{\includegraphics[scale=0.050]{images/case5_chem/0.30_0.05/all_merged_phi3_1.jpeg}}{495}{9.887}{10} }  \vspace{-0.1cm}
        \subfloat[]{\scalebarbackground{\includegraphics[scale=0.050]{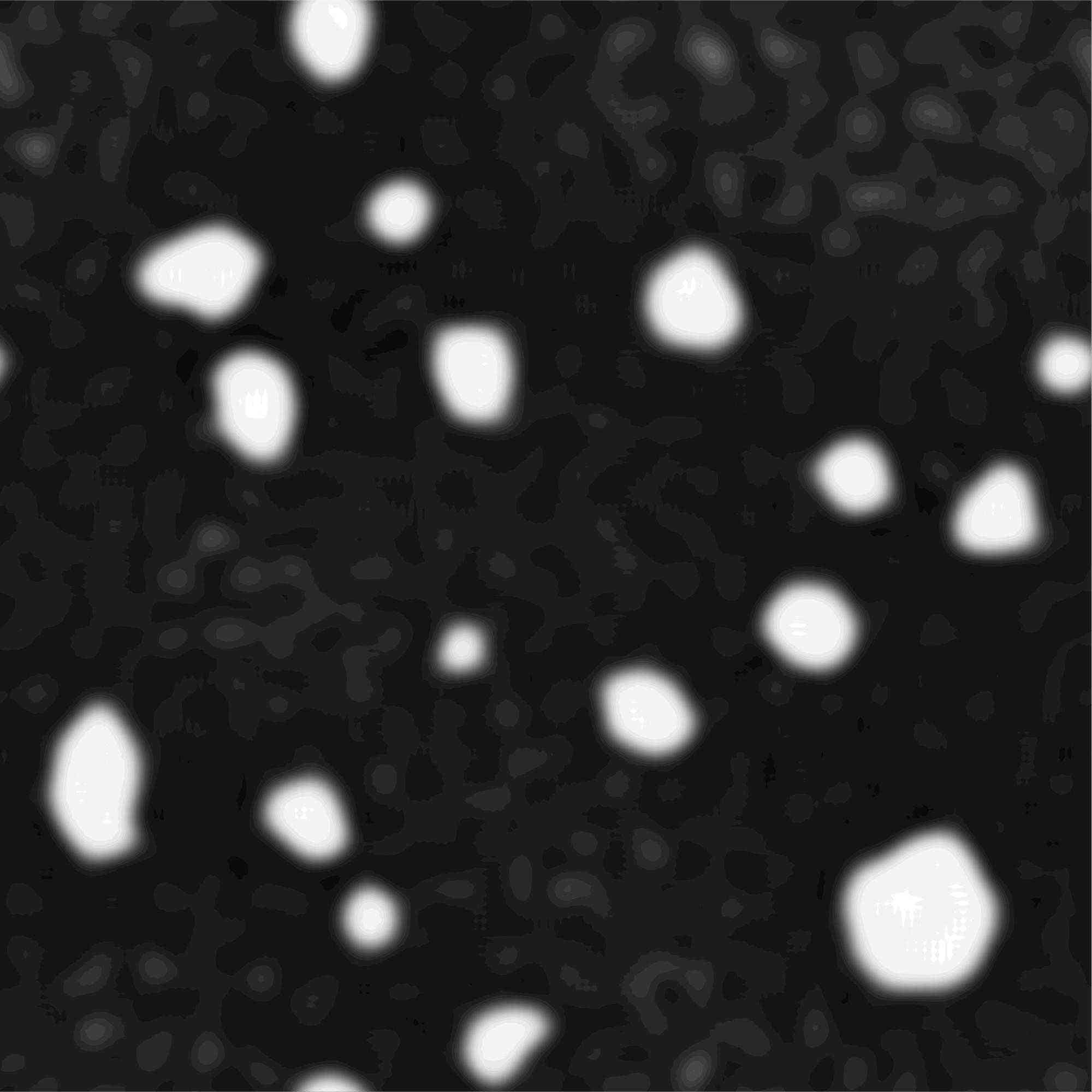}}{495}{9.887}{10} }  \vspace{-0.1cm}
        \subfloat[]{\scalebarbackground{\includegraphics[scale=0.050]{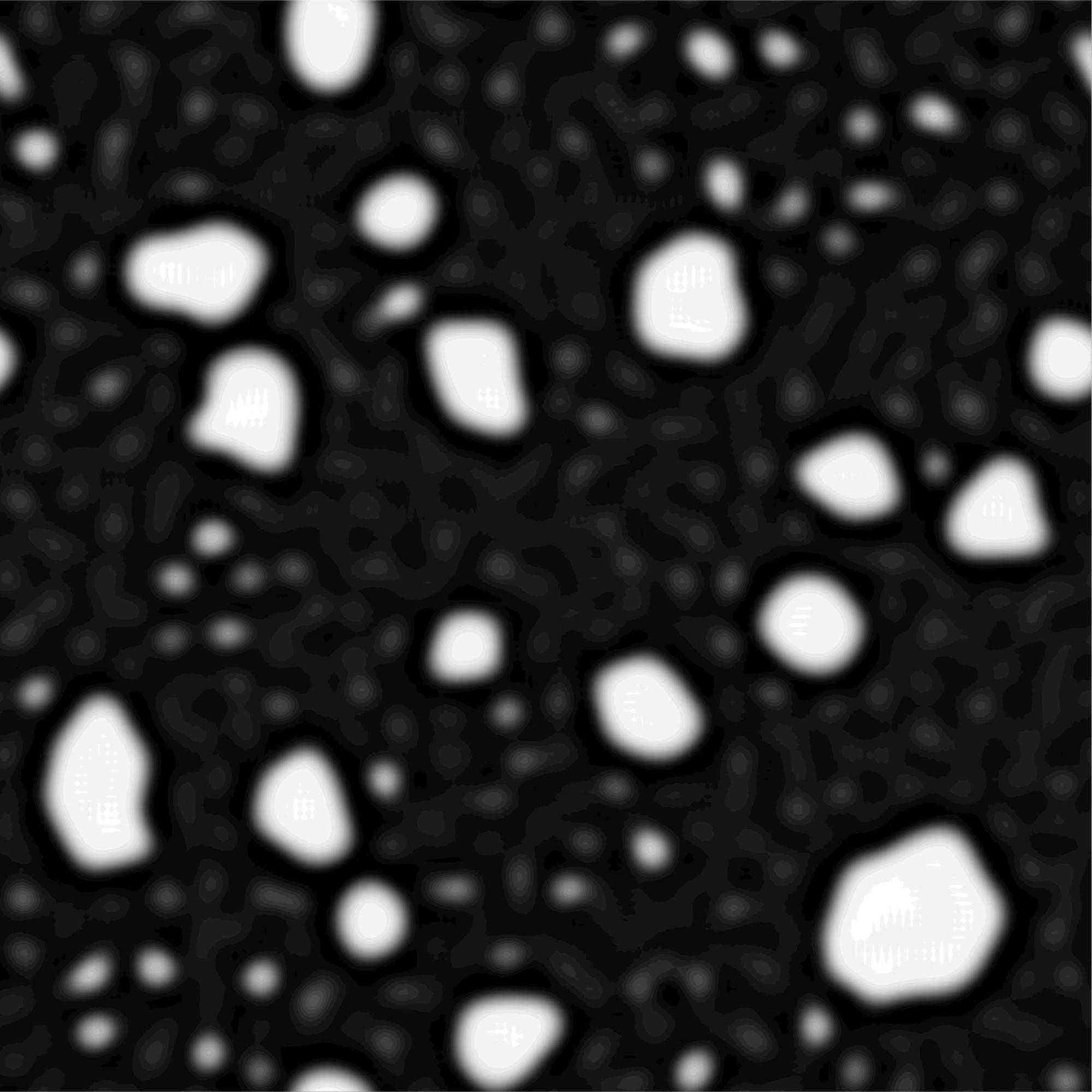}}{495}{9.887}{10} }  \vspace{-0.1cm}
        \subfloat[]{\scalebarbackground{\includegraphics[scale=0.050]{images/case5_chem/0.65_0.05/phi3_00100_dark.jpeg}}{495}{9.887}{10} } \\  \vspace{-0.1cm}        
        %% 0.70
        \raisebox{1.5cm}{$x_{Al} = 0.70$}
        \subfloat[]{\scalebarbackground{\includegraphics[scale=0.050]{images/case5_chem/0.30_0.05/all_merged_phi3_1.jpeg}}{495}{9.887}{10} }  \vspace{-0.1cm}
        \subfloat[]{\scalebarbackground{\includegraphics[scale=0.050]{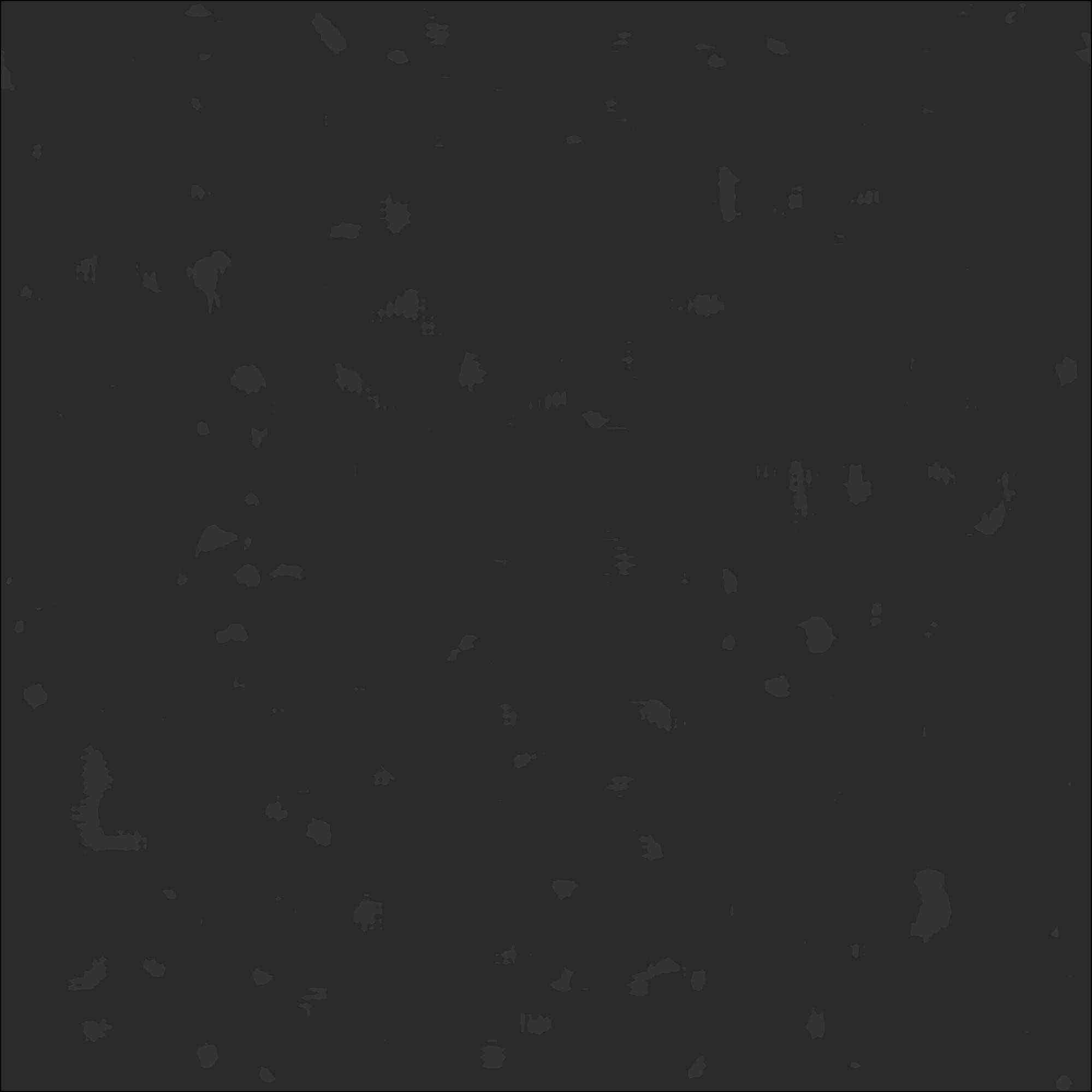}}{495}{9.887}{10} }  \vspace{-0.1cm}
        \subfloat[]{\scalebarbackground{\includegraphics[scale=0.050]{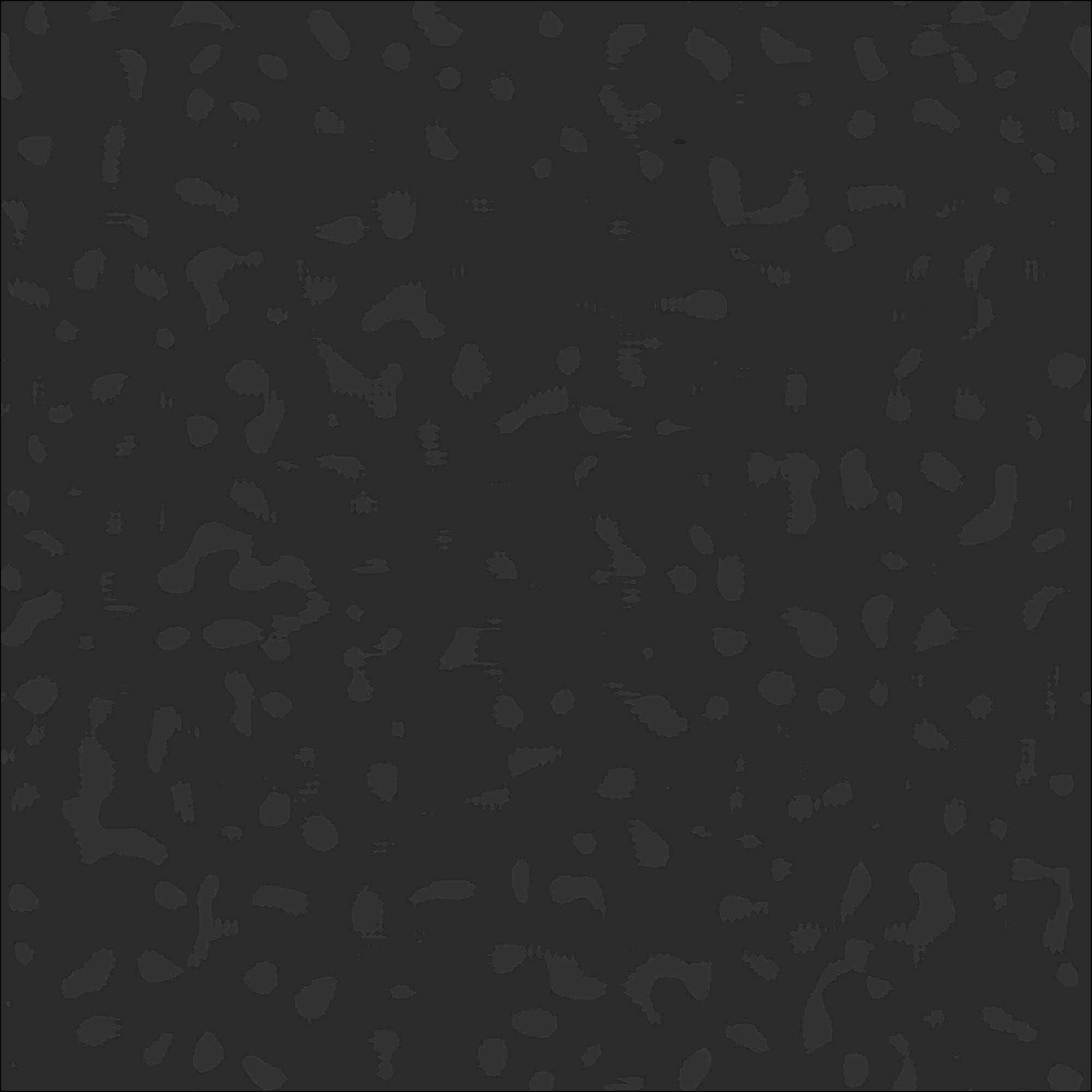}}{495}{9.887}{10} }  \vspace{-0.1cm}
        \subfloat[]{\scalebarbackground{\includegraphics[scale=0.050]{images/case5_chem/0.70_0.05/phi3_00100.jpeg}}{495}{9.887}{10} } \\  \vspace{-0.1cm}
        
        \caption{Evolution of microstructure of Ti$_{1-x-0.05}$Al$_x$Zr$_{0.05}$N system for different Al content during annealing at T=1200$^\circ$C. Here, each column correspond to the same time, and each row correspond to an specific Al composition shown to the left of each set. The evolution of the microstructures over time demonstrates that AlN and TiN phases segregate first, and ZrN appears later in $x_{Al}\leq0.65$ range. In the case of $x_{Al}=0.7$, ZrN segregates first in the AlN matrix and TiN forms in the grain boundaries of ZrN.} 
        \label{fig:time_evolution_2}
    \end{figure*}

\begin{figure*}
        \centering
        \subfloat[$x_{Al}=0.25$]{\includegraphics[scale=0.085]{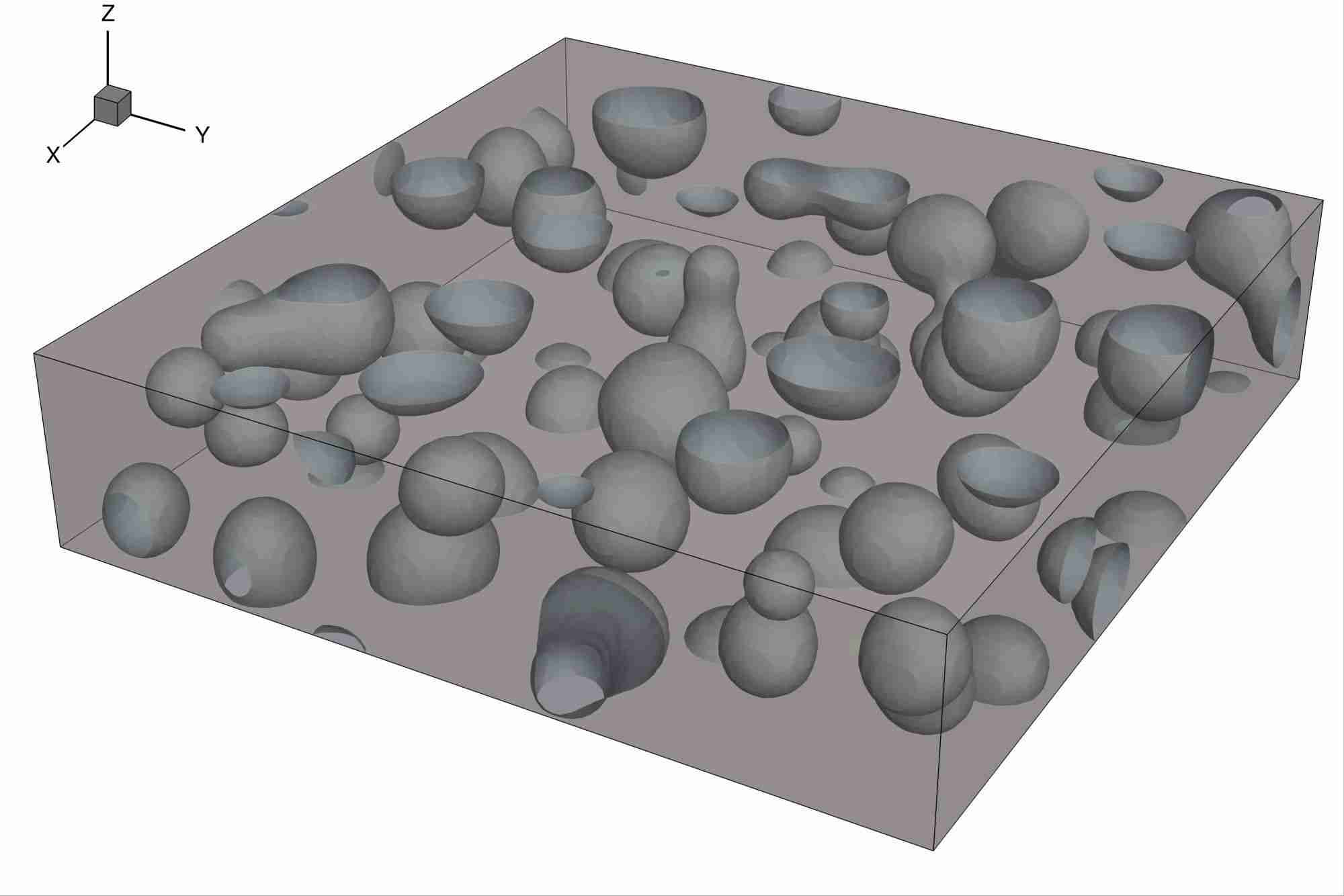}} \vspace{-0.16cm}
        \subfloat[$x_{Al}=0.50$]{\includegraphics[scale=0.085]{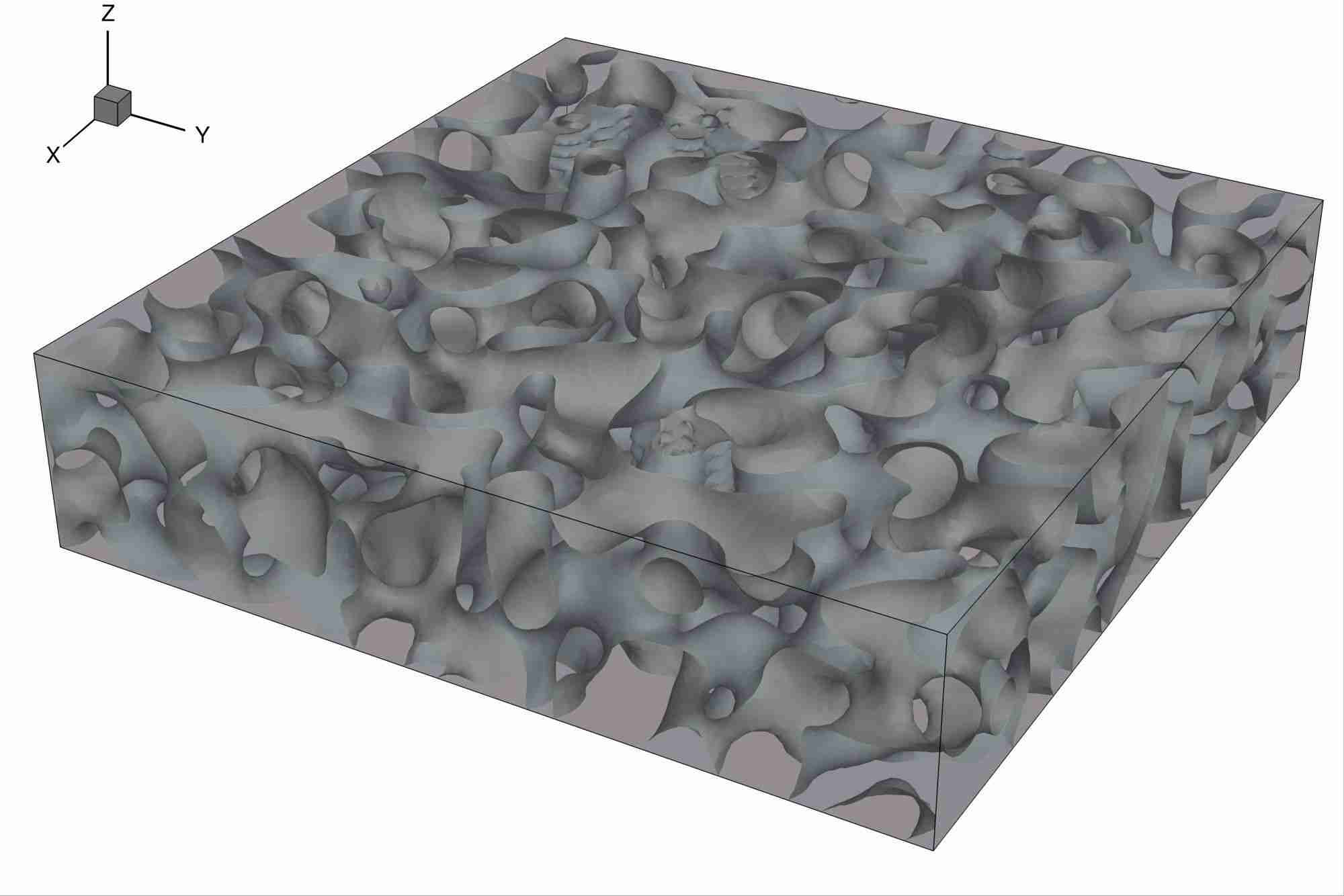}} \\ \vspace{-0.16cm}
        \subfloat[$x_{Al}=0.30$]{\includegraphics[scale=0.085]{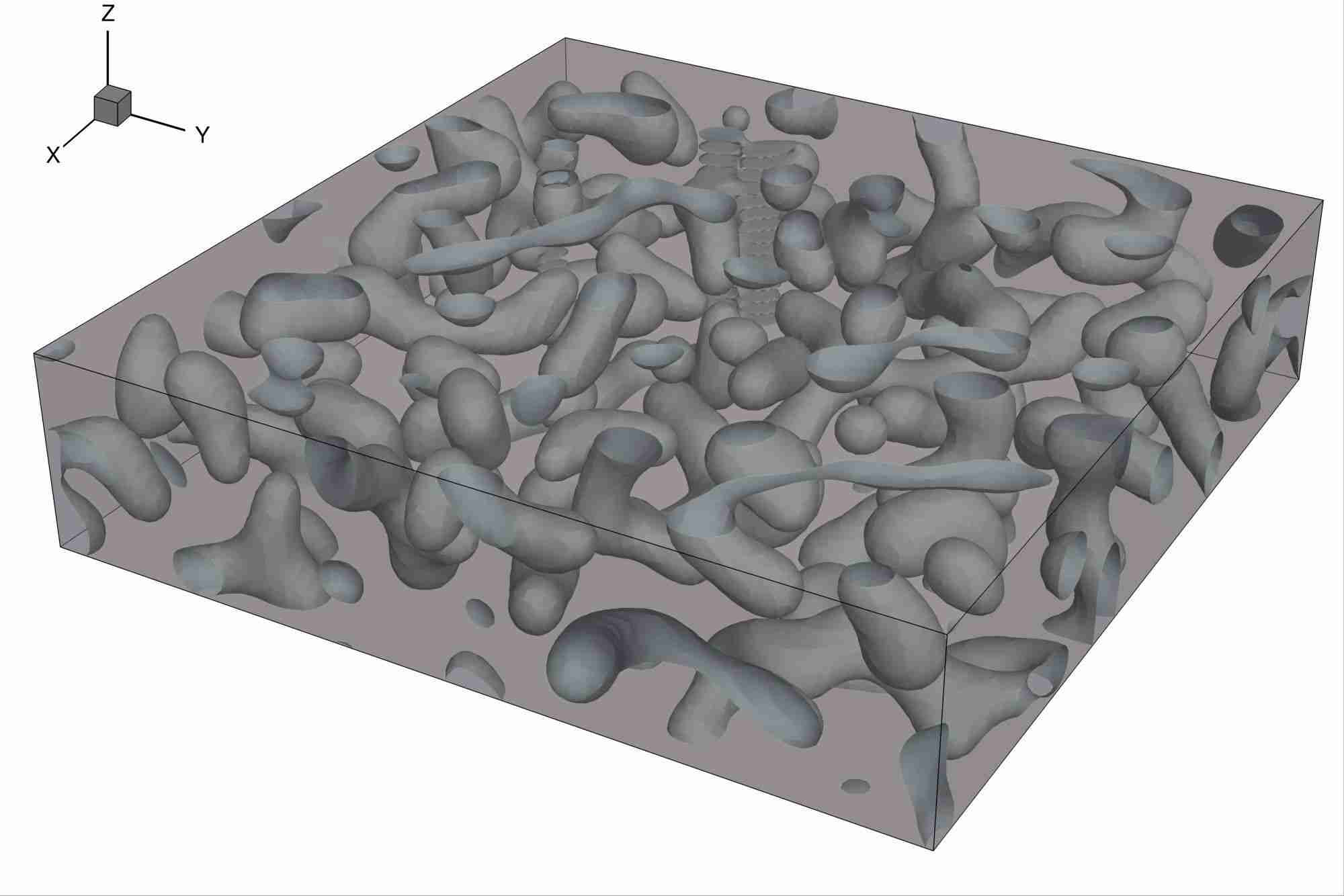}} \vspace{-0.16cm}
        \subfloat[$x_{Al}=0.55$]{\includegraphics[scale=0.085]{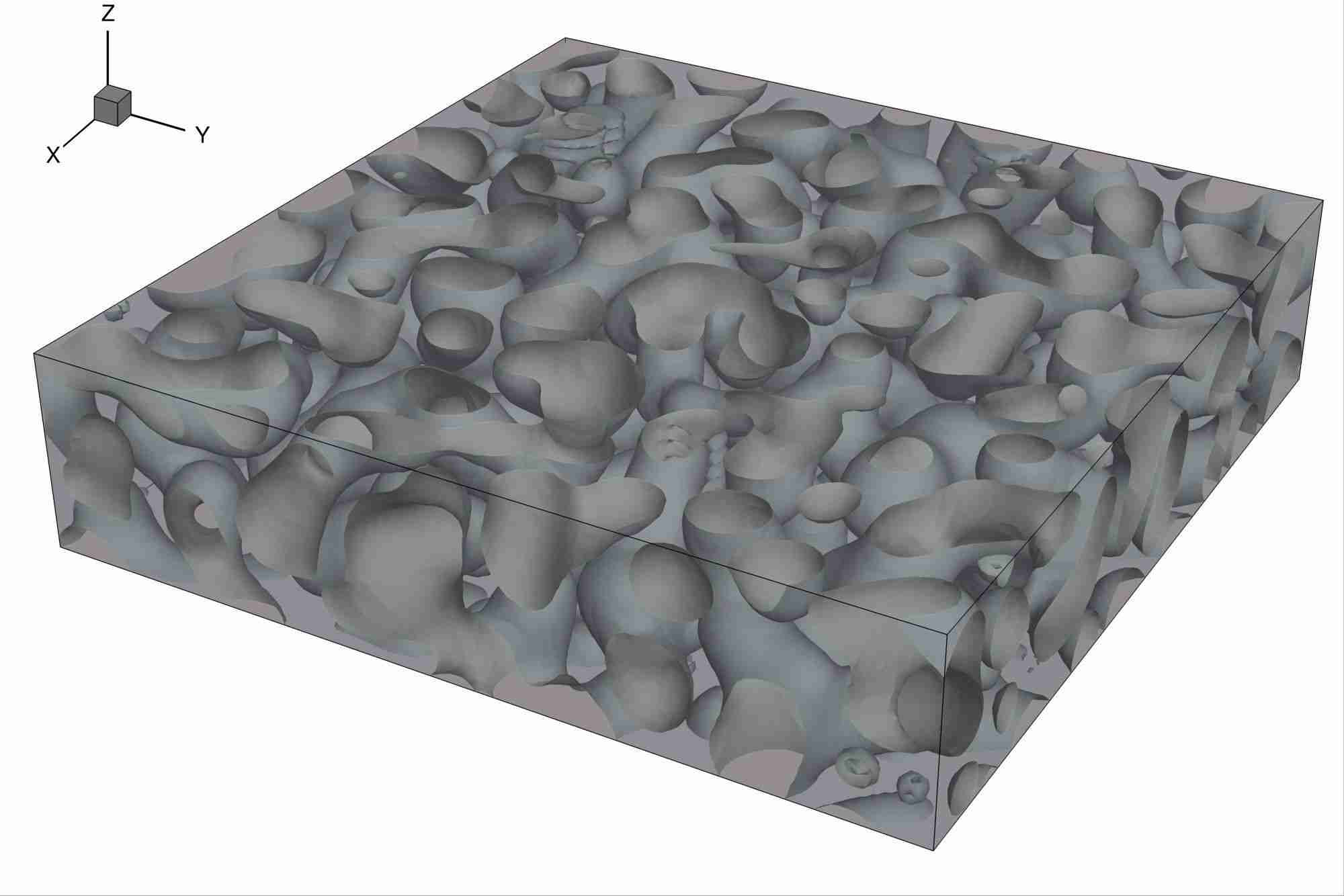}} \\ \vspace{-0.16cm}
        \subfloat[$x_{Al}=0.35$]{\includegraphics[scale=0.085]{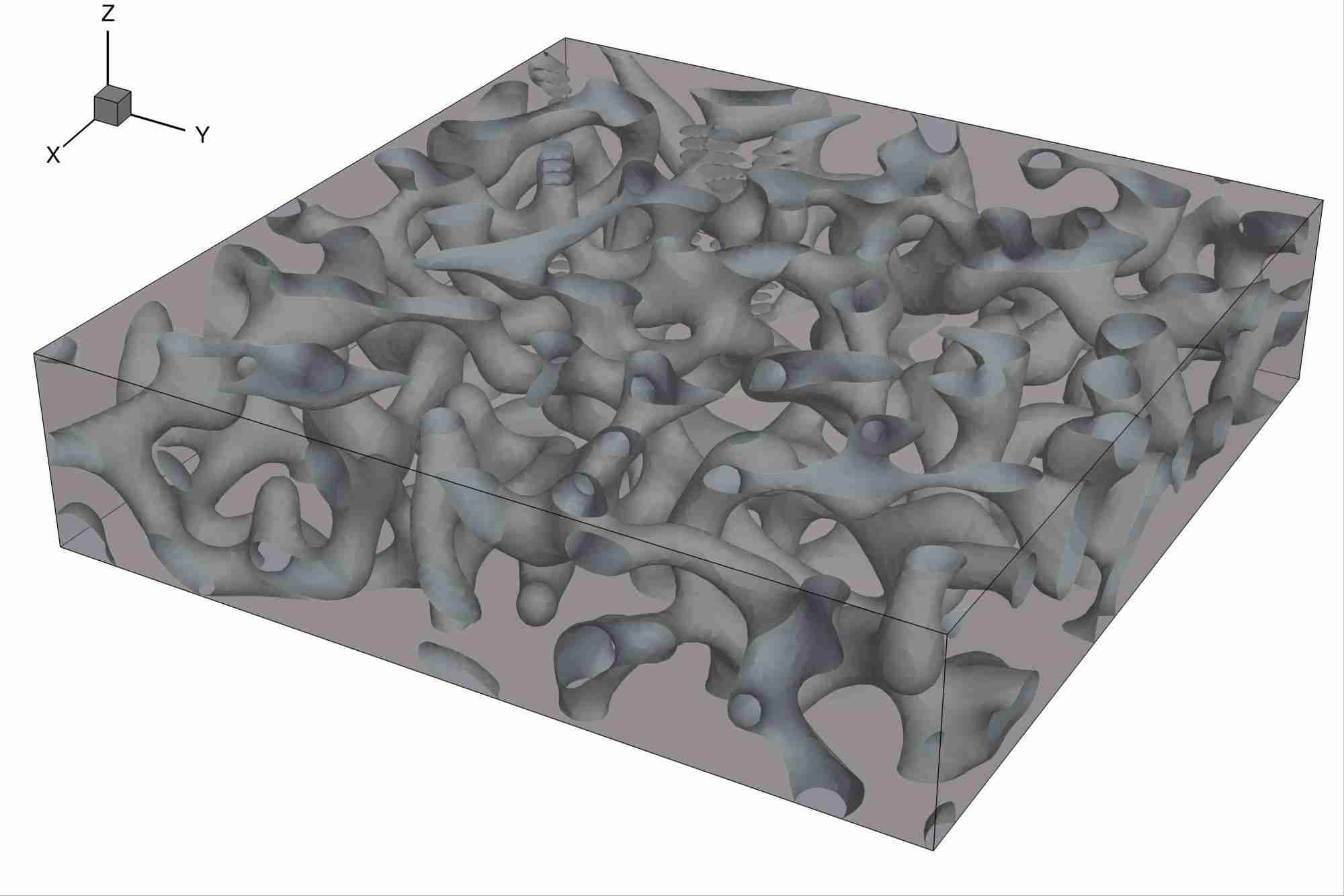}}  \vspace{-0.16cm}
        \subfloat[$x_{Al}=0.60$]{\includegraphics[scale=0.085]{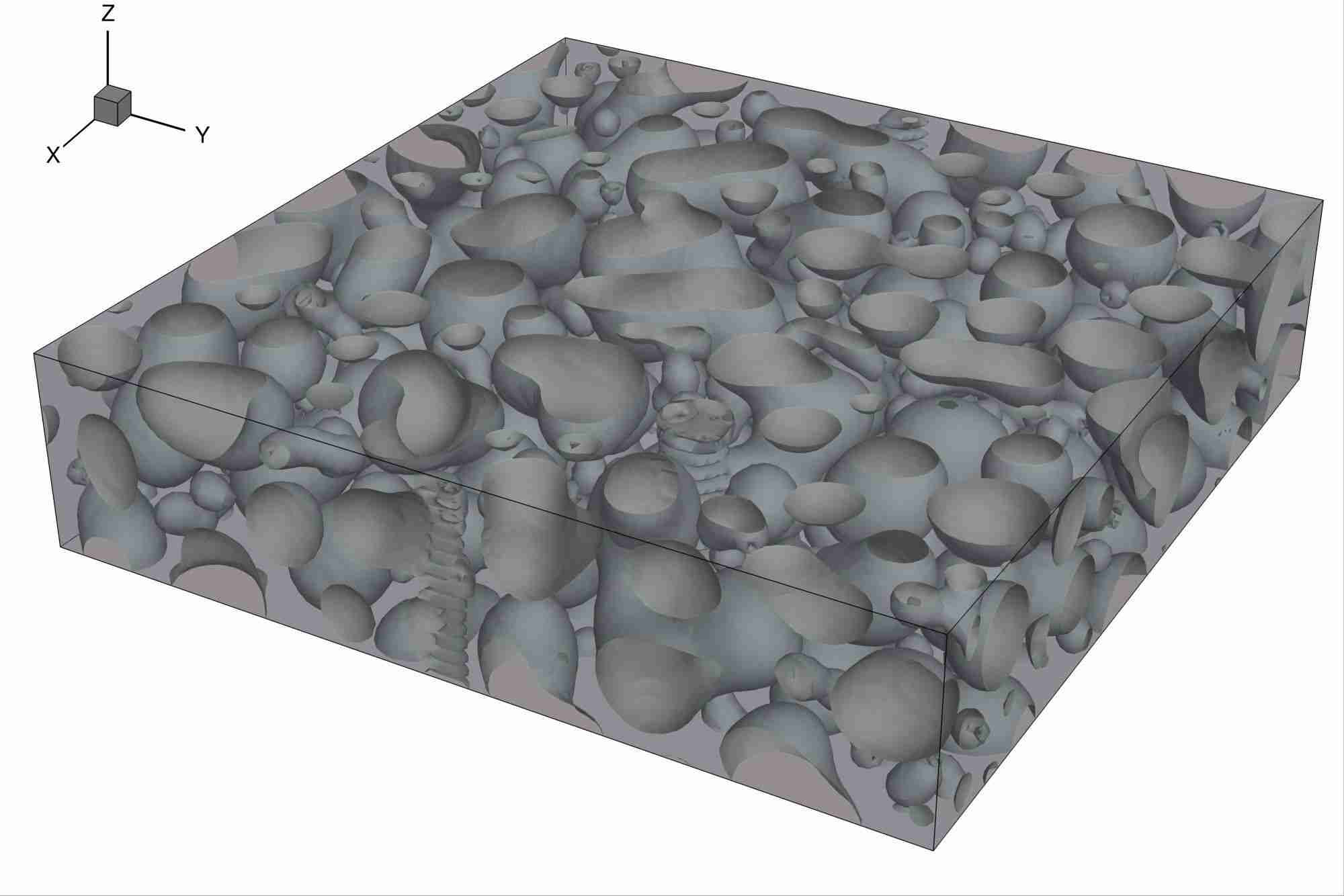}} \\ \vspace{-0.16cm}
        \subfloat[$x_{Al}=0.40$]{\includegraphics[scale=0.085]{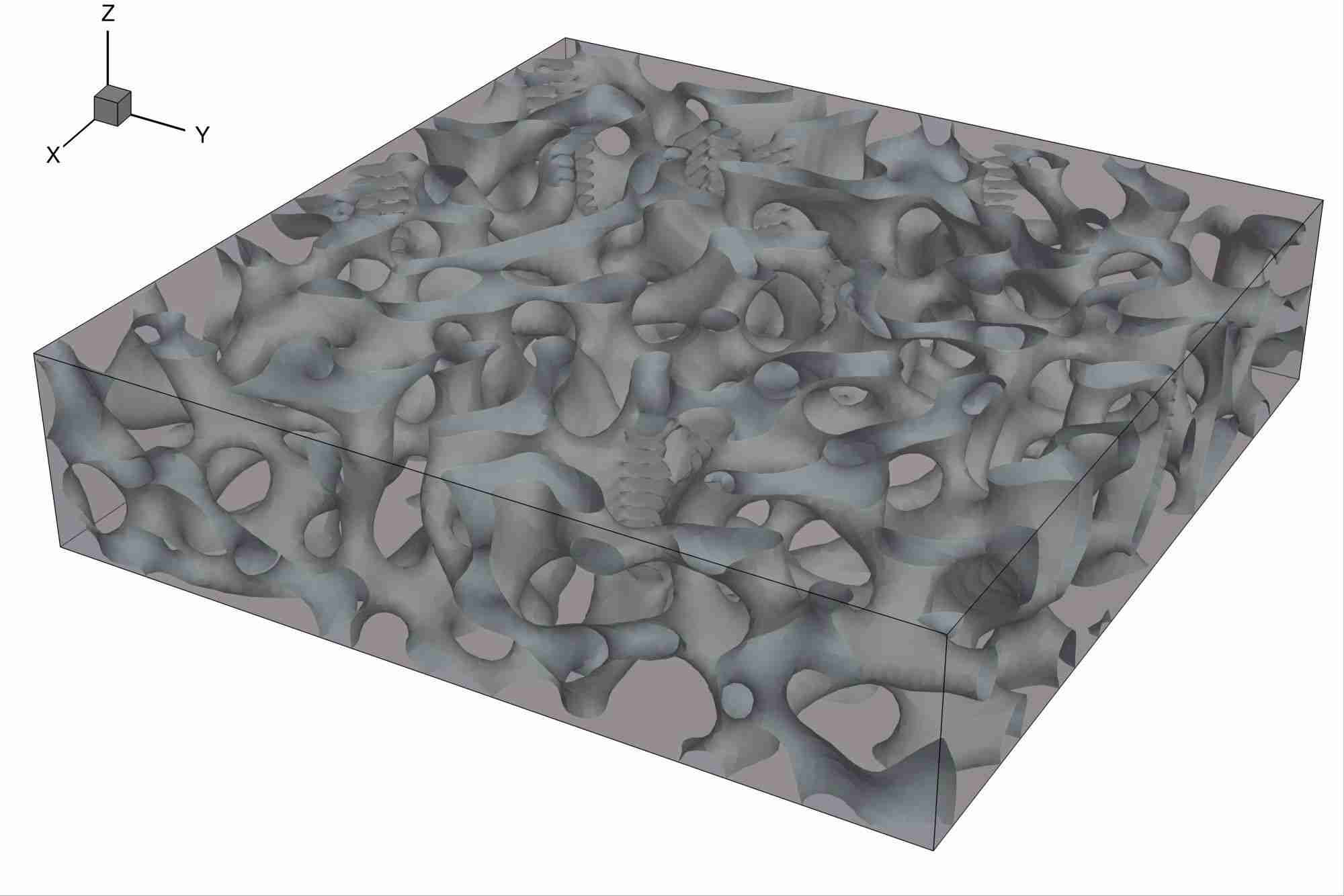}} \vspace{-0.16cm}
        \subfloat[$x_{Al}=0.65$]{\includegraphics[scale=0.085]{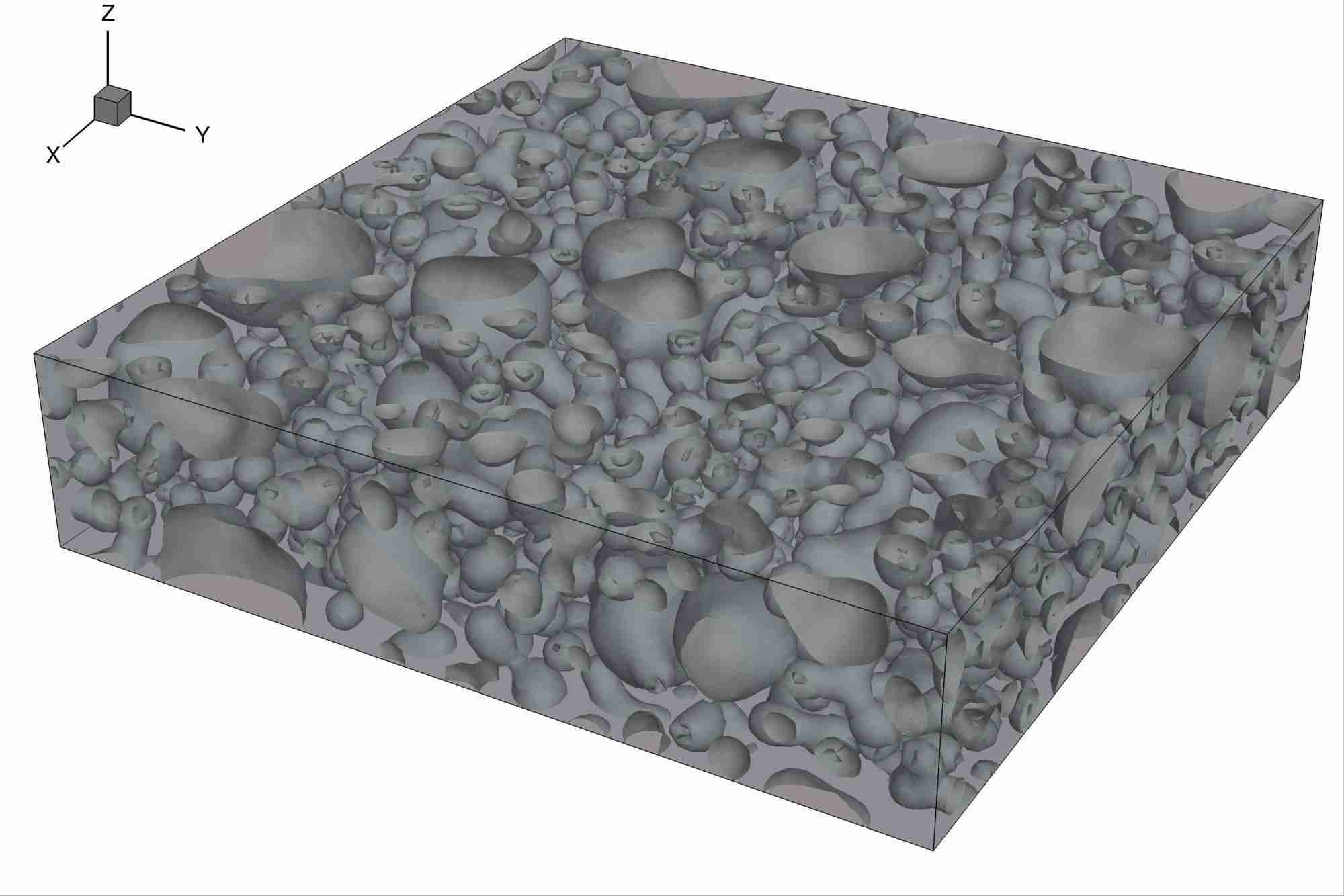}} \\ \vspace{-0.16cm}
        \subfloat[$x_{Al}=0.45$]{\includegraphics[scale=0.085]{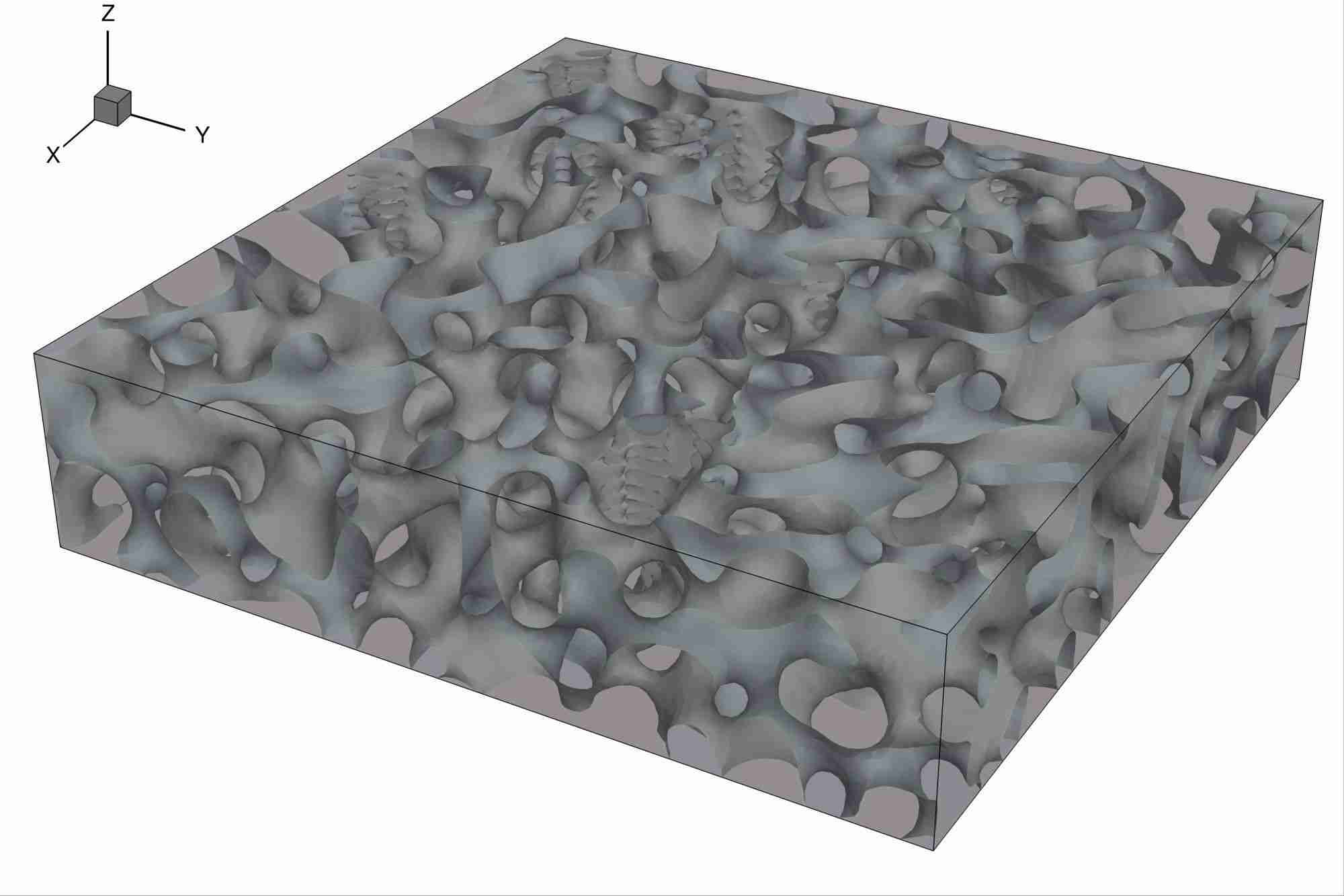}}  \vspace{-0.16cm}
        \subfloat[$x_{Al}=0.70$]{\includegraphics[scale=0.085]{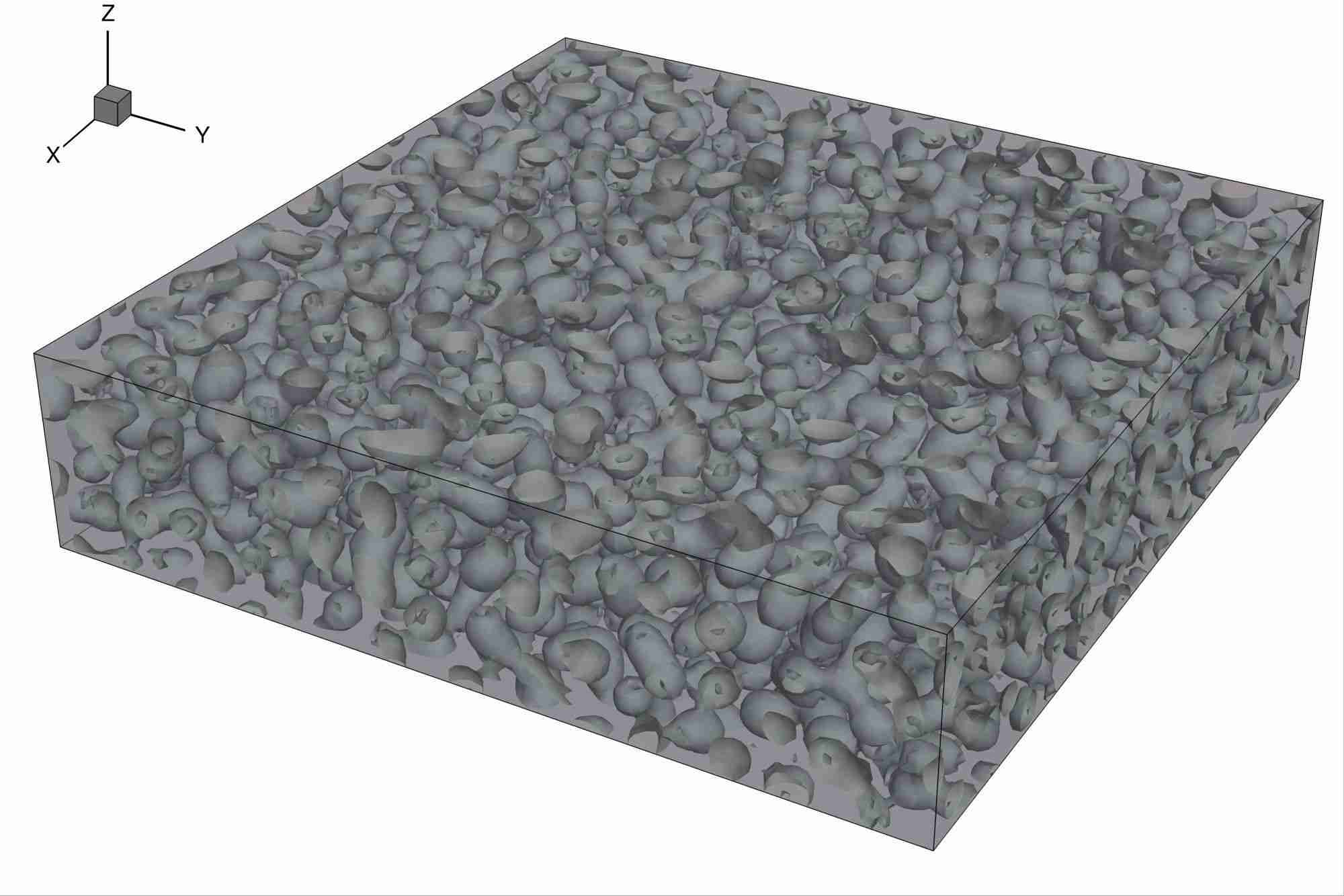}}  \vspace{-0.16cm}
        \caption{3D topological distribution of the interfaces for different Al content in Ti$_{1-x-0.05}$Al$_x$Zr$_{0.05}$N for $0.25\leq x_{Al}\leq 0.70$ range during chemical-only regime.}
        \label{fig:3D_mic_chemical}
    \end{figure*}

\begin{figure*}
        \centering
        \subfloat[$x_{Al}=0.25$]{\includegraphics[scale=0.085]{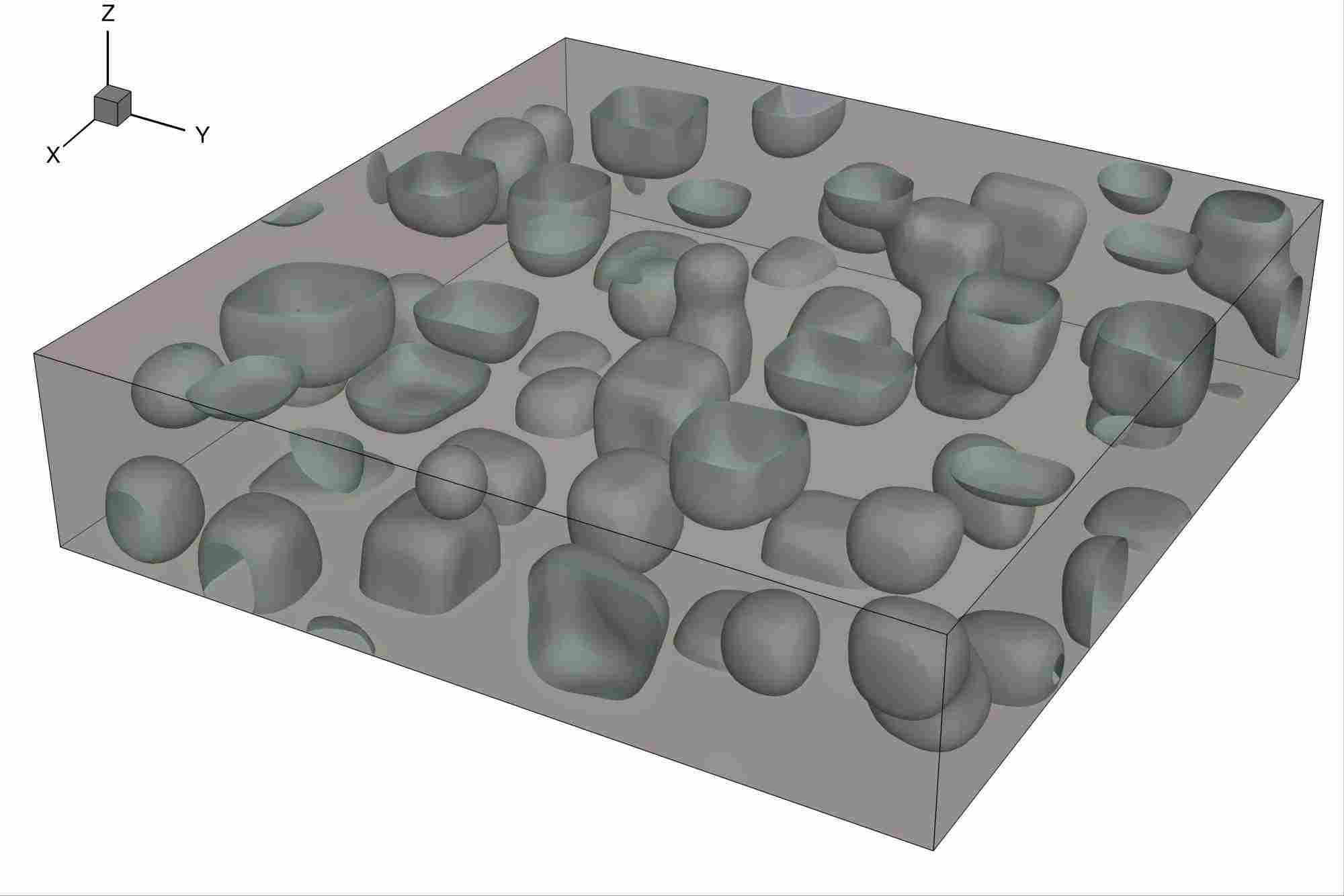}} \vspace{-0.16cm}
        \subfloat[$x_{Al}=0.50$]{\includegraphics[scale=0.085]{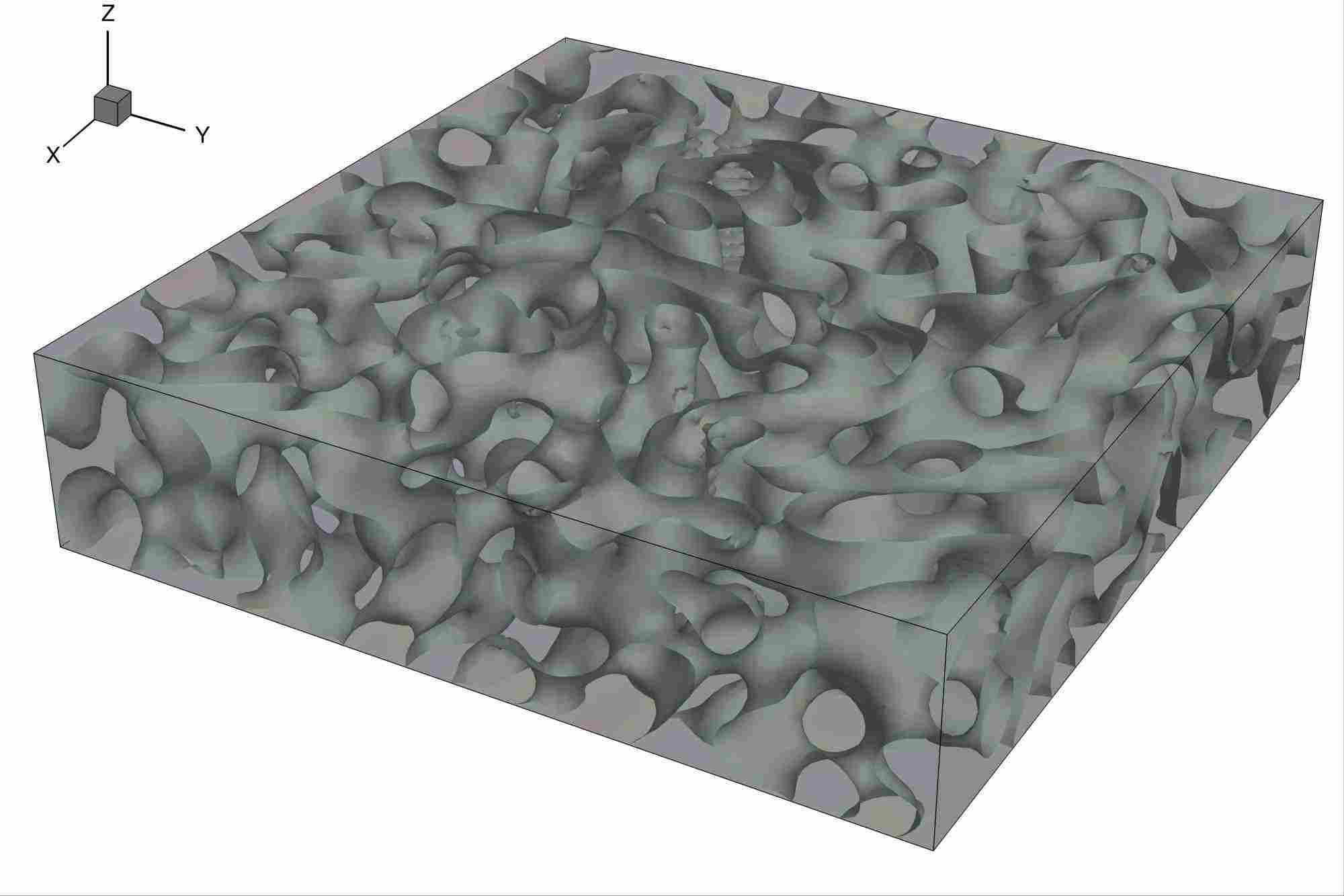}} \\ \vspace{-0.16cm}
        \subfloat[$x_{Al}=0.30$]{\includegraphics[scale=0.085]{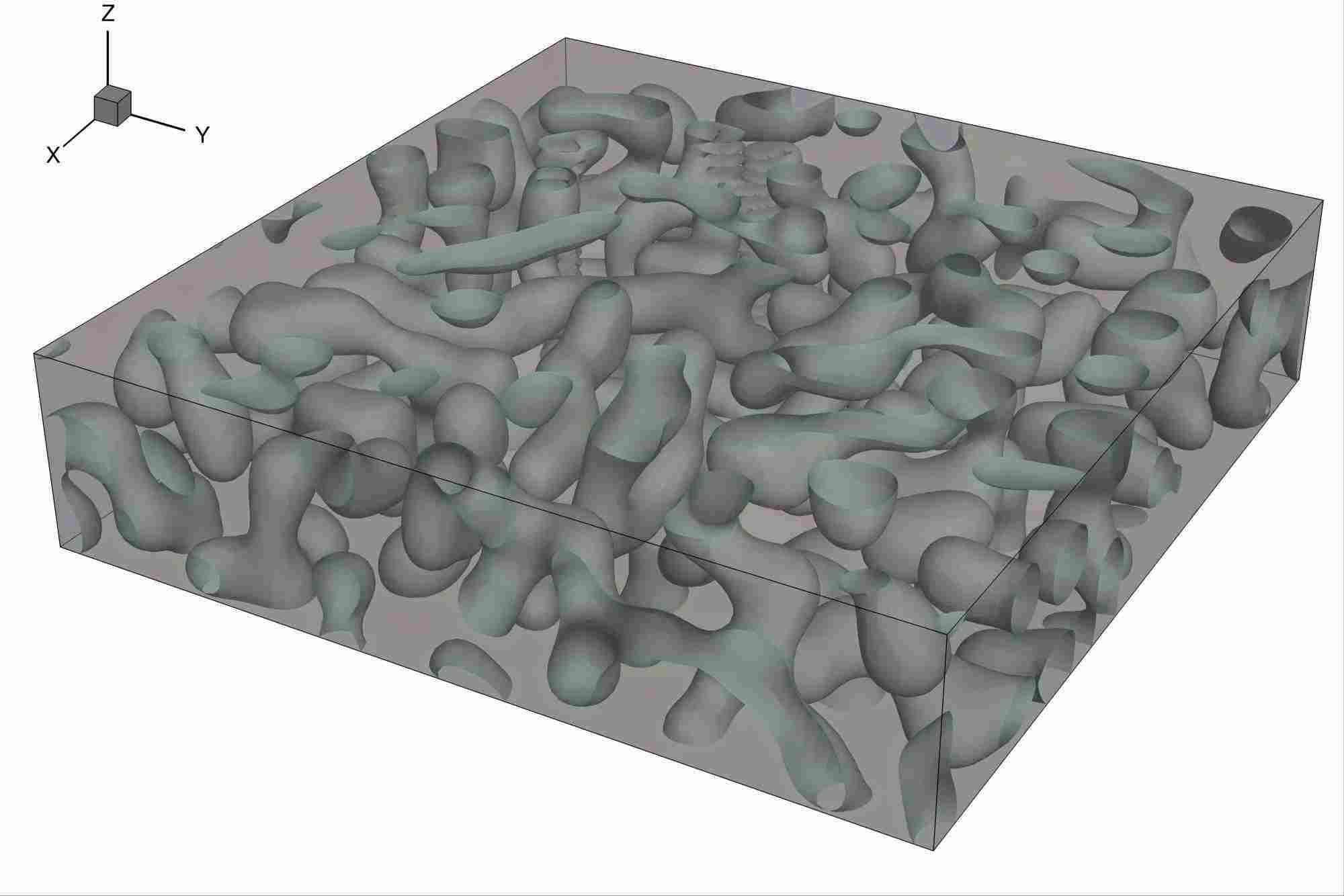}} \vspace{-0.16cm}
        \subfloat[$x_{Al}=0.55$]{\includegraphics[scale=0.085]{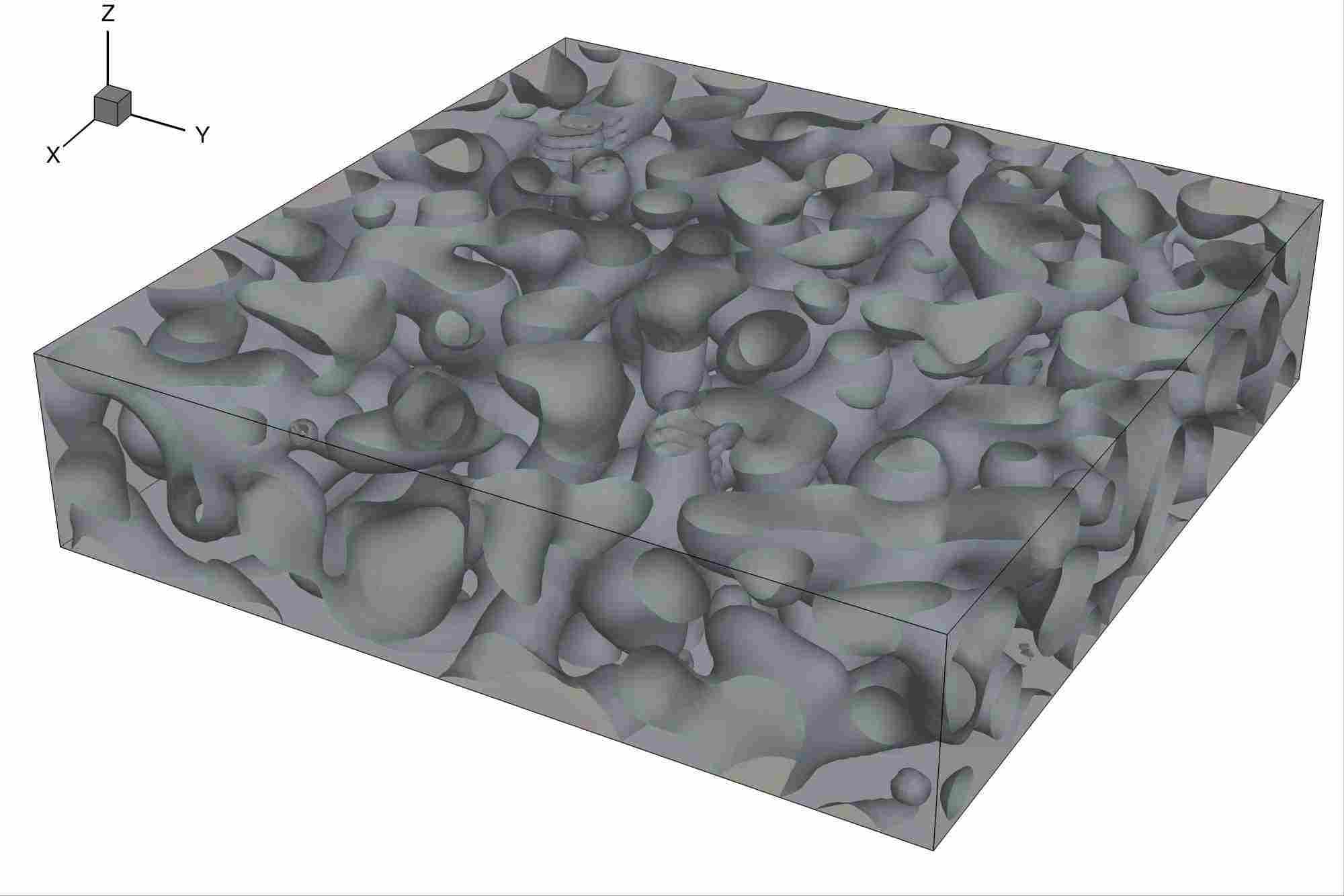}} \\ \vspace{-0.16cm}
        \subfloat[$x_{Al}=0.35$]{\includegraphics[scale=0.085]{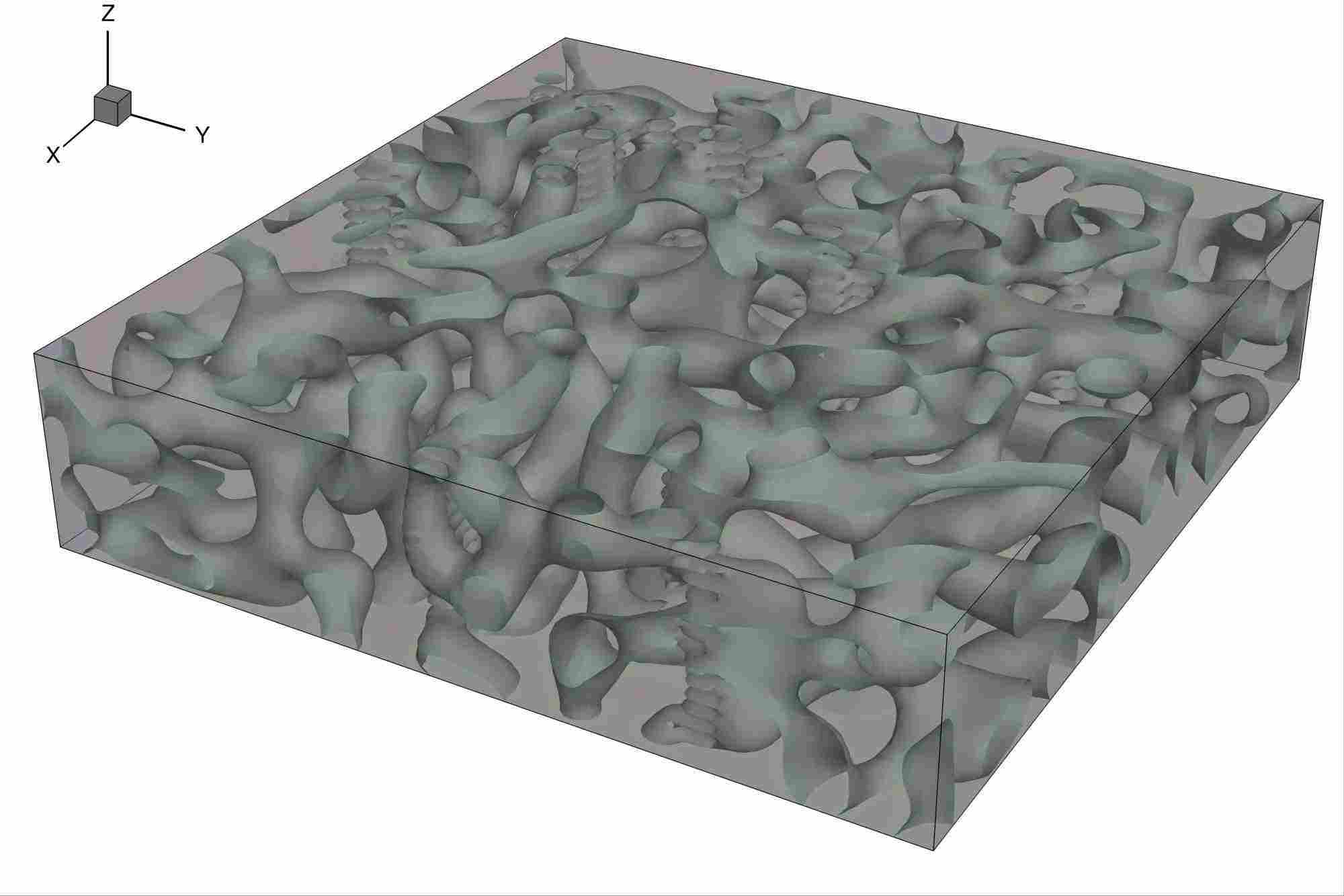}}  \vspace{-0.16cm}
        \subfloat[$x_{Al}=0.60$]{\includegraphics[scale=0.085]{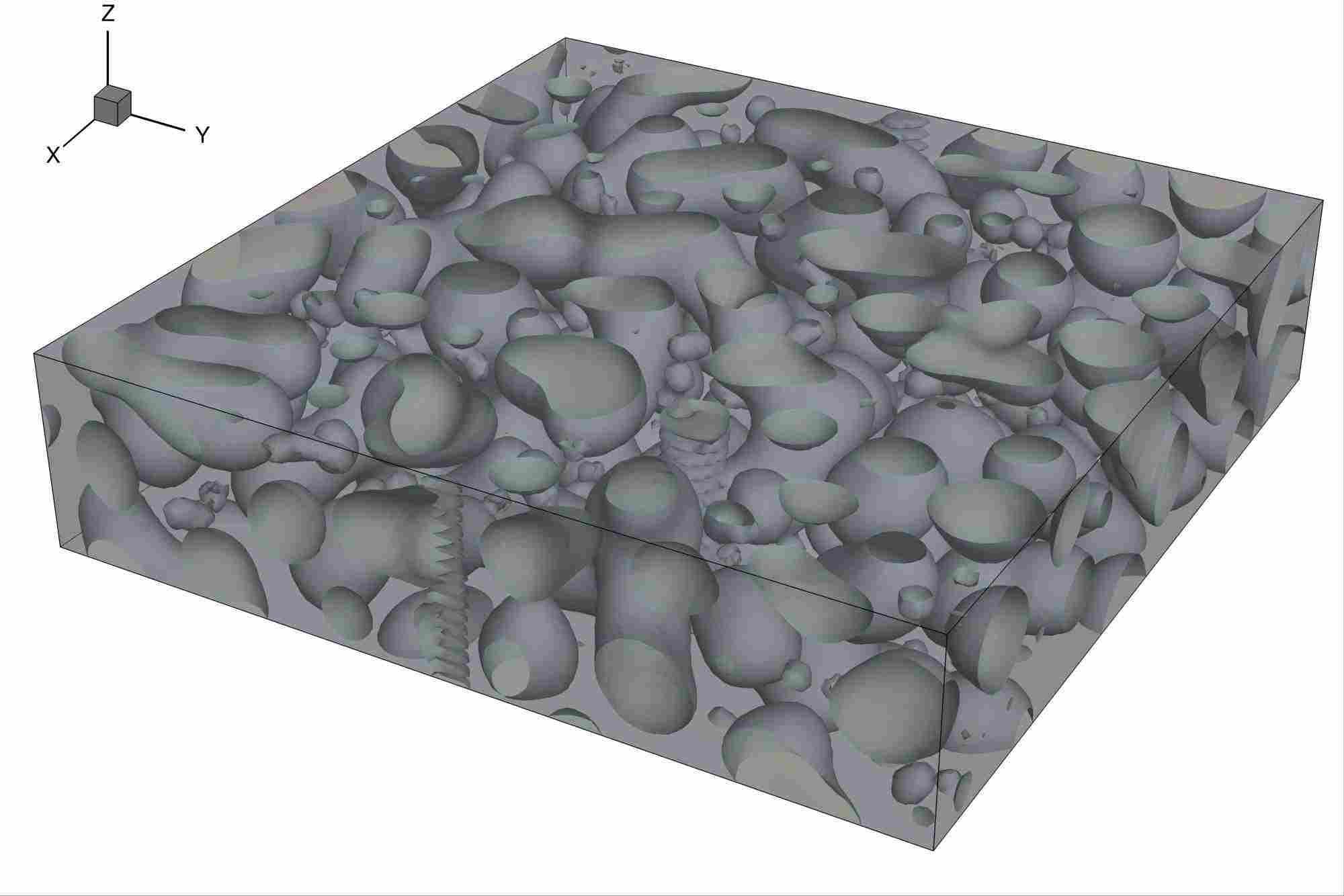}} \\ \vspace{-0.16cm}
        \subfloat[$x_{Al}=0.40$]{\includegraphics[scale=0.085]{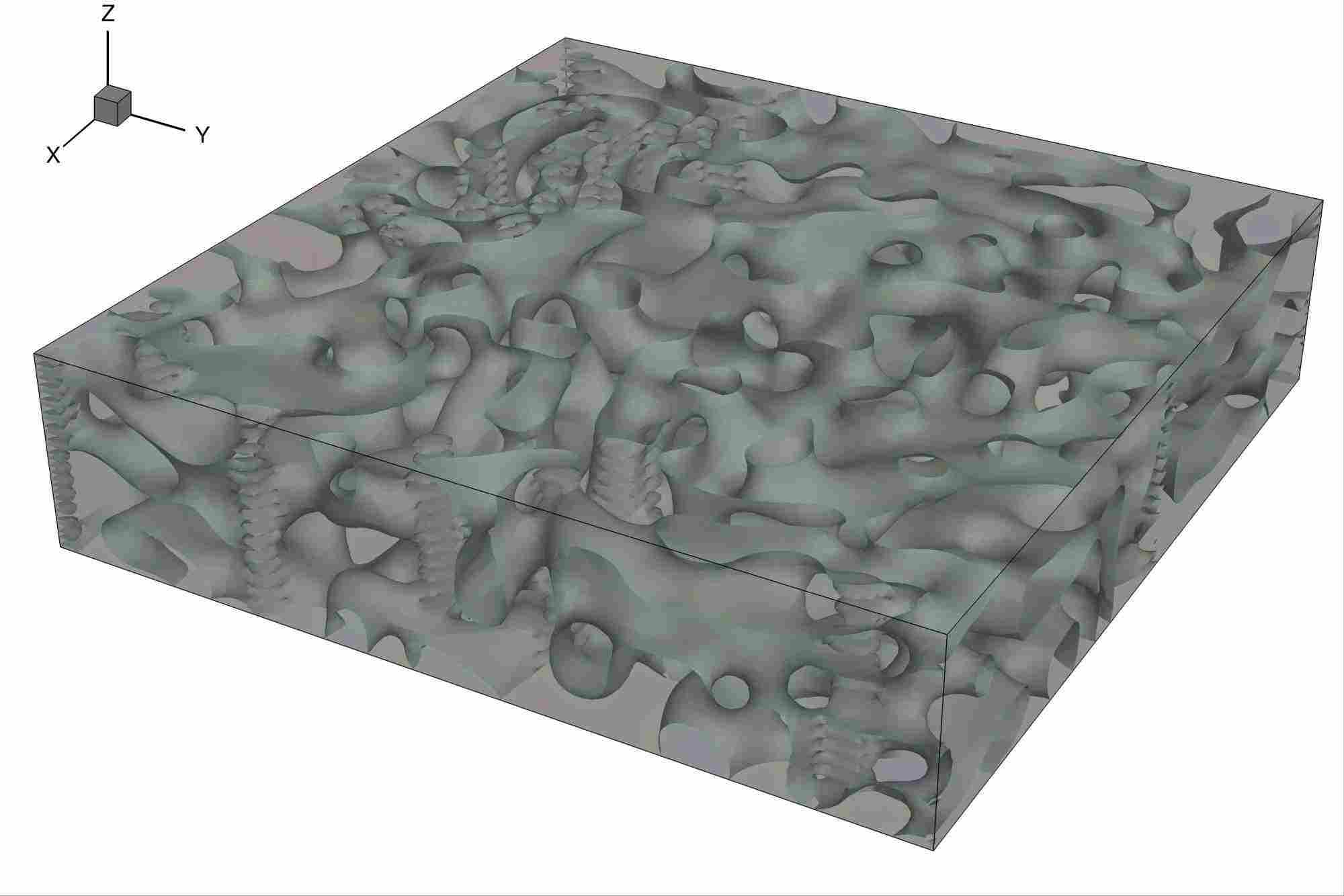}} \vspace{-0.16cm}
        \subfloat[$x_{Al}=0.65$]{\includegraphics[scale=0.085]{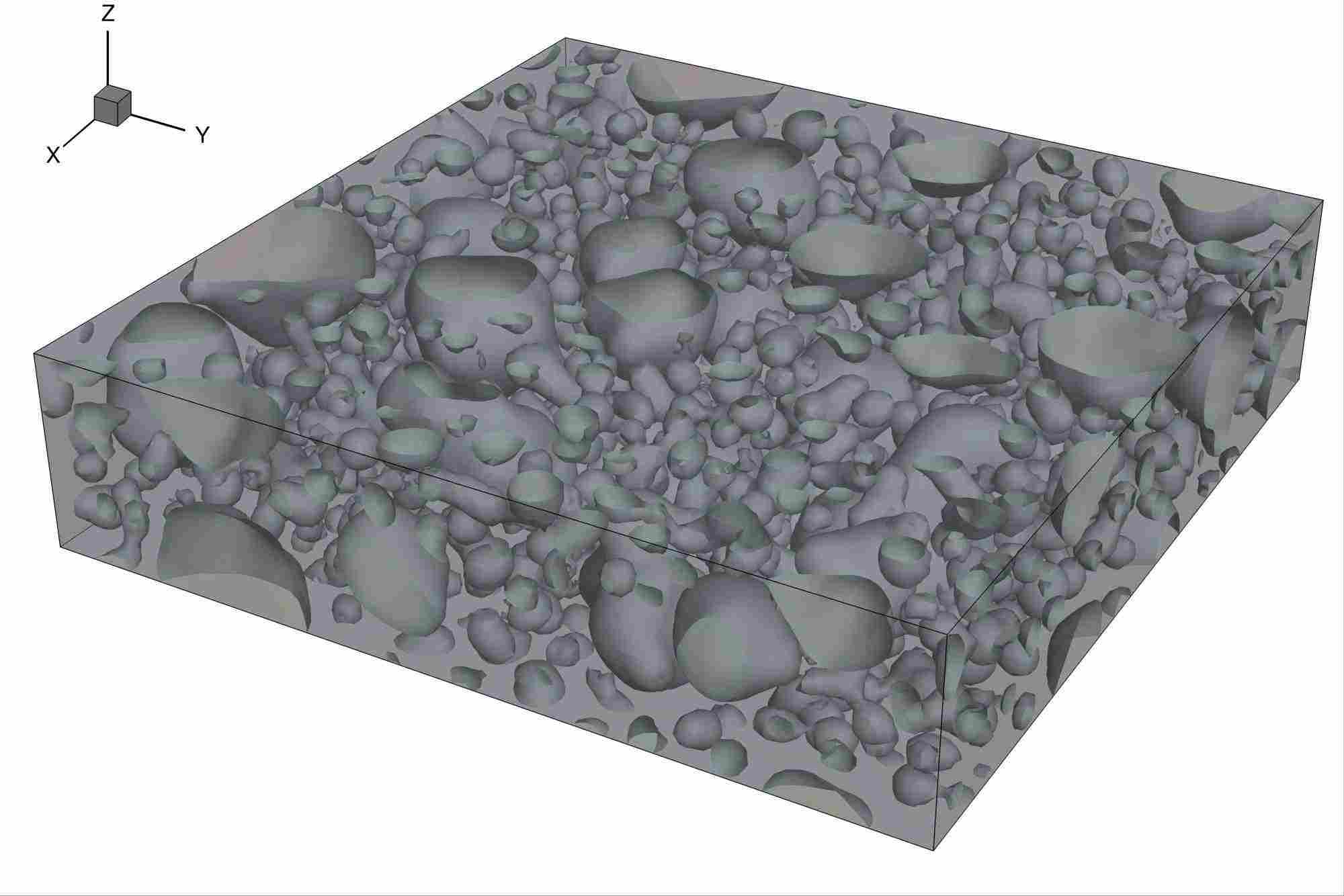}} \\ \vspace{-0.16cm}
        \subfloat[$x_{Al}=0.45$]{\includegraphics[scale=0.085]{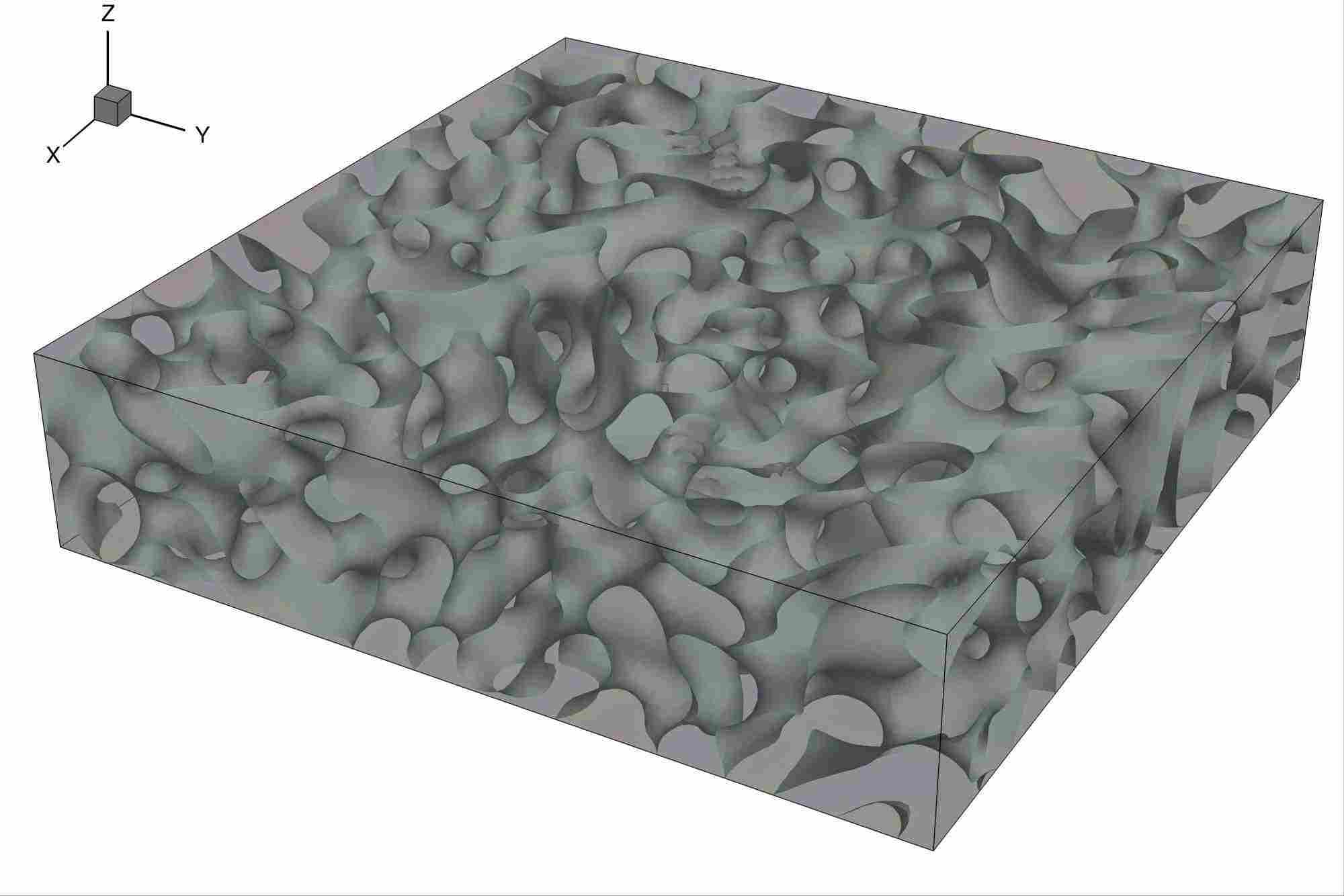}}  \vspace{-0.16cm}
        \subfloat[$x_{Al}=0.70$]{\includegraphics[scale=0.085]{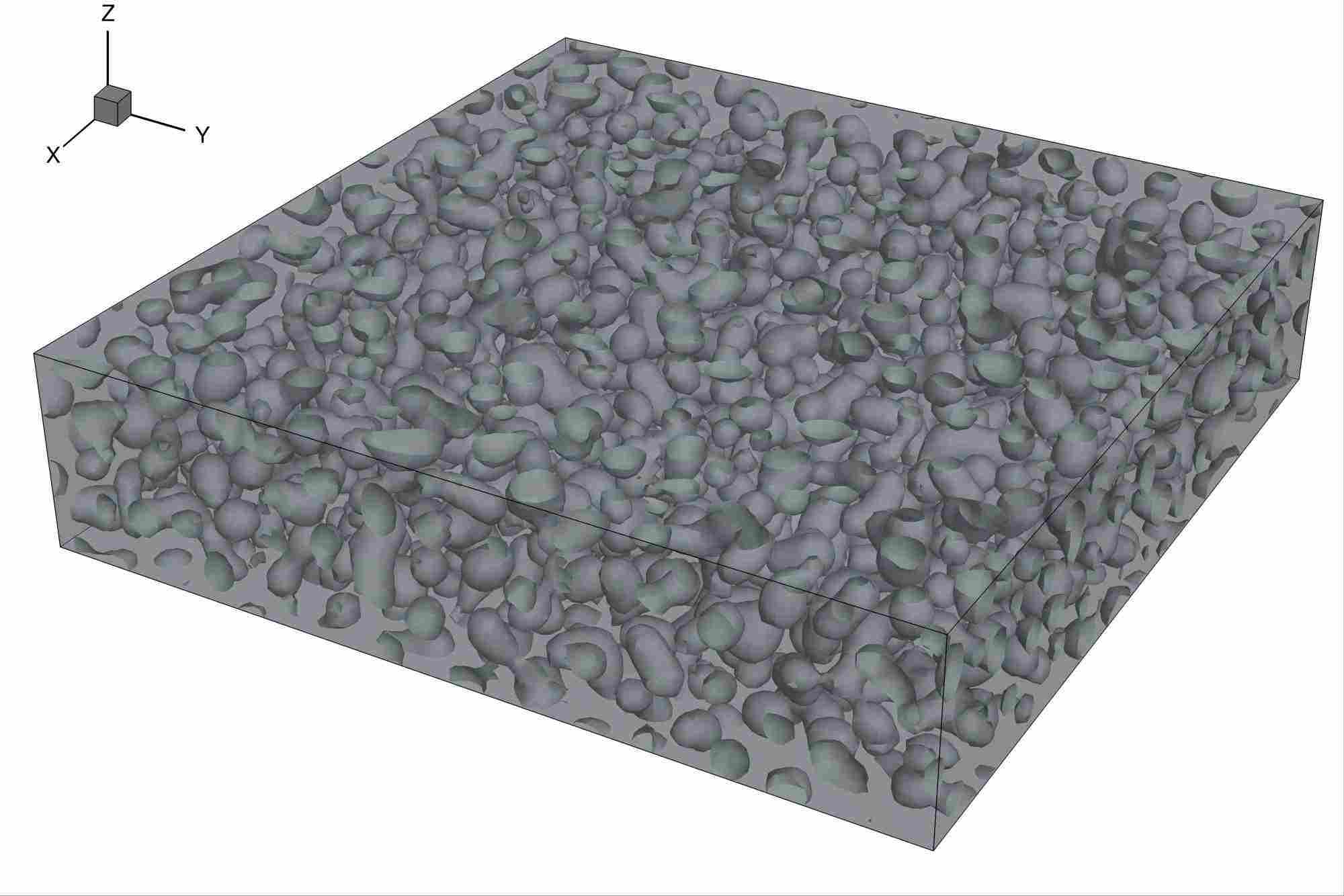}}  \vspace{-0.16cm}
        \caption{3D topological distribution of the interfaces for different Al content in Ti$_{1-x-0.05}$Al$_x$Zr$_{0.05}$N for $0.25\leq x_{Al}\leq 0.70$ range during elastochemical regime.}
        \label{fig:3D_mic_elastochemical}
    \end{figure*}

    In addition to the results in the body of the paper, the supplementary results are provided in this section. The results are provided in the below order: 
    
    \begin{itemize}
        
        \item Evolution of the microstructure of Ti$_{1-x-0.05}$Al$_{x}$Zr$_{0.05}$N for $0.25\leq x_{Al}\leq0.45$ range (Refer to Fig.~\ref{fig:time_evolution_1}).
        \item Evolution of the microstructure of Ti$_{1-x-0.05}$Al$_{x}$Zr$_{0.05}$N for $0.50\leq x_{Al}\leq0.70$ range (Refer to Fig.~\ref{fig:time_evolution_2}). 
        \item 3D microstructure of Ti$_{1-x-0.05}$Al$_{x}$Zr$_{0.05}$N system for $0.25\leq x_{Al}\leq0.70$ range during chemical growth (Refer to Fig.~\ref{fig:3D_mic_chemical}).  
        \item 3D microstructure of Ti$_{1-x-0.05}$Al$_{x}$Zr$_{0.05}$N system for $0.25\leq x_{Al}\leq0.70$ range during elastochemical growth (Refer to Fig.~\ref{fig:3D_mic_elastochemical}). 
    \end{itemize}

    %%%%%%%%%%%%%%%%%%
    %%%%%%%%%%%%%%%%%%   
    \newpage
    \clearpage
    %%%%%%%%%%%%%%%%%%
    %%%%%%%%%%%%%%%%%% 
    
    \newpage
    \clearpage
    
\end{document}